\documentclass[aps,pre,twocolumn,showpacs,epsfig,epsf,amssymb,showkeys,tightenlines,floatfix]{revtex4-1}
\usepackage{graphicx}
\usepackage{amssymb}
\usepackage{amsmath}
\usepackage{subfigure}
\usepackage{relsize}
\usepackage{booktabs}
\begin{document}
%\widetext

\title{
Heirarchical and synergistic self-assembly in composites of model Wormlike micellar-polymers and nanoparticles  
results in nanostructures with  diverse morphologies.}

\author{Shaikh Mubeena$^1$, Apratim Chatterji$^{1,2}$}
\email{apratim@iiserpune.ac.in}
\affiliation{
1. Dept. of Physics, IISER-Pune, Dr. Homi Bhabha Road,  Pune-411008, India.\\
2. Center for Energy Science, IISER-Pune, Dr. Homi Bhabha Road,  Pune-411008, India.
}
\date{\today}
\begin{abstract}
 
         Using Monte carlo simulations, we investigate the self-assembly of model nanoparticles inside  a 
matrix of model equilibrium polymers (or matrix of Wormlike micelles) as a function of the polymeric matrix 
density and the excluded volume parameter between polymers and nanoparticles. 
In this paper, we show morphological transitions in the system architecture 
via synergistic self-assembly of nanoparticles and the equilibrium polymers. In a synergistic self-assembly, 
the resulting morphology of the system is a result of the interaction between both nanoparticles and the polymers, 
unlike the polymer templating method. We report the morphological transition of nanoparticle aggregates from 
percolating network-like structures to non-percolating clusters as a result of the change in the excluded volume 
parameter between nanoparticles and polymeric chains. In parallel with the change in the self-assembled structures 
of nanoparticles, the matrix of equilibrium polymers also shows a transition from a dispersed state to a percolating 
network-like structure formed by the clusters of polymeric chains. We show that the shape anisotropy of the 
nanoparticle clusters formed is governed by the polymeric density resulting in rod-like, sheet-like or other 
anisotropic nanoclusters. It is also shown that the pore shape and the pore size of the porous network of 
nanoparticles can be changed by changing the minimum approaching distance between nanoparticles and polymers. 
We provide a theoretical understanding of why various nanostructures with very different morphologies are obtained.

\end{abstract}
\maketitle
\section{Introduction}

      Nanoparticle assembly inside a polymeric matrix is a powerful route to form hybrid materials with the desired material,
 magnetic, optical and electronic properties by a suitable choice of system parameters 
~\cite{1,2,3,4,5,6,7,8,9,lin2005self,rozenberg2008polymer}. The synergistic interaction between nanoparticles and 
the polymer matrix gives rise to a variety of hybrid systems and nanostructures having a wide range of applications 
like in cosmetics ~\cite{arraudeau1989composition,tatum1988organoclay,gadberry1997organoclay,patel2006nanoclays,care3skin,corcorran2004particle,laufer1996objective,lochhead2007role,katz2007nanotechnology,miller2006nanomaterials,raj2012nanotechnology}, food ~\cite{sorrentino2007potential,de2009nanocomposites,rhim2013bio,arora2010nanocomposites,ray2006potential,sozer2009nanotechnology}, pharmacy ~\cite{sambarkar2012polymer,ray2006polymer,mourino2016polymer,dwivedi2013application}, novel functional materials ~\cite{ingrosso2010colloidal,segala2012ruthenium,striccoli2009nanocrystal,mazumdar2015polymer,nicolais2004metal,qi2017novel}, ultraviolet lasers ~\cite{bloemer1998transmissive,jakvsic2004silver, scalora1998transparent,kedawat2014fabrication}, hybrid nanodiodes ~\cite{gence2010conjugated,park2004hybrid,das2012review}, DNA templated electronic junctions \cite{turberfield2003dna, park2004electronic}, quantum dots and thin wires ~\cite{alam2013synthesis}. 
%{ shenhar2003nanoparticles, boal2000self, goubault2005self, ditsch2005controlled, colfen2003higher, oaki2005morphological, he2005vertically, he2005polymer, qi2000control, sun2004one, baia2005gold, monson2003dna, lynch2005preparation, lu2005preparation, sohn2003directed, von1999preparation, yun2005tunable, spatz1999micellar, spatz2000ordered, glass2003block, yun2005electrically, cheyne2005hierarchical}. 

%Due to their economic importance and ease of tailoring them with desired properties, nanocomposites have gained a considerable attention over the last few decades. 

        The investigation of nanoparticle assembly inside a polymeric matrix is important for fabrication of complex 
nanodevices as well as is of fundamental interest to explore the equilibrium and non equilibrium phase behaviour of the
 system ~\cite{dijkstra1999phase,ilett1995phase,lekkerkerker1992phase,poon2002physics,asakura1954interaction,aarts2002phase,bolhuis2002influence,gast1983polymer}.
 The colloid-polymer mixtures have been widely studied theoretically and experimentally and an equilibrium phase 
diagram is predicted using Asakura-Oosawa model ~\cite{asakura1954interaction,lekkerkerker1992phase,gast1983polymer,de1981interactions,fleer2007simple,poon2002physics} 
for their varying size ratios and volume fractions. Experimental studies confirmed the predicted equilibrium 
colloid-polymer phase diagram ~\cite{poon2002physics,de1981interactions,fleer2007simple} and also observed 
non-equilibrium phase behaviour ~\cite{poon1997non}. An increase in the polymer density in the colloid-polymer system
 leads to its demixing into two or more phases~\cite{zhang2013phase}. In a deeply quenched system, the colloid rich phase
 may get kinetically arrested (on reaching the volume fraction $\approx 0.57-0.59$) during the phase separation process,
 suppressing the demixing and producing a non-equilibrium gel state ~\cite{zhang2013phase}. For non-adsorbing 
colloid-polymer mixtures with the polymer to colloid size ratio of $\approx 0.08$, these non-equilibrium gel phases
 are observed for colloid volume fraction $\leq 0.3$ above a threshold of polymer concentration ~\cite{starrs2002collapse}. 
This limiting value of colloid volume fraction depends on polymer concentration. Above this threshold polymer 
concentration, these non-equilibrium phases show a period of latency (observation period over which the morphology is 
maintained) before sedimentation of colloids under gravity. The latency period increases with increase in the polymer
 concentration and may lead to "freezing of structures" at sufficiently high concentrations ~\cite{starrs2002collapse}. 
Experimental studies have reported a variety of non-equilibrium phases including clusters, gels and glasses 
~\cite{joshi2014dynamics}. These non-equilibrium states are a challenge for theoretical studies and forms an active 
area of research.
       
       The discovery of the mesoporous material MCM-41 by Mobil Oil Corporation 
~\cite{kresge1992ordered} widely popularised the method of 
assembling nanostructures inside polymeric matrices and thus giving rise to the "bottom-up" approach which revolutionised 
the whole industry of nanofabrication ~\cite{ seul1995domain, tang2006self, van2006symmetry, bedrov2005molecular, 
shay2001thermoreversible,fejer2007helix,glotzer2007anisotropy,lee2004self}. Polymeric and other supramolecular 
matrices have long 
been explored as templates for fabricating nanostructures due to low cost and ease of tailoring nanomaterials with 
the desired properties. Later, realising the importance of synergistic interactions between nanoparticles (NPs) and 
the matrix, a myriad of synergistically assembled nanocomposites have also been generated. The NPs can be incorporated 
in a matrix by both in situ ~\cite{xu2017situ,sun2008situ} or ex-situ ~\cite{guo2014comparison} methods. However, due 
to high surface interactions between NPs, it is difficult to disperse NPs inside the matrix. Therefore, in situ method 
is preferably used to produce a homogeneously dispersed NP-matrix composite. Nanorods 
~\cite{luo2006formation,abyaneh2016direct,cao2004nanostructures}, nanowires and nanobelts 
~\cite{wang2013nanowires}, nanocombs and nanobrushes ~\cite{umar2008formation,lao2006formation,singh2010synthesis},
 nanosheets ~\cite{pandey2016proton, vyborna2017morphological}, nanoporous ~\cite{xu2013nanoporous} 
and mesoporous structures ~\cite{ying1999synthesis}, nanotubes, spherical and complex morphological nanostructures 
have been reported in the literature. However, to exploit NP properties for nanodevice fabrication, the control over 
their morphological structures is still a big challenge to the researchers. This paper reports transformation in the 
nanostructure morphology with the change in the polymeric matrix density and the excluded volume parameter between
 nanoparticles and polymers.
% and shows that these morphological changes in the system architecture are the consequences 
%of the competition between the repulsive interactions present in the system.

       Generally, the (thermotropic) polymeric matrices have a value of elastic constants of the order 
of $\approx 10 pN$ ~\cite{sharma2009self}. But there exists a class of polymers known as lyotropic systems 
showing interesting phase behaviour due to their low value of elastic constants ($\approx 1 pN$) 
~\cite{ramos2000light,sallen1995surface}. Wormlike micelles (WLM) is one of the examples of such systems. 
When the concentration of the surfactant molecules in presence of suitable salts becomes higher than a 
critical micellar concentration, the surfactant molecules then aggregate to form micelles. 
For appropriate shape and size of the surfactant molecules, they can aggregate into long thread-like 
polymeric Wormlike micelles ~\cite{berret2006rheology}. They possess an extra mode of relaxation by scission 
and recombination reactions giving rise to an exponential distribution of their polymeric length 
~\cite{berret2006rheology,turner1990relaxation,cates1990nonlinear}. Moreover, the average spacing of 
WLMs in a hexagonal H1 phase is 5-10 nm, which is much larger than that of a thermotropic polymeric matrix 
~\cite{ramos2000light,sharma2009self}. Thus a low value of elastic constant, extra mode of relaxation and 
characteristic mesoscopic length scales makes the Wormlike micellar matrix  different from a 
thermotropic polymeric matrix. In this paper, we study the hierarchical self organization of NPs in a model 
of self-assembling polymeric matrix  that mimics the characteristic behaviour of a Wormlike micellar matrix. 
Our generic model of self assembling polymers represents the class of polymers known as  equilibrium 
polymers, WLMs at mesocopic length scales are just an example of equilibrium polymers. Equilibrium polymers are 
those whose bond energies are order of the thermal 
energies $k_BT$, and thereby chains can break or rejoin due to thermal fluctuations.

        We study the effect of the addition of NPs with diameters of the order of the diameter of equilibrium 
polymer (EP) chains on the polymeric self-assembly, which in turn affects the NP self-organization. 
The parameters which are systematically varied in our investigations are (1) the volume 
fractions of the polymeric matrix and (2) the minimum distance of approach between NPs and the matrix polymers 
(EVP - Excluded Volume Parameter). We show that for a low value of EVP, a uniformly mixed state of polymers and 
NPs is observed for all the densities of equilibrium polymers considered here. An increase in the value of the 
EVP leads to the formation of clusters of polymeric chains which joins to form a percolating network-like structure. 
The network of nanoparticles breaks into non-percolating clusters of NPs at some higher value of the EVP. We are able 
to present these morphological transformations in a diagram. There exist reports of a sudden decrease in the measured 
conductivity of polymer-nanocomposites in the literature ~\cite{li2010nanocomposites, ambrosetti2010insulator}. 
This decrease in the conductivity is attributed to the transition of NPs from non-percolating to percolating 
state on an decrease in NP volume fraction ~\cite{li2010nanocomposites, ambrosetti2010insulator}. 
In an attempt to explain these morphological transformations, we propose that these morphological transitions 
are due to the competition between the repulsive interaction between nanoparticles and the polymeric matrix and 
the repulsion between polymeric chains themselves. We try to explain this by analyzing the behaviour of 
total excluded volume in the system. We thus present how the system parameters can be used to tailor the 
mesoporous nanostructures formed in the system. This has a great potential to optimize the shape and size of
nano-structures obtained by in-situ methods and there improve the design and performance of fuel cells and 
batteries (especially Li-ion batteries), drug delivery, optoelectronic devices and other device-properties which depend
on the nano-structure morpholology.

     The following section presents the potentials  used to model NPs and equilibrium polymeric matrix and 
gives the details of the computational method. In the 3rd section, we briefly summarize our results which 
were published previously. In the next section, we present our new results which is divided into five
sub-sections where we prove the robustness of our results and then give a detailed qualitative and quantitative 
analysis of the model system. We investigate the variation in EP number density and the minimum 
approaching distance 
between polymers and NPs and its effect on the system behaviour. We present our Conclusions in the last section.

\section{Model and method}

\subsection{Modeling micelles (Equilibrium Polymers)}
\begin{figure}
\centering
\includegraphics[scale=0.2]{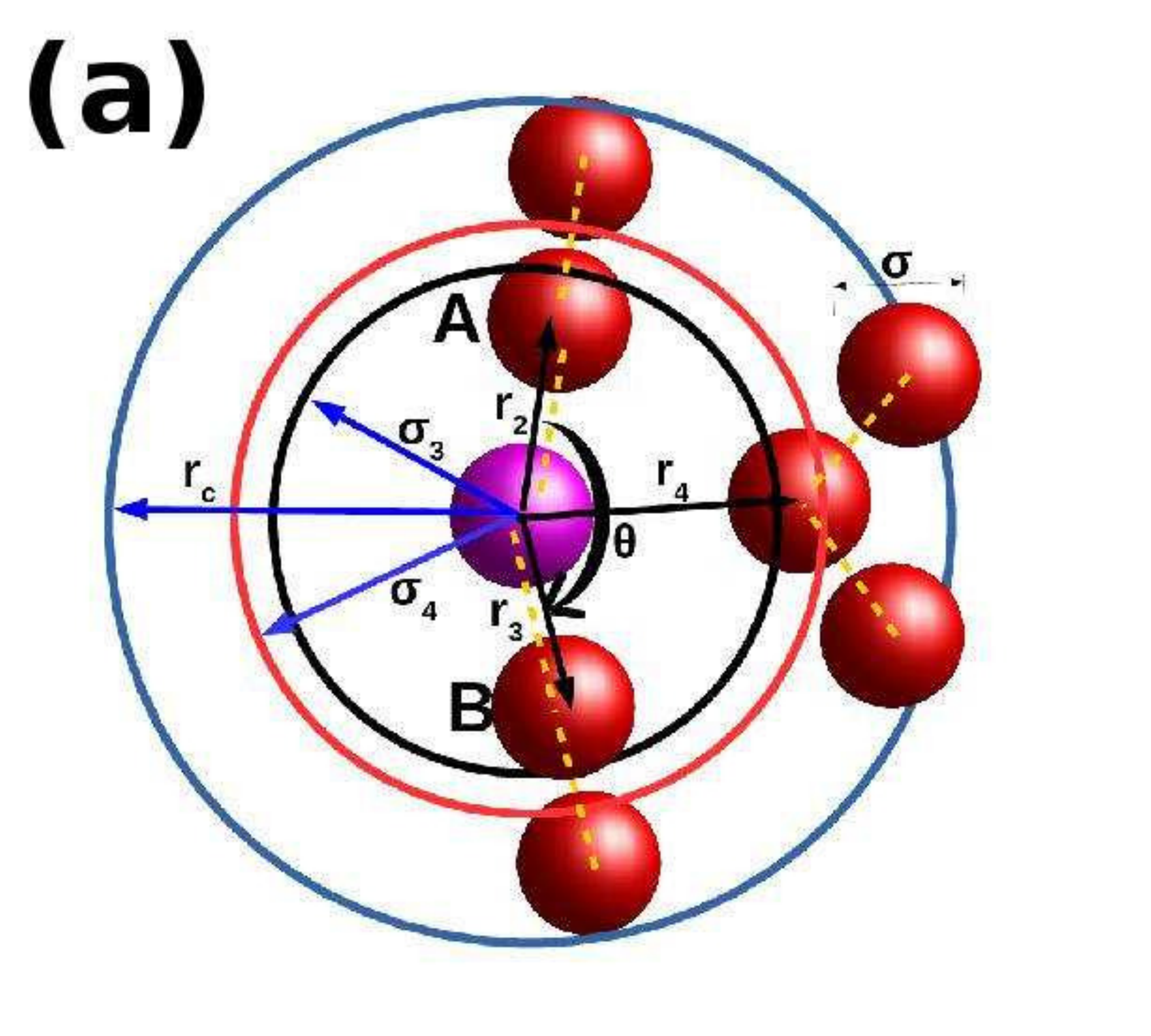}
\includegraphics[scale=0.2]{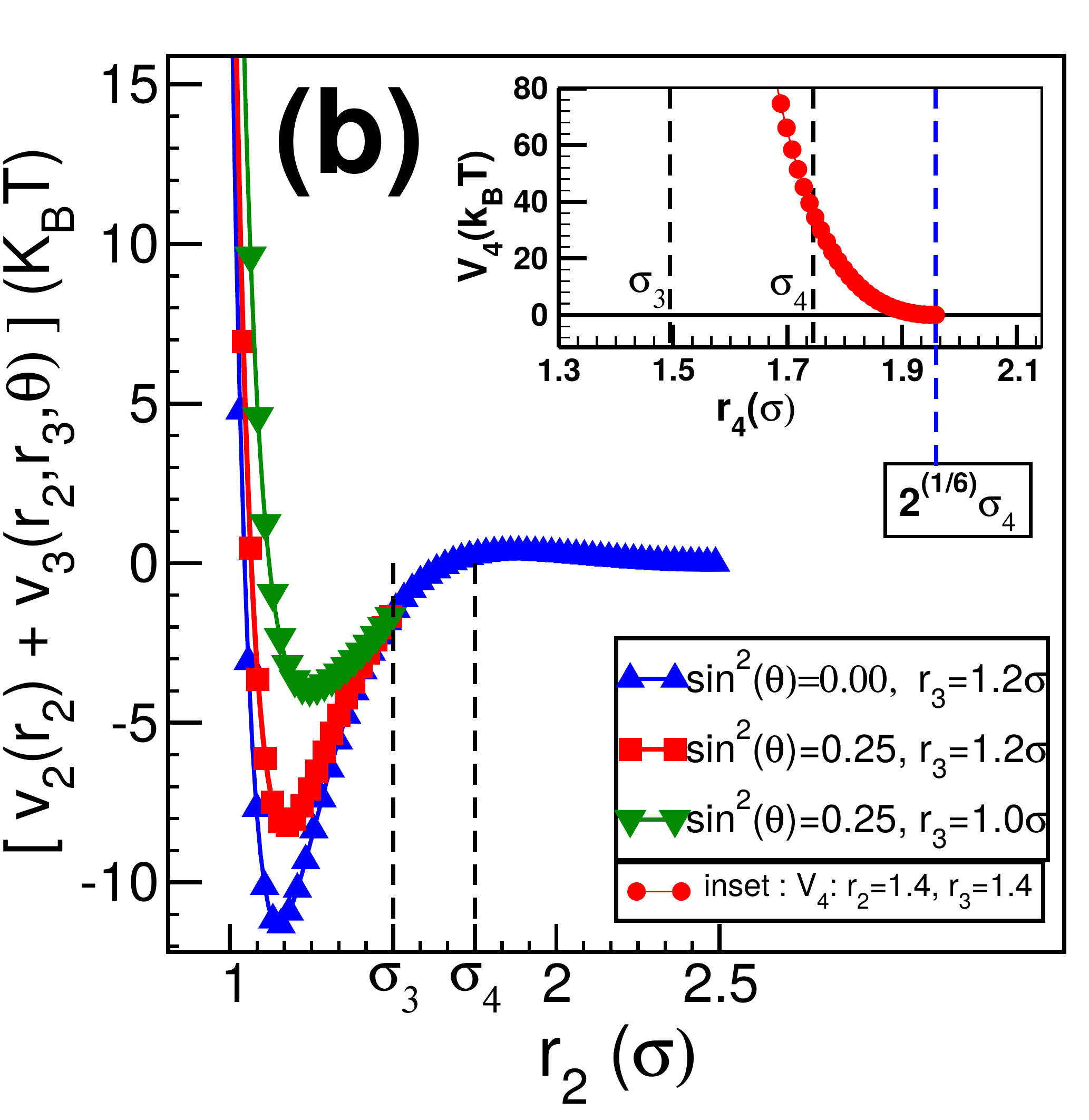}
\caption{(colour online) (a) The figure shows the schematic representation of the range 
of different interactions ($V_2, V_3, V_4$) of the model potentials. The red spheres denote the 
monomers of size $\sigma$ which self assemble to form equilibrium polymers. 
All the distances are shown with respect to the monomer shown in pink. 
These monomers are acted upon by two body potential $V_2$ having a cutoff range of $r_c$. A three-body potential 
$V_3$ is acting on a triplet with a central monomer bonded by two other monomers (joined by a dashed line) at 
distances of $r_2$ and $r_3$, forming an angle $\theta$ at the central monomer (shown in pink) and having a 
cutoff range of $\sigma_3$. There is also a purely repulsive four-body potential $V_4$ which is a  shifted 
Lennard-Jones potential introduced to prevent branching. The distance $2^{1/6} \sigma_4$ is defined as the cutoff 
distance for $V_4$. (b) The figure shows the  plots of $V_2+V_3$ for different values of 
versus $r_2$ for fixed values of $\sin^2 (\theta)$. $\sigma_3 =1.5 \sigma$ is the cutoff distance 
for potential $V_3$.  The inset shows a suitably shifted Lennard-Jones potential $V_4$ with a cutoff 
distance of $2^{1/6}\sigma_4$,  where $\sigma_4= 1.75 \sigma$.}
\label{fig:model}
\end{figure}

         The model we use is a modified version of a previous model presented by Chatterji and Pandit 
\cite{rahul,chatterji2003statistical} and also has been used in \cite{mubeena2015hierarchical}. In this model, 
WLMs are represented by a chain of beads (called monomers here which are shown as spheres in Fig.\ref{fig:model}(a)) 
which interact with the given potentials to form equilibrium polymers (Wormlike micelles) without branches. 
Each single bead or a 
monomer in the model is representative of an effective volume of a group of amphiphilic molecules. Thus, the WLMs 
are modelled at a mesoscopic scale ignoring all the chemical details. The diameter of a monomer $\sigma$ is taken 
as the unit of length in the system. The potential is the combination of three terms $V_2$, $V_3$ and $V_4$:

\begin{itemize}
\item {
We have a two body Lennard-Jones (L-J) potential modified with an exponential term.\\
   \begin{equation}
V_2 = \epsilon [ (\frac{\sigma}{r_2})^{12} - (\frac{\sigma}{r_2})^6 + \epsilon_1 e^{-a r_2/\sigma}]; 
\, \forall  r_2 \leq r_c,
\label{eq1}
\end{equation}
where $r_2$ is the distance between two interacting monomers with $\epsilon=110k_BT$ and the cutoff distance of 
the potential is at $r_c=2.5\sigma$. The exponential term in the above potential creates a maximum at 
$\approx 1.75\sigma$ which acts as a potential barrier for two monomers approaching each other. 
 Two monomers are defined as bonded when the distance $r_2$ between them becomes $\leq \sigma_3=1.5\sigma$, 
such that the interaction energy is negative. This is explained schematically in Fig.\ref{fig:model}(a) and 
the plot of the potential is also shown in Fig.\ref{fig:model}(b) (full line-blue). 
We set $\epsilon_{1}=1.34\epsilon$ and $a= 1.72$.
}
\item{
To model semiflexibility of chains, we use a three-body potential $V_3$,
\begin{equation}
V_3 = \epsilon_3 (1 - \frac{r_{2}}{\sigma_3})^2(1 - \frac{r_{3}}{\sigma_3})^2 \sin^2(\theta); 
\, \forall r_{2},r_{3} \leq \sigma_3. 
\label{eq2}
\end{equation}

      Here $r_{2}$ and $r_{3}$ are the distances of the two monomers A and B (refer Fig.\ref{fig:model}(a)) 
from the central monomer (coloured pink) which forms a triplet with an angle $\theta$ subtended by 
$\vec{r_{2}}$ and $\vec{r_{3}}$. Here $\epsilon_{3}=6075k_{B}T$. The prefactors to $\sin^2(\theta)$ are 
necessary to ensure that the potential and force goes smoothly to zero at the cutoff distance 
$\sigma_3=1.5\sigma$ for $V_3$.  This three-body potential modeling semiflexibility acts only if a monomer 
has two bonded neighbors at distances of $\leq \sigma_3$. Fig.\ref{fig:model}(b) shows the plots 
of $V_2+V_3$ for $r_2=\sigma$ and different values of $r_3$ and $\theta$. This potential penalizes angles $
< 180^\circ$. Configurations with $\theta = 0^{circ}$ are prevented by excluded volume interactions of the monomers.
}

\item{ Monomers interacting by potentials $V_2$ and $V_3$ self-assemble to form linear polymeric chains 
with branches. A fourth  monomer can get bonded with a monomer which already has two bonded neighbours to 
form a branch. To avoid branching, it is necessary to repel any monomer which can potentially form a branch. 
Therefore, we introduce a four-body potential term which is a shifted Lennard-Jones repulsive potential $V_{4}$ 
which repels  any branching monomer at  distances $\approx \sigma_4$, where  $\sigma_4=1.75\sigma > \sigma_3$. 
 The explicit form of the potential $V_4$ is, 
\begin{equation}
V_4 = \epsilon_4 (1 - \frac{r_{2}}{\sigma_3})^2(1 - \frac{r_{3}}{\sigma_3})^2 \times V_{LJ}(\sigma_4,r_4);
  \forall     r_4   \leq  2^{1/6}\sigma_4
\label{eq3}
\end{equation}
 Here $r_{2}$ and $r_{3}$ are the distances of the two monomers from central monomer(coloured pink) which forms 
a triplet, while $r_{4}$ is the distance of the fourth monomer which tries to approach the central monomer which 
already has two bonds, as shown in Fig.~\ref{fig:model}(a). We set the cutoff distance for this potential to be 
$2^{1/6}\sigma_4$ with  a large value of  $\epsilon_4=2.53 \times 10^5 k_BT$. The value of $\epsilon_4$ is kept 
large enough to ensure enough repulsion in case $(1 - \frac{r_{2}}{\sigma_3})^2 << 1$ and/or 
$(1 - \frac{r_{3}}{\sigma_3})^2 << 1$. These terms are necessary to ensure that if $r_2$ or $r_3$ becomes 
$\geq \sigma_3$, $V_4$ and the corresponding force goes smoothly to zero at the cutoff of $\sigma_3$. 
}
\end{itemize}

%@@@@@@@@@@@@@@@@@@@@@@@@@@@@@@@@ describe pot @@@@@@@@@@@@@@@@@@@@@@@@@@@@@@@@@@

%@@@@@@@@@@@@@@@@@@@@@@@@@@@@@@@@@@@@@@@@@@@@@@@@@@@@@@@@@22222222222222222

      All the three potentials are summarised in Fig.\ref{fig:model}(a). In this figure, the red spheres denote the 
micellar monomers having size $\sigma$. These monomers are acted upon by two body potential $V_2$ having a cutoff range 
of $r_c$. The potential is shown in Fig.\ref{fig:model}(b) indicated by the legends $\sin^2\theta=0$ 
(triangle-blue line). A three-body potential $V_3$ is acting on a triplet with a central monomer bonded by two other 
monomers at distances $r_2$ and $r_3$ (bonded monomers are joined by dashed lines). The vectors $\vec{r_2}$ and 
$\vec{r_3}$ subtends an angle $\theta$ at the central monomer and the potential $V_3$ has a cutoff distance of 
$\sigma_3$ (Fig.\ref{fig:model}(b)). The three-body potential $V_3$ modifies the two body potential $V_2$ such 
that the resultant potential energy $V_2+V_3$ gets shifted to a higher value of energy depending on the value 
of $\theta$, $r_2$ and $r_3$. This is illustrated in the graph for $\sin^2\theta=0.25$ and two different values 
of $r_3=1.2\sigma$ and $\sigma$ keeping the value of $r_2$ fixed at $r_2=\sigma$. For a given value of $\theta$, 
the potential energy $V_2+V_3$ is lower for higher value of $r_3$.  In addition to $V_2$ and $V_3$, there is also 
a four-body potential $V_4$ which is a shifted Lennard-Jones potential introduced to prevent branching and 
having a cutoff distance $2^{1/6}\sigma_4$. The behaviour of potential energy $V_4$ is shown in the inset 
of Fig.\ref{fig:model}(b). The vertical lines indicate $\sigma_3=1.5\sigma$ and $\sigma_4=1.75\sigma$. 
The vertical line (in blue) in the inset figure shows the cutoff distance of $V_4$ at $2^{1/6}\sigma_4$.

\subsection{Modeling Nanoparticles}

\begin{figure}
\includegraphics[scale=0.3]{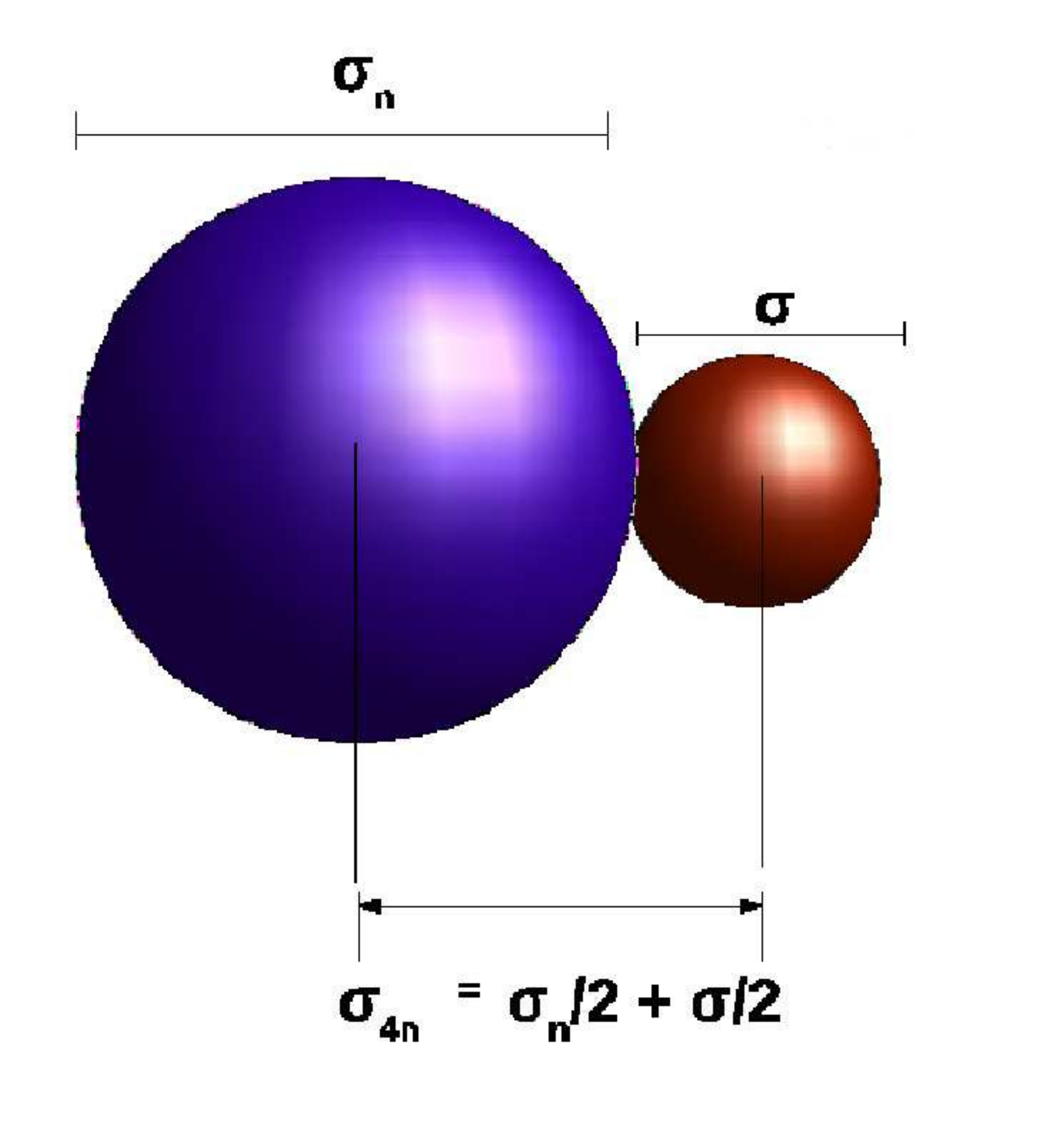}
\caption{ The figure is a schematic diagram to explain the role of the parameter $\sigma_{4n}$ by
which we control the excluded volume of the monomers when they interact with nan-particles.
 The parameter $\sigma_{4n}$ is shown as the minimum approaching distance between nanoparticles and 
monomers and calculated as $\sigma_{4n} = \sigma/2 + \sigma_n/2 $. For monomers of diameter $\sigma$ 
and nanoparticles of diameter $\sigma_n=1.5\sigma$, the minimum allowed value of $\sigma_{4n}$ is 
 $(\sigma_{4n})_{min}=1.25\sigma$.  We use  $\sigma_{4n}$ as a parameter and we choose  values 
$\sigma_{4n} \geq 1.25\sigma$ keeping $\sigma_n=1.5\sigma$ fixed.}
\label{sig4}
\end{figure}

       To explore the self-organization of NPs inside Wormlike micellar matrix (or EP matrix), we need to 
introduce a suitable model for NPs. The NP-NP interactions are modelled by a Lennard-Jones potential for 
particles of size $\sigma_n$,
\begin{equation}
V_{2n} = \epsilon_n[(\frac{\sigma_{n}}{r_n})^{12} - (\frac{\sigma_{n}}{r_n})^6],      \forall     r_n   \leq   r_{cn}
\end{equation}
 where, ${r_n}$ is the distance between the centres of two interacting NPs. The potential $V_{2n}$ goes
 smoothly to zero at the cutoff distance $r_{cn}=2\sigma_n$.
The interaction between NPs and micellar monomers at a distance r is modelled by a repulsive WCA 
(Weeks-Chandler-Anderson) potential given by,
\begin{equation}
V_{4n} = \epsilon_{4n}[(\frac{\sigma_{4n}}{r})^{12} - (\frac{\sigma_{4n}}{r})^6],  \forall r \leq 2^{1/6}\sigma_{4n},
\end{equation}
The strength of attraction for the above potential is kept to be $\epsilon_{4n}=30k_BT$. The value of 
$\sigma_{4n}$ here defines the minimum approaching distance between a NP and a monomer. Therefore, it is used 
as EVP between micelles and NPs. Since the monomers and NPs of diameters of $\sigma$ and $\sigma_n$ are 
considered to be nearly impenetrable, the EVP cannot be less than $\sigma/2+\sigma_n/2$. Thus, there exists 
a lower bound for the variable $\sigma_{4n}$ and it can only have values $\sigma_{4n} \geq \sigma/2+\sigma_n/2$.
This is explained in Fig.\ref{sig4}. In summary, the model has a fixed value of monomer diameter $\sigma$ and 
NP diameter $\sigma_n$ ($=1.5\sigma$) with the minimum approaching distance between them $\sigma_{4n}$ used 
as a variable along with monomer number density $\rho_m$ . All the other values are kept constant. 
The polymeric matrix thus modelled is equally applicable for a Wormlike micellar matrix or any other
equilibrium polymeric matrix.

\subsection{Method}

       To study the system behaviour,  we randomly place model monomers and NPs with a fixed volume fraction 
in a box of size $V=30\times 30\times 60\sigma^3$ with periodic boundary conditions 
where the unit of energy is set by choosing $k_BT=1$. 
We used Monte Carlo method (Metropolis algorithm) to evolve the system towards equilibrium. For a high-density 
regime, the Metropolis Monte Carlo moves seemed to be inefficient to equilibrate the system. Therefore, the 
system is first allowed to equilibrate for $10^5$ iterations with a few (100-200) NPs inside it so that monomers 
can self-assemble in the form of chains in the presence of the seeding of NPs. Once the chains are formed in 
the presence of seeded NPs ($~10^5$ iterations) then, a semi-grand canonical Monte Carlo (GCMC) scheme is 
switched on for the rest of the iterations. According to this scheme, every 50 Monte Carlo Steps (MCSs), 
300 attempts are made to add a NP at a random position or remove a randomly chosen NP in the system. 
Each successful attempt of adding or removing a NP is penalised by an energy gain or loss of $\mu_n$, which 
sets the chemical potential of NP system. A higher value of $\mu_n$ leads to the lower number of NPs getting 
introduced in the system. The system is evolved with the GCMC scheme for $2\times 10^6$ to $4\times 10^6$ 
iterations and the thermodynamic quantities considered are averaged over ten independent runs. 
For the lowest value of $\sigma_{4n}$ used, i.e. $\sigma_{4n} = 1.25 \sigma$, a rapid introduction of 
a large number of NPs into the box was observed as  soon as GCMC was switched on. This led to a large 
number of NPs ($\approx 10^4$ or more) in the simulation box and shorter runs. After the number of 
NPs in the simulation box had relatively stabilized,  the acceptance rate of of GCMC attempts 
was $\sim 5\%$, with a  slightly lower rate of removal of NPs from the simulation box, which resulted
in a net $~0.1\%$ addition of particles in the box. For the higher values of $\sigma_{4n}$, the rate of 
GCMC acceptance is $10-20\%$ and also reaches to $30\%$ in case of $\sigma_{4n}=3.5\sigma$. 
Similarly, the acceptance rate of attempts to remove NPs also remains approximately same resulting in 
the net $0.02-0.07\%$ rate of addition of NPs. The rate with which NPs are added decreases with 
increase in $\sigma_{4n}$.

%\section{Results }
\section{Previous results}

 In our previous communication ~\cite{mubeena2015hierarchical}, we used the same model potentials described 
here to show that monomers interacting with the potentials self-assemble to form long cylindrical chains 
akin to WLMs. These polymeric chains show an exponential distribution of chain length, thus, confirming 
the characteristic behaviour of equilibrium polymer (or Wormlike micellar) matrix ~\cite{turner1990relaxation}. 
The self-assembled polymers at low densities were observed to be in a disordered (isotropic) state with 
relatively small chains. With an increase in the monomer number density, the short chains or monomers 
start joining up to form longer chains. These chains then start getting parallel to neighbouring chains,
 and the system shows a transition to a nematically ordered state at a high value of density. 
We observe a  isotropic to nematic first order transition at monomer number density 
$\rho_m\approx 0.13\sigma^{-3}-0.14\sigma^{-3}$.   

          The behaviour of the mixture of the model NPs and equilibrium polymeric chains was investigated 
at a high monomer number density $\rho_m=0.126\sigma^{-3}$ with the change in $\sigma_{4n}$. The kinetically 
arrested assembly of NPs shows a transition from a system spanning percolating network-like structure 
to non-percolating clusters with an increase in EVP ($\sigma_{4n}$). The clusters of NPs were observed 
to be having rod-like structures.

\section{Present investigation}

       In this paper, the self-assembly of the equilibrium polymeric chains + NP system is investigated 
as a function of the $\sigma_{4n}$ (EVP) and $\rho_m$. Throughout the paper, the diameter of NPs $\sigma_n$, 
the quantity $\sigma_4$ and the NP chemical potential are kept fixed at $1.5\sigma$, $1.75 \sigma$  and $-8k_BT$,
 respectively, unless otherwise mentioned. All the distances mentioned here refer to the centre-to-centre 
distances of the particles involved. To investigate the behaviour of NPs and wormlike micellar system, 
four different values of the number of monomers $N_m$ are considered viz. $N_m=2000$, $4000$, $5000$ and $6800$ 
corresponding to the monomer number densities $\rho_m=0.037\sigma^{-3}, 0.074\sigma^{-3}$, $0.093\sigma^{-3}$ and 
$0.126\sigma^{-3}$, respectively. 

\begin{figure}
\centering
\includegraphics[scale=0.35]{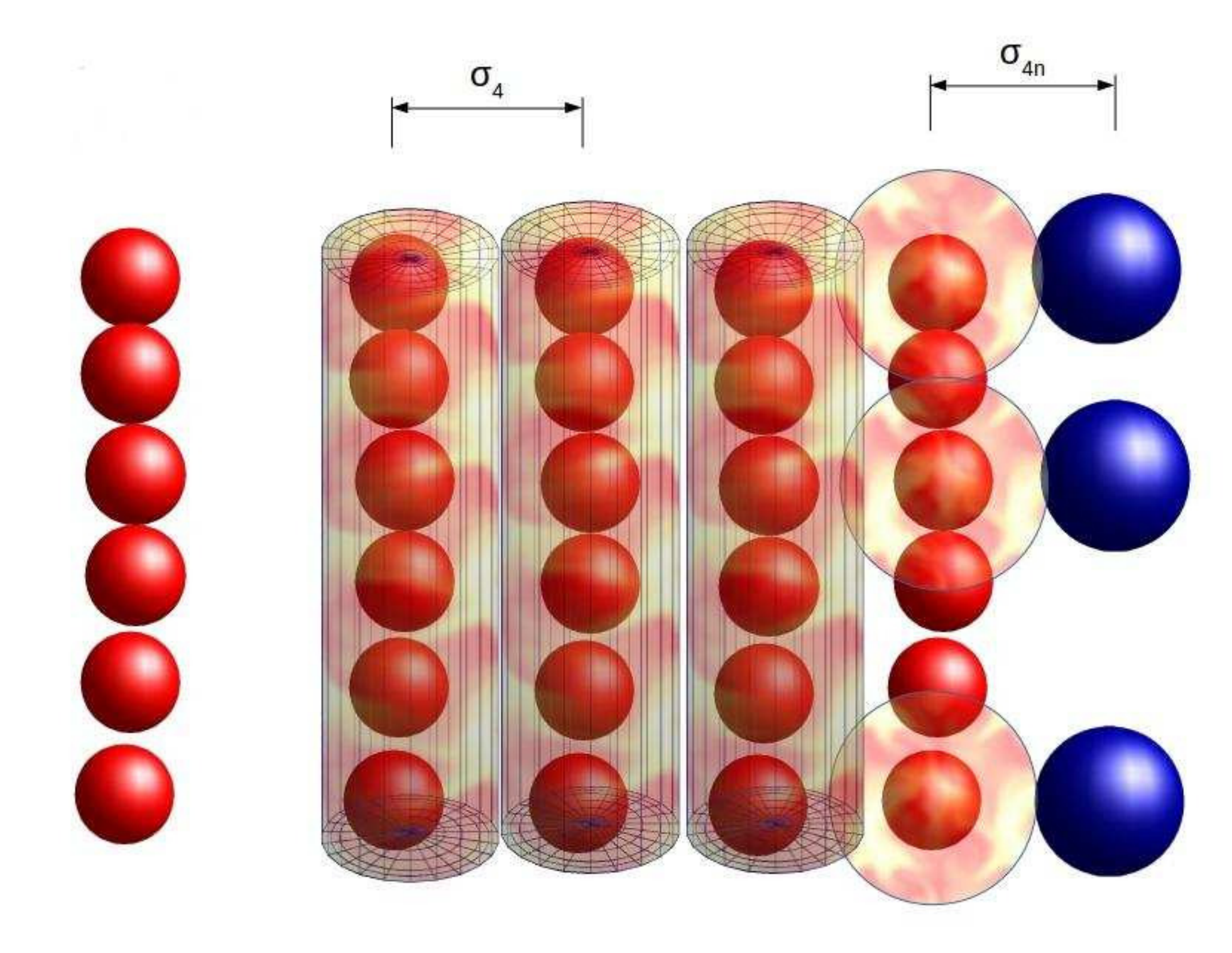}
\caption{(colour online) The figure explains the calculation of the effective volume of monomers. 
Red colored particles represents monomers, whereas the nanoparticles are shown in  blue. Two neighbouring polymeric 
chains with distance between monomers centers $ <  \sigma_{4}=1.75 \sigma$ are considered as the cylinders of 
radius $\sigma_4$. Monomers whose position is within the range $ 2^{1/6}\sigma_{4n}$ of repulsive potential $V_{4n}$ 
are assumed as spheres of radius $(\sigma_{4n}-\sigma_n/2)$, where $\sigma_n$ is the radius of 
nanoparticles. }
\label{eff_vol}
\end{figure}

\begin{figure*}
\centering
\includegraphics[scale=0.19]{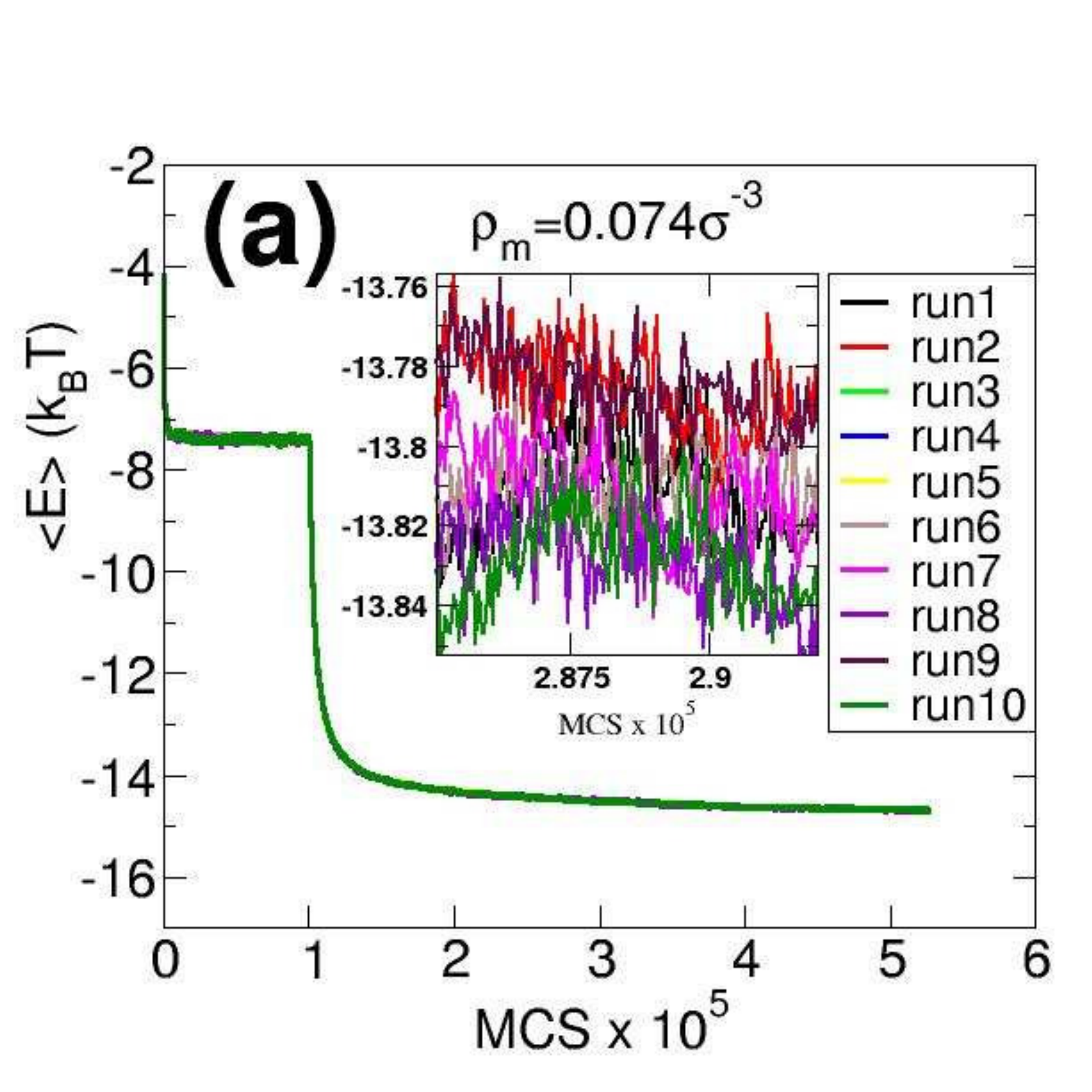}
\includegraphics[scale=0.19]{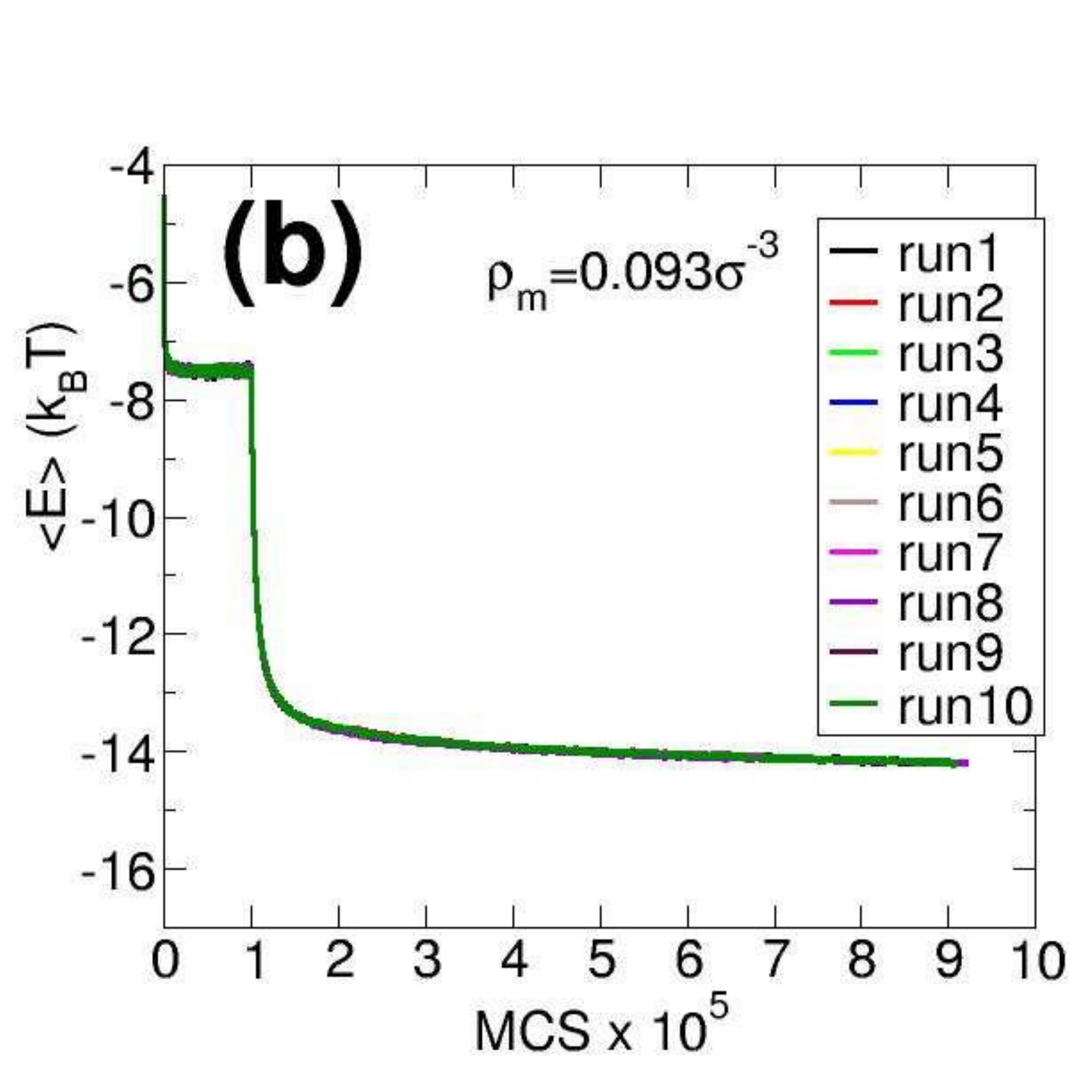}
\includegraphics[scale=0.2]{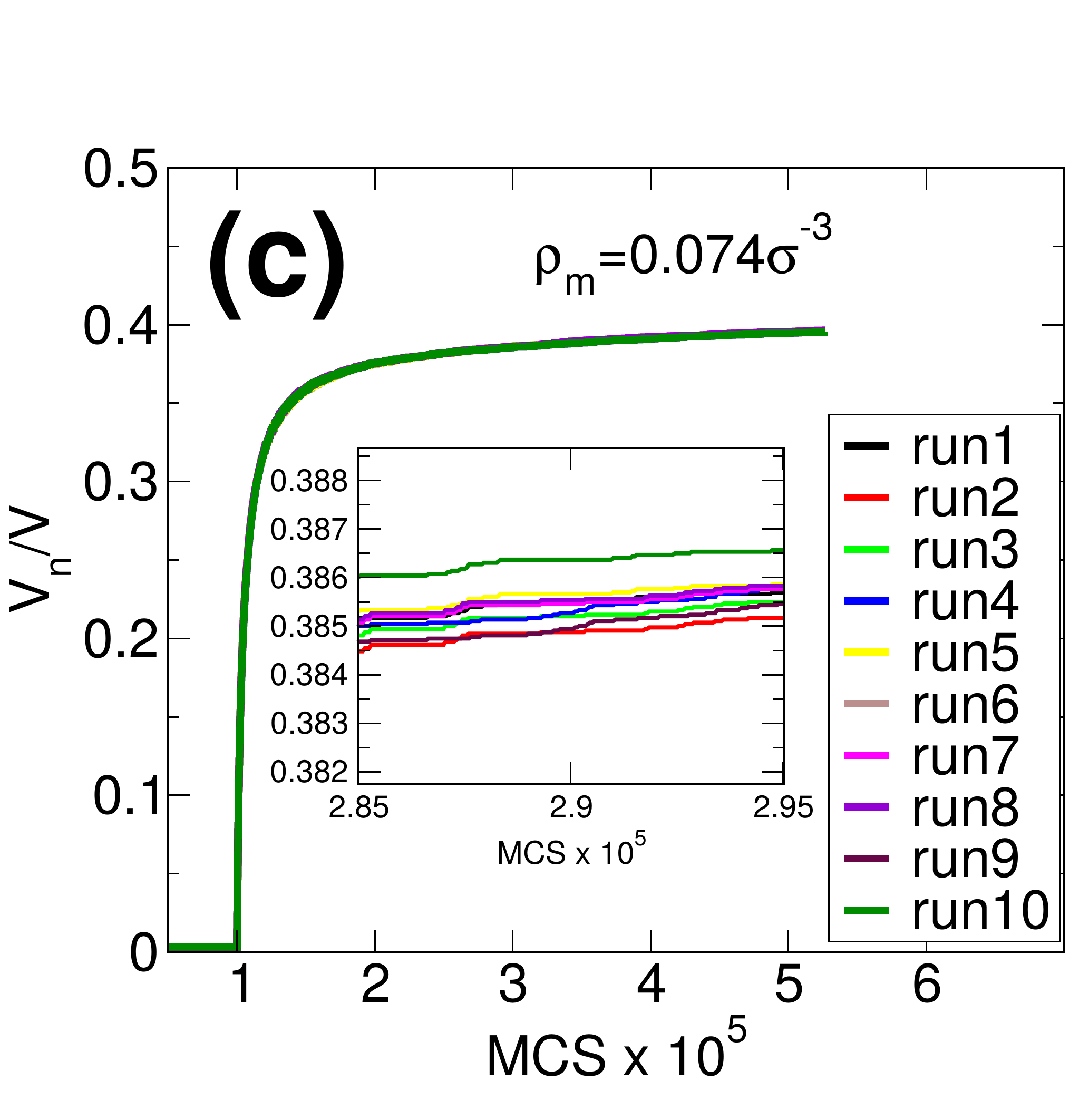}
\includegraphics[scale=0.2]{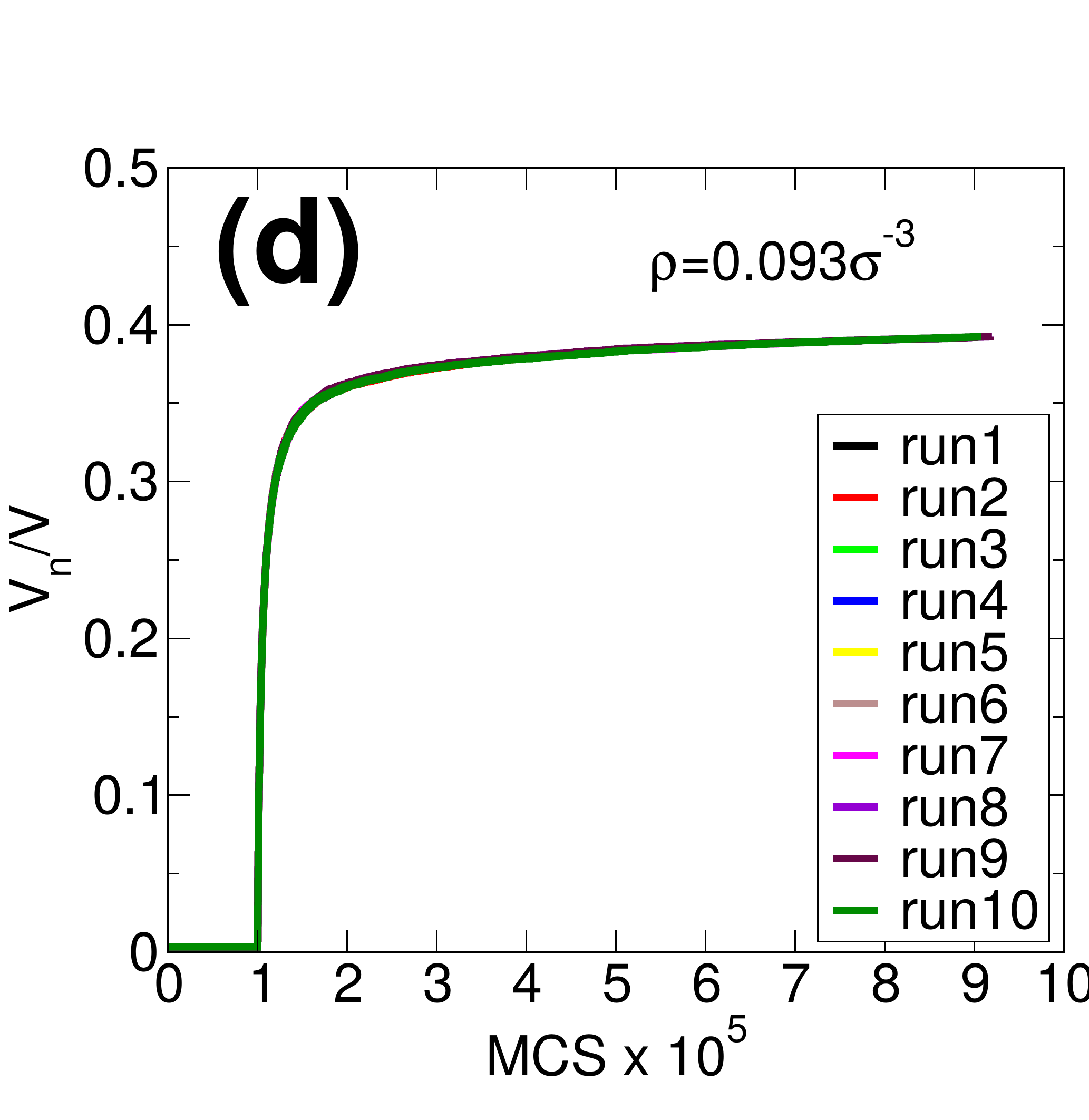}
\caption{(colour online) The subplots (a),(b)  shows the plots of average energy per particle (both monomers and NPs)
 for two different values of monomer number density $\rho_m$.  The subplots (c),(d) shows nanoparticle volume fraction
$V_n/V$ versus MCSs, $V_n$ is the volume of NPs and $V$ is the volume of simulation box.
 The parameter  $\sigma_{4n}$ is kept fixed at $1.25 \sigma$. 
Each figure shows multiple graphs generated from $10$  independent runs initialized with different initial 
configurations which can be clearly seen in the magnified parts of the graphs shown in the inset of figures
(a) and (c).  The system is evolved with MC steps for a first $10^5$ iterations and then subjected to the grand-canonical
MC (GCMC) scheme on nanoparticles. The data show a sudden jump in their values when we switch on GCMC.}
\label{runs}
\end{figure*}

\begin{figure*}
\centering
\includegraphics[scale=0.2]{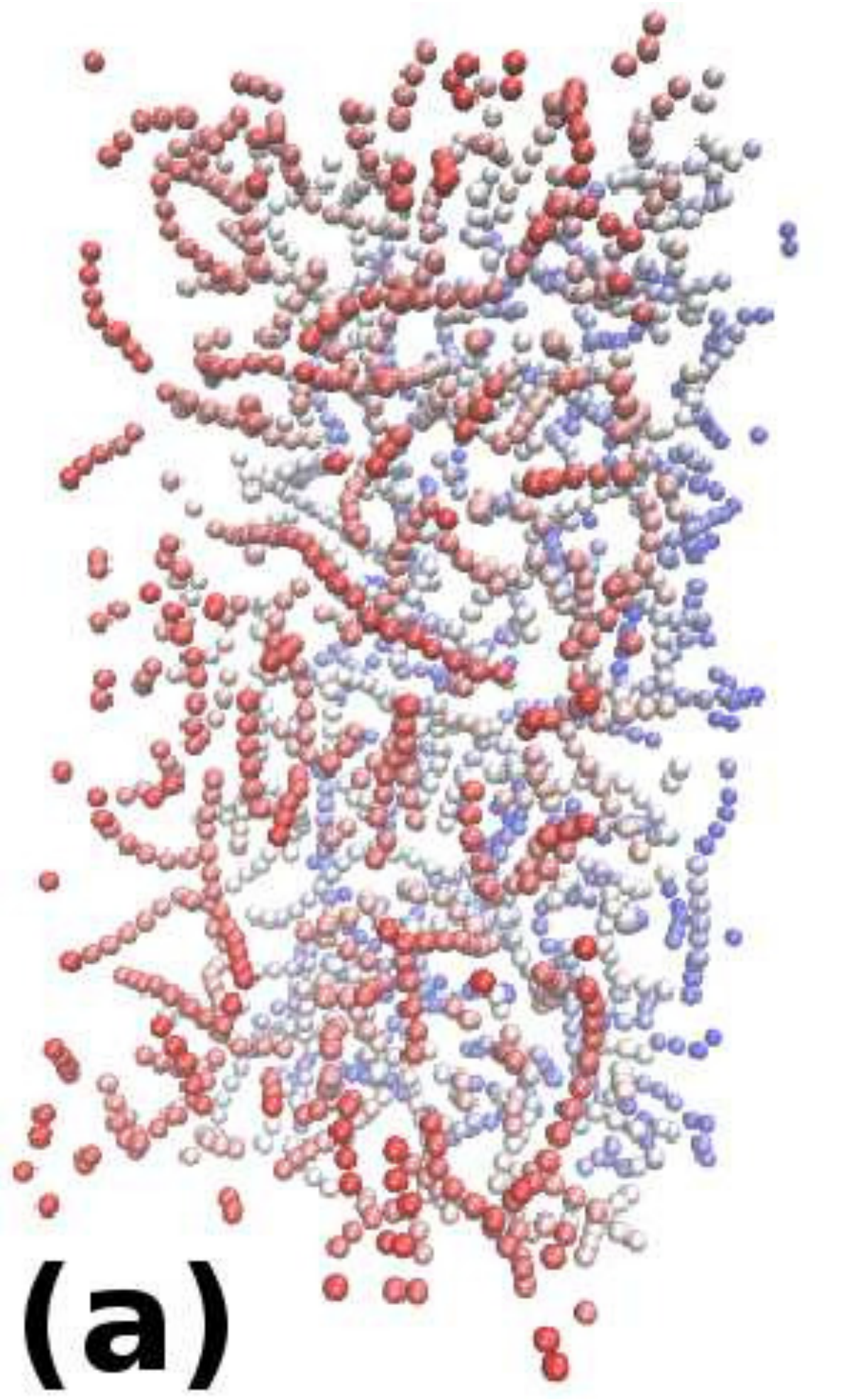}
\hspace{2cm}
\includegraphics[scale=0.2]{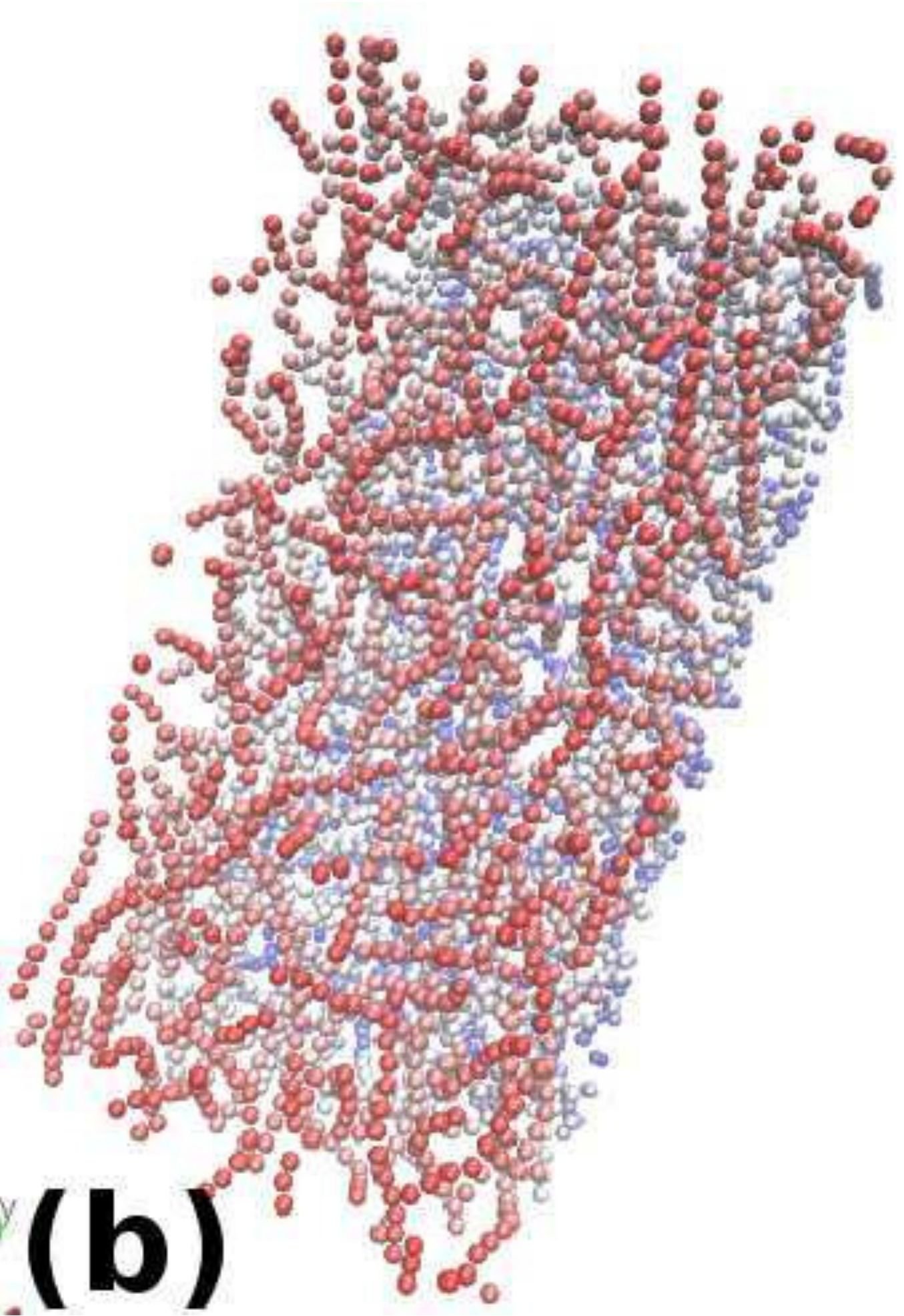}
\hspace{2cm}
\includegraphics[scale=0.2]{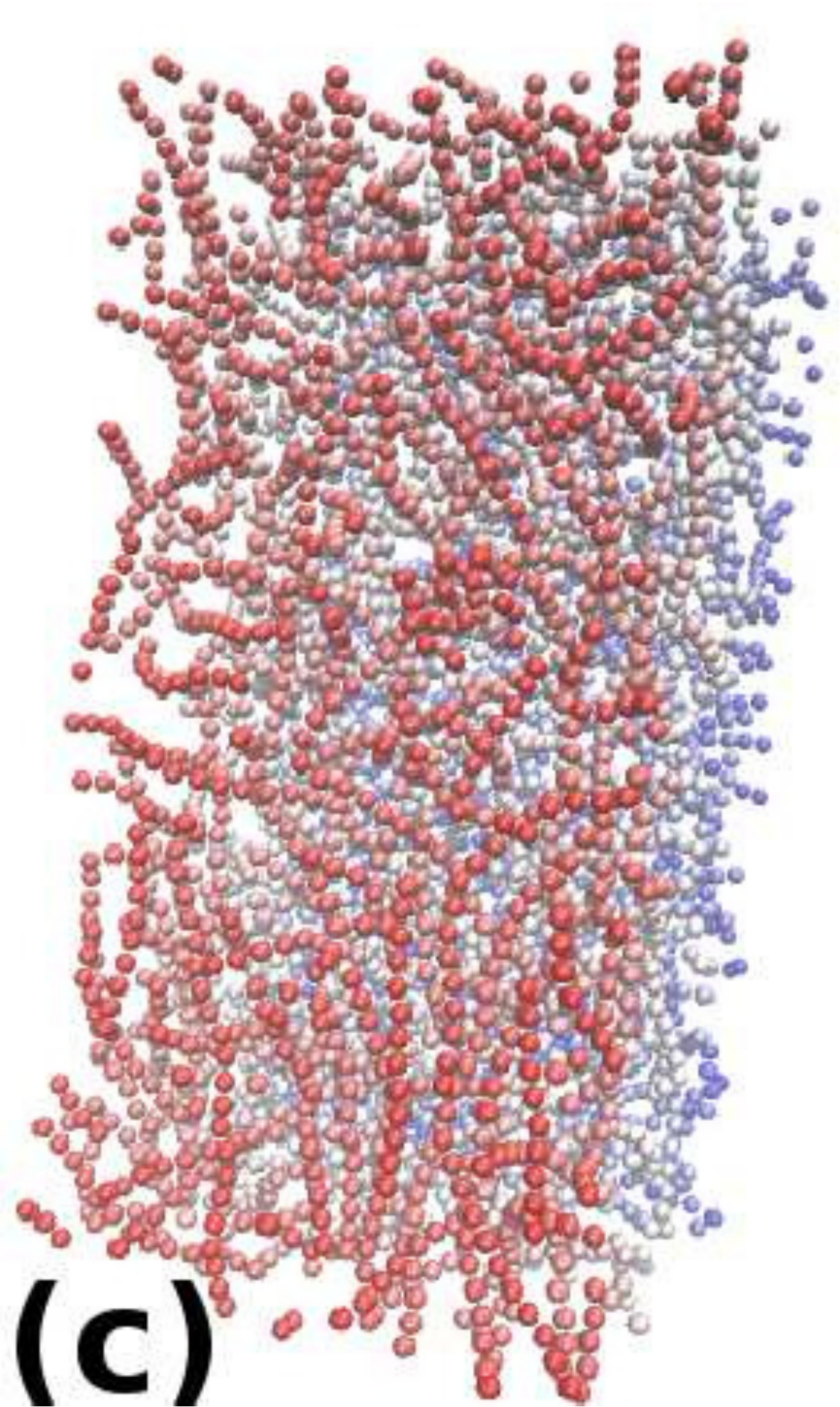}
\hspace{2cm}
\includegraphics[scale=0.2]{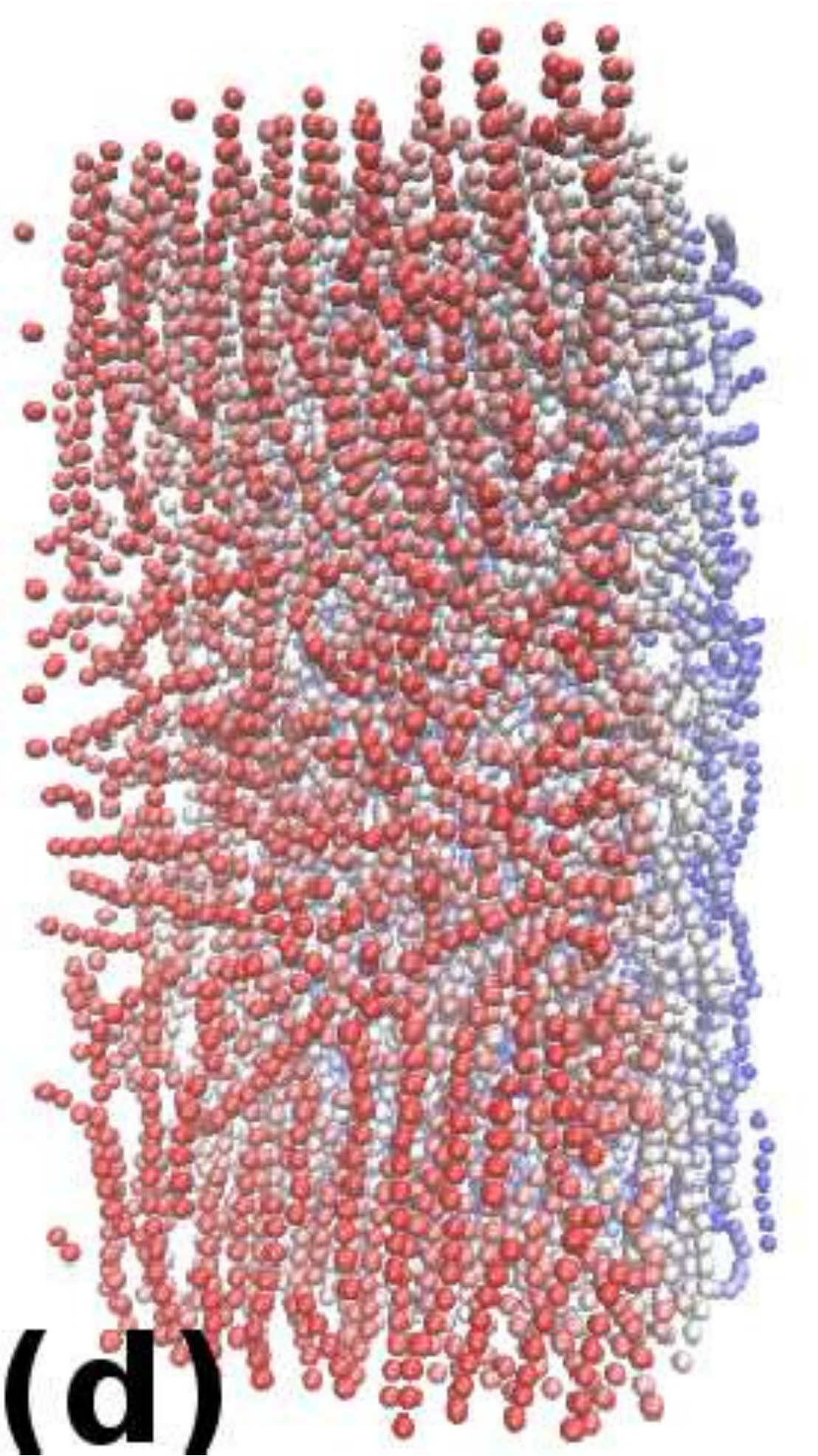} \\
\includegraphics[scale=0.2]{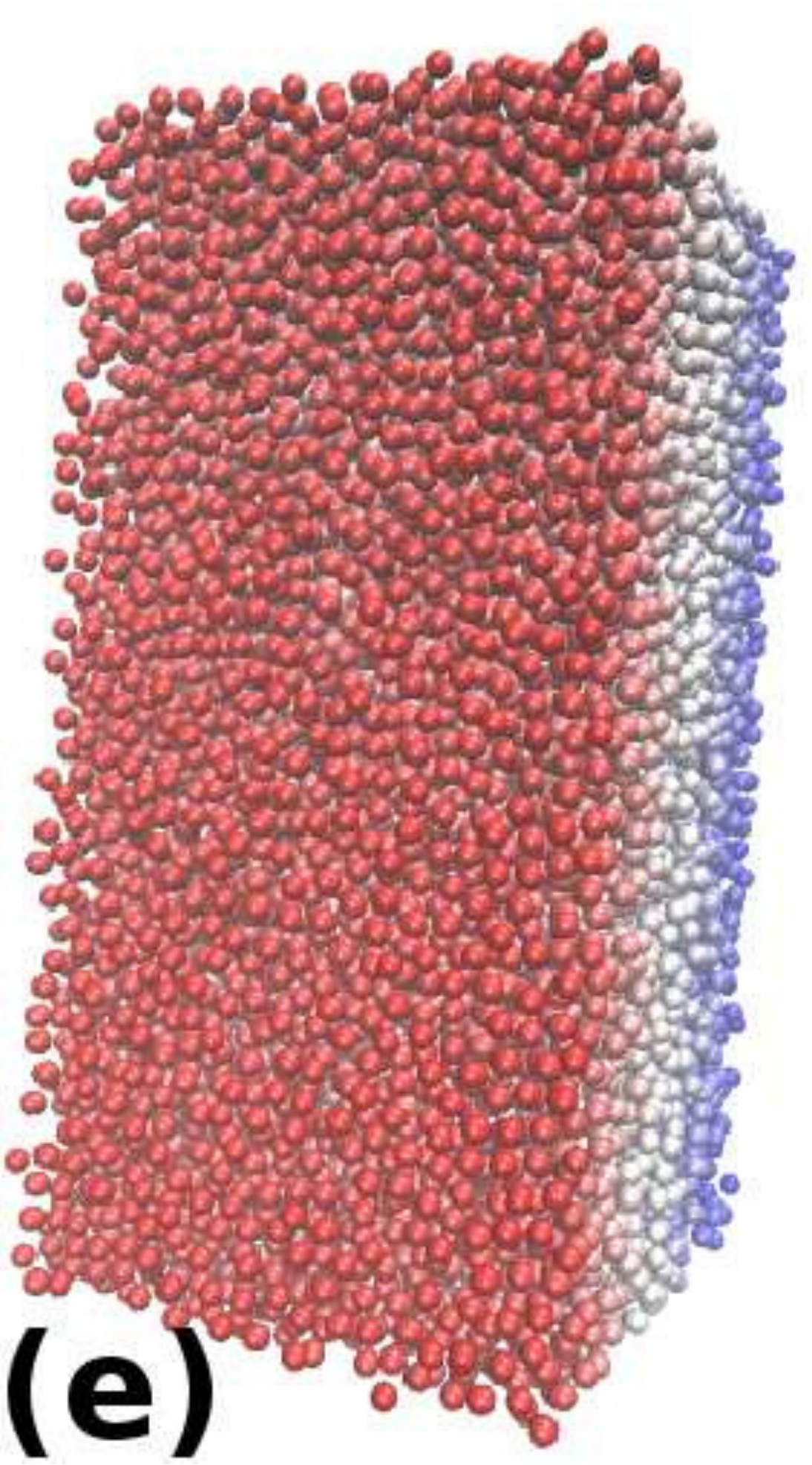}
\hspace{2cm}
\includegraphics[scale=0.2]{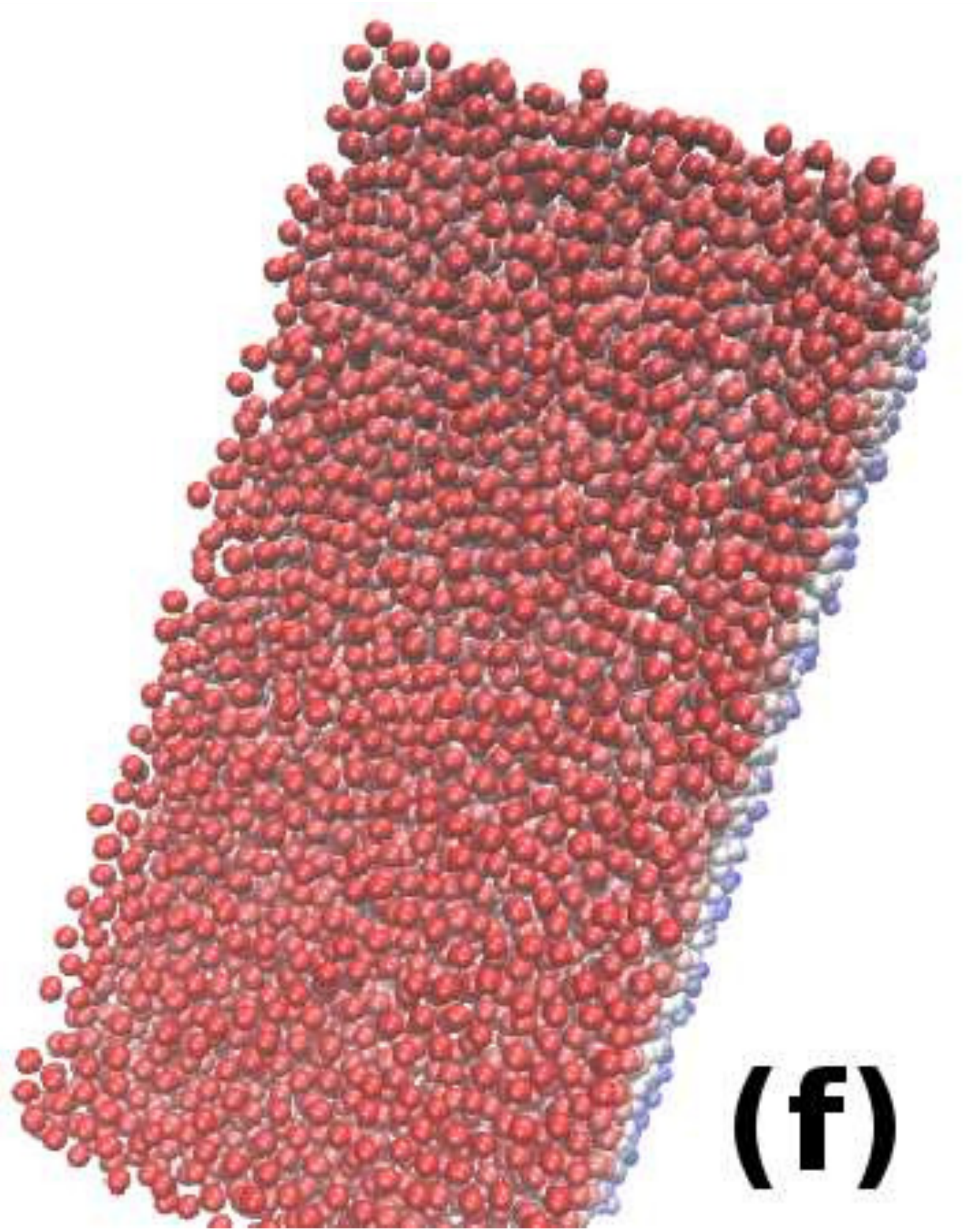}
\hspace{2cm}
\includegraphics[scale=0.2]{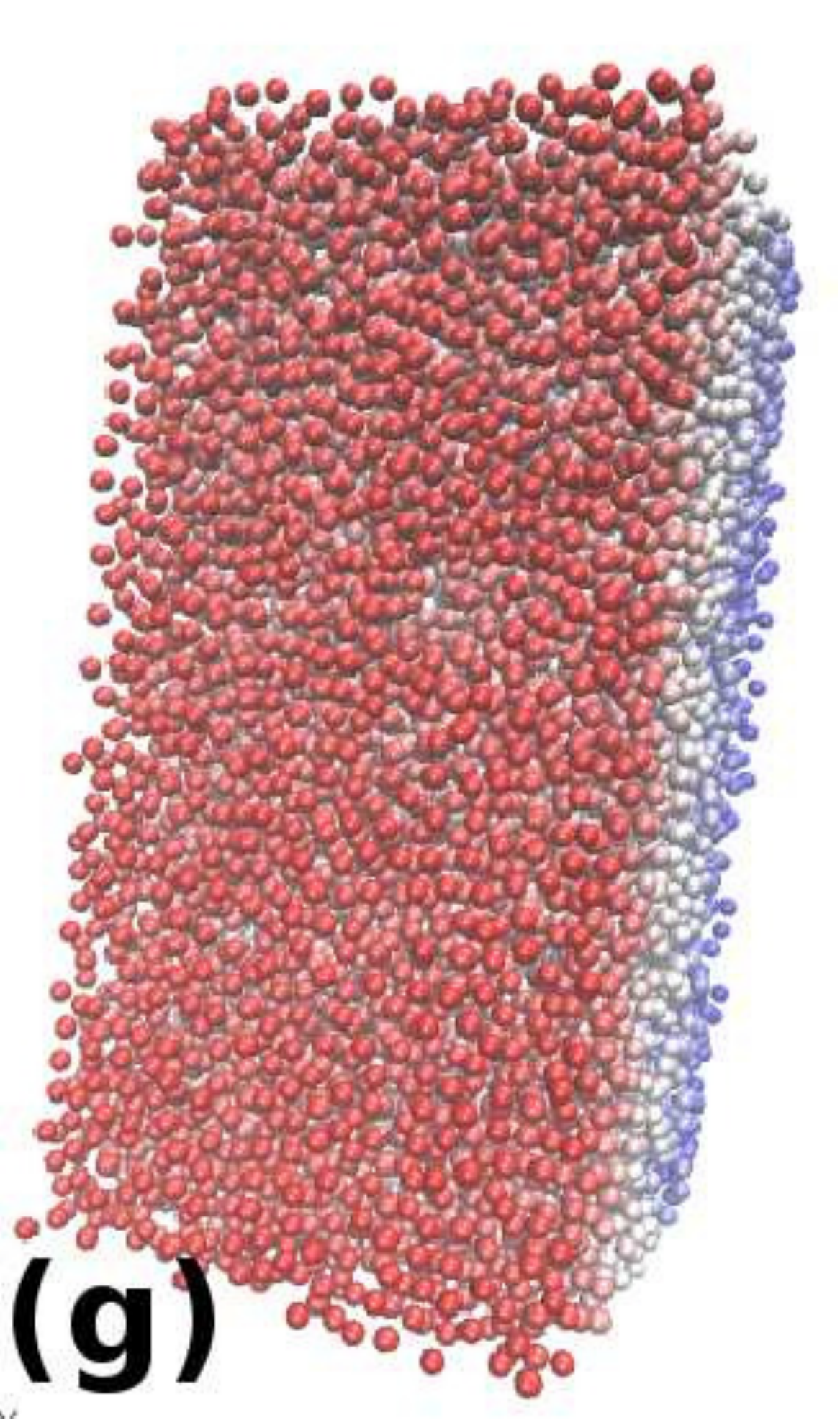}
\hspace{2cm}
\includegraphics[scale=0.2]{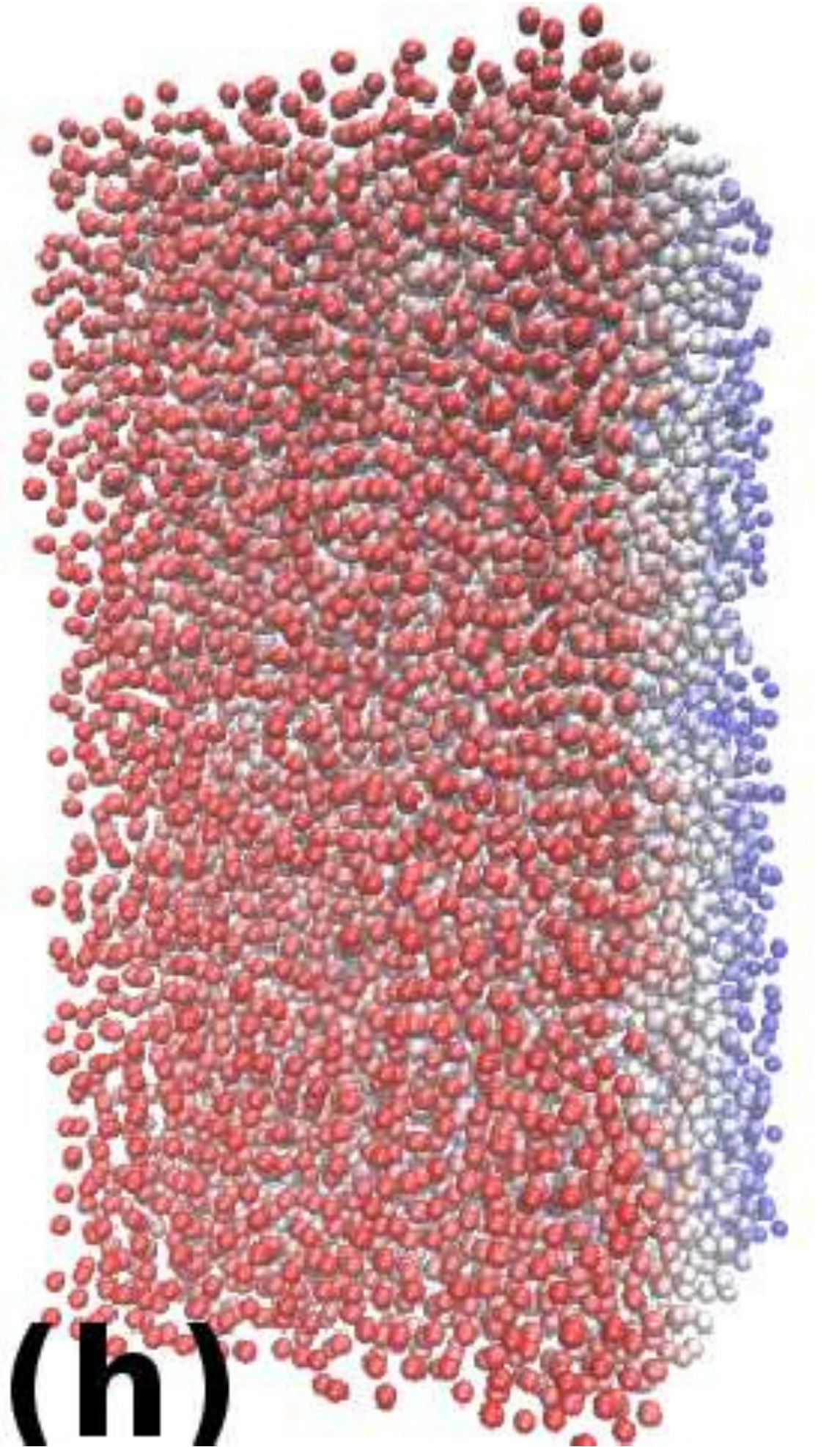}
\caption{(colour online) The upper row of snapshots shows snapshots of only the equilibrium polymers (micellar chains)
 and the  lower row shows only the nanoparticles  of the NP+polymer system. Different snapshots are for 
different values of monomer densities;  (a),(e) is for $\rho_m = 0.037\sigma^{-3}$, (b),(f) is for $\rho_m 0.074 
=\sigma^{-3}$, (c),(g) is for $\rho_m =0.093\sigma^{-3}$ and (d),(h)  is for $\rho_m 0.126\sigma^{-3} $. 
The value of $\sigma_n $ and $\sigma_{4n}$ is kept fixed at $1.5\sigma$ and $1.25\sigma$, respectively. 
All the snapshots show that the nanoparticle-monomer system are in  uniformly mixed state and clustering 
of micellar chains is not observed. All the figures have a gradient in their colour varying from red to blue 
to identify and distinguish between  the particles which are near the front plane (red) and the ones 
which are closer to the the back of the box (blue).}
\label{crystal}
\end{figure*}

\subsection{Calculation of effective Volume}

       We realize that the system behaviour can be better characterized by considering the total 
excluded volume of clylindrical polymer chains and NPS due to the repulsive interactions $V_4$ and $V_{4n}$ instead of 
only considering the volume fraction of spherical monomers. 
We define the total volume excluded by a chain due to the repulsive potentials from $V_4$ and 
and $V_{4n}$ as the effective volume of the chain. The effective volume of a monomer is 
the effective volume of the chains divided by the total number of monomers. 
The calculation of the effective volume of WLMs is explained in Fig.\ref{eff_vol}.  The figure shows 
that any two micellar chains (red spheres) at a distance of the cutoff distance ($2^{1/6}\sigma_4$)
 (range of the repulsive interaction ($V_4$)) from each other are considered as non-overlapping 
cylinders of diameter $\sigma_4$ (shaded cylinders). Correspondingly, the effective volume of the 
monomer in the chain is defined to be $v_1=\pi (\sigma_{4}/2)^{2}\sigma$. If a monomer is at a distance 
less than the cutoff distance $2^{1/6}\sigma_{4n}$ (for $V_{4n}$) from a nanoparticle, then the monomer 
is considered to be a sphere of radius $(\sigma_{4n}-\sigma_n/2)$ (shaded sphere) and the effective volume of 
the monomer is $v_2=4\pi/3(\sigma_{4n}-\sigma_n/2)^{3}$. Any monomer that is not involved in any of 
the repulsive interactions $V_4$ and $V_{4n}$ has the volume $v_3=4\pi/3(\sigma/2)^3$. 

To calculate this effective volume using a suitable algorithm, first, we identify  and label the monomers 
which are part of a chain.  All monomers within a distance of $1.5\sigma$ (cutoff distance for $V_3$) from each 
other are considered as bonded and thus form part of a chain. We do not observe branching in the chains.
Then all the chains that are involved in the repulsive interactions from other chains or nanoparticles 
are identified and the effective volume of micelles is calculated according to the scheme discussed above 
and  illustrated in Fig.\ref{eff_vol}. If a monomer interacts with a monomer of a neighbouring chain as 
well as a NP simultaneously,  then, the higher of the two values of  effective volume  ($v_1$ or $v_2$)
is considered.  The effective volume fraction of micelles, thus calculated, depicts the actual excluded 
volume fraction because of the repulsive potentials $V_4$ and $V_{4n}$ in addition to the volume fraction 
of monomers ($\rho_m 4/3\pi(\sigma/2)^3$). If there are $n_1$ monomers interacting with other monomers 
with potential $V_4$, $n_2$ monomers having repulsive interaction $V_{4n}$ with a nanoparticle and 
$n_3$ monomers out of range of the potentials $V_4$ and $V_{4n}$ then, $n_1v_1+n_2v_2+n_3v_3$ 
depicts the effective volume of micelles. This effective volume of monomers is not only dependent 
upon the value of $\sigma_{4}$ and $\sigma_{4n}$ but also on the arrangement of the constituent 
particles. The effective volume thus calculated will help in understanding the behaviour of the 
system. For monomers situated in between a micellar chain and NPs, the volume may get slightly 
overestimated when $\sigma_{4n} - \sigma_n > \sigma_4$. This is because of the over-calculation of 
the volume in case of overlapping of the shaded (red) spheres and shaded monomer chains (cylinders) 
as shown in the Fig.\ref{eff_vol}. However, this overestimation does not change the interpretation of results.

\subsection{Investigation for systems with $\boldsymbol{(\sigma_{4n})_{min}=1.25\sigma}$:
                Formation of a dispersed state of polymeric chains}

\begin{table*}
 \hspace{-1cm}$\boldsymbol{\mathlarger{\mathlarger{\mathlarger{\rho_m=0.037\sigma^{-3}}}}}$ \hspace{4cm}   $\boldsymbol{\mathlarger{\mathlarger{\mathlarger{\rho_m=0.126\sigma^{-3}}}}}$\\
\hspace{-1cm} $\overbrace{\boldsymbol{initial}  \hspace{1cm} \rightarrow \hspace{1cm} \boldsymbol{final} }$ \hspace{2cm}  $\overbrace{\boldsymbol{initial} \hspace{1cm} \rightarrow \hspace{2cm} \boldsymbol{final}}$\\
\hspace{0.1cm}  \hspace{3cm}  \hspace{7cm}  $\overbrace{\mu_n=-8k_BT \hspace{1cm} \mu_n=4k_BT}$
\end{table*}
\begin{figure*}
\centering
\includegraphics[scale=0.25]{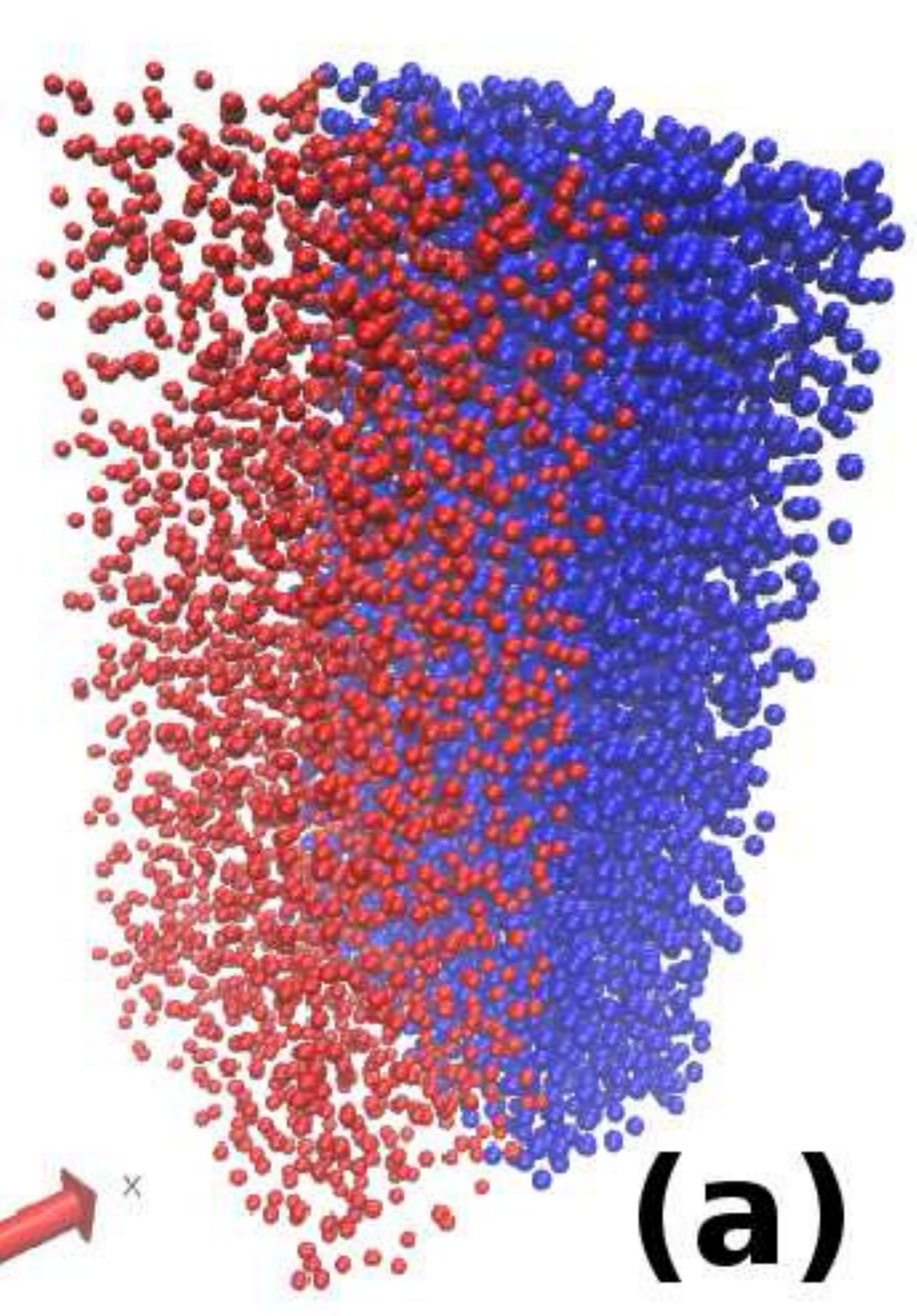}
\includegraphics[scale=0.23]{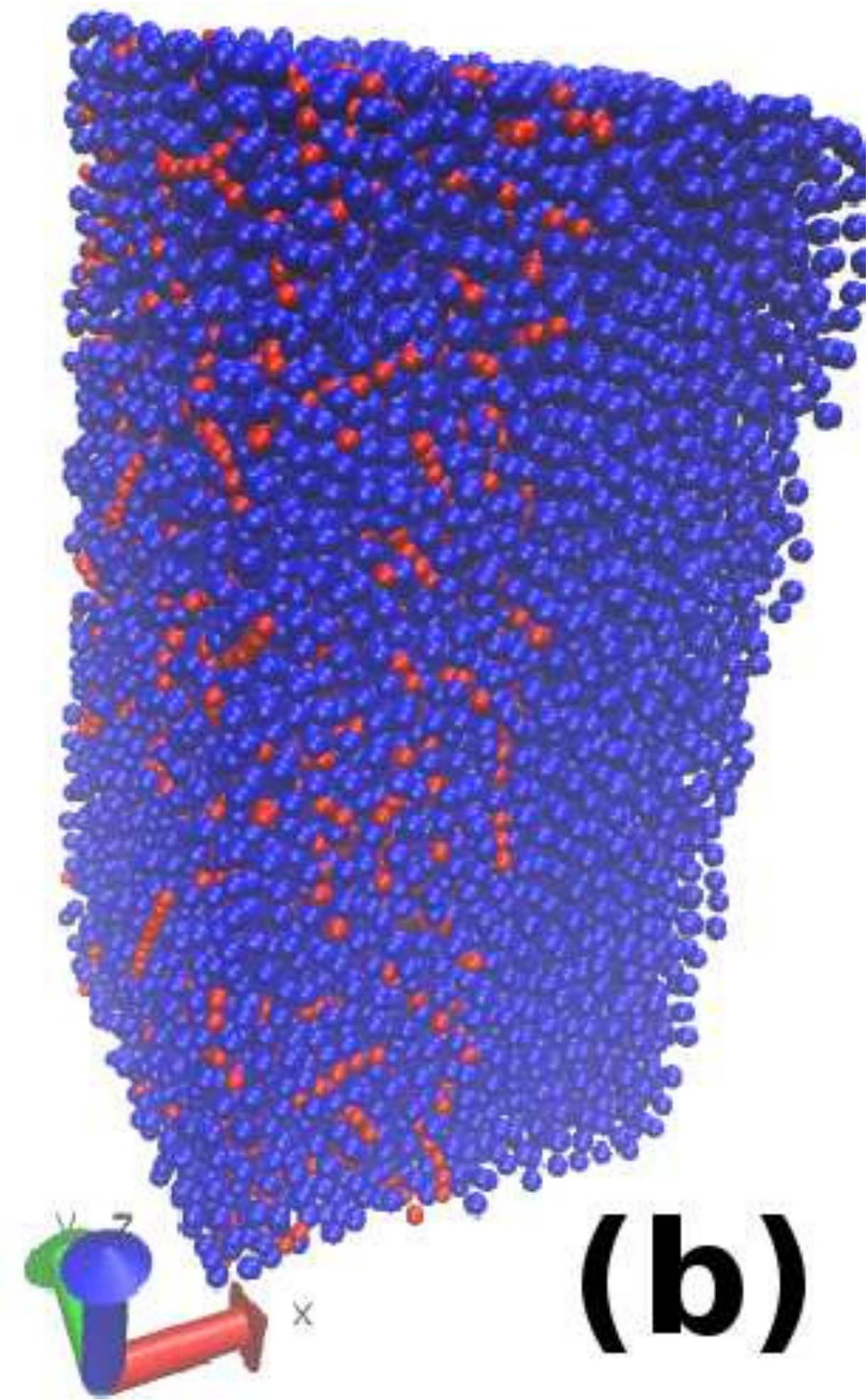}
\includegraphics[scale=0.23]{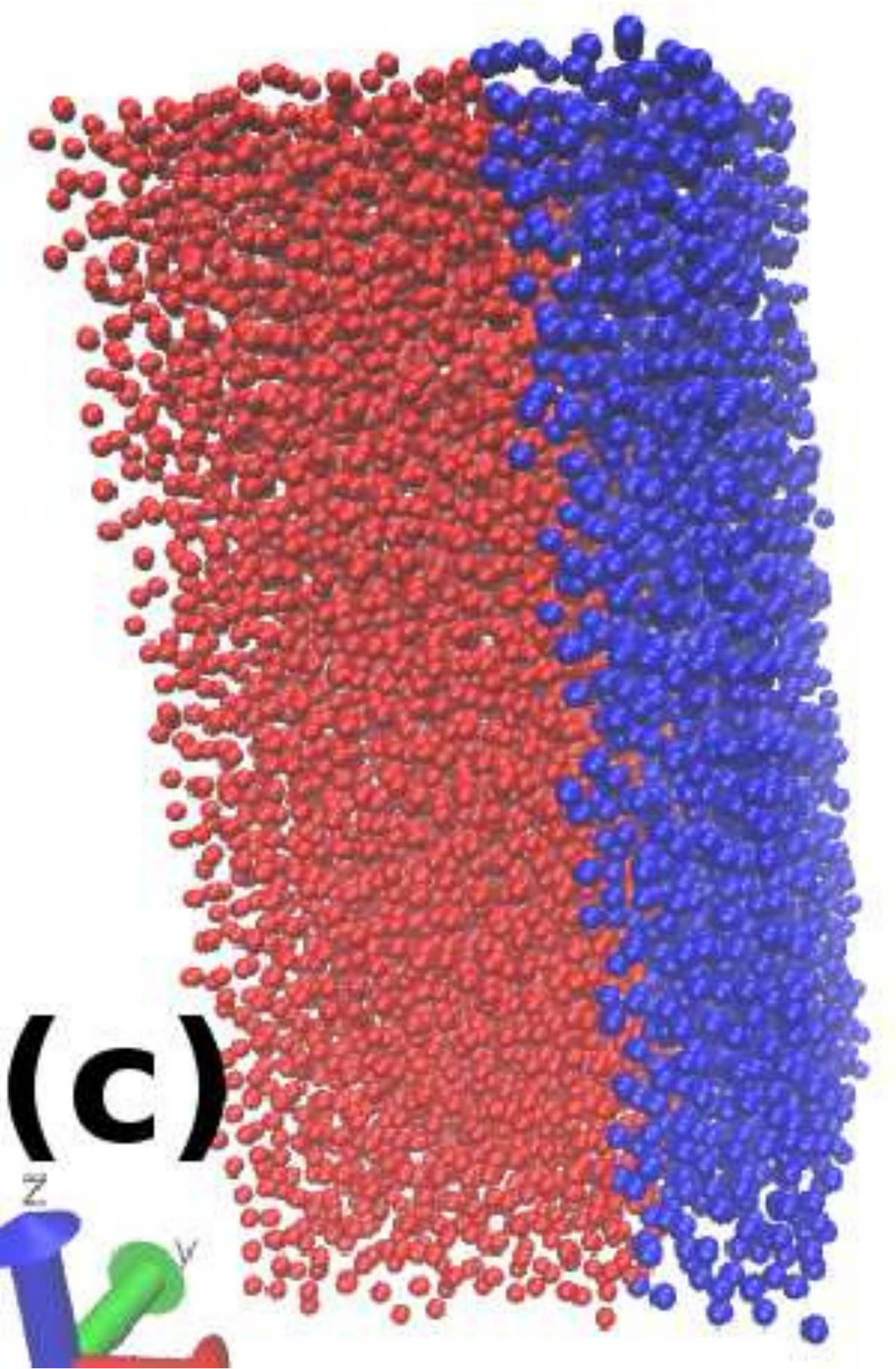}
\includegraphics[scale=0.23]{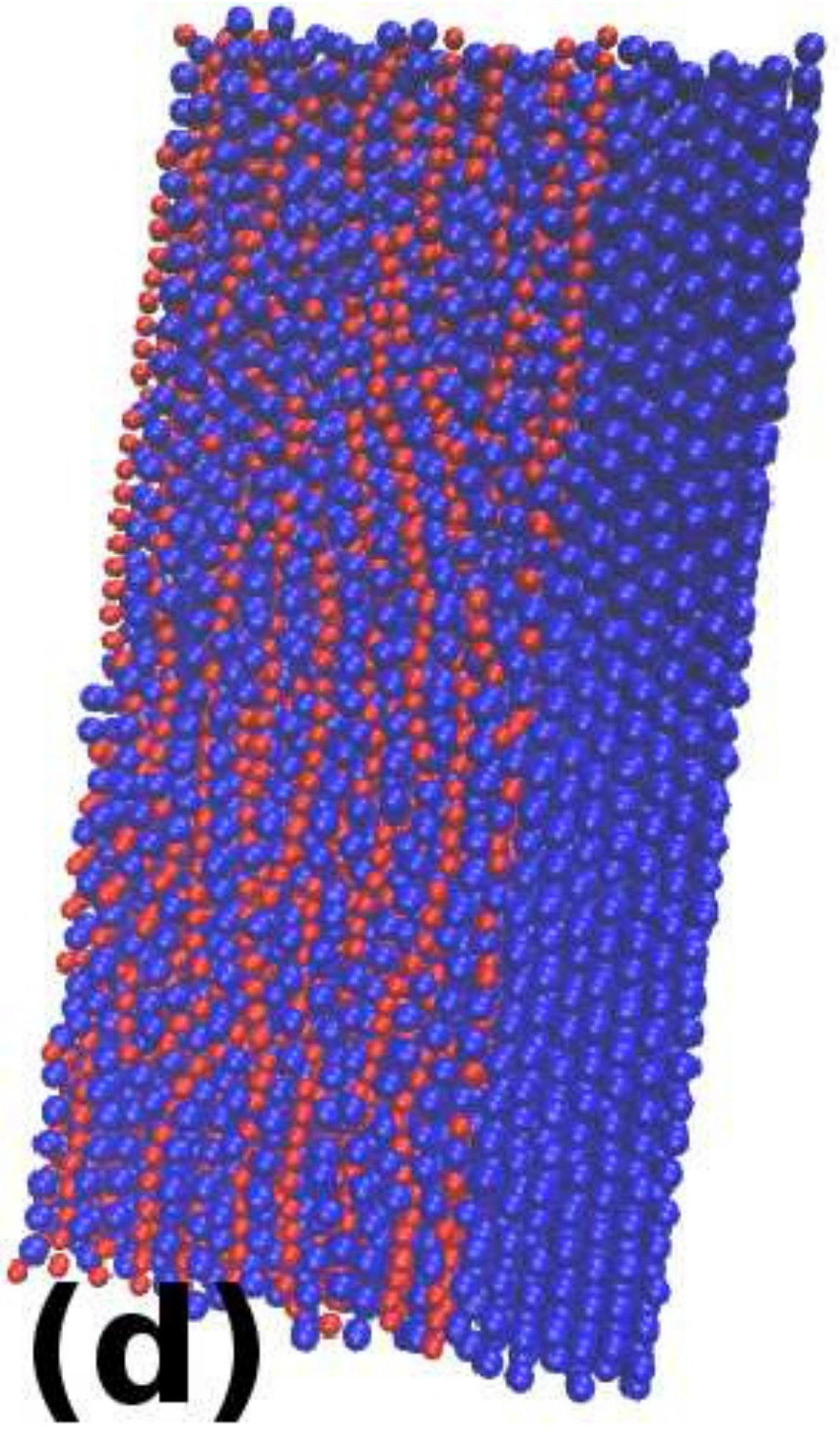}
\includegraphics[scale=0.23]{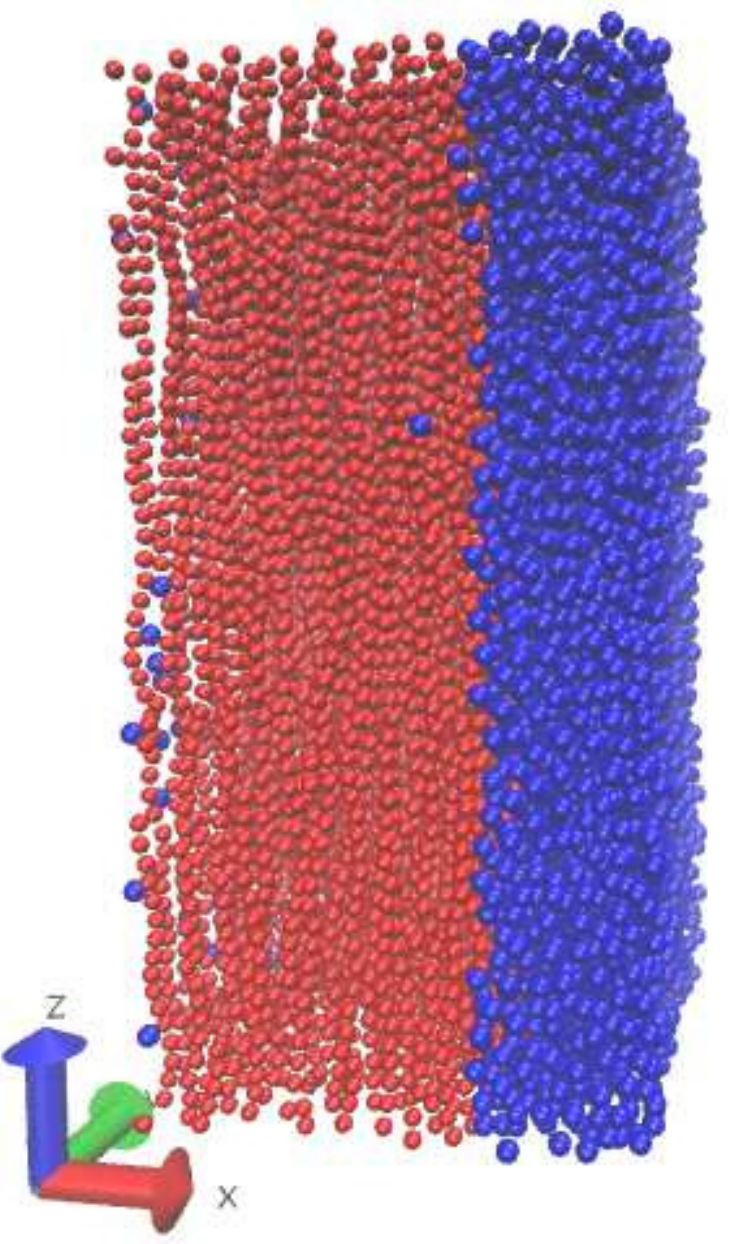}
\caption{(colour online) The figure shows the snapshots of the system for monomer number densities 
$ \rho_m=0.037\sigma^{-3} $ (figures (a) and (b)) and $ 0.126 \sigma^{-3} $ (figures (c), (d) and (e)). 
Red particles indicate monomers while nanoparticles are shown in blue. For each $\rho_m$, the two 
snapshots in (a) and (c) represent the initial configuration of monomers and NPS, where NPs and monomers
are kept on two sections of the simulation box.
The snapshots in (b) and (d) show the microstates after $4\times 10^6$ iterations (with GCMC switched 
on after $10^5$ iterations).  In both the cases of densities, the final state deviates from the initial 
state forming a uniformly mixed 
states of micellar chains + nanoparticles,  this coexists with a state which occupies part of the box
which has only densely packed nanoparticles (without micellar chains). The snapshot in figure (e) shows 
the final state of the system initialized as in (c) but with a higher value of chemical potential 
$\mu_n=4k_BT$. The states reached after the runs for $\mu_n=-8k_BT$ (figure (d)) and $\mu_n=4k_BT$ 
(figure (e)) are different. The state in (e) shows a phase separated state.}
\label{diff_init1}
\end{figure*}

           To study the effect of change of micellar density, the system is simulated with different monomer
 number densities at $\sigma_{4n}=1.25\sigma$ (minimum allowed value of $\sigma_{4n}$; refer Fig.\ref{sig4}). 
To check the robustness of the results, for each number density of monomers, the system is observed over 
ten independent runs each starting with different initial random configurations. First, the system is evolved 
by applying Monte Carlo technique for $10^5$ iterations and then GCMC scheme is switched on for the rest 
of the iterations. Fig.\ref{runs} shows the evolution of the average energy of the constituent particles 
($\langle E\rangle$) and the NP volume fraction $V_n/V$ with MCSs, for two different values of monomer 
number densities. The NP volume $V_n$ is given by $V_n=N_n  4/3\pi {\sigma_n}^{3}$ where, $N_n$ is the 
number of NPs in the simulation box. Both the figures for energy per unit constituent particle 
(Fig.\ref{runs}(a) and (b)) and the volume fraction of NPs graphs (Fig.\ref{runs}(c) and (d)) show 
data for multiple runs generated from 10 independent runs. All these graphs from different independent 
runs overlap and look indistinguishable from each other. Therefore, magnified parts of the graphs are 
shown in the insets of figures (a) and (c).  in the number of NPs with MCS. The inset figures clearly 
show different graphs for different 
independent runs indicated by different colours, for a small range of iterations. All the graphs show a 
jump on switching to the GCMC scheme at $10^5$ of MCSs, indicating an increase in the NP volume 
fraction in the system. The insets clearly show the energy fluctuations and gradual increase
in the number of NPs with MCS.

          All the independent runs for each $\rho_m$ produce similar configurations after 
(2 to 4)$\times 10^5$ iterations and the energy and NP volume fraction graphs converge to the same value. 
For each $\rho_m$, a very long run (~ $4\times 10^6$ iterations) was given to check if there is any change 
in the morphology of the system within thermal fluctuations. The system is found to maintain its morphology 
across all independent runs.  After $\approx 2\times 10^5$ iterations, the system morphology remains 
relatively unchanged for the length of the runs varying from $4\times 10^5$ to $4\times 10^6$ of iterations. 
Once the energy graphs show a very slow variation in its values and no further change in the polymer-NP 
morphology is observed, we show the representative snapshots for each of the monomer number densities $
\rho_m$ in Fig.\ref{crystal}. The monomers and the NPs are shown separately in the upper row and the 
lower row, respectively. All the snapshots have a gradient in colour varying from red (front plane) to 
blue (rear plane) along one of the box lengths, for a better visualisation of a three-dimensional figure. 
The figure shows the snapshots for (a) $\rho_m=0.037\sigma^{-3}, (b) 0.074\sigma^{-3}, (c) 0.093\sigma^{-3}$ 
and (d) $0.126\sigma^{-3}$. No clustering of micellar chains is observed, i.e. any two neighbouring chains 
are always found with NPs in between. In other words, the NP-monomer system forms a uniformly mixed state.

 An increase in the nematic ordering of polymeric chains and their chain length can be recognized from 
the figures as a result of the change in $\rho_m$ in Fig.\ref{crystal}. The snapshot in Fig.\ref{crystal}(a) 
shows smaller chains in a disordered state and the chain length and order increases for 
Figs.\ref{crystal}(b) and \ref{crystal}(c) and finally the system forms a nematic state with longer and 
aligned chains in Fig.\ref{crystal}(d). All the systems shown in the figure may show different polymer 
arrangements, but for all the micellar densities a uniformly mixed state of polymeric chains and NPs 
is observed.
       
All the independent runs discussed till now had the systems initialized in a state where the positions 
of both the monomers and 200-300 seed nanoparticles are randomly chosen. To ensure that our conclusions 
are not dependent on initial conditions, we gave additional runs initialized with all monomers on one 
side and all the NPs on the other side of the box. The representative snapshots from these runs after 
$4\times 10^6$ iterations, for $\rho_m=0.037\sigma^{-3}$ [Fig.\ref{diff_init1}(a) and (b)] and 
$\rho_m=0.126\sigma^{-3}$ [Fig.\ref{diff_init1}(c), (d) and (e)] are shown. Figure \ref{diff_init1}(a) 
[and (c)] shows the initial state with $\rho_m=0.037\sigma^{-3}$ with $N_m=2000$ [$\rho_m=0.126\sigma^{-3}$ 
with $N_m=6800$] and 5000 [5000] NPs in the box. Figures \ref{diff_init1} (b) and (d) show the snapshots 
after $4\times 10^6$ MCSs for systems initialized as shown in Figs. (a) and (c), respectively. The 
Fig.\ref{diff_init1}(e) shows the snapshot initialized with the same configuration shown in (c) but having 
the value of $\mu_n=4k_BT$. For both the densities of monomers considered for $\mu=-8k_BT$, the final state 
is different from the initial state forming a uniformly mixed state of monomer chains and NPs (left side 
of the box) coexisting with NPs without monomers (right side of the box) (figures (b) and (d)). The runs 
were tested with different initial numbers of NPs, but all the runs resulted in a mixed state of polymeric 
chains and NPs (as shown in the left part of the boxes in Figs.\ref{diff_init1}(b) and \ref{diff_init1}(d)). 
These mixed states are found to be coexisting with NPs shown in the right sides of the boxes. We expect 
that if the system reaches  equilibration, the systems will evolve to a completely mixed state as in 
Figs.\ref{crystal}(e), (f), (g) and (h) for $\sigma_{4n}=1.25\sigma$ and $\mu_n=-8k_BT$. But
in our studies, the density of monomers and NPs are locally very high, which leads to very packed structures
which is unable to relax to equilibrium. 
However, keeping the initial state similar to Fig.\ref{diff_init1}(c) and increasing the value 
of $\mu_n$ from $\mu_n=-8k_BT$ to $\mu_n=4k_BT$ for $\rho_m=0.126\sigma^{-3}$ resulted in a phase separated 
state as shown in Fig.\ref{diff_init1}(e).

\begin{figure}
\centering
\includegraphics[scale=0.26]{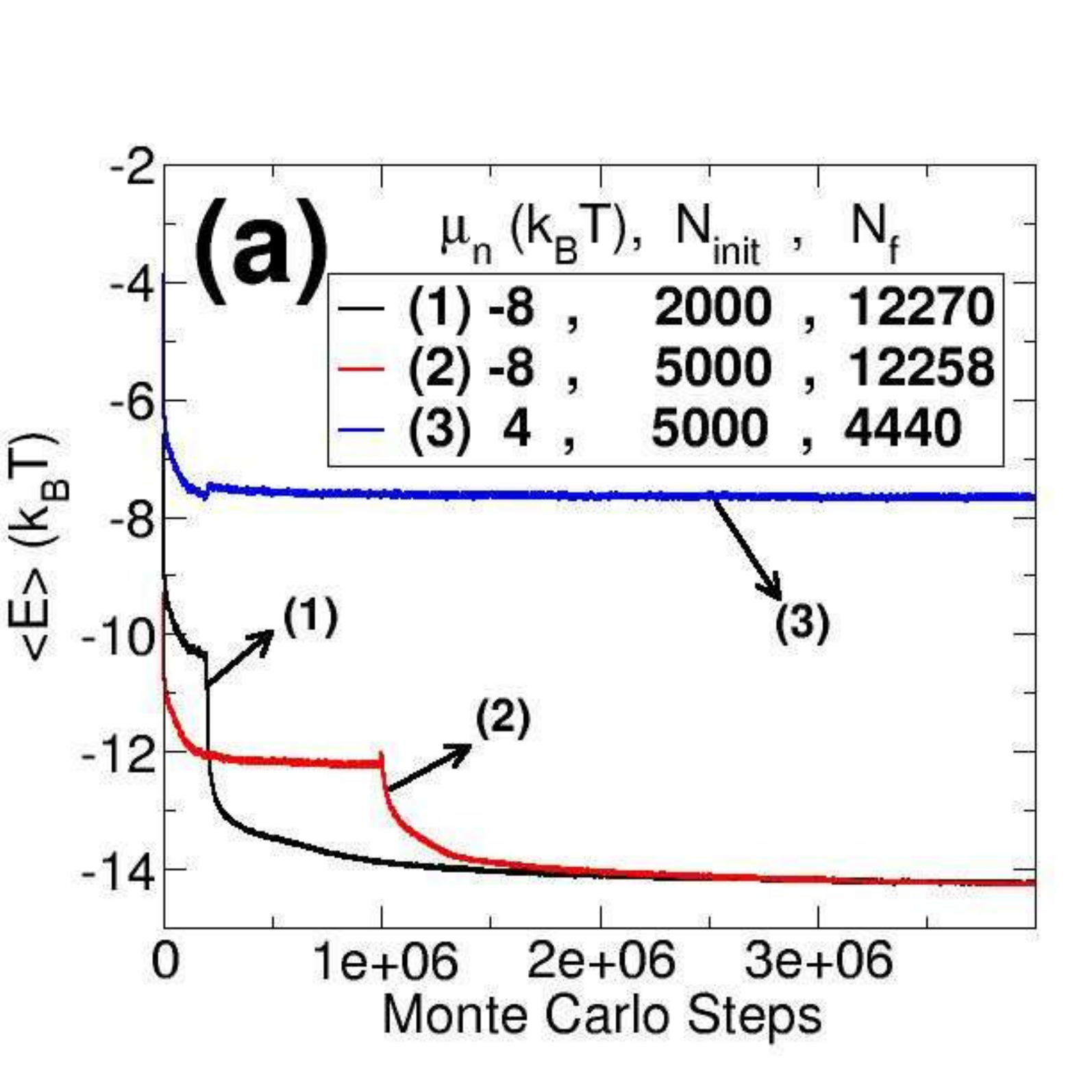}
\includegraphics[scale=0.2]{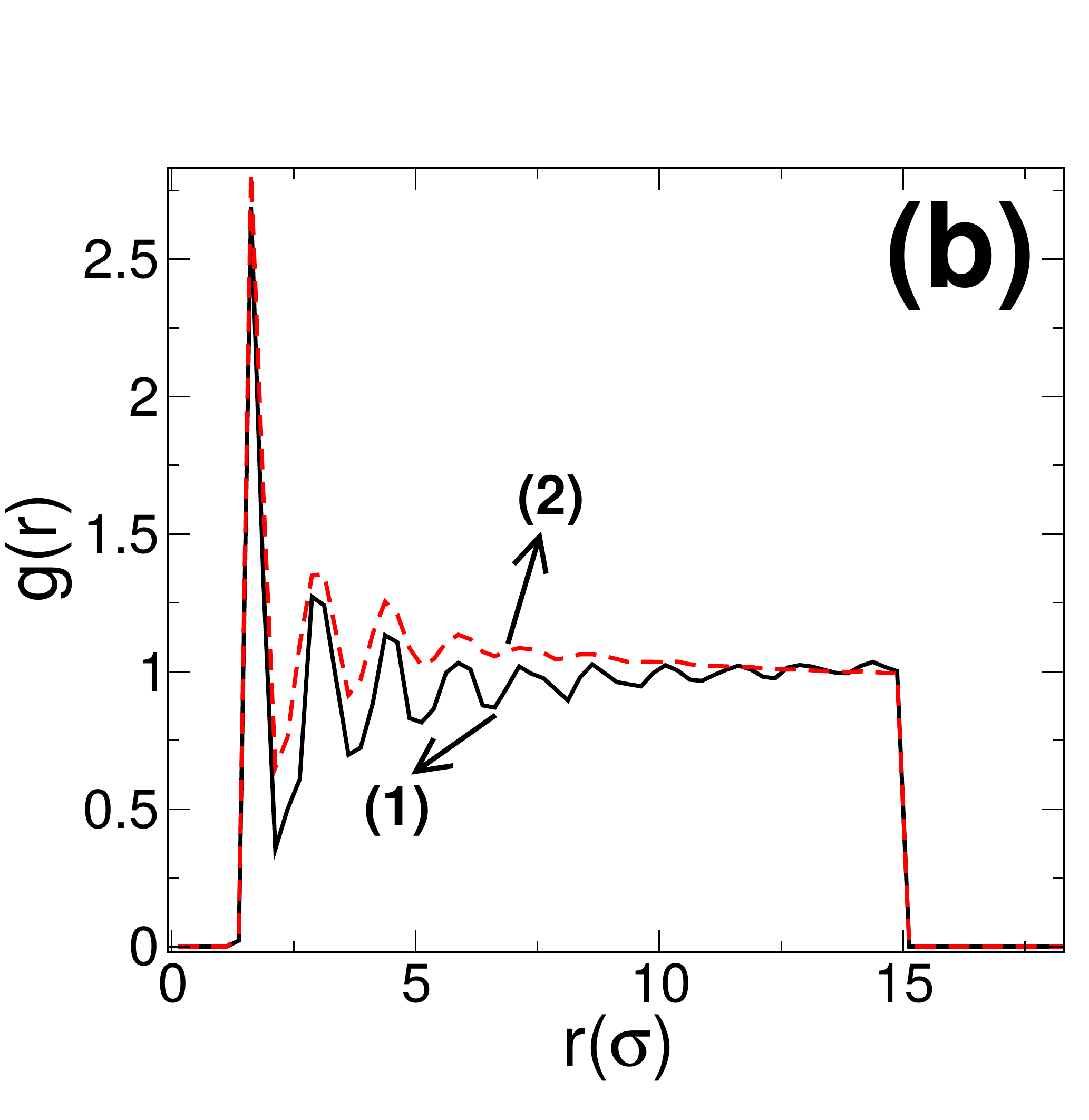}
\caption{(colour online) (a) The figure compares the average potential energy $\langle E \rangle$ of the constituent 
particles (monomers and NPs) versus MCSs of the systems that are initialized as shown in 
Fig.\ref{diff_init1}(c). Lines (1) and (2) correspond to data  with  $\mu_n=-8k_BT$ and 
line (3) corresponds to data with $\mu_n=4k_BT$. The graphs (2) and (3) 
corresponds to the systems represented in Figs.\ref{diff_init1}(d) and (e), respectively, at the end 
of the run. The graphs (1) and (2) differs in their number of nanoparticles in the initial configuration 
$N_{init}$ as well as in the 
value of iterations/MCS  after which the GCMC scheme is switched on. The two graphs (1) and (2) illustrates that the 
results are independent of the value of $N_{init}$ and the number of iterations after which the GCMC scheme is 
switched on, as for both cases $\langle E \rangle$ converges to the same value within statistical fluctuations. On 
switching on to GCMC scheme, graphs (1) and (2) jumps to a lower value of $\langle E \rangle$. This is indicative of the 
rapid addition of nanoparticles in the system. The line (3) jumps to a slightly higher value 
of energy indicating the removal of nanoparticles. The legends $N_f$ indicates their final value of 
the number of nanoparticles after $4 \times 10^6$ MCS. 
(b) The figure shows the pair correlation function $g(r)$ for the NPs in left two-thirds
of the box represented by line-2 and the NPs in the right one-third of the box (line-1) corresponding
to the configuration shown in the Fig.5(d). The pair correlation function shows the presence of a long-range 
correlations in the position of NPs corresponding to a crystalline ordered state in the right section of the box.}
\label{diff_init2}
\end{figure}

\begin{figure*}
\centering
\includegraphics[scale=0.2]{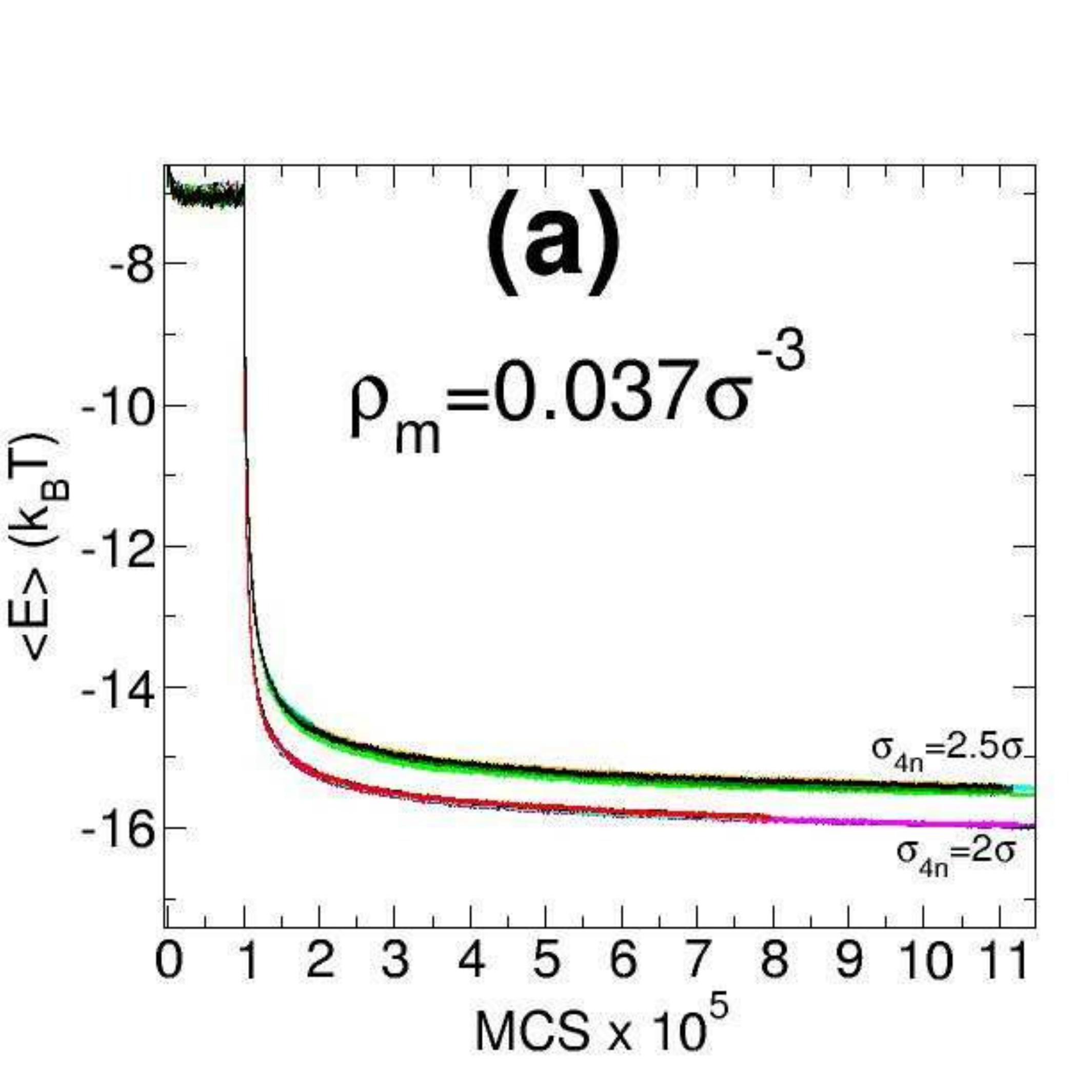}
\includegraphics[scale=0.2]{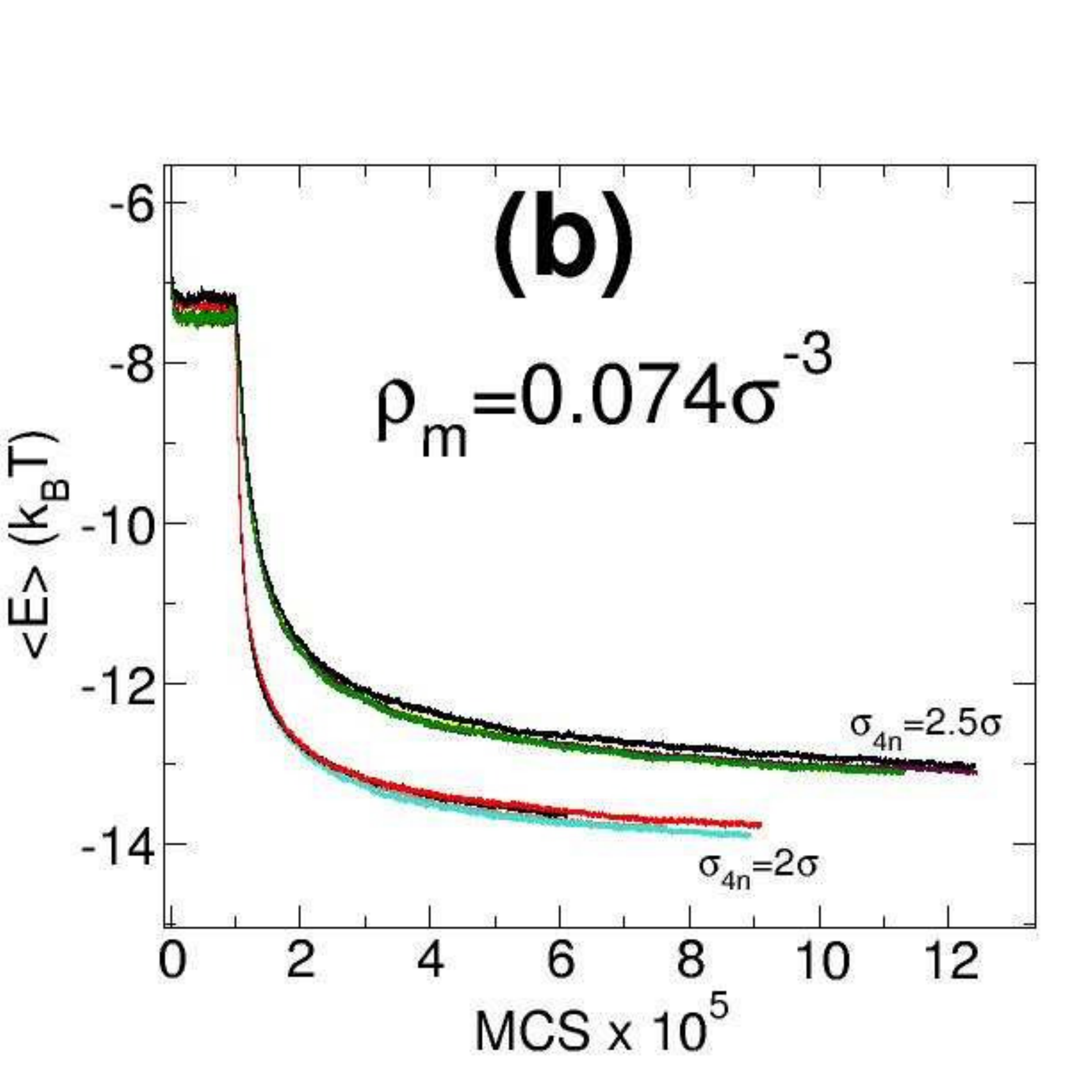}
\includegraphics[scale=0.2]{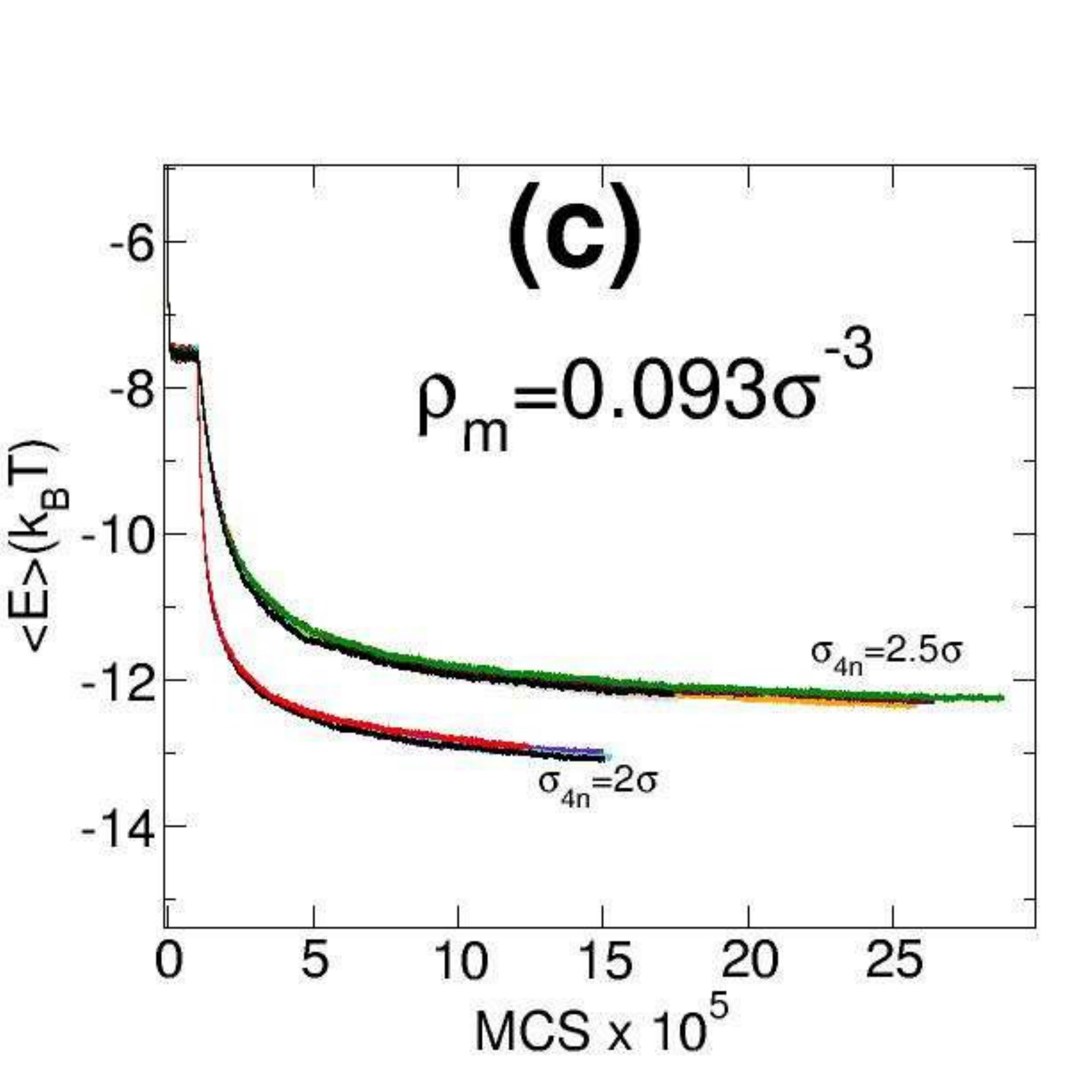}
\includegraphics[scale=0.2]{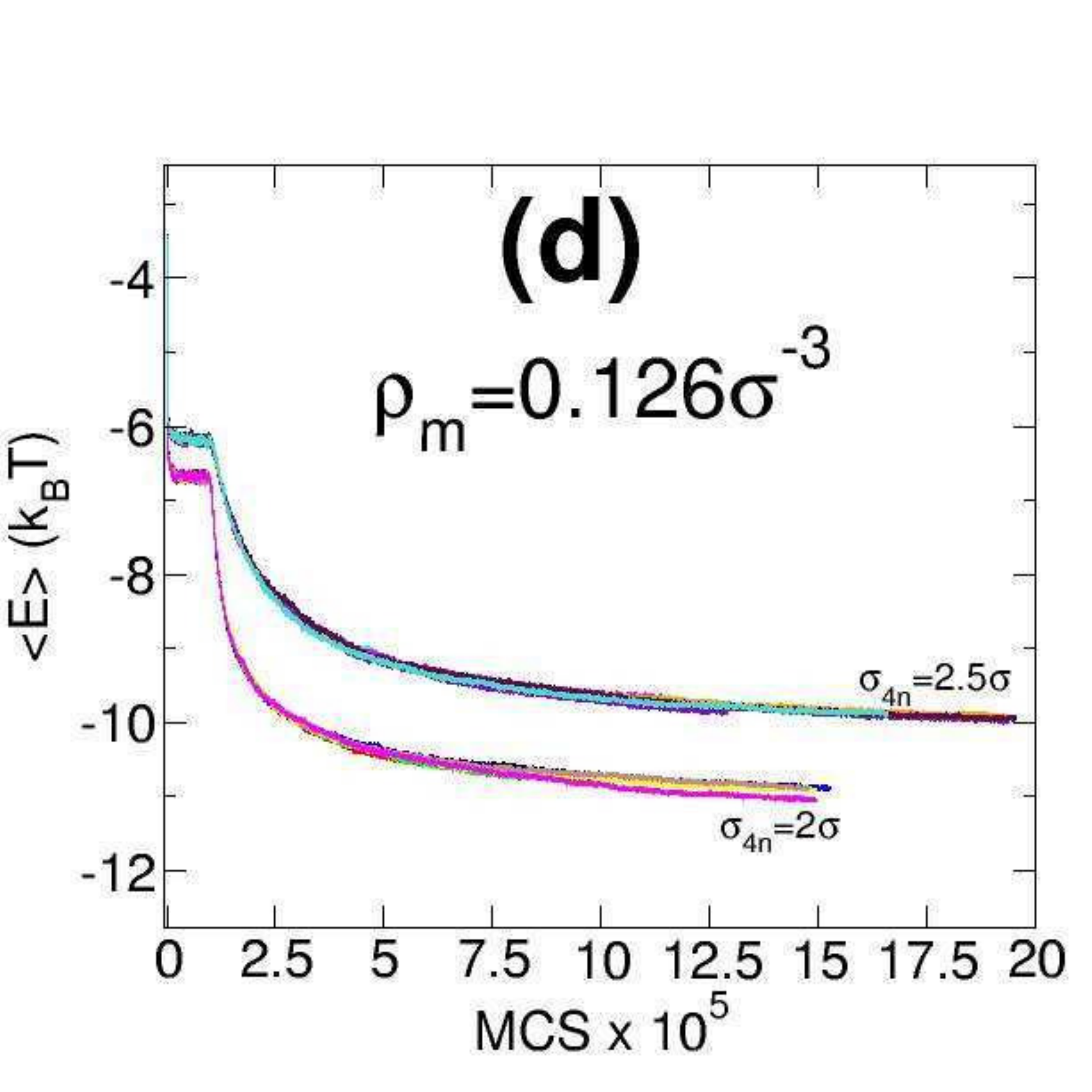}
\caption{(colour online) Plots of average potential energy $\langle E \rangle$ per particle (monomer and NP)
 of the system for different values of monomer 
number densities (a) $ \rho_m=0.037\sigma^{-3}, (b) 0.074\sigma^{-3} , (c) 0.093\sigma^{-3} $ and 
(d) $ 0.126\sigma^{-3} $. Each figure shows the energy graphs for two different values of 
$ \sigma_{4n}=2\sigma $ and $ 2.5\sigma$. For each value of $\sigma_{4n}$, there are ten graphs overlapping 
each other and converging to the same values. The Monte Carlo scheme (without GCMC) is applied for 
first $10^5$ iterations. Then the GCMC scheme is switched on which leads to a further decrease in the energy 
values indicated by jumps in the values of $\langle E \rangle$ in each graph.}
\label{energies}
\end{figure*}

        Thus, it is observed that for $\mu_n=-8k_BT$, the system initialized as shown in 
Fig.\ref{diff_init1}(c) always leads to form a mixed state while, for $\mu_n=4k_BT$, the system remains 
in a phase separated state. This difference in the behaviour of the systems with different values of 
$\mu_n$ can be confirmed by the plots shown in Fig.\ref{diff_init2}(a). The figure shows three graphs 
(1), (2) and (3). For all the graphs the state of initialization is as shown in Fig.\ref{diff_init1}(c) 
but with a different initial number of nanoparticles $N_{init}$ as indicated in the figure. Graphs 
(1) and (2) shows the evolution of energy with MCSs for $\mu_n=-8k_BT$ initialized with 2000 and 5000 NPs, 
respectively. Both the graphs show convergence to the same value of energy after $4\times 10^6$ iterations 
with a jump to lower energy values on switching on the GCMC scheme at $10^5$ and $10^6$ iterations, 
respectively. This jump to lower energy indicates the addition of NPs into the box on switching \textit{on} 
the GCMC scheme. Irrespective of (i) the number of MCSs at which the GCMC scheme is switched \textit{on} 
and (ii) the initial number of NPs $N_{init}$, all the systems are observed to evolve to form a mixed state 
and converges to the same value of energy for $\mu_n=-8k_BT$ (two of which constitutes the graphs (1) and 
(2)). Graph (3) shows the evolution of energy for the system initialized with the same state as in 
Fig.\ref{diff_init1}(c) with 5000 NPs in the box, but having a higher value of $\mu_n=4k_BT$. 
On switching \textit{on} the GCMC scheme, the graph shows a jump at $10^5$ MCSs to a higher value of 
energy indicating removal of NPs from the box. The system evolves to form an equilibrated phase 
separated state as shown in \ref{diff_init1}(e) after removal of $\approx 550$ NPs from the box. 
Since all the systems initialized with the configuration shown in \ref{diff_init1}(c) and 
$\mu_n=-8k_BT$ evolves to form a mixed state irrespective of the initial number of NPs in the system, 
it can be concluded that for $\mu_n=-8k_BT$, a phase separated state is not a thermodynamically 
preferred state. Therefore, we can now claim from  the snapshots shown in Fig.\ref{crystal} that the mixed 
state will remain the thermodynamically preferred state as well, for $\sigma_{4n}=1.25\sigma$. 
We remind the reader that we do not claim that the microstates corresponding to the the snapshots of 
Fig.\ref{crystal} are in equilibrium that the values of both the average energy as well $V_n/V$ are evolving.

      For the value of $\mu_n=-8k_BT$, one more difference can be observed for different values of 
$\rho_m$ which can be seen in Figs.\ref{diff_init1}(b) and \ref{diff_init1}(d). Both the figures are 
initialized with similar configurations (unmixed state) and number of NPs $N_{init}=5000$, but with 
different values of monomer number density $\rho_m=0.037\sigma^{-3}$ [Fig.\ref{diff_init1}(a)] and 
$\rho_m=0.126\sigma^{-3}$ [Fig.\ref{diff_init1}(c)]. In case of higher density of monomers, the phase 
of NPs which is devoid of monomers (in the right part of the box in Fig.\ref{diff_init1}(d)) seems to 
have a long-range crystalline order compared to the lower density case [refer the right part of the box 
in Fig.\ref{diff_init1}(b)].  The figure \ref{diff_init2}(b) 
compares the pair correlation function for the NPs in left section box  in Fig.\ref{diff_init1}(d)
as well as the right section where the NPs dont have monomers interspersed between them. 
The correlation graphs for NPs in the mixed state (in the left part of the box in Fig.\ref{diff_init1}(d)) 
depicted in graph (2) shows very few and low peaks which die out quickly compared to the graph for NPs 
(in the right part of the box of Fig.\ref{diff_init1}(d)) as depicted by graph (1). The graph (1) has very 
sharp peaks and shows a longer range of correlation hence, confirming the observed ordered structure of NPs. 
We expect that these two coexisting states as shown in Fig.\ref{diff_init1}(b) and (d) are stuck in a 
metastable state and they will form a fully mixed state after equilibration.

 All the results for ten independent runs and the runs shown in Fig.\ref{diff_init1} indicate that the
 mixed state of NP and micellar chains could be the thermodynamically preferred state of the system for 
the value of $\mu_n=-8k_BT$. The systems considered here are quite dense and show a very slow increase 
in the number of nanoparticles even after a very long run (Fig.\ref{energies}(c,d)).  Therefore, we assume 
that the structures discussed in the previous sections (see Fig.\ref{crystal}) are most probably kinetically 
arrested states that are relaxing slowly to equilibrium. The rest of the paper considers the systems 
initialized with a randomly mixed state of micelles and NPs and investigates the effect of the 
change in $\sigma_{4n}$ for each value of $\rho_m$.

\subsection{$\sigma_{4n}>1.25\sigma$: Polymer and NP clusters with different morphologies }

 We next investigate the effect of the change in the value of $\sigma_{4n}$ for different monomer densities 
by varying $\rho_m$ and $\sigma_{4n}$. The results from these runs were further substantiated with ten 
independent runs for each value of $\rho_m$ and $\sigma_{4n}$. The average energy of the system for 
ten independent runs for two different values of $\sigma_{4n}=2.5\sigma$ and $2\sigma$ is shown in 
Fig.\ref{energies} for values of densities (a) $0.037\sigma^{-3}, (b) 0.074\sigma^{-3}, (c) 0.0931\sigma^{-3}$ 
and $(d) 0.126\sigma^{-3}$. The system is evolved with Monte Carlo steps without GCMC for the first 
$10^5$ iterations and then the GCMC scheme is switched on which is marked by a rapid lowering in the 
energy values in Fig.\ref{energies}. This lowering of energy values corresponds to the rapid addition 
in the number of nanoparticles in the system but, the rate of addition slows down after 
$(2-5) \times 10^5$ iterations. In each figure, each visible line shows multiple lines overlapping 
each other, generated by ten independent runs. All the ten independent runs converge to the same value 
of energy. Here, we would like to emphasise again that all the systems are initialized 
with random positions of monomers and 200-300 seed nanoparticles and with the grand-canoninal Monte Carlo
steps NPs get introduced in the midst of the micellar chains. 
This method of initialization is a reminiscent of the in situ method of preparation of 
nanoparticles inside the matrix of polymers, where the NPs start nucleating out from a chemical solution 
on suitable addition of a reactant.

\begin{figure*}
\centering
\includegraphics[scale=0.2]{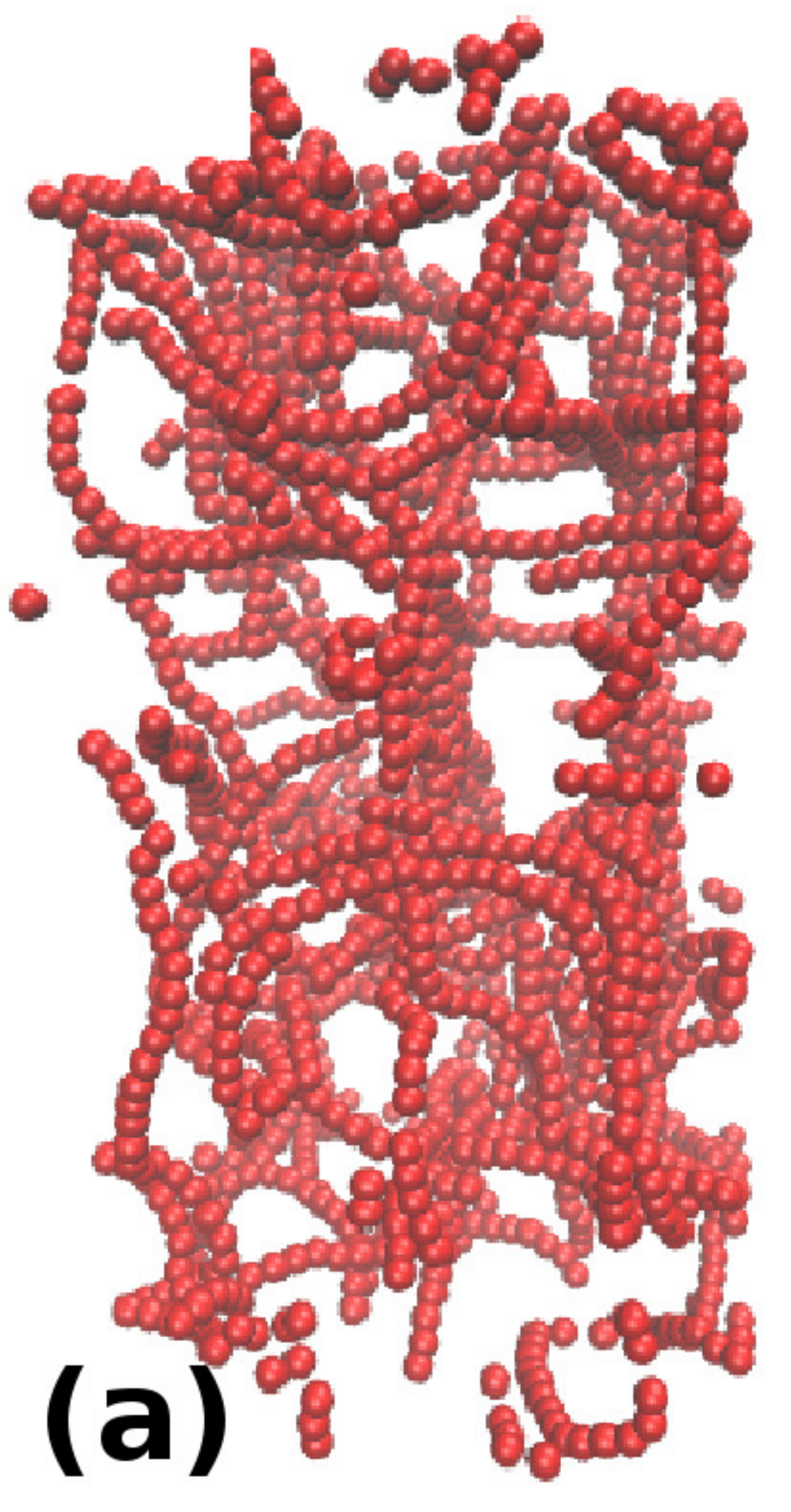}
\hspace{2cm}
\includegraphics[scale=0.2]{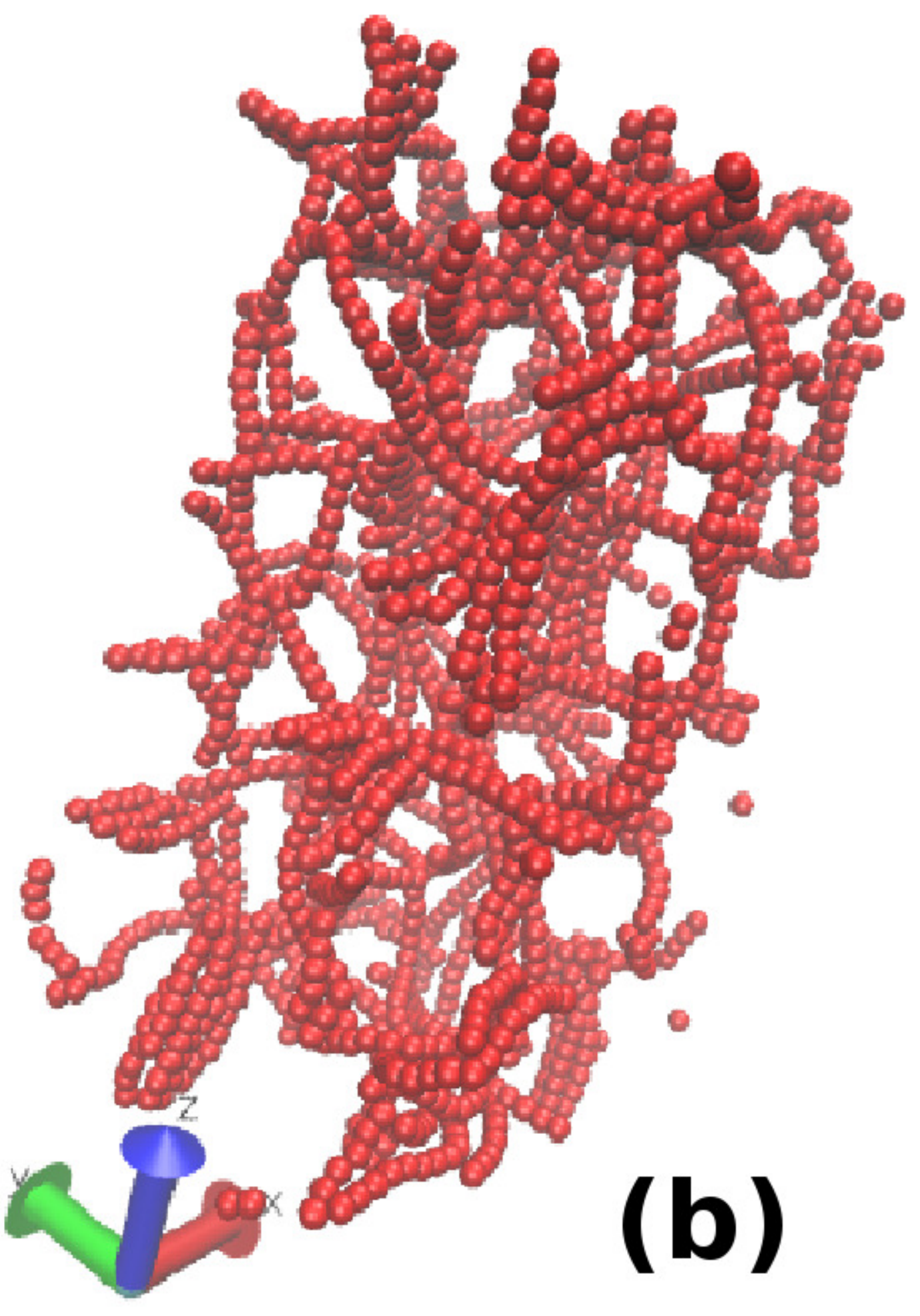}
\hspace{2cm}
\includegraphics[scale=0.2]{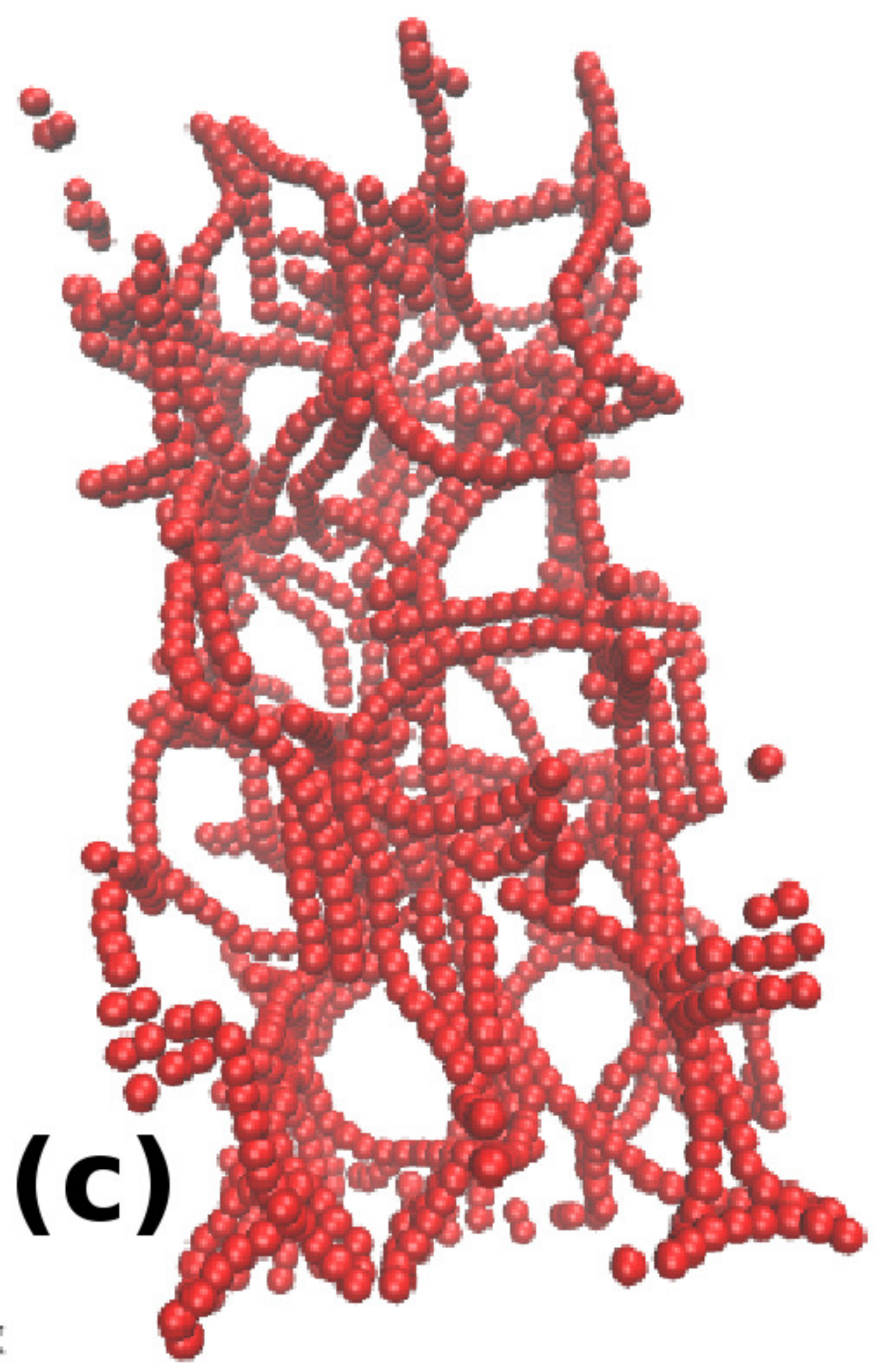}
\hspace{2cm}
\includegraphics[scale=0.2]{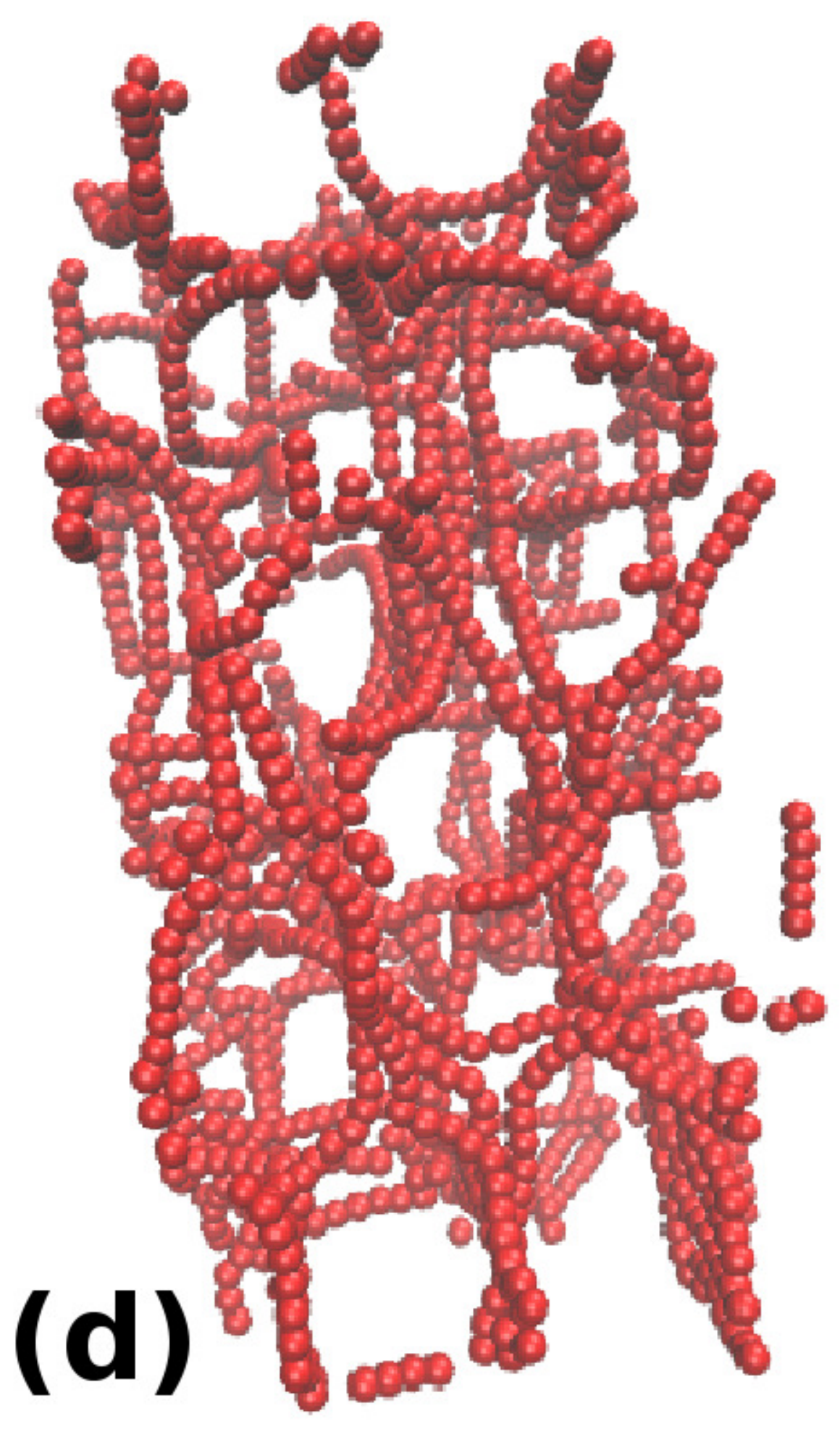}\\
\includegraphics[scale=0.2]{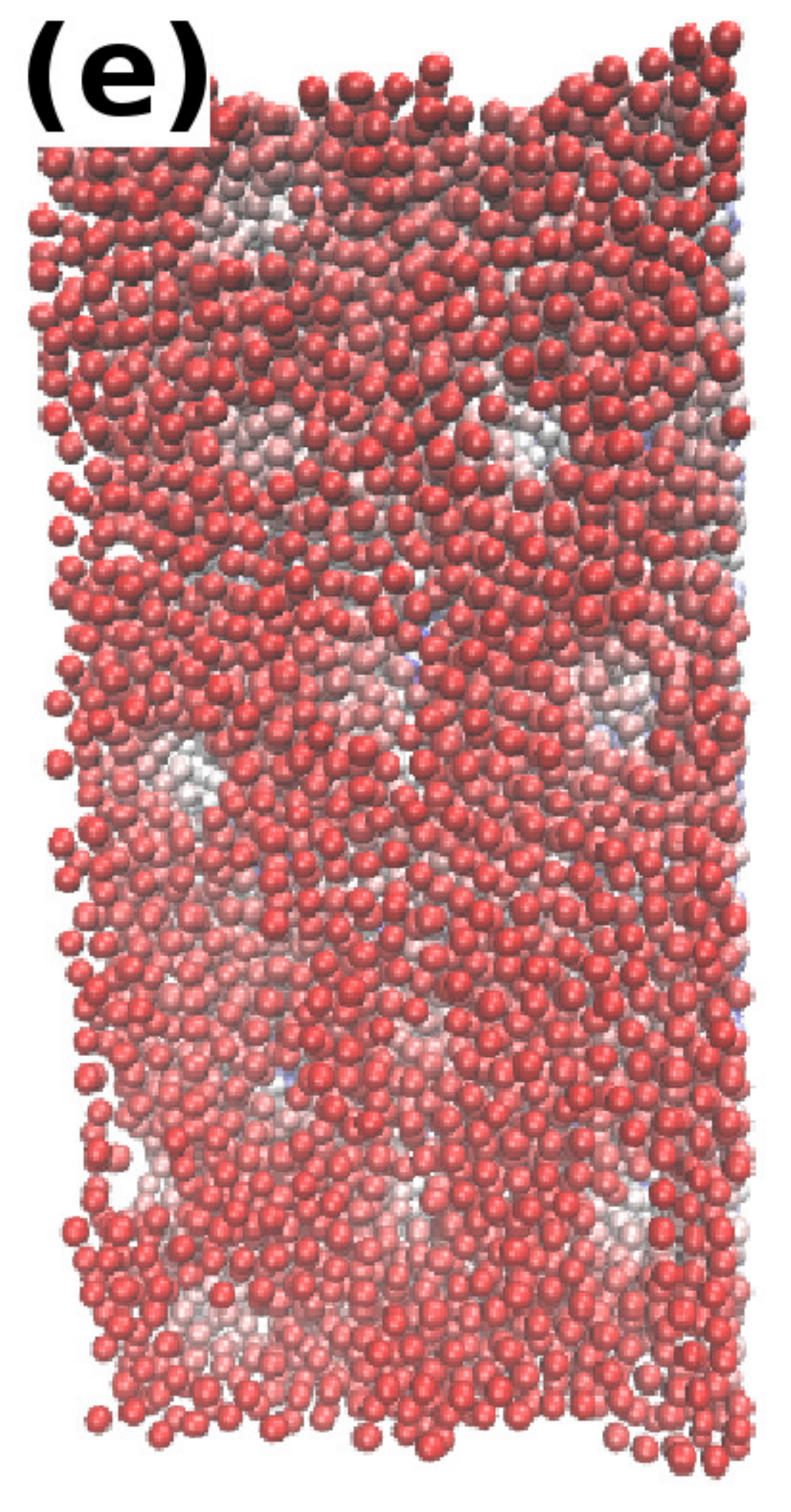}
\hspace{2cm}
\includegraphics[scale=0.2]{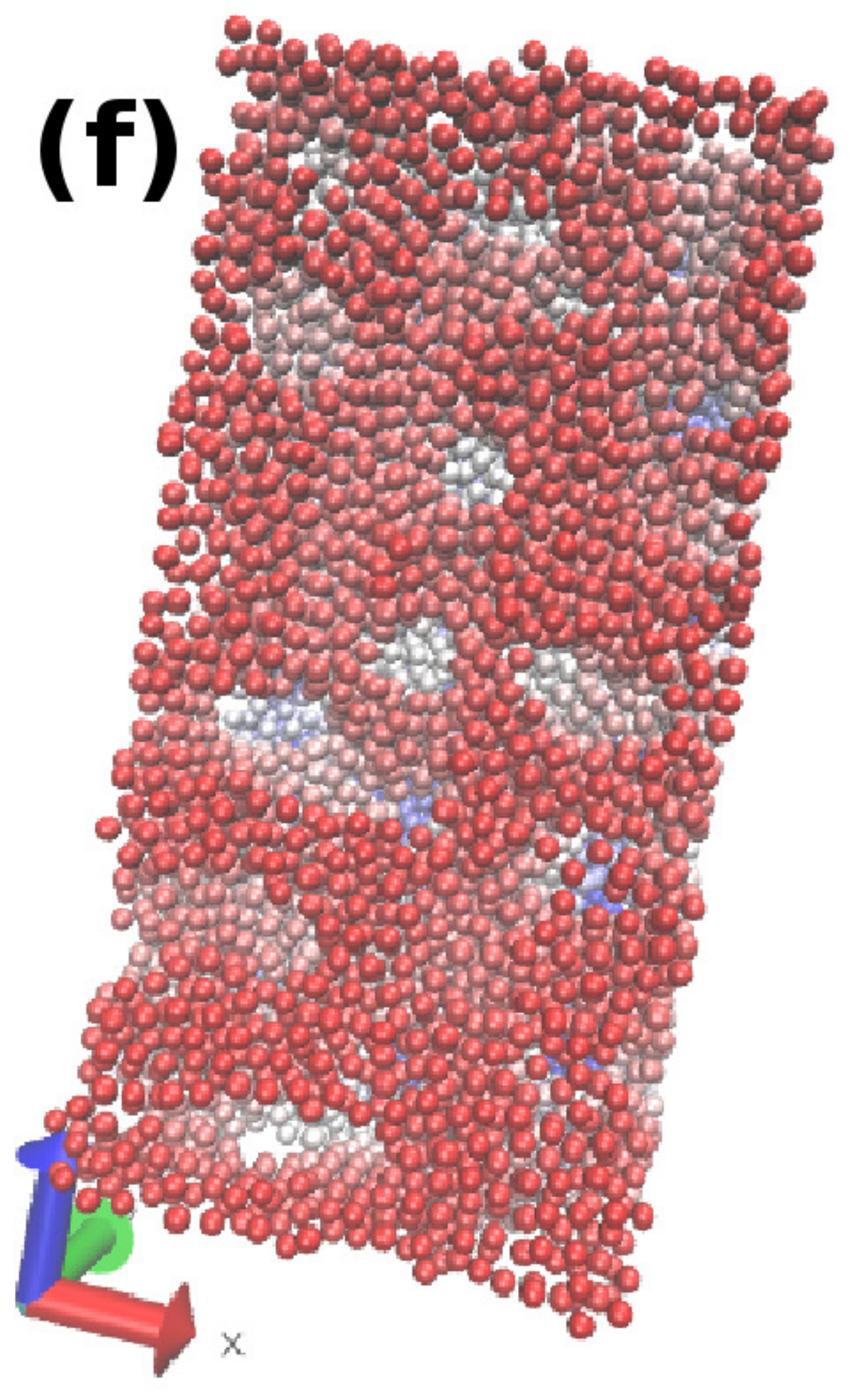}
\hspace{2cm}
\includegraphics[scale=0.2]{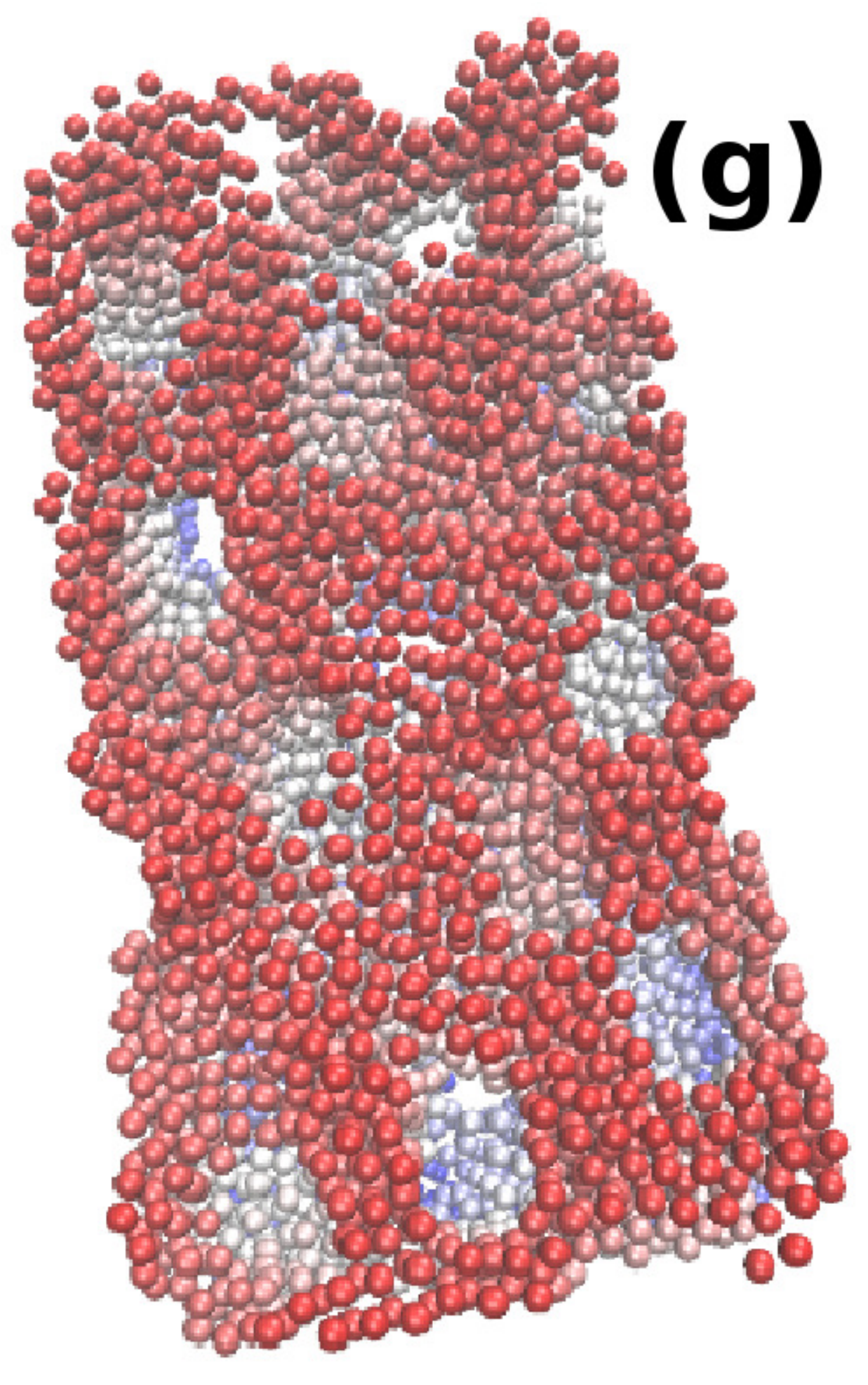}
\hspace{2cm}
\includegraphics[scale=0.2]{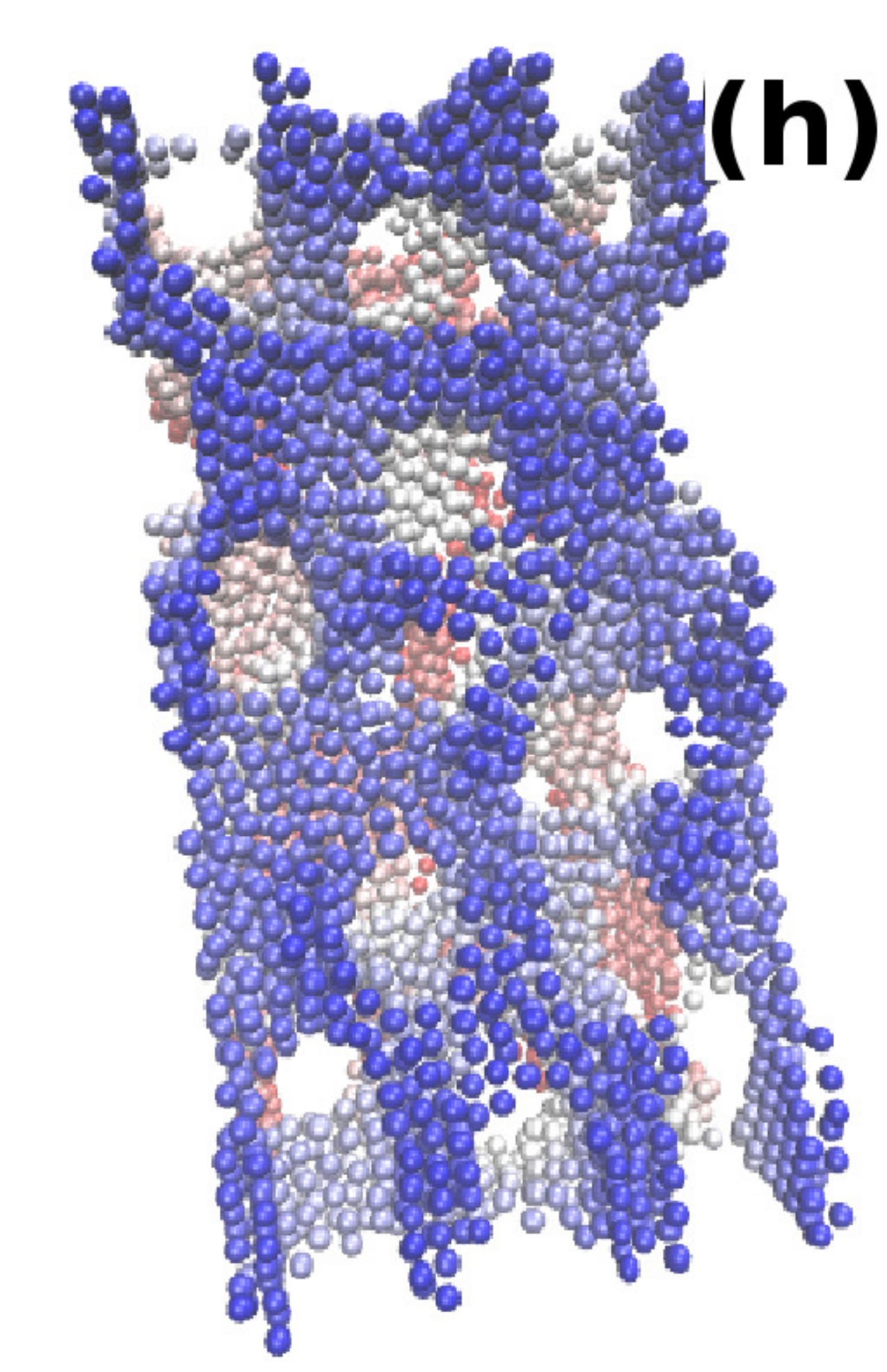}\\
\caption{(colour online)  The figure shows the snapshots for NPs and monomers from the NP + micellar-polymer 
system for the lowest value of  monomer number density$\rho_m=0.037\sigma^{-3}$. The snapshots in 
the upper row show only the micellar monomers while only the nanoparticles are shown in the lower row. 
The snapshots from (a) to (d) [and correspondingly  (e) to (h)] are for 
$\sigma_{4n} = 2.0\sigma, 2.5\sigma, 3\sigma$ and $3.5\sigma$, respectively. There is a gradient in 
colour along one of the shorter axis of the box for the snapshots (e) to (h), to help reader differentiate 
the particles present near the front plane from those at the rear. 
All the snapshots indicate the formation of the network-like 
structure of aggregates of nanoparticles and micellar chains inter-penetrating each other.}
\label{low_dens}
\end{figure*}

\begin{figure*}
\centering
\includegraphics[scale=0.2]{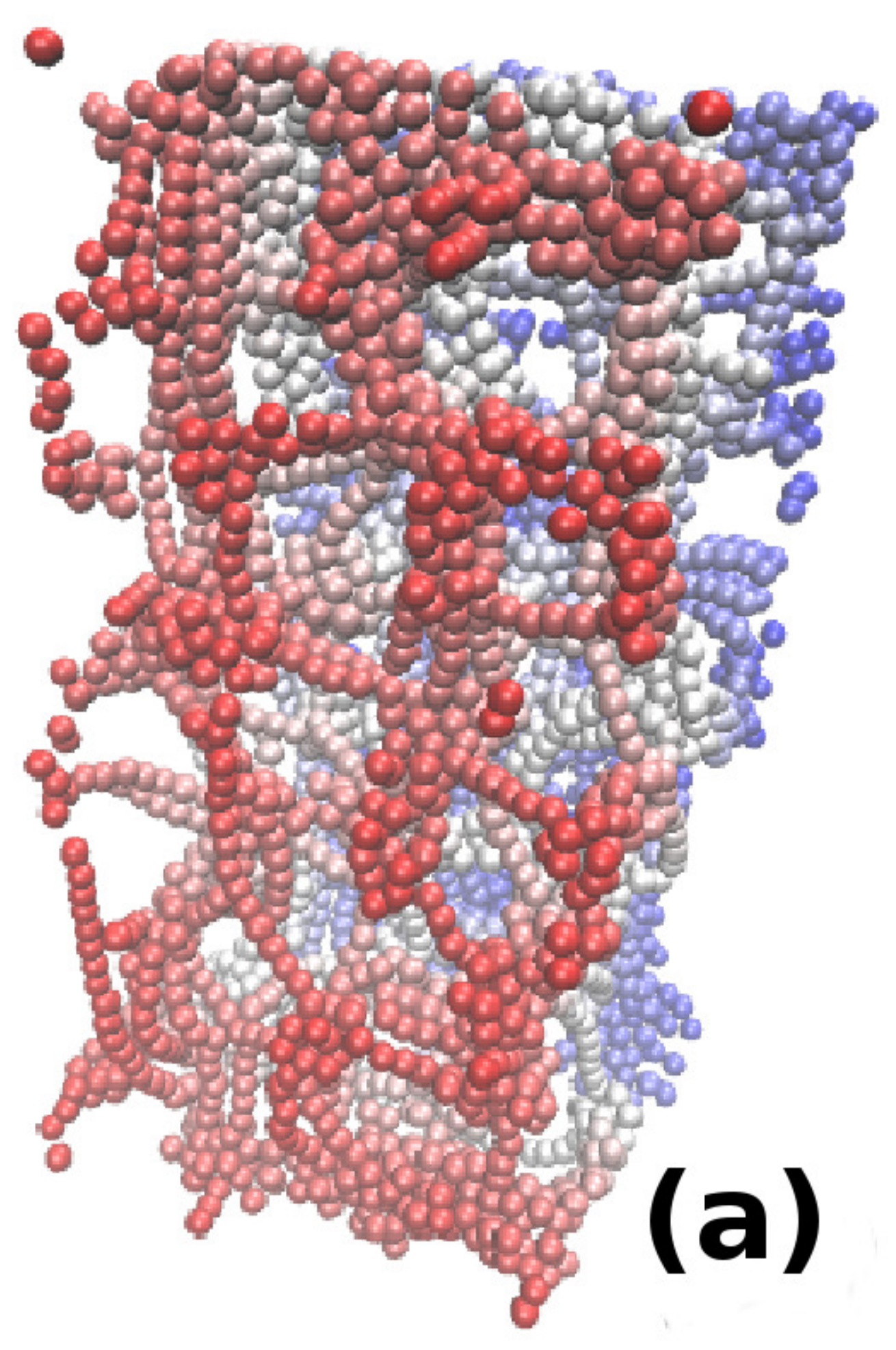}
\hspace{2cm}
\includegraphics[scale=0.2]{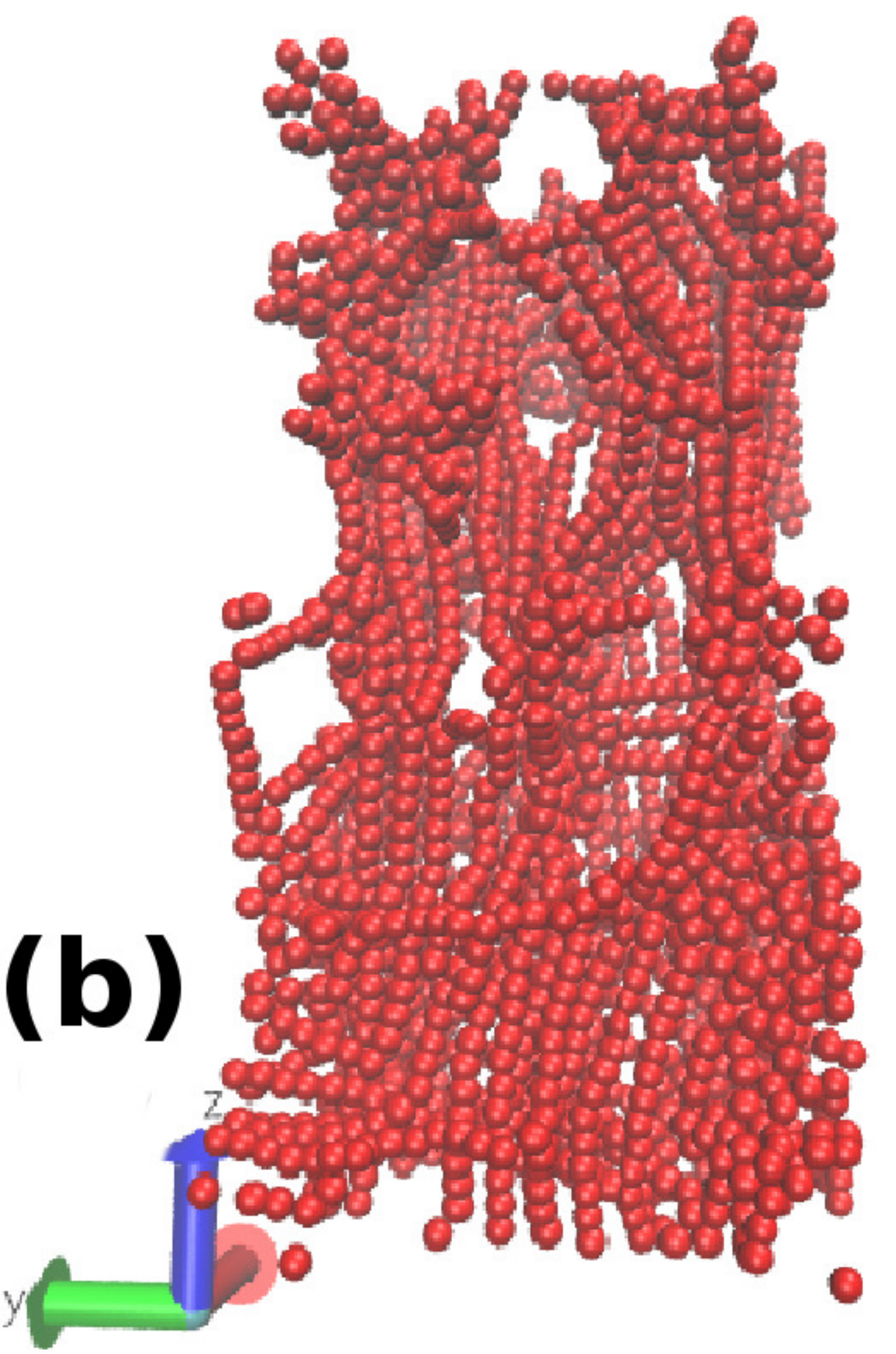}
\hspace{2cm}
\includegraphics[scale=0.2]{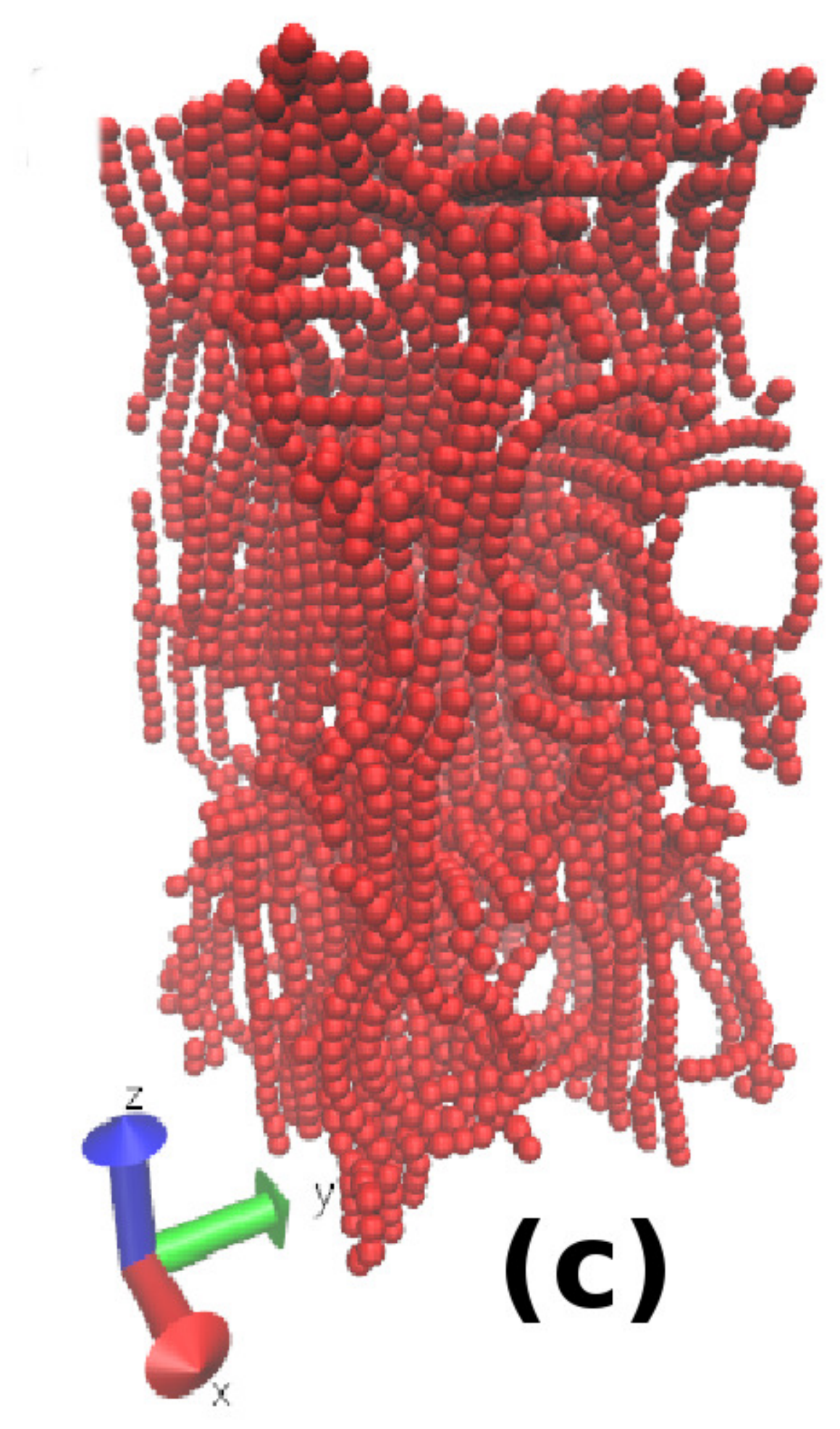}
\hspace{2cm}
\includegraphics[scale=0.2]{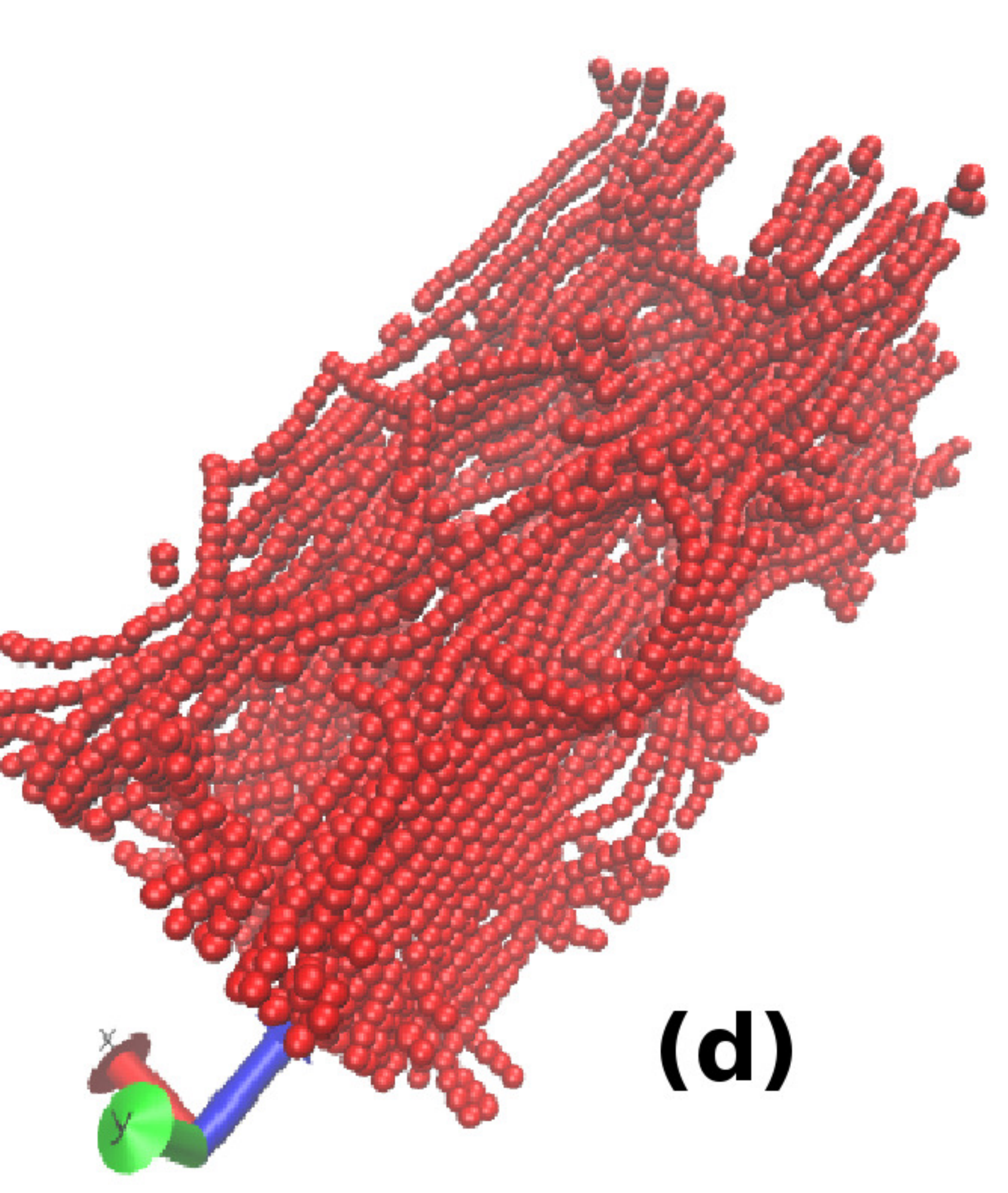}\\
\includegraphics[scale=0.2]{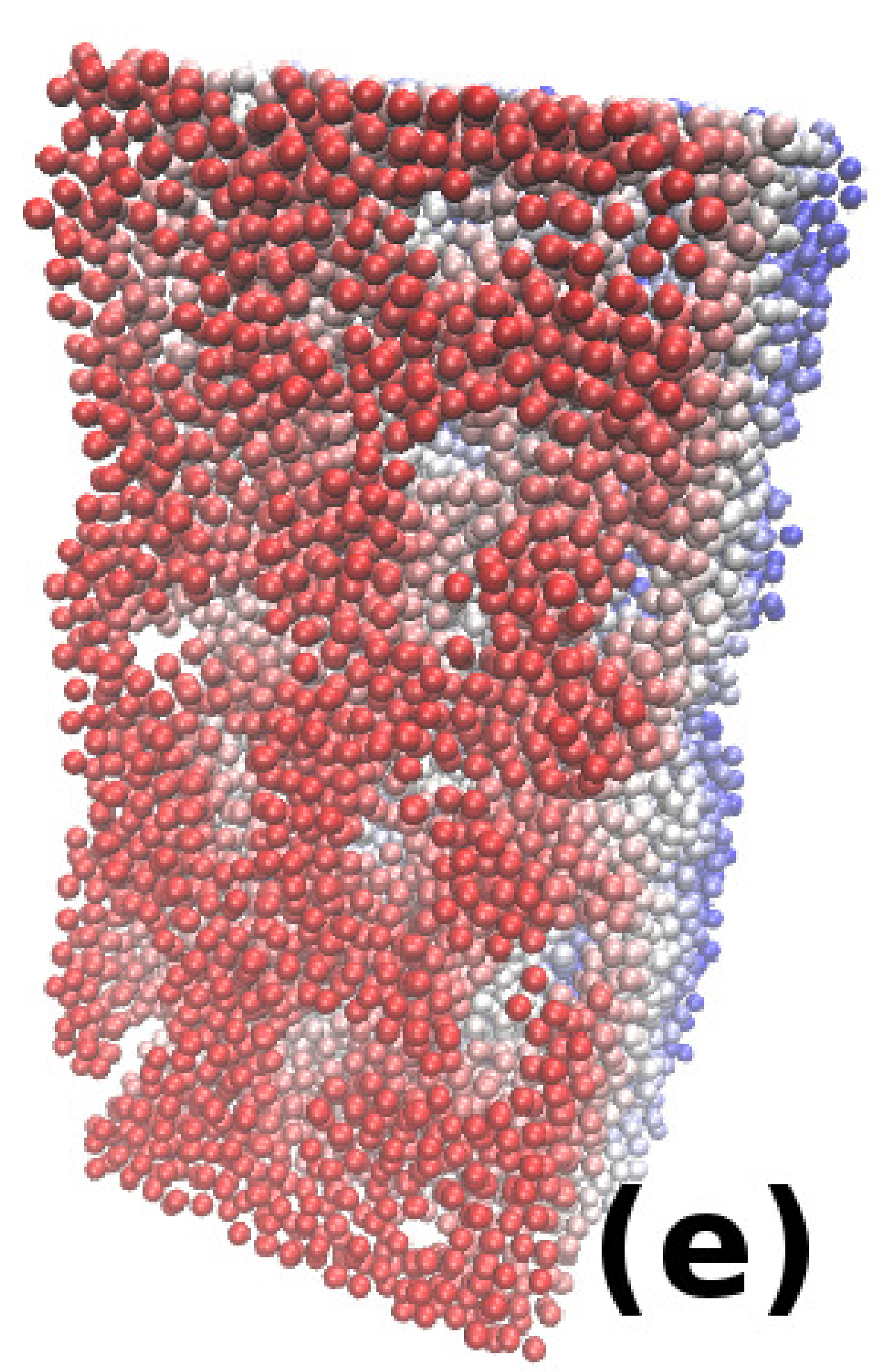}
\hspace{2cm}
\includegraphics[scale=0.2]{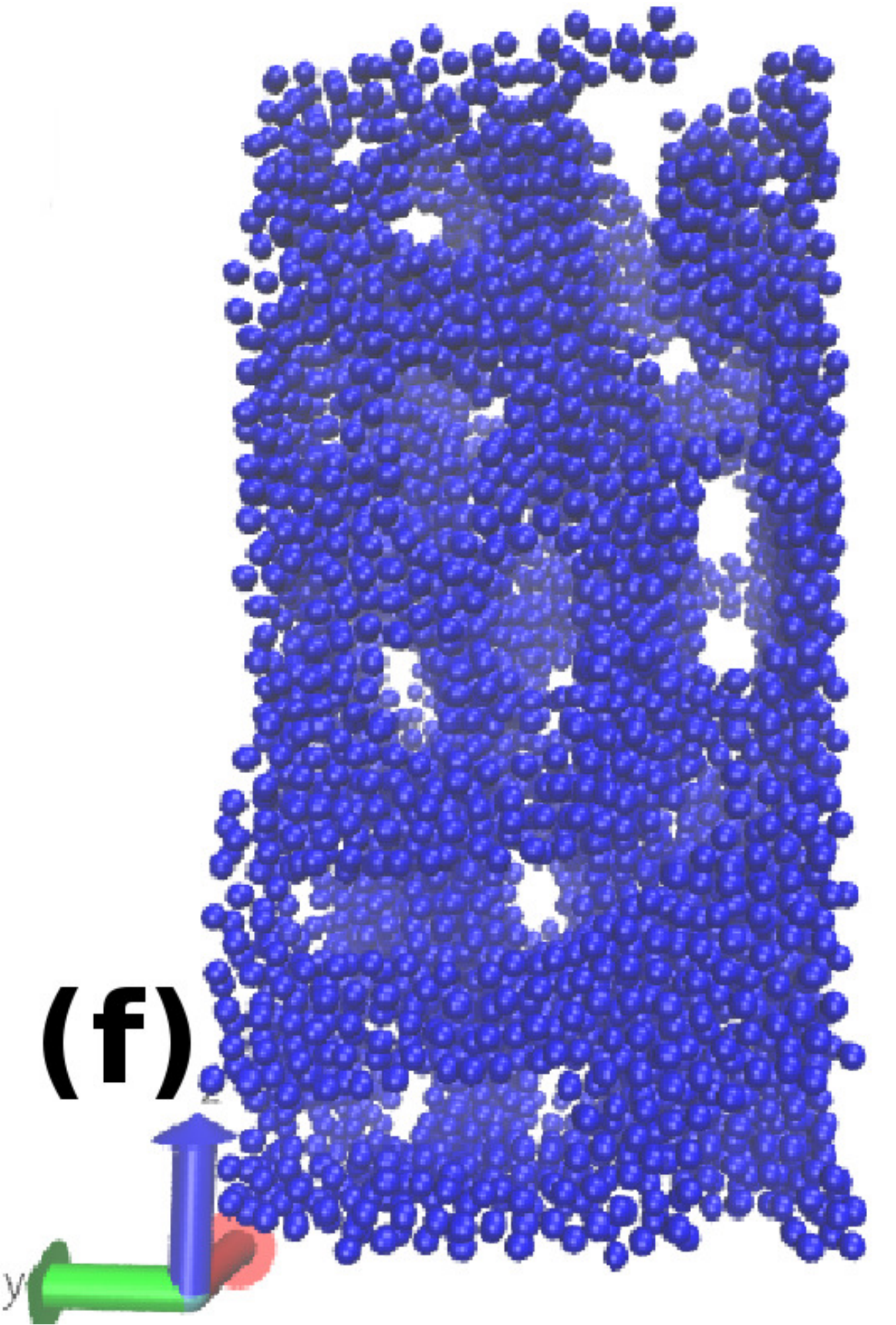}
\hspace{2cm}
\includegraphics[scale=0.2]{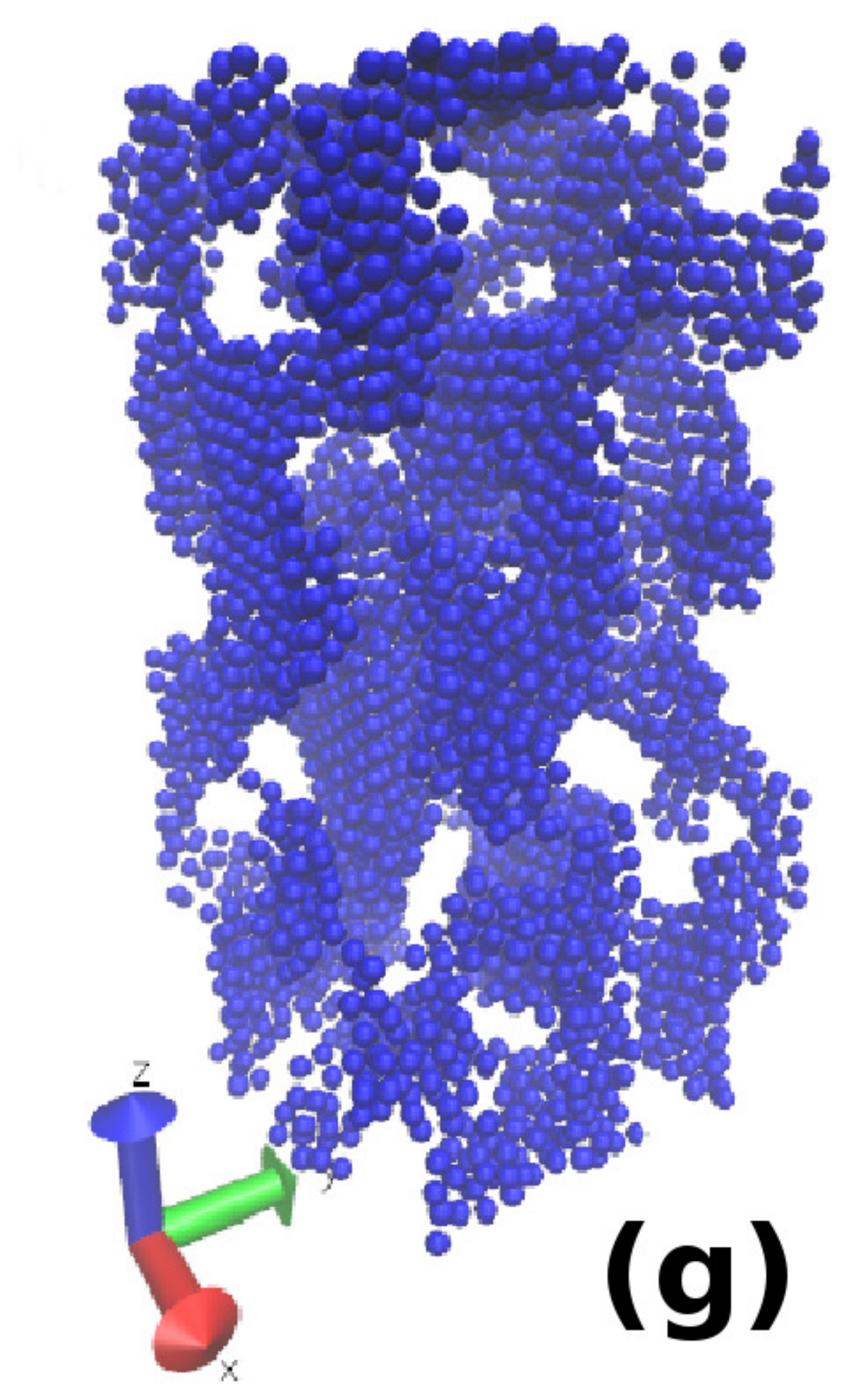}
\hspace{2cm}
\includegraphics[scale=0.2]{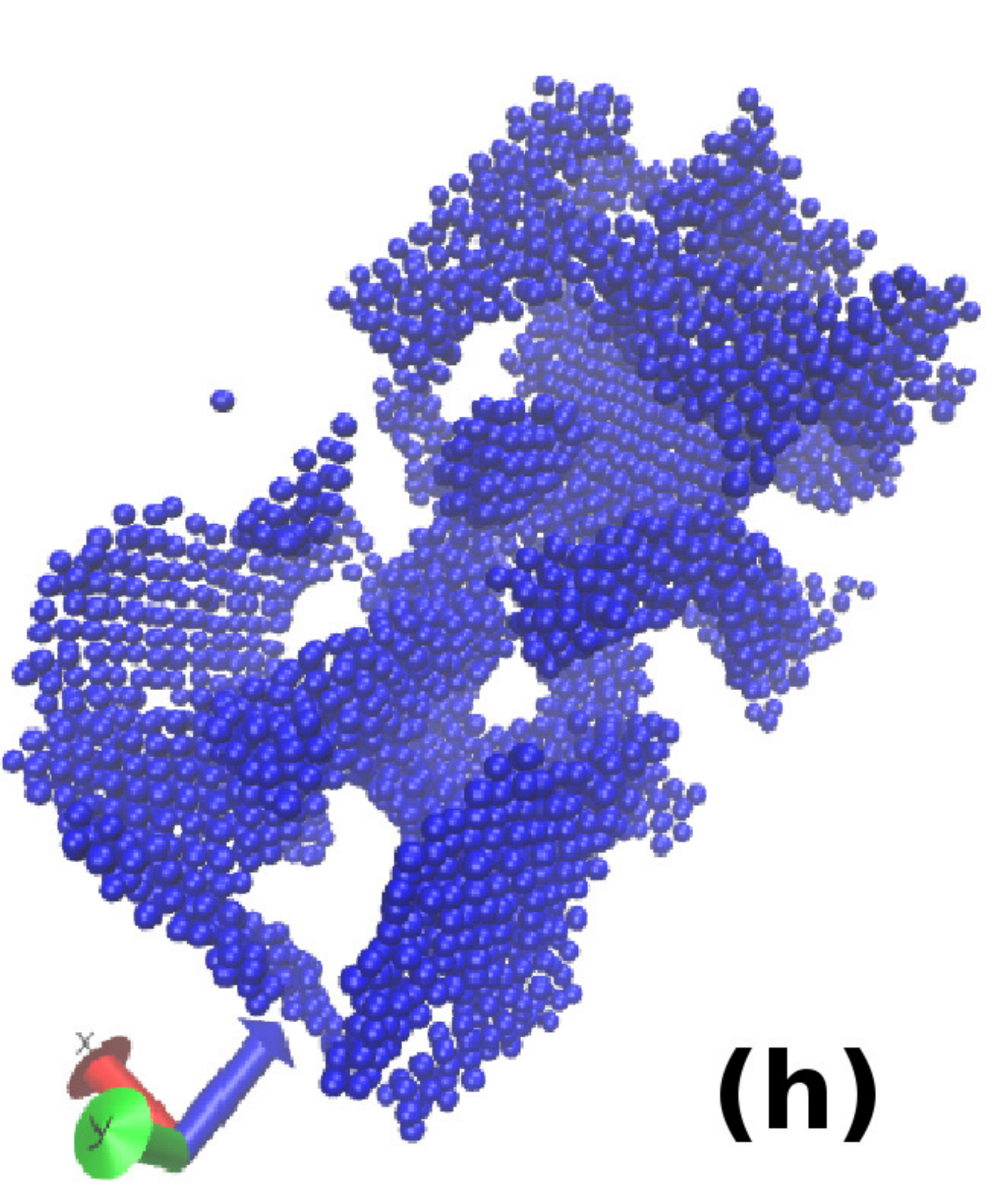}\\
\caption{(Colour online)  The figure shows the snapshots of NPs and monomers from the NP + micellar-polymer system 
for the value of monomer number density $ \rho_m= 0.074\sigma^{-3} $. The snapshots in the upper row 
show only the micellar monomers while, the lower row only shows the nanoparticles. Snapshots from 
(a) to (d) [or correspondinly (e) to (h)] are for  $\sigma_{4n}=1.75\sigma, 2.25\sigma, 2.75\sigma$ and 
$3.25\sigma$, respectively. The snapshots (a) and (e) have a gradient in colour varying from red to blue 
along one of the shorter axis of the simulation box to help the reader diferentiate the particles at the 
front plane and those closer to the rear. Snapshots (e), (f), and (g)  are forming network-like structures of 
nanoparticles, while the nanoparticle network breaks down into non-percolating clusters in the figure 
(h) by forming individual sheetlike nano-structures.}
\label{int_dens1}
\end{figure*}

\begin{figure*}
\centering
\includegraphics[scale=0.2]{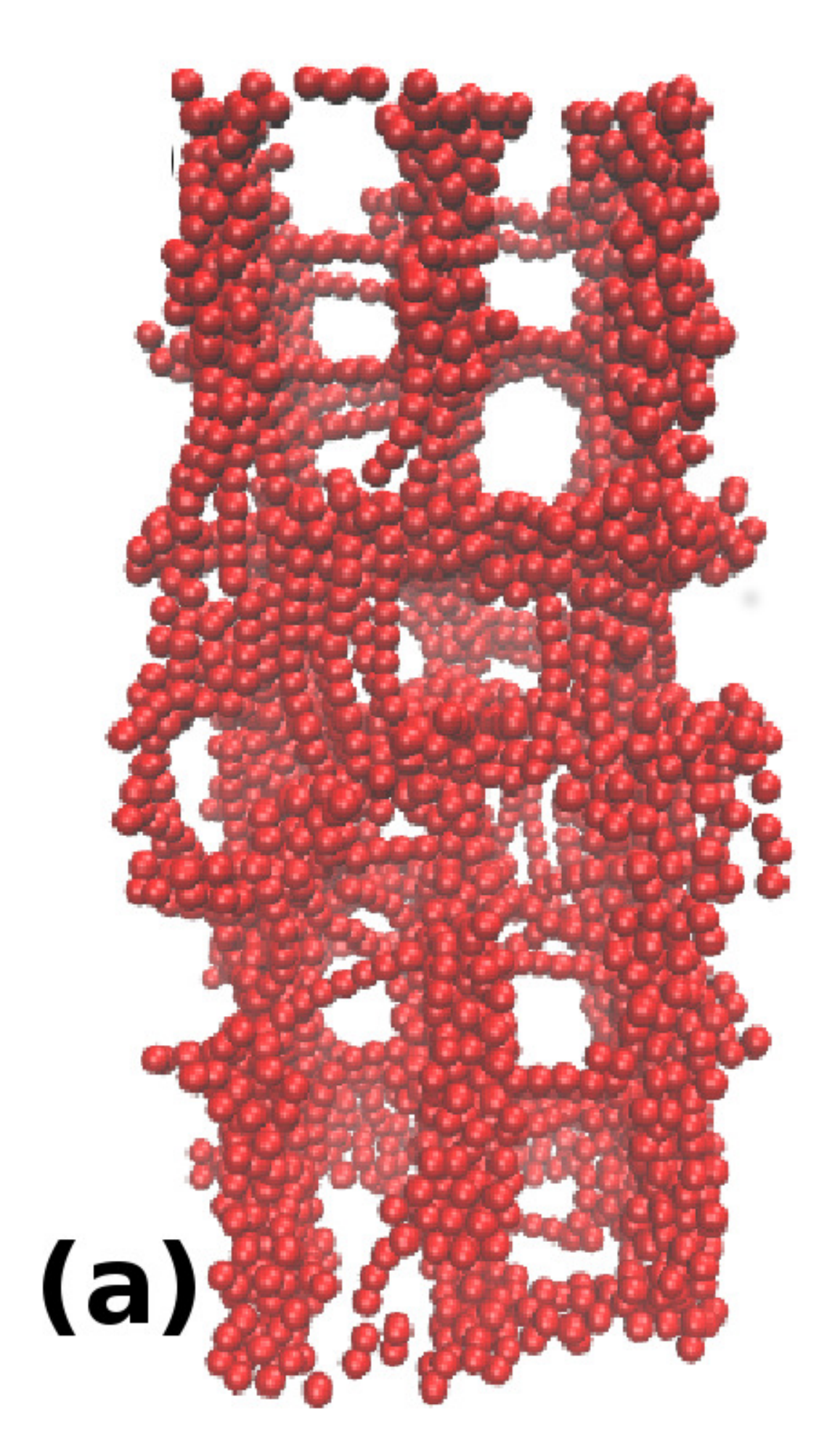}
\hspace{2cm}
\includegraphics[scale=0.2]{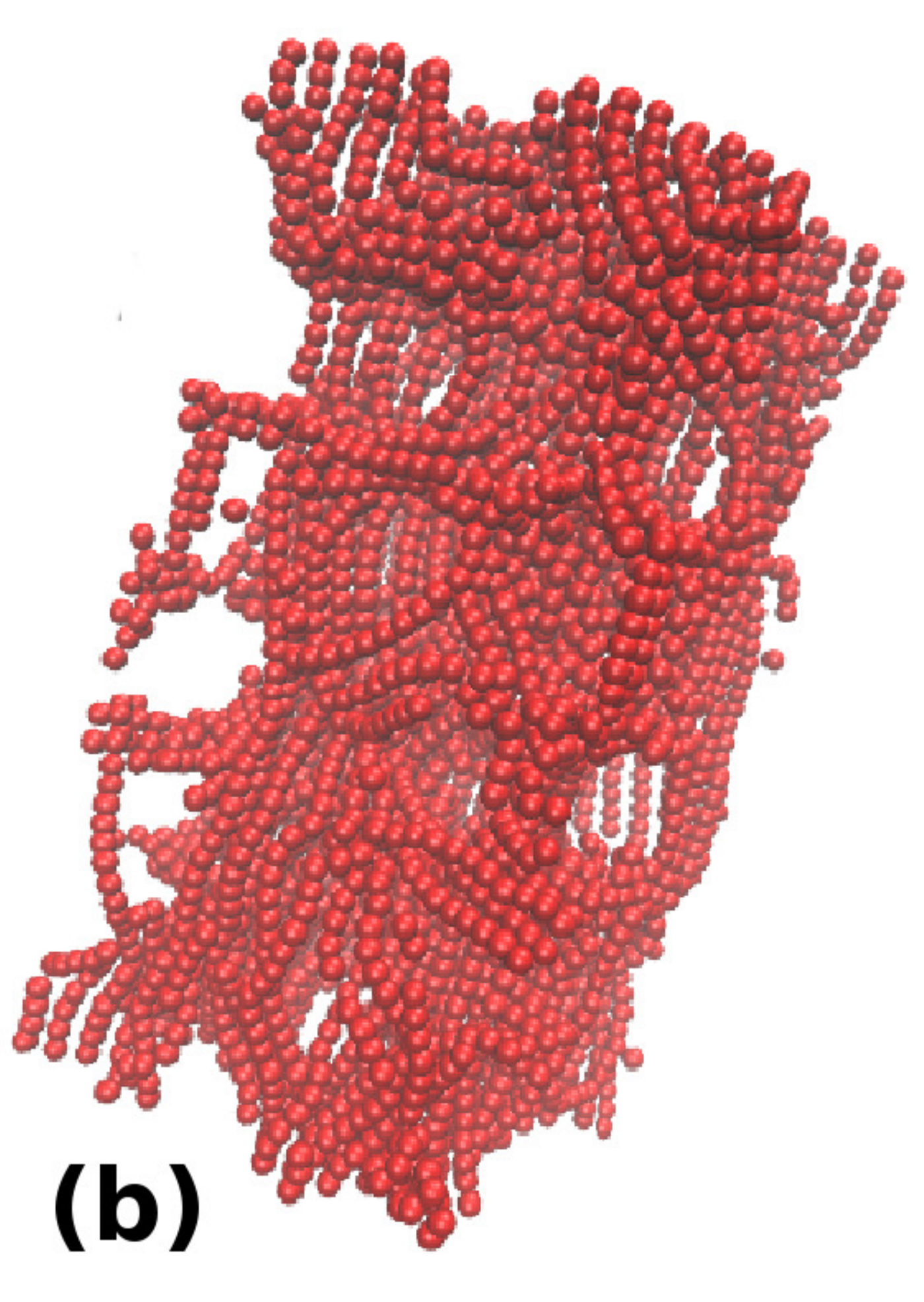}
\hspace{2cm}
\includegraphics[scale=0.2]{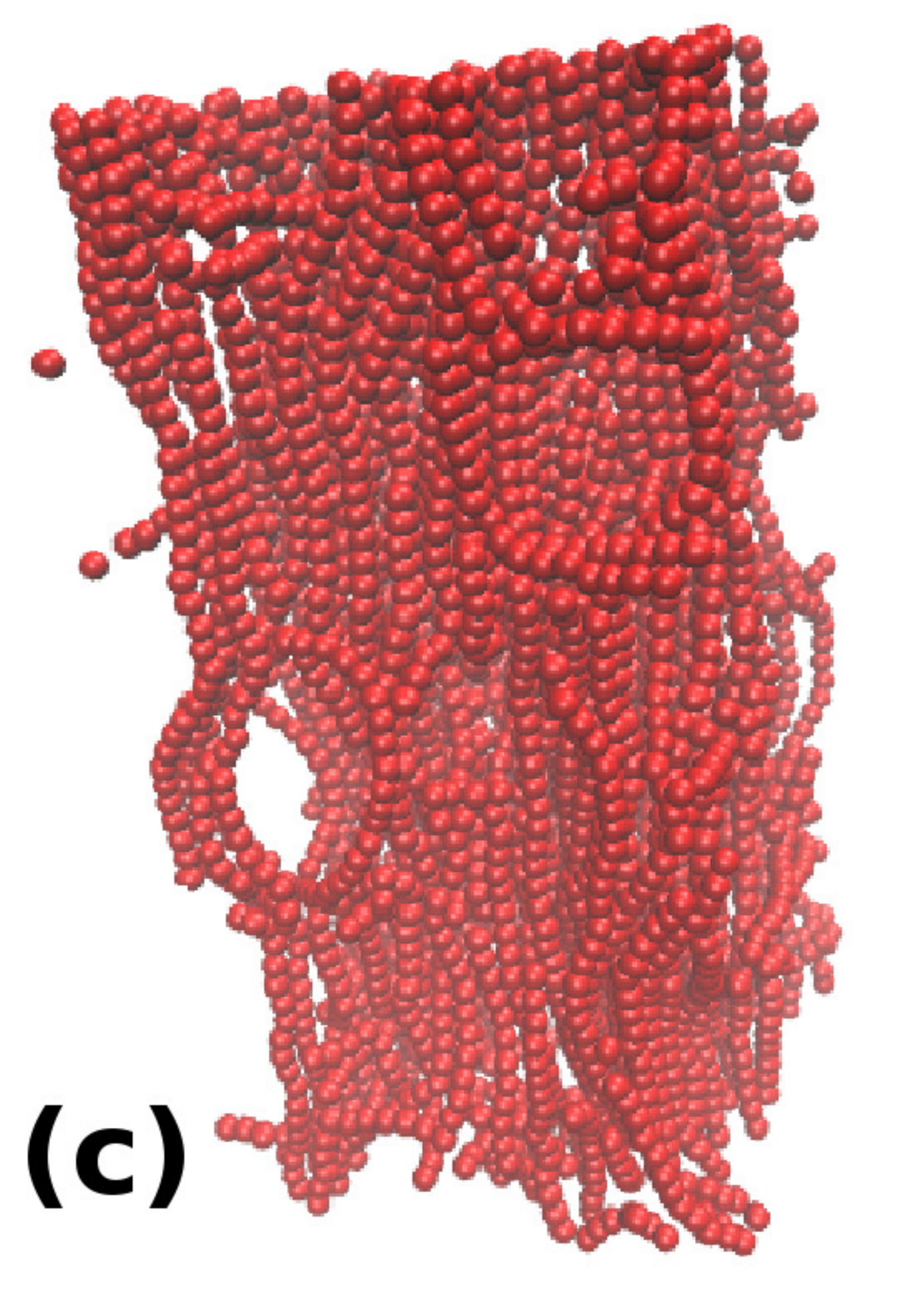}
\hspace{2cm}
\includegraphics[scale=0.2]{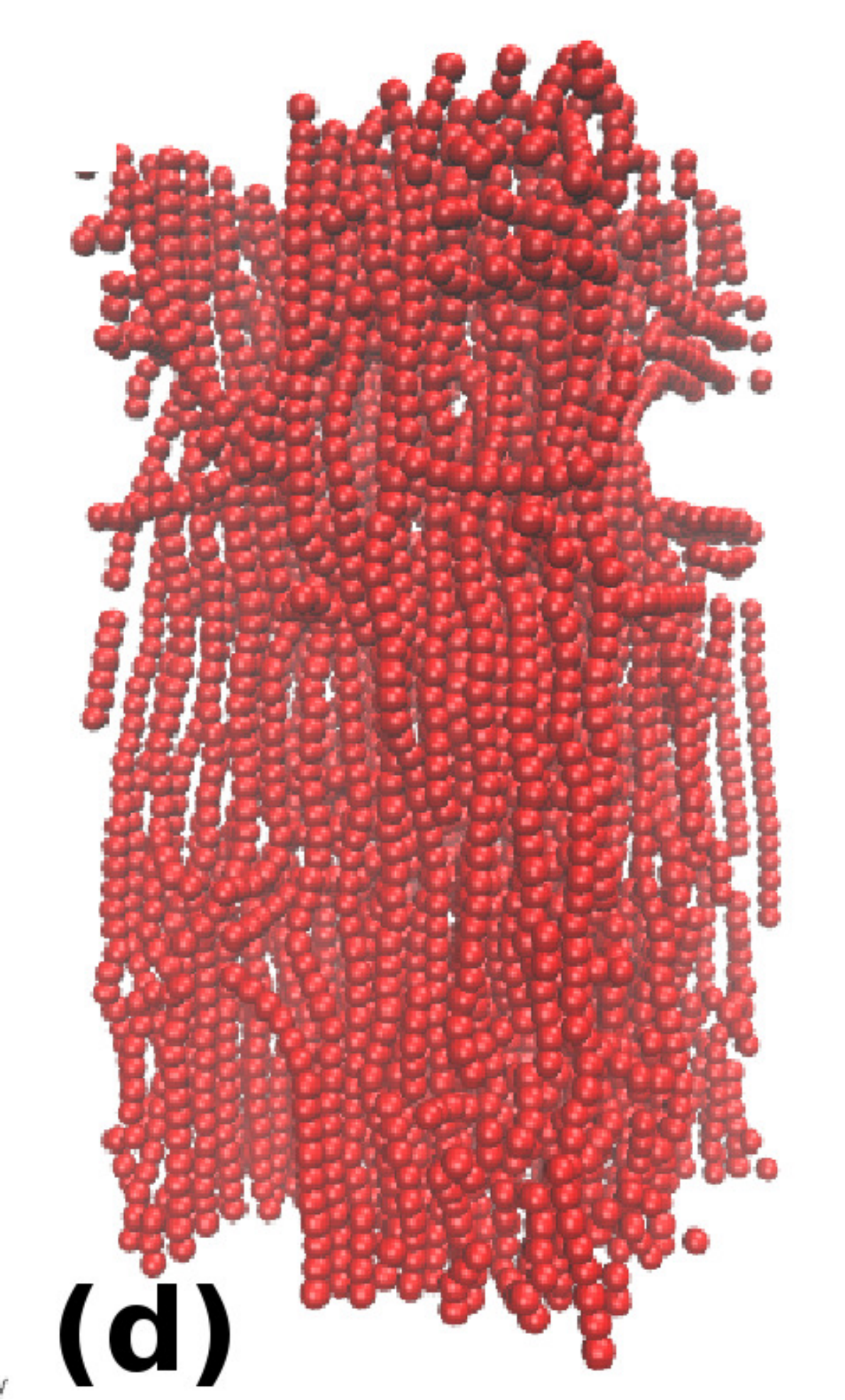}\\
\includegraphics[scale=0.2]{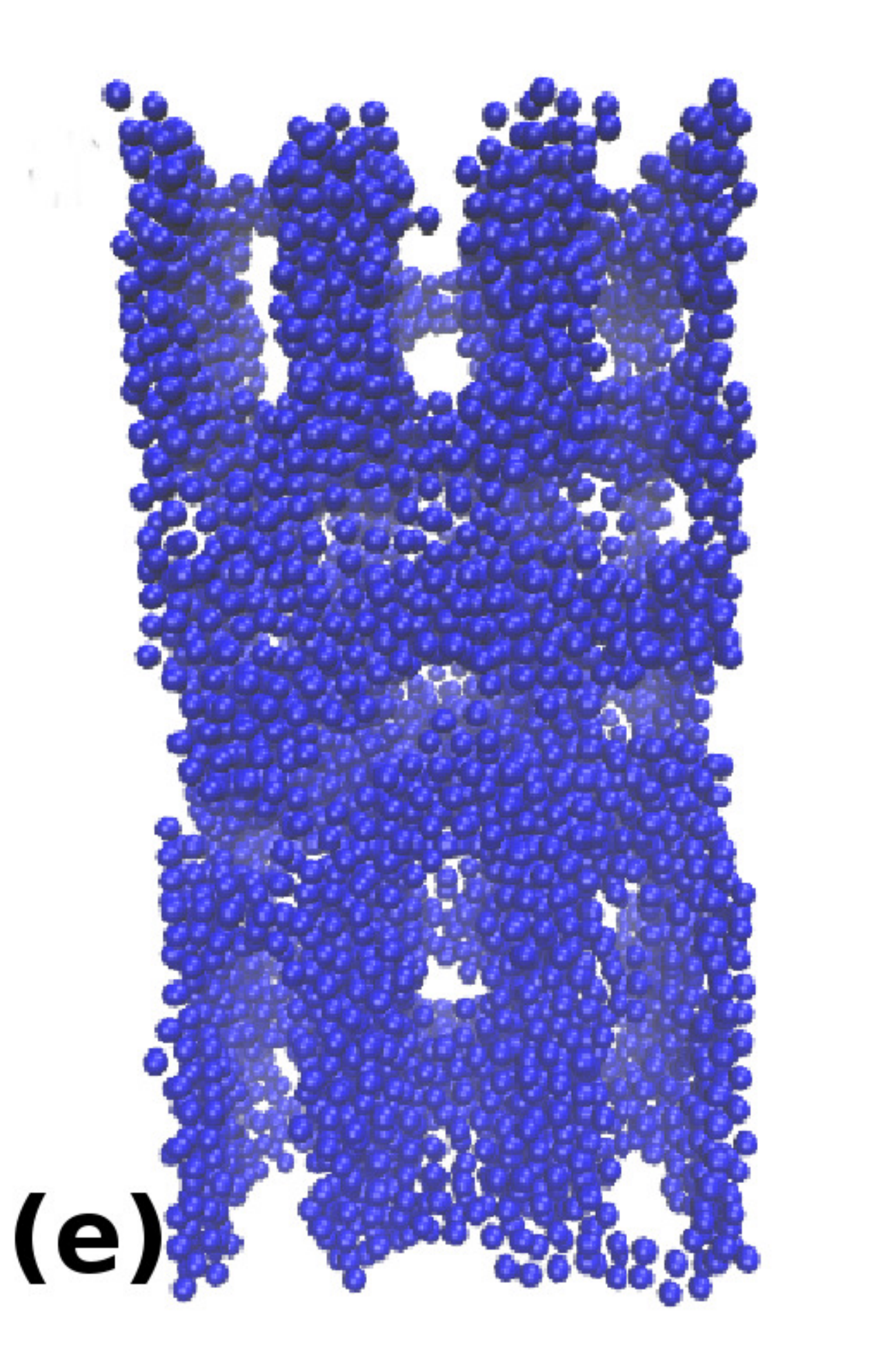}
\hspace{2cm}
\includegraphics[scale=0.2]{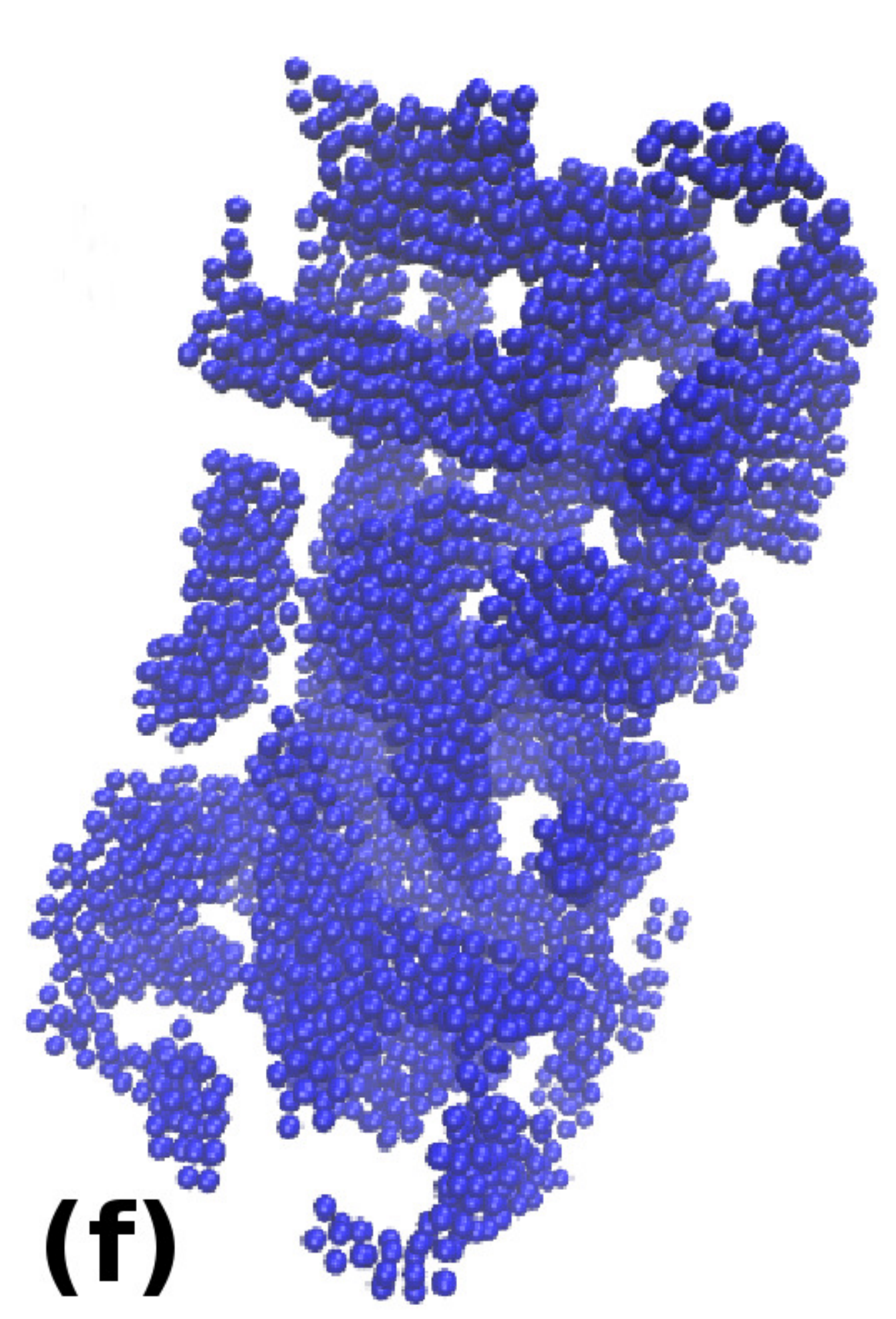}
\hspace{2cm}
\includegraphics[scale=0.2]{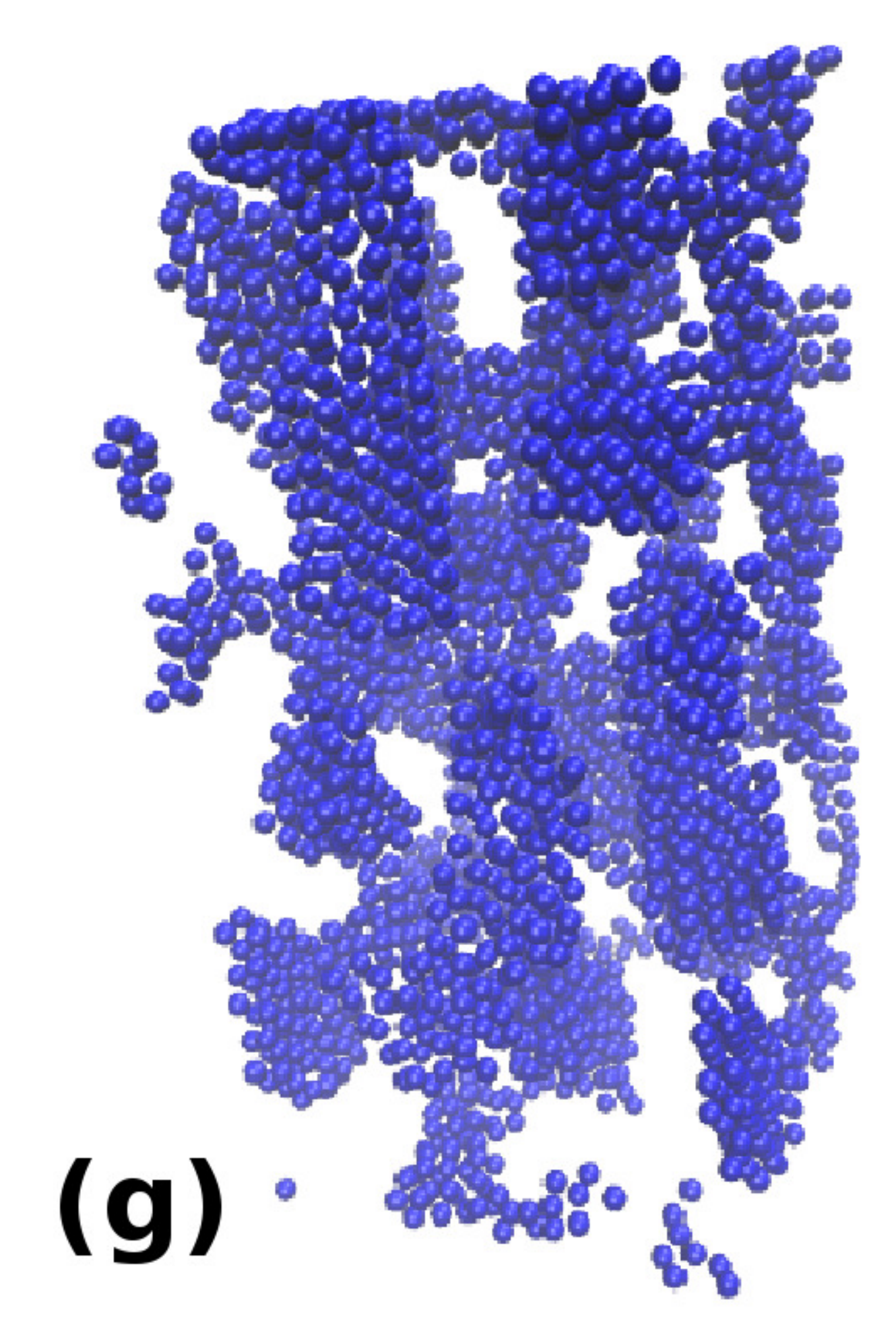}
\hspace{2cm}
\includegraphics[scale=0.2]{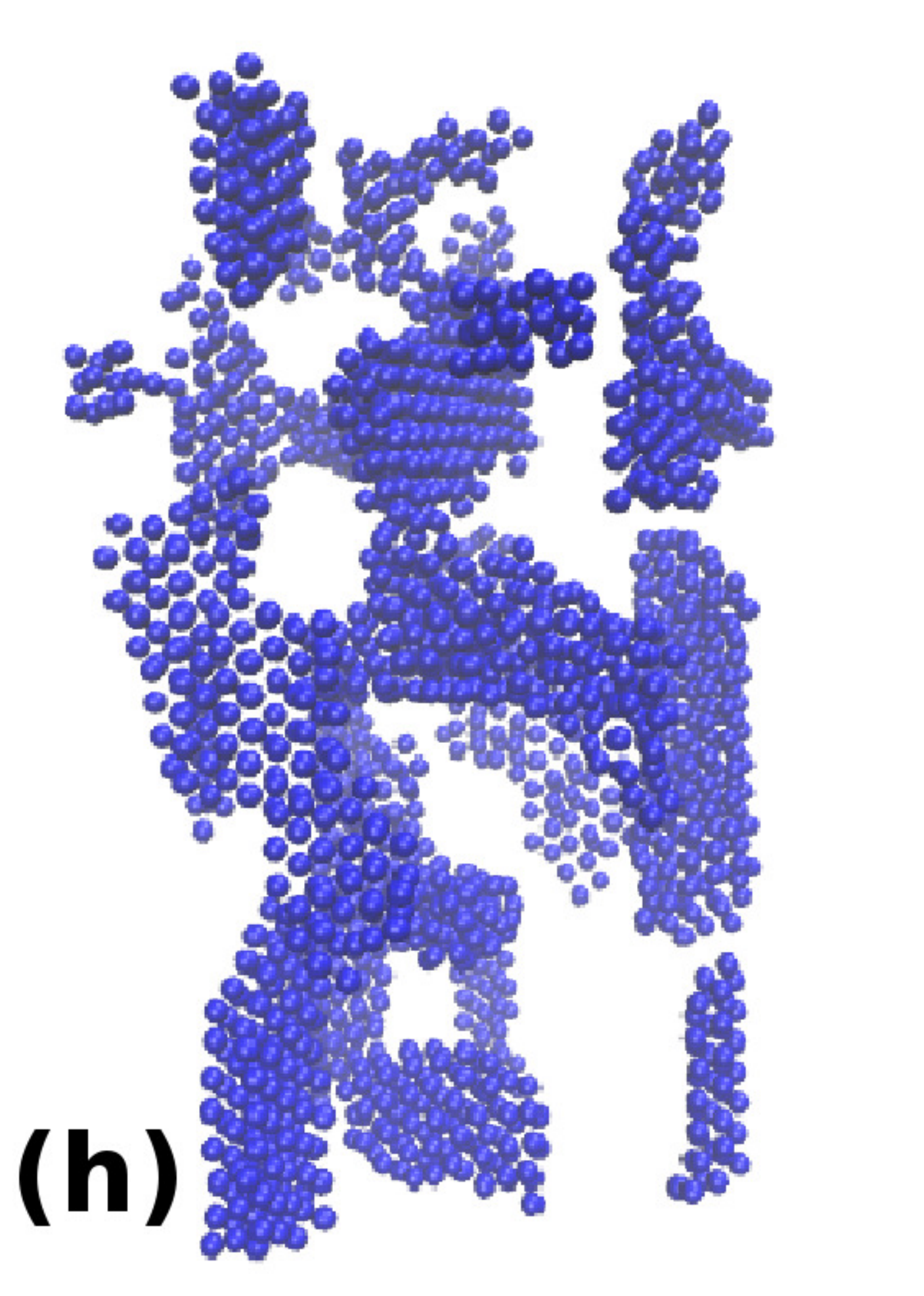}
\caption{( colour online). The figure shows the snapshots of NPs and monomers from the NP + micellar-polymer 
system for the value of number density of monomers $ \rho_m=0.093\sigma^{-3} $. Snapshots in the upper 
row show only micellar monomers (red) while the snapshots in the lower row only show nanoparticles (blue). 
Snapshots from (a) to (d) [or correspondingly (e) to (h)] are for $\sigma_{4n}=  1.75\sigma, 2.25\sigma, 
2.5\sigma $ and $3\sigma $, respectively. With the increase in the value of $\sigma_{4n} $ from left 
to right, the snapshots of nanoparticles show that the network gradually breaks and form individual 
sheet-like clusters for a high value of $\sigma_{4n}$, as shown in the snapshot (h).}
\label{int_dens2}
\end{figure*}

\begin{figure*}
\centering
\includegraphics[scale=0.2]{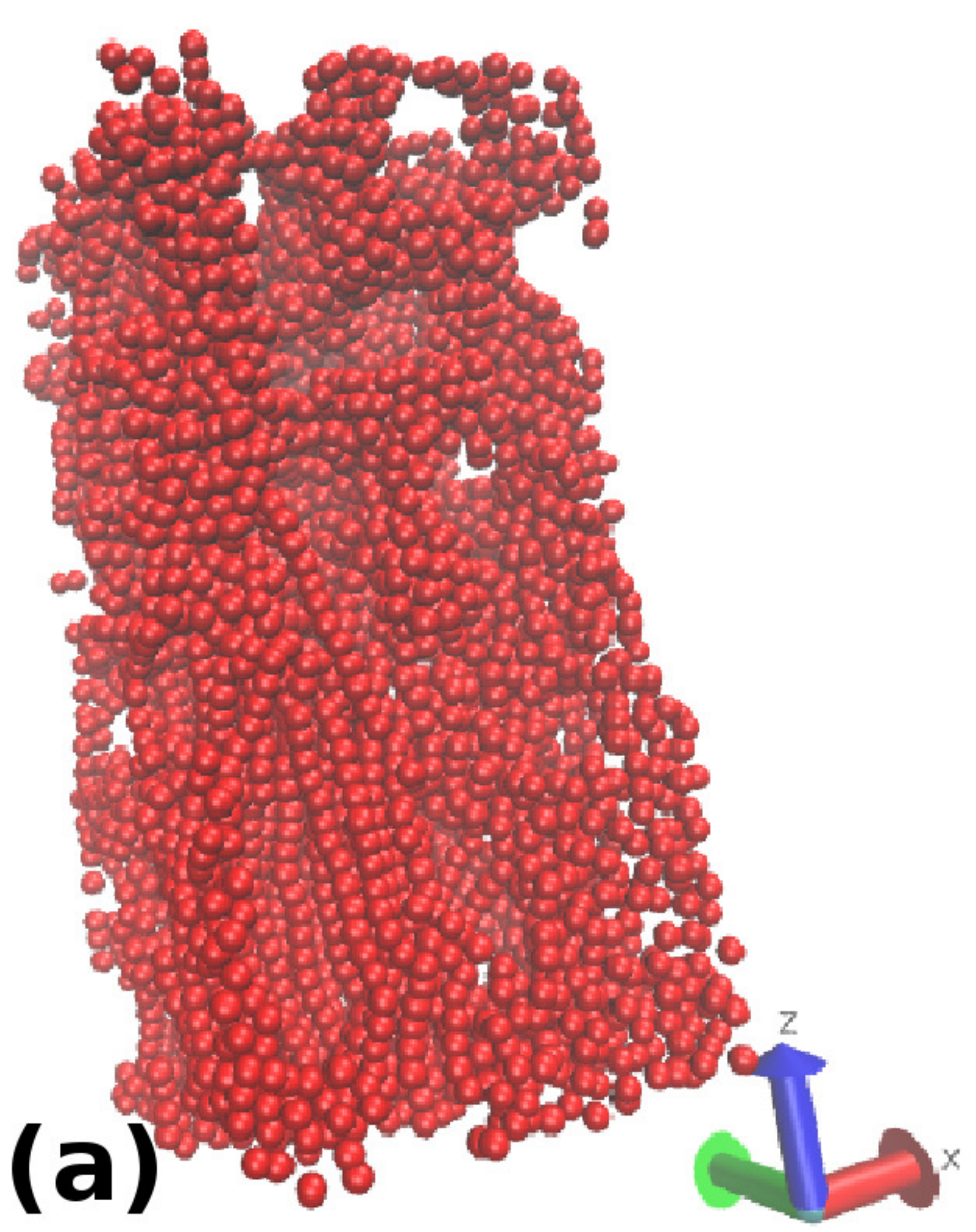}
\hspace{2cm}
\includegraphics[scale=0.2]{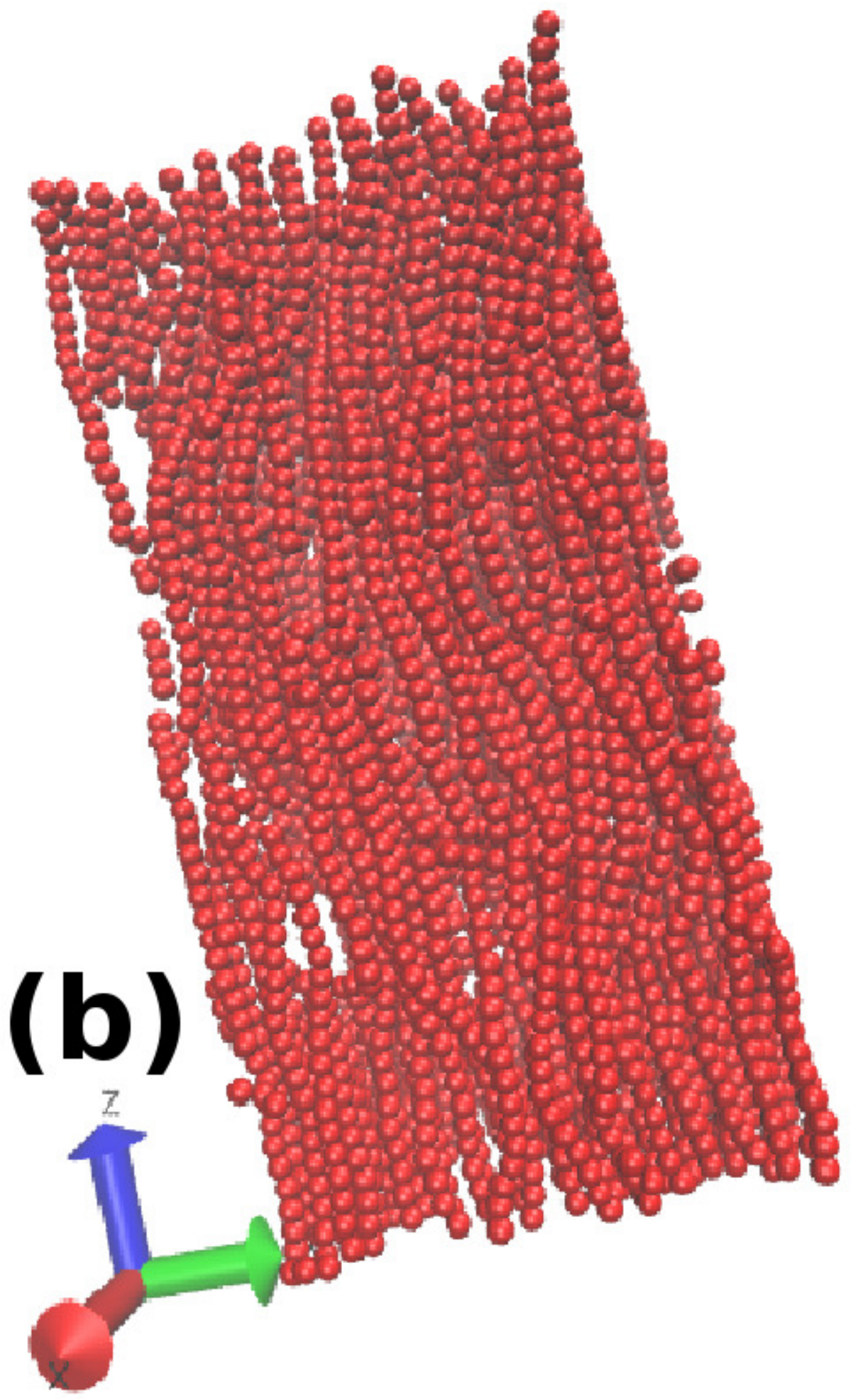}
\hspace{2cm}
\includegraphics[scale=0.2]{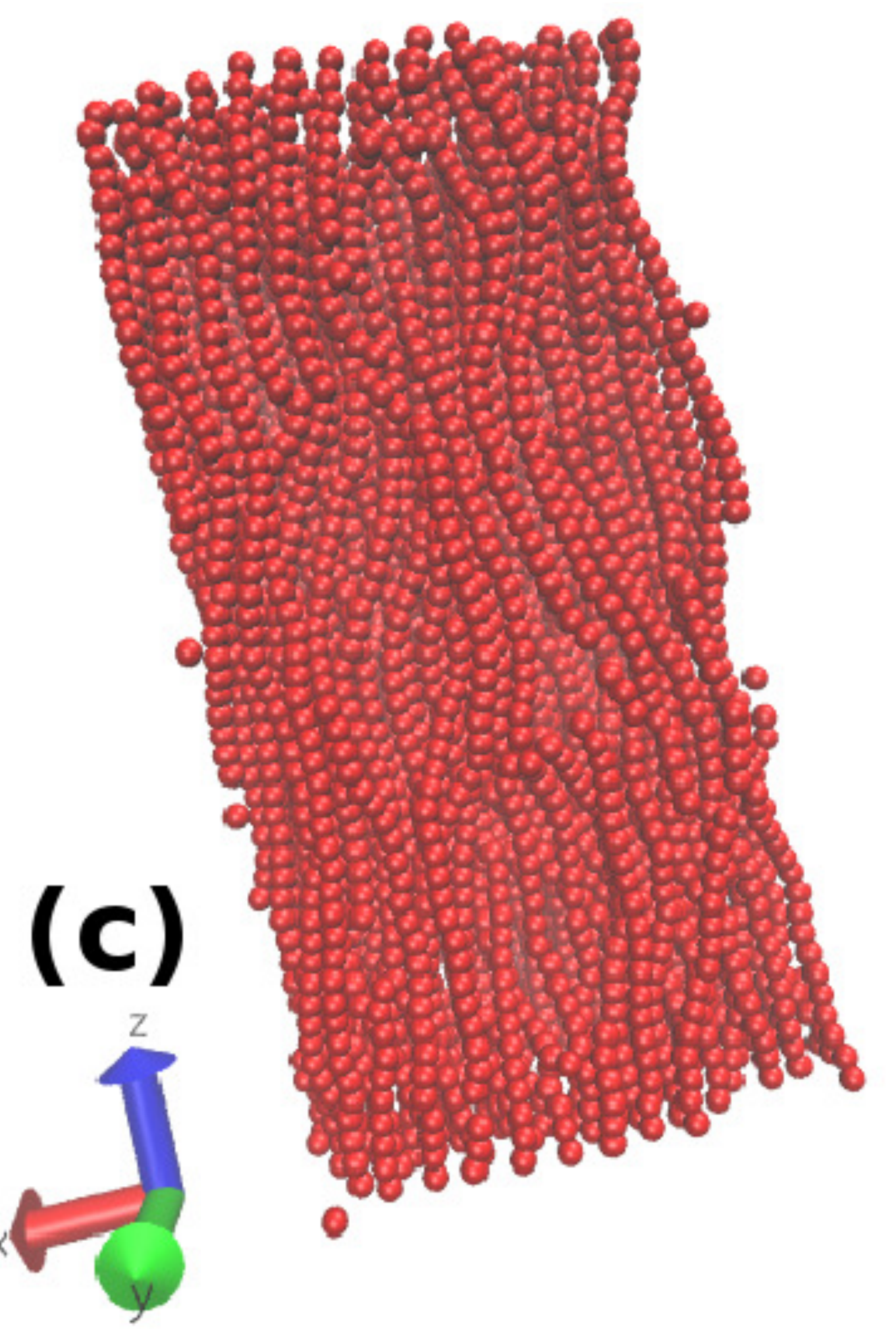}
\hspace{2cm}
\includegraphics[scale=0.2]{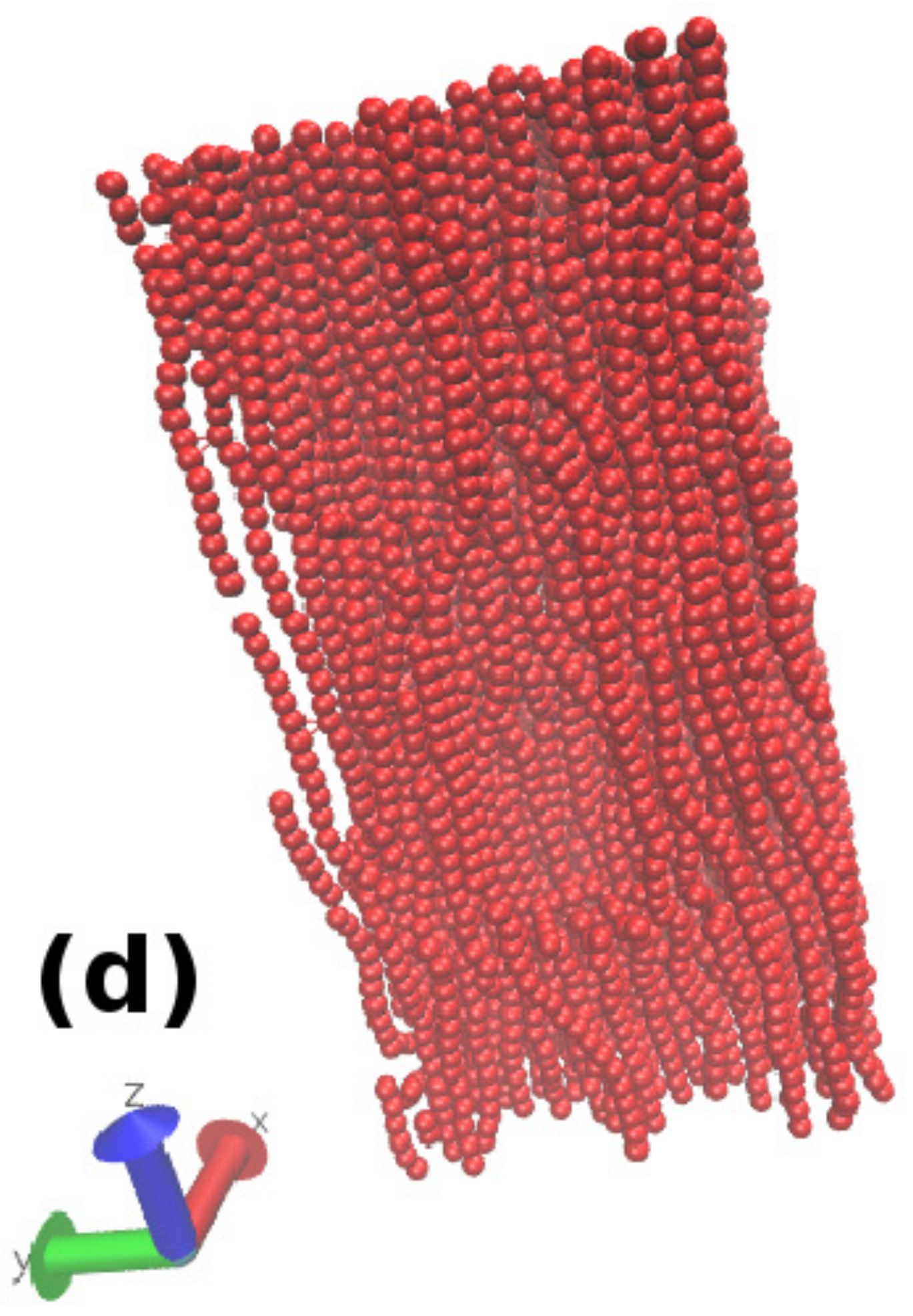} \\
\includegraphics[scale=0.2]{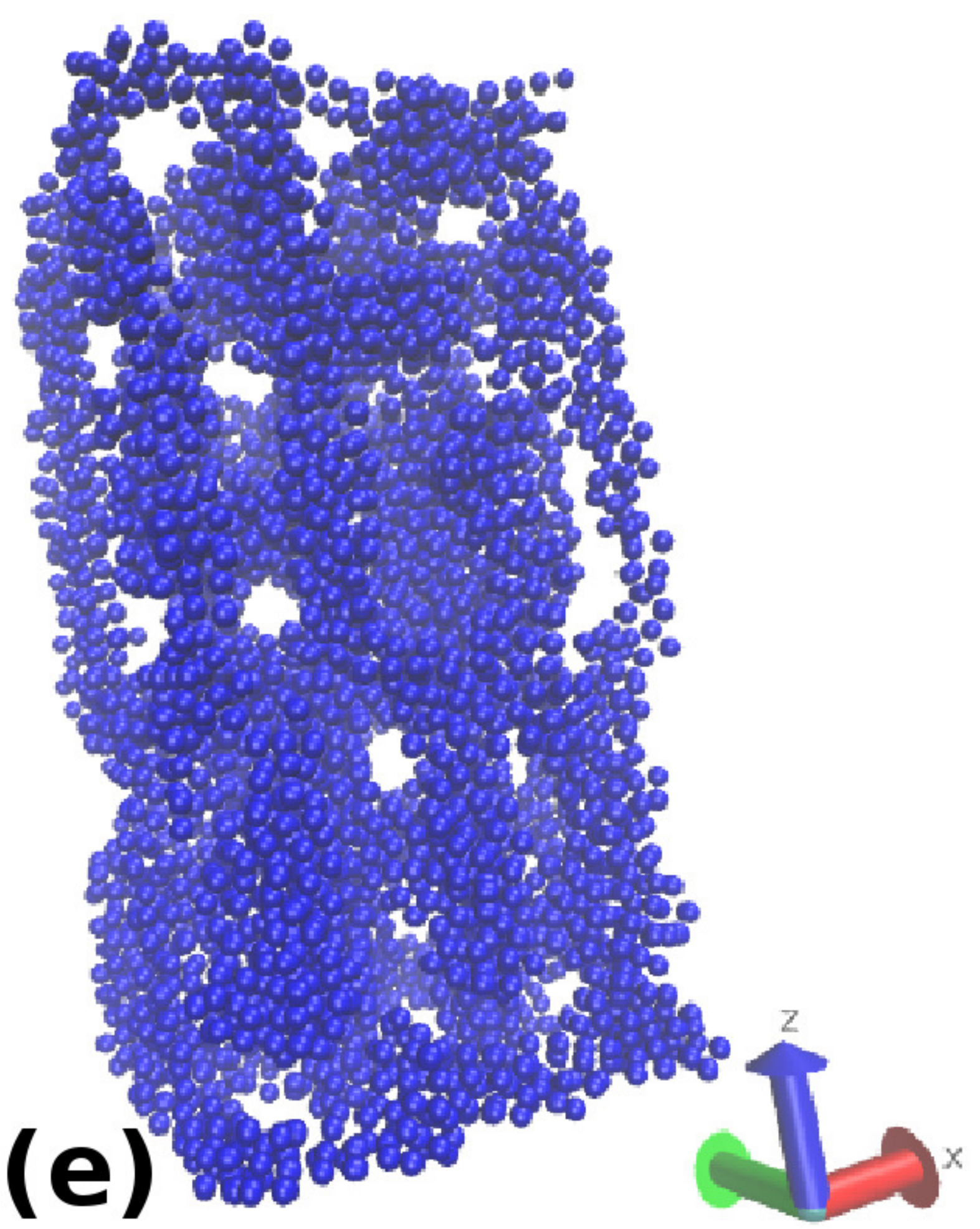}
\hspace{2cm}
\includegraphics[scale=0.2]{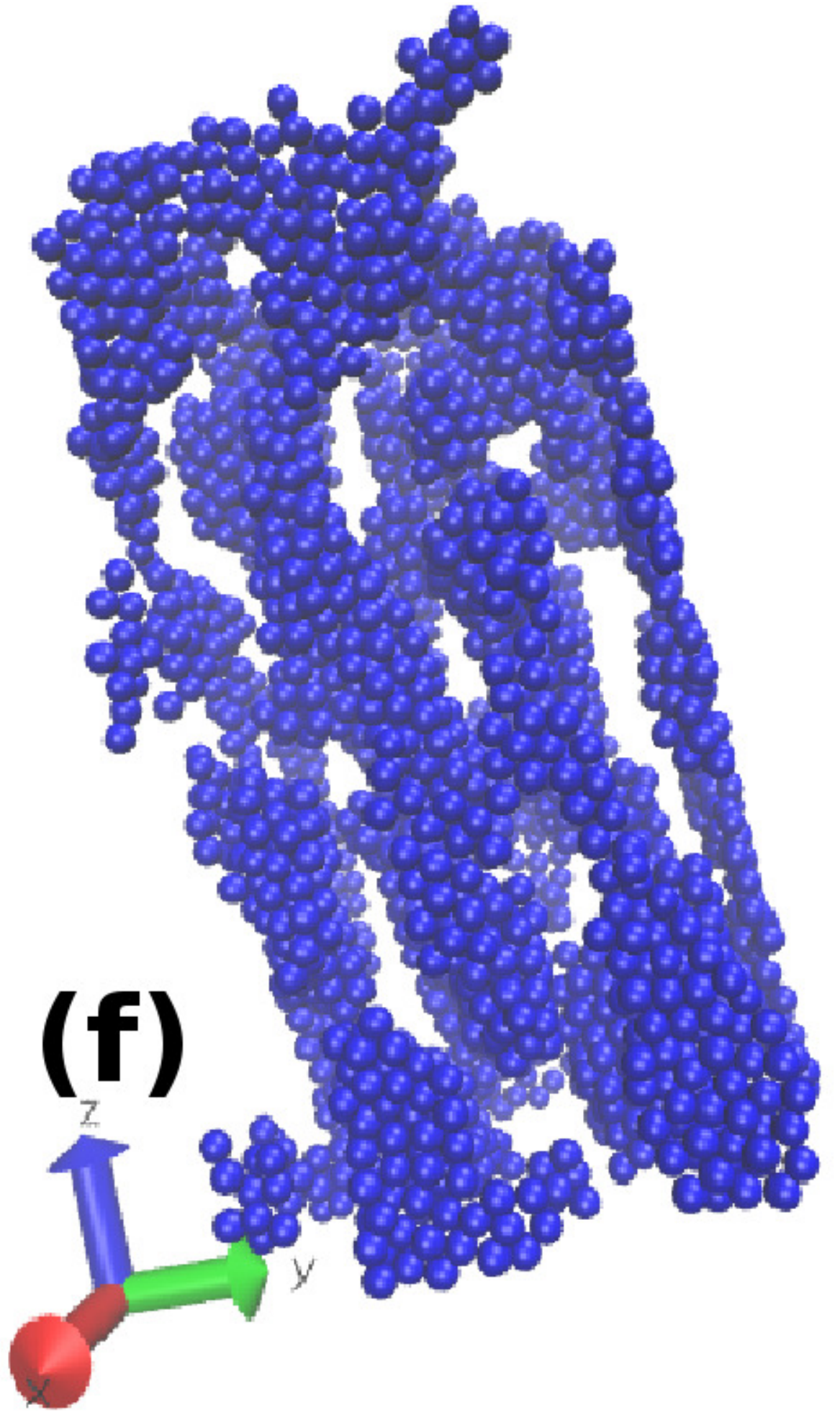}
\hspace{2cm}
\includegraphics[scale=0.2]{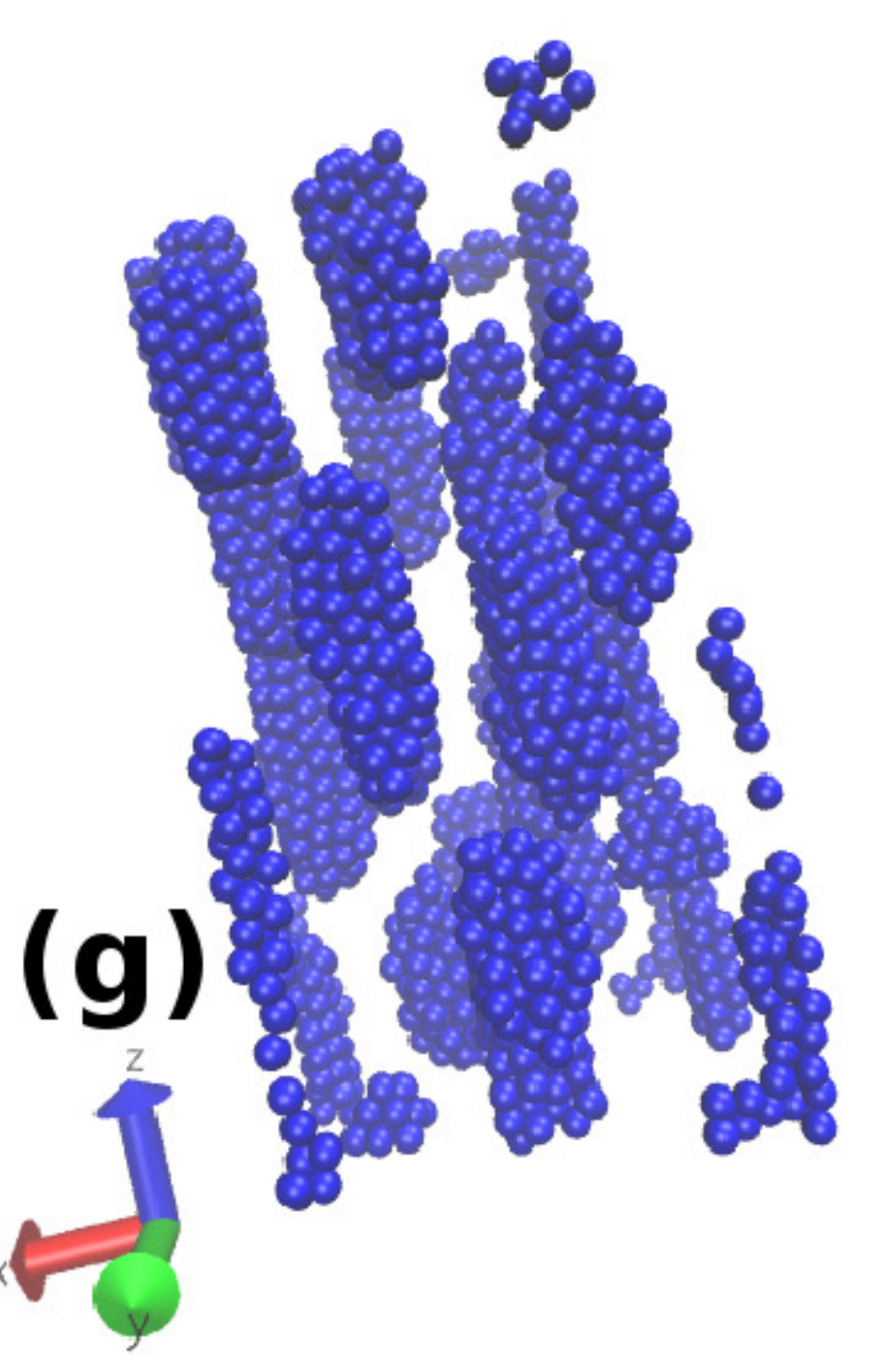}
\hspace{2cm}
\includegraphics[scale=0.23]{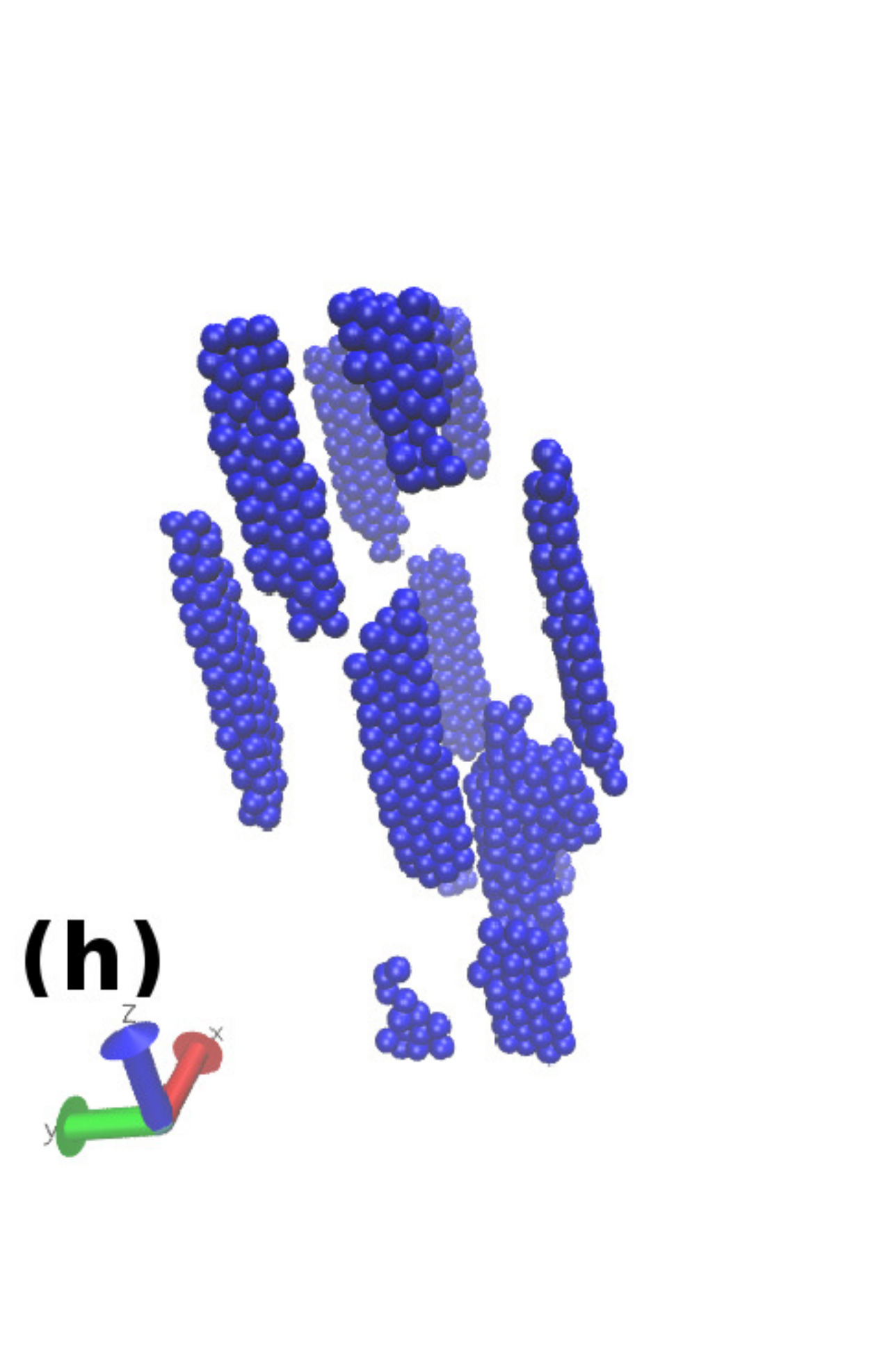}
\caption{(colour online). Figure shows the snapshots of NPs and monomers from the NP + micellar-polymer system 
for monomer number density $\rho_m= 0.126\sigma^{-3} $. The snapshots in the upper row show only 
micellar monomers while, only the NPs are shown in the lower row. Snapshots (a) to (d) (or (e) to 
(h)) correspond to $\sigma_{4n}=1.75\sigma$, $2\sigma$, $2.25\sigma$ and $2.5\sigma$, respectively. 
With increase in the value of $\sigma_{4n}$, the figure shows the nanoparticle networks in (e) and 
(f)  breaking into individual rod-like nanostructures as shown in the snapshots (g) and (h).}
\label{high_dens}
\end{figure*}

After evolving the system for $(2-4)\times 10^6$ iterations, the representative snapshots for four 
different values of $\rho_m=0.037\sigma^{-3}$, $0.074\sigma^{-3}$, $0.093\sigma^{-3}$ and 
$0.126\sigma^{-3}$ are shown in figures \ref{low_dens}, \ref{int_dens1}, \ref{int_dens2} and 
\ref{high_dens}, respectively. 
For each NP+micelles system, the monomers and NPs are shown separately 
in the upper and lower rows, respectively, in figures \ref{low_dens}, \ref{int_dens1}, 
\ref{int_dens2} and \ref{high_dens}. 
Each figure shows the snapshots for four different values of $\sigma_{4n} $ increasing  from (a) to (d) 
for monomers (in red) (or (e) to (h) for NPs in blue). Only for snapshots of NPs in Figs.\ref{low_dens}, 
\ref{int_dens1}(a) and \ref{int_dens1}(e), there exists a gradient in colour (varying from red to blue 
along one of the length of the simulation box) from front plane to the rear plane. This helps to identify particles 
lying in different planes and thereby clearly see the pores in NP aggregates. These pores are occupied by monomers.

In each of the figures, one can see in the leftmost snapshot (snapshots (a) and (e), which are for a lower 
value of $\sigma_{4n}$) a network-like structure of micellar chains or NPs which spans the system. An 
increase in the value of $\sigma_{4n}$ leads to a decrease in number of nanoparticles that gradually 
breaks the network connections, thus, gradually breaking the network of NPs as shown in 
Figs.\ref{int_dens1}(g), \ref{int_dens2}(f,g) and \ref{high_dens}(f).  With further increase in 
$\sigma_{4n}$, these networks break into non-percolating clusters as shown for $\sigma_{4n}=3.25\sigma, 
3\sigma$ and $2.25\sigma$ in figures \ref{int_dens1}(h), \ref{int_dens2}(h) and \ref{high_dens}(g), 
respectively. We do not observe the breaking of networks into individual NP clusters for 
$\rho_m=0.037\sigma^{-3}$ in Fig.\ref{low_dens} for the range of values of $\sigma_{4n}$ considered here. 
Therefore with an increase in $\rho_m$, the value of $\sigma_{4n}$ at which the network breaks into 
non-percolating clusters gets shifted to a lower value of $\sigma_{4n}$. From the snapshots 
\ref{int_dens1}(h), \ref{int_dens2}(h) and \ref{high_dens}(g), it can be clearly seen that different 
densities of micelles lead to different shape-anisotropy of nanoparticle clusters. It forms sheet-like 
structures for $\rho_m=0.074\sigma^{-3}$ [Fig.8(h)] and $0.0931\sigma^{-3}$ [Fig.9(h)] while rods are 
formed for higher density of monomers [Fig.10(h)]. We give the reasons of how the sheet like structures
are formed at the end of the paper.

\begin{figure}
\includegraphics[scale=0.33]{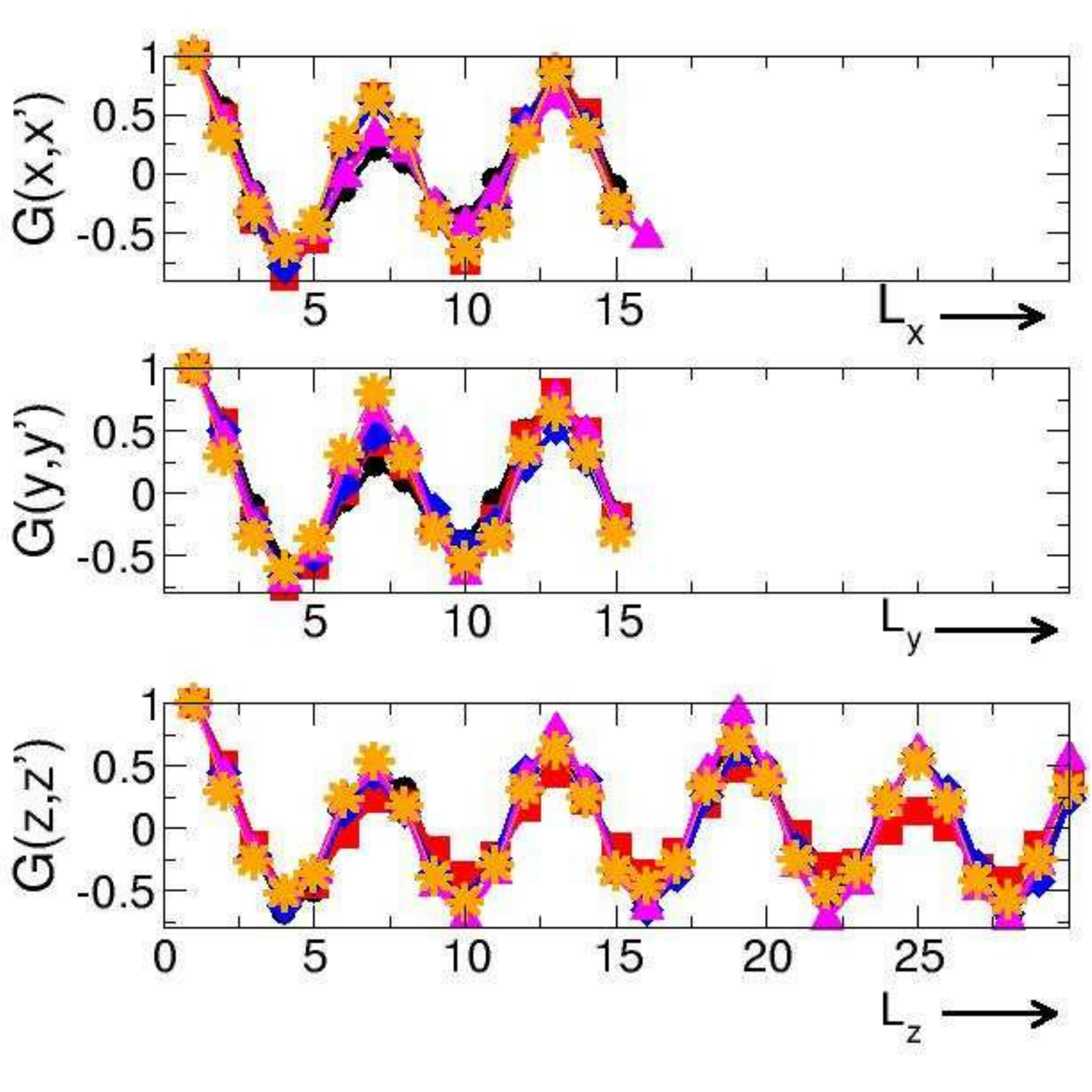}
\caption{(colour online) The figure shows the average of the local density-density correlation function for monomers 
for $\sigma_{4n}=1.5\sigma$ and different values of micellar densities indicated by different symbols. 
The density correlations along $x, y$ and $z$-axis are shown separately from top to bottom. 
The different symbols are for $\rho_m=0.037\sigma^{-3}$ (circle-black), $0.074\sigma^{-3}$ (square-red), 
$0.093\sigma^{-3}$(diamond-blue), and $0.126\sigma^{-3}$ (triangle-magenta). The symbols (star-orange) 
indicate the correlation plots for $\rho_m=0.037\sigma^{-3}$ but $\sigma_{4n}=3\sigma$.}
\label{dens_corr}
\end{figure}

It is observed that not only the non-percolating clusters of NPs have different shapes but also the clusters 
which constitute the NP-networks differ in their morphologies. The snapshots in figures \ref{low_dens}, 
\ref{int_dens1}(e) (\& \ref{int_dens2}(e)) and \ref{high_dens}(e) show networks of NP aggregates, 
network of sheet-like clusters of NPs and network of rod-like clusters of NPs, respectively. 
Thus the density of micelles governs the morphology of NP structures with the anisotropy of NP clusters 
increasing with increase in micellar density viz. clusters to sheets to rods. We also observe that 
as soon as the value of $\sigma_{4n}$ changes from $1.25\sigma$ (Fig.\ref{crystal}) to $1.5\sigma$ 
(Fig.\ref{low_dens}(e), \ref{int_dens1}(e), \ref{int_dens2}(e), \ref{high_dens}(e) ), the micellar chains 
aggregate to form clusters which joins to form a system spanning network such that the networks of micellar 
chains and NPs are inter-penetrating each other. These networks at $\sigma_{4n}=1.5\sigma$ seem to have 
the same periodicity in their structure irrespective of the micellar density.  

\subsection{Polymeric chains of the matrix:\\
               quantitative analysis of micro-structure}

To confirm the above observations, we plot the density correlation function for monomers which is shown in 
Fig.\ref{dens_corr}. It shows the density correlation function for monomers along the x, y, and z-axes of the 
box for $\sigma_{4n}=1.5\sigma$.  The function is calculated using the expression,
\begin{equation}
G(x_i,x'_i) = \frac{\Big\langle(\rho(x_i)-\langle\rho(x'_i)\rangle)(\rho(x_i)-\langle\rho(x'_i)\rangle)\Big\rangle}{\Big\langle(\rho(x_i)-\langle\rho(x_i)\rangle)^2\Big\rangle}
\label{eq:corr}
\end{equation}

where, $\rho(x_i)$ is the density of the monomers in cubic box of size $\sigma^3$ at $x_i$ at a particular $y-z$ plane.
The angular brackets in the numerator correspond to averages taken at different $y-z$ planes as well as over different
independent runs. The term $\langle\rho(x_i)\rangle$ is the mean density of monomers in the simulation box.
The term in the denominator normalizes the function from 1 to -1. The different symbols indicate the 
different values of micellar densities in Fig.\ref{dens_corr}. All the plots for different densities 
overlap each other and are indistinguishable from each other. They have the same spatial period for all the densities 
and along different the axes. This indicates the periodic nature of the network-like structures in all 
three directions and this remains unaltered by the change in monomer number density. This is observed not only 
for the value of $\sigma_{4n}=1.5\sigma$, but also for all the values of $\sigma_{4n}$ considered for 
$\rho_m=0.037\sigma^{-3}$ (refer Fig.\ref{low_dens}).

\begin{figure}
\includegraphics[scale=0.3]{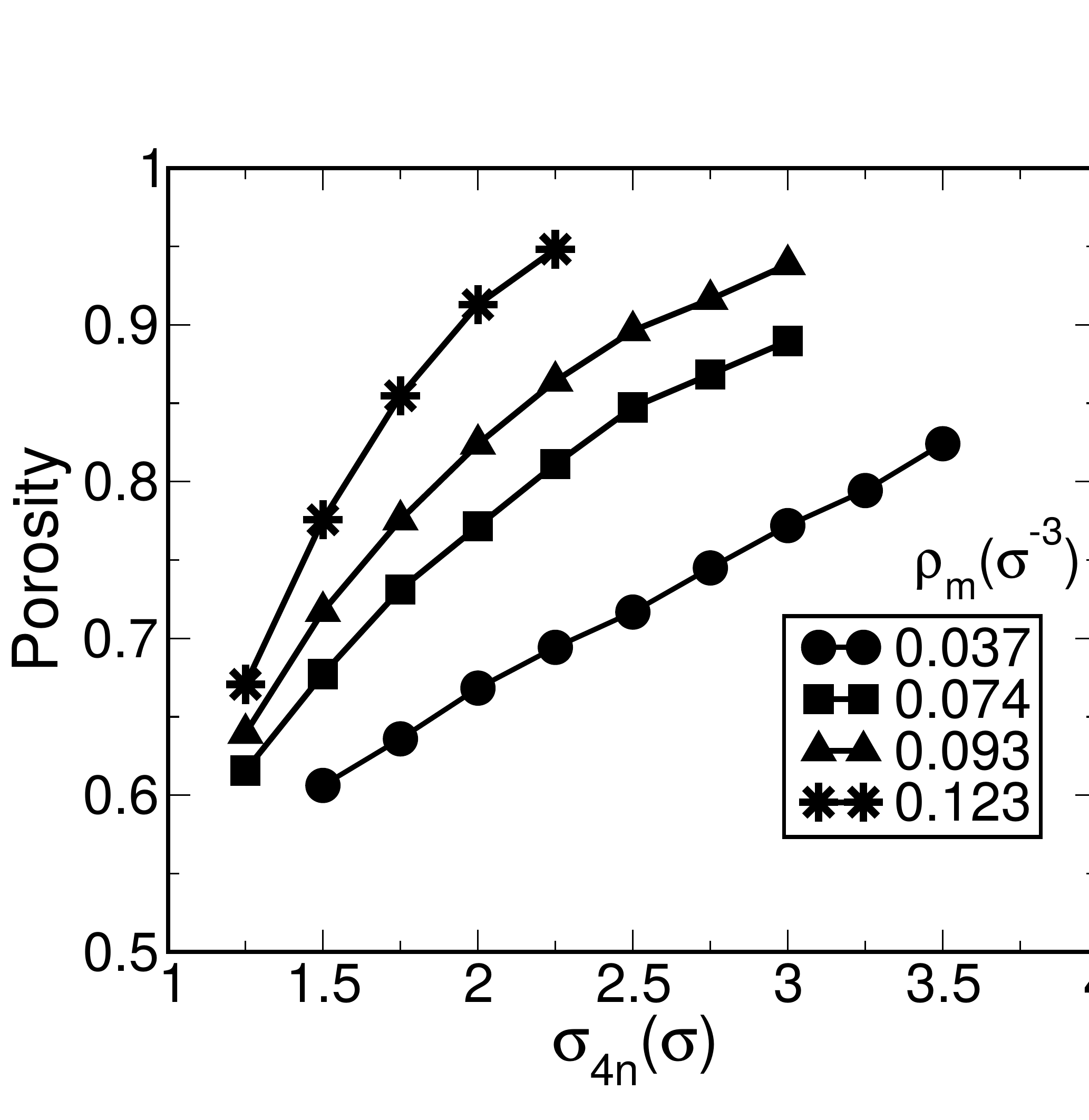}
\caption{(colour online) The figure shows the variation in the porosity of the network of nanoparticles. 
Porosity is calculated by subtracting the nanoparticle volume fraction from $1$, i.e.
it represents  the variation in the volume fraction of the void (currently occupied by monomers, but which 
can be dissolved away) inside a porous network of NPs due to the change in $\sigma_{4n}$ values. 
The different symbols indicate the different values of micellar densities. 
It can be seen that the porosity of the networks increases with either increase in the value of 
$\sigma_{4n}$ or increase in monomer number density $\rho_m$, as expected.  }
\label{porosity}
\end{figure}

\begin{figure}
\centering
\includegraphics[scale=0.3]{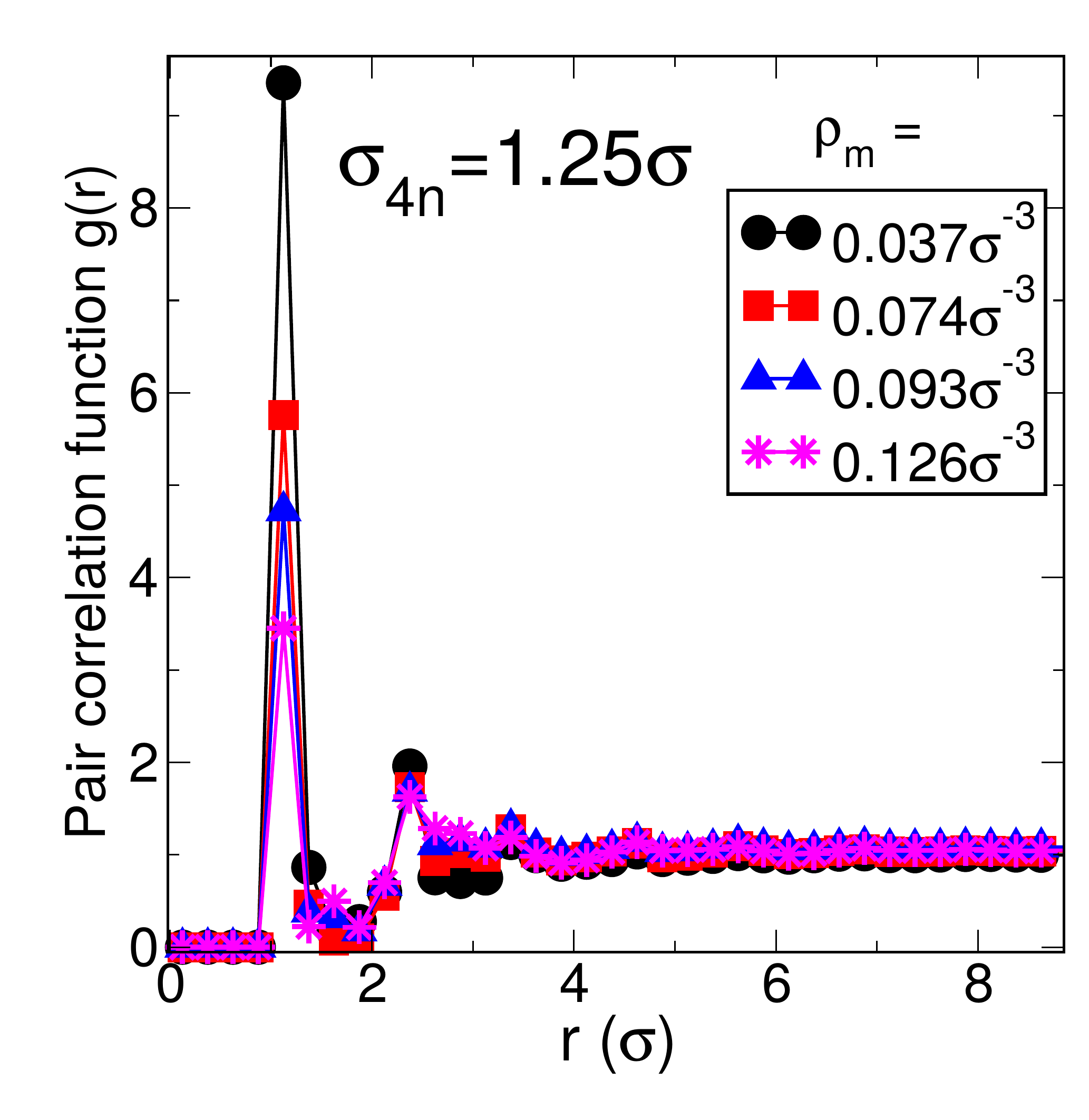}
\caption{ The figure shows the pair correlation function $g(r)$ for monomers at $\sigma_{4n}=1.25\sigma$.  
It shows four different plots for $ \rho_m= 0.037\sigma^{-3}, 0.074\sigma^{-3} , 0.093\sigma^{-3} $ 
and $ 0.126\sigma^{-3} $ which are indicated by different symbols in the figure. The absence of a peak 
around $1.75\sigma$ indicates that polymeric chains are out of the range of repulsive interaction $V_4$ 
from each other and hence depicts the dispersed state of chains, with NPs in between adjacent chains.}
\label{corr_mono_125}
\end{figure}

\begin{figure*}
\centering
\includegraphics[scale=0.2]{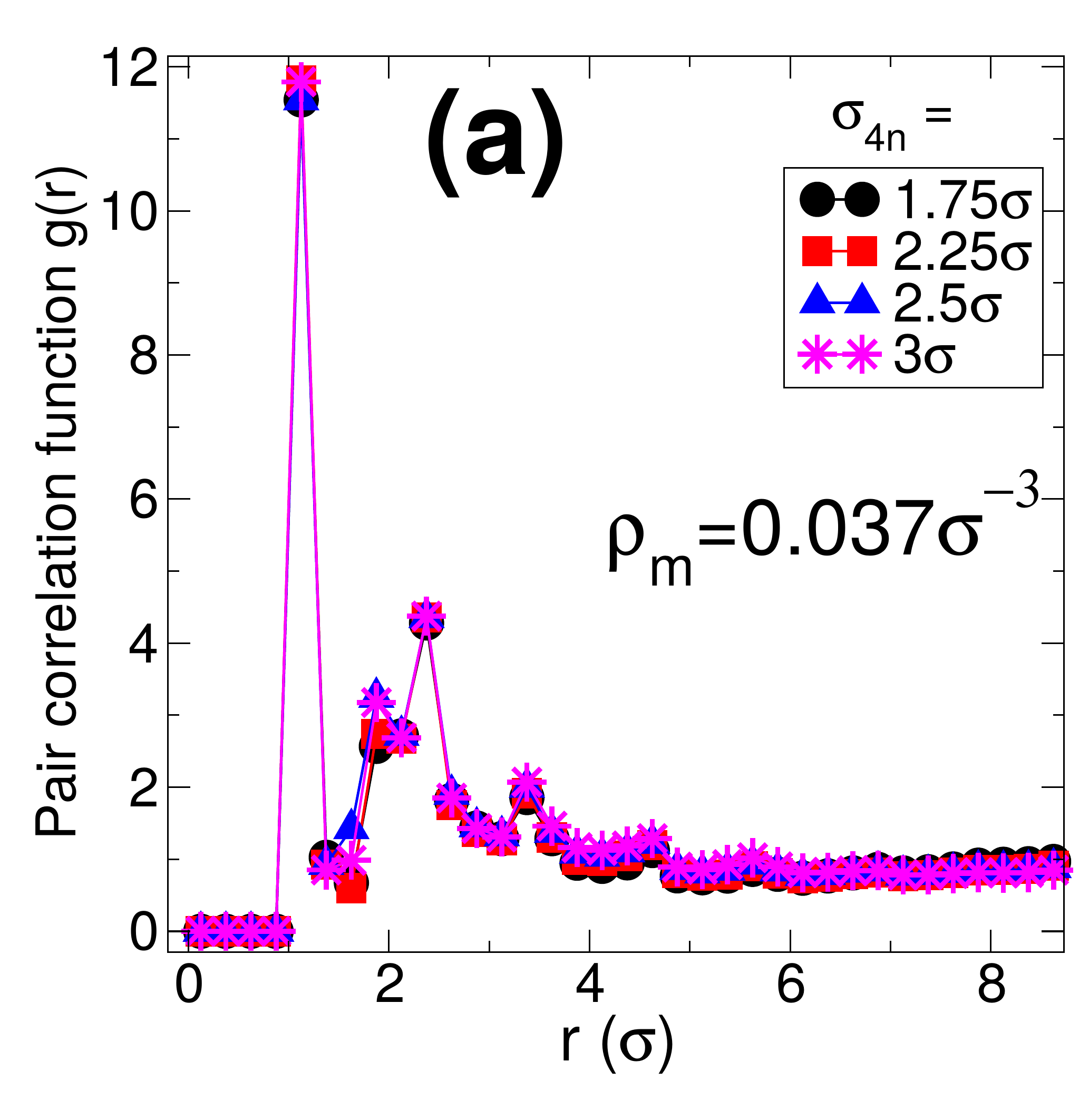}
\includegraphics[scale=0.2]{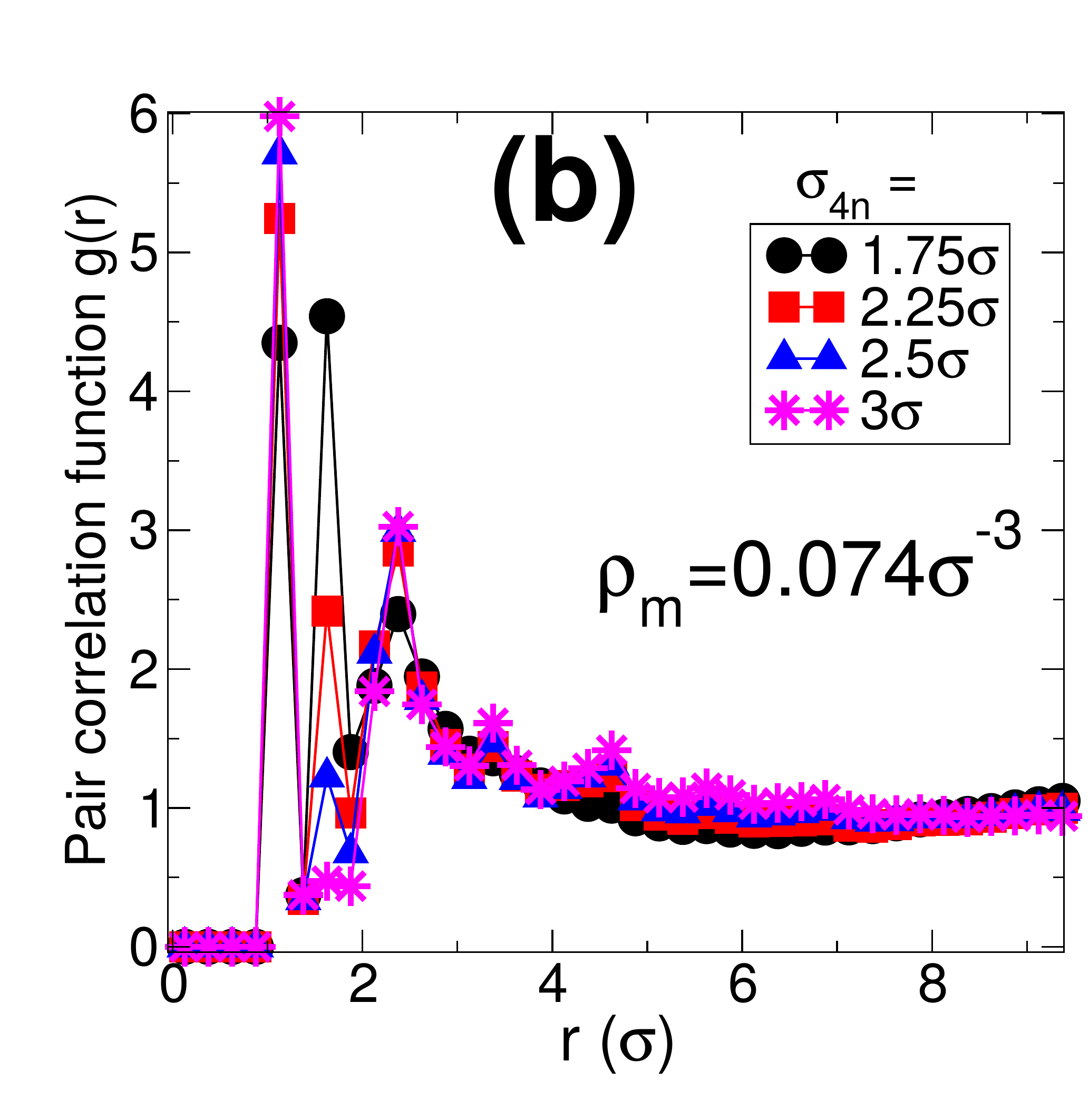}
\includegraphics[scale=0.2]{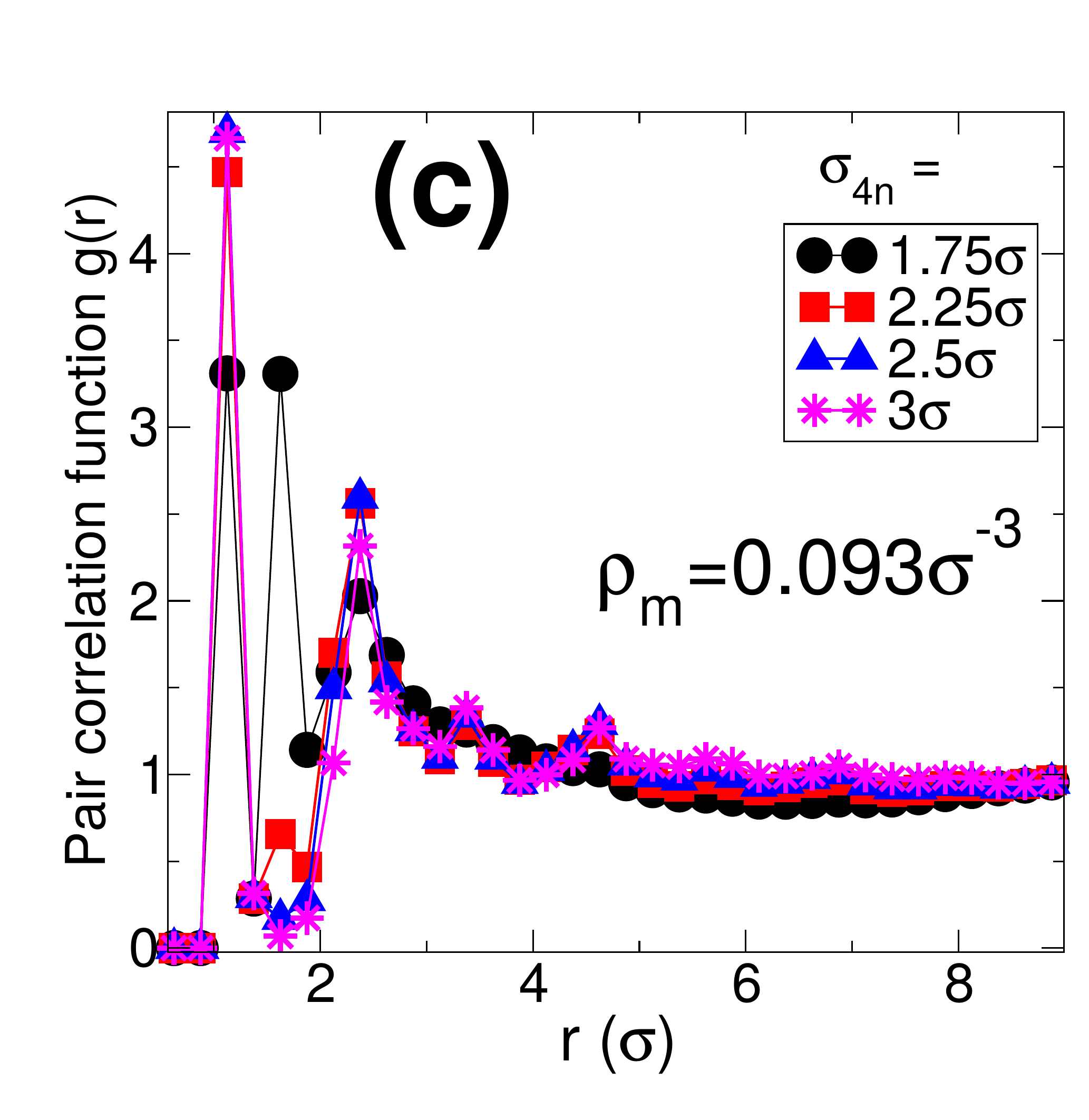}
\includegraphics[scale=0.2]{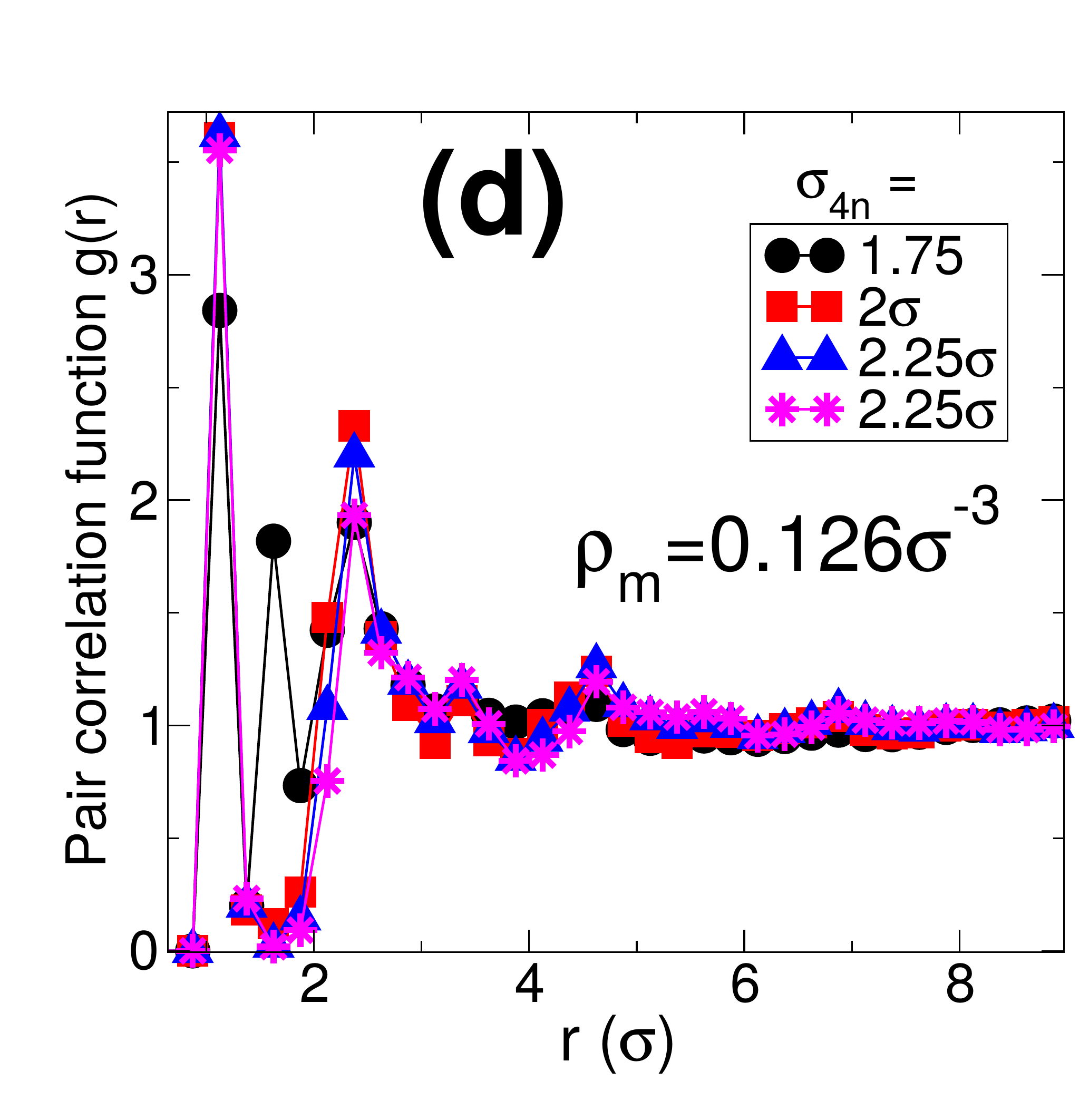}
\caption{(colour online ). The  figure shows the pair correlation function for micellar monomers 
for $\sigma_{4n}=1.25\sigma$ and micellar densities (a) $\rho_m= 0.037\sigma^{-3}$, (b) 
$\rho_m= 0.074\sigma^{-3}$, (c) $0.093\sigma^{-3}$ and (d) $0.126\sigma^{-3} $. Each figure shows the 
correlation function for four different values of $\sigma_{4n} > 1.25\sigma$ as indicated in the graph. 
Contrary to the observations from $g(r)$ data shown in Fig.\ref{corr_mono_125} (for $\sigma_{4n}=1.25\sigma$), 
all the figures here show the appearance of peaks around $1.75\sigma$. The appearance of a peak around 
$1.75\sigma$ indicates that the distance between adjacent monomer chains $\approx 1.75\sigma$ (the range of
$V_4$). Therefore, chains are forming clusters that joins to form a network-like structure as shown in 
Fig.\ref{low_dens}, \ref{int_dens1}, \ref{int_dens2}, \ref{high_dens} for $\sigma_{4n}>1.25\sigma$. 
With increase in the value of $\sigma_{4n}$, the height of the peak around $1.75\sigma$ decreases and 
finally vanishes for a higher value of $\sigma_{4n}$, except for the lowest density shown in (a) for 
$\rho_m=0.037\sigma^{-3}$. For the lowest density, the micellar chains forms networks of clusters of 
micellar chains for all the values of $\sigma_{4n}>1.25\sigma$ as shown in Fig.\ref{low_dens}. }
\label{corr_mono}
\end{figure*}

For the lowest density of micelles considered here , i.e. $\rho_m=0.037\sigma^{-3}$, the morphological changes 
from network of NPs to sheets to rod-like structures is not observed for $\sigma_{4n}>1.25\sigma$. 
The effect of an increase in the value of $\sigma_{4n}$ is observed to decrease the number density of NPs 
without changing the periodicity of the networks. The plot of density correlation function for all the 
values of $\sigma_{4n}$ considered for $\rho_m=0.037\sigma^{-3}$, has the same periodicity as shown by 
the plots in figure. \ref{dens_corr}. To confirm this, the plots shown in figure \ref{dens_corr} shows 
graphs for $\rho_m=0.037\sigma^{-3}$ for two different values of $\sigma_{4n} =1.5\sigma $ 
(circle-black symbols) and $3\sigma$ (star-orange symbols). 

Furthermore, the snapshots in figure \ref{low_dens} show that as $\sigma_{4n}$ increases (from left to right), 
the pore size increases and the wall thickness (or the branch thickness of the network) is seen to be 
decreasing. Hence, we conclude that the thickness of the walls of the NP networks decreases as a result of 
an increase in $\sigma_{4n}$.  This behaviour affects the porosity of the network. We define porosity as 
the volume fraction of the void inside the NP network. The porosity is calculated by subtracting the NP 
volume fraction from 1. The figure \ref{porosity} shows the variation of porosity versus $\sigma_{4n}$ 
quantifying the observed behaviour of the porosity with increase in $\sigma_{4n}$. For a given value of 
$\sigma_{4n}$, an increase in micellar density decreases the available volume for NPs, thereby, 
increasing the porosity of the NP network. Therefore, the figure \ref{porosity} also shows an increase 
in porosity with the increase in micellar density $\rho_m$ for a fixed value of $\sigma_{4n}$, as expected.  
The non-linear behaviour of the plot for $\rho_m =0.123 \sigma^{-3}$ is related to the drop in the number 
of NPs which can get introduced in the box as one increases $\sigma_{4n}$.
The nanoparticle porous structure can be obtained and used after removing by dissolving away the polymeric matrix.

In figures \ref{low_dens}, \ref{int_dens1}, \ref{int_dens2} and \ref{high_dens}, we observe the clustering 
of NPs and micellar chains into different morphologies for values of $\sigma_{4n} \geq 1.5\sigma$ whereas, 
Fig.\ref{crystal} shows that for $\sigma_{4n}=1.25\sigma$ (the minimum possible value of $\sigma_{4n}$), 
we obtain a uniformly mixed state of self-assembled micellar chains and NPs. To get an insight into the spatial 
arrangement of micellar chains, we calculate the pair correlation function g(r) of monomers corresponding to
the snapshots shown in figures \ref{crystal}, \ref{low_dens}, \ref{int_dens1}, \ref{int_dens2} and \ref{high_dens}. 
The plots in Fig. \ref{corr_mono_125} corresponds to correlation function for the snapshots shown in 
Fig.\ref{crystal} while the plots in figures \ref{corr_mono}(a), \ref{corr_mono}(b), \ref{corr_mono}(c) and 
\ref{corr_mono}(d) correspond to the snapshots shown in figures \ref{low_dens}, \ref{int_dens1}, 
\ref{int_dens2} and \ref{high_dens}, respectively. Different symbols represents $g(r)$ for different 
values of $\sigma_{4n}$.

In the correlation function of monomers, a first peak is expected to occur around the value of the diameter
$\sigma$ of the  monomer (which is set as 1) followed by peaks at its multiples indicating the distance 
between bonded monomers along a chain. If there are chains that are situated at a distance $r <= 1.75\sigma$  
(i.e. having the repulsive interaction $V_4$) from other chains, then peaks around $1.75\sigma$ and its 
multiples are expected. We do not observe any peak around $1.75\sigma$ in Fig.\ref{corr_mono_125} for all 
the values of micellar densities. This indicates that the micellar chains in all the snapshots shown in 
Fig.\ref{crystal} (for $\sigma_{4n}=1.25\sigma$), are situated at a distance $r > 1.75\sigma$. 
This is consistent with the observation that no two micellar chains in Fig.\ref{crystal} are found to exist 
without NPs in between. With a NP of size $1.5\sigma$ between two monomer chains, the monomer chains have 
no possibility to remain within the range of repulsive potential ($r \leq 1.75\sigma$). Thus, it confirms 
the observation that, for all the snapshots shown in Fig.\ref{crystal}, no clustering of micellar chains 
is observed and a uniformly mixed state of micellar chains and NPs is formed.

In contrast to the plots in Fig.\ref{corr_mono_125}, there appears a peak around $1.75\sigma$ in 
Figs.\ref{corr_mono}(a), \ref{corr_mono}(b), \ref{corr_mono}(c) and \ref{corr_mono}(d). Except for the 
plots in Fig.\ref{corr_mono}(a), it is observed that the height of this peak decreases with the increase 
in the value of $\sigma_{4n}$ and finally vanishes for a higher value of $\sigma_{4n}$. The appearance 
of a peak around $1.75\sigma$ is consistent with the observed clustering of micellar chains as shown in 
the snapshots in Figs.\ref{low_dens}, \ref{int_dens1}, \ref{int_dens2} and \ref{high_dens} where, they 
form network-like structures. As the value of $\sigma_{4n}$ increases, the NP network gradually starts 
breaking and hence giving more space for micellar chains to be relatively further away from each other. 
This explains the decrease in the peak height around $1.75\sigma$ with increase in $\sigma_{4n}$. When the 
NP network breaks to the extent that most of the micellar chains get enough volume to be at a distance 
$r > 1.75\sigma$, this peak disappears and a minimum is observed at that point. The value of $\sigma_{4n}$ 
at which this happens can be seen to be decreasing with increase in micellar density viz. $3\sigma$ and 
$2.25\sigma$ for $\rho_m= 0.093\sigma^{-3}$ and $0.126\sigma^{-3}$, respectively. 

However, this behaviour is not observed for $\rho_m=0.037\sigma^{-3}$ [refer Fig.\ref{corr_mono}(a)] 
where for all the values of $\sigma_{4n}$ the peaks are present around $1.75\sigma$. This is because, 
with an increase in $\sigma_{4n}$, the network-like structure is not observed to be breaking as shown 
in Fig.\ref{low_dens}. Therefore, for all the values of $\sigma_{4n}$, the polymeric chains show the 
formation of clusters of chains that is reflected in the form of peaks around $1.75\sigma$ for all 
the values of $\sigma_{4n}$ in Fig.\ref{corr_mono}(a).

\begin{figure}
\includegraphics[scale=0.22]{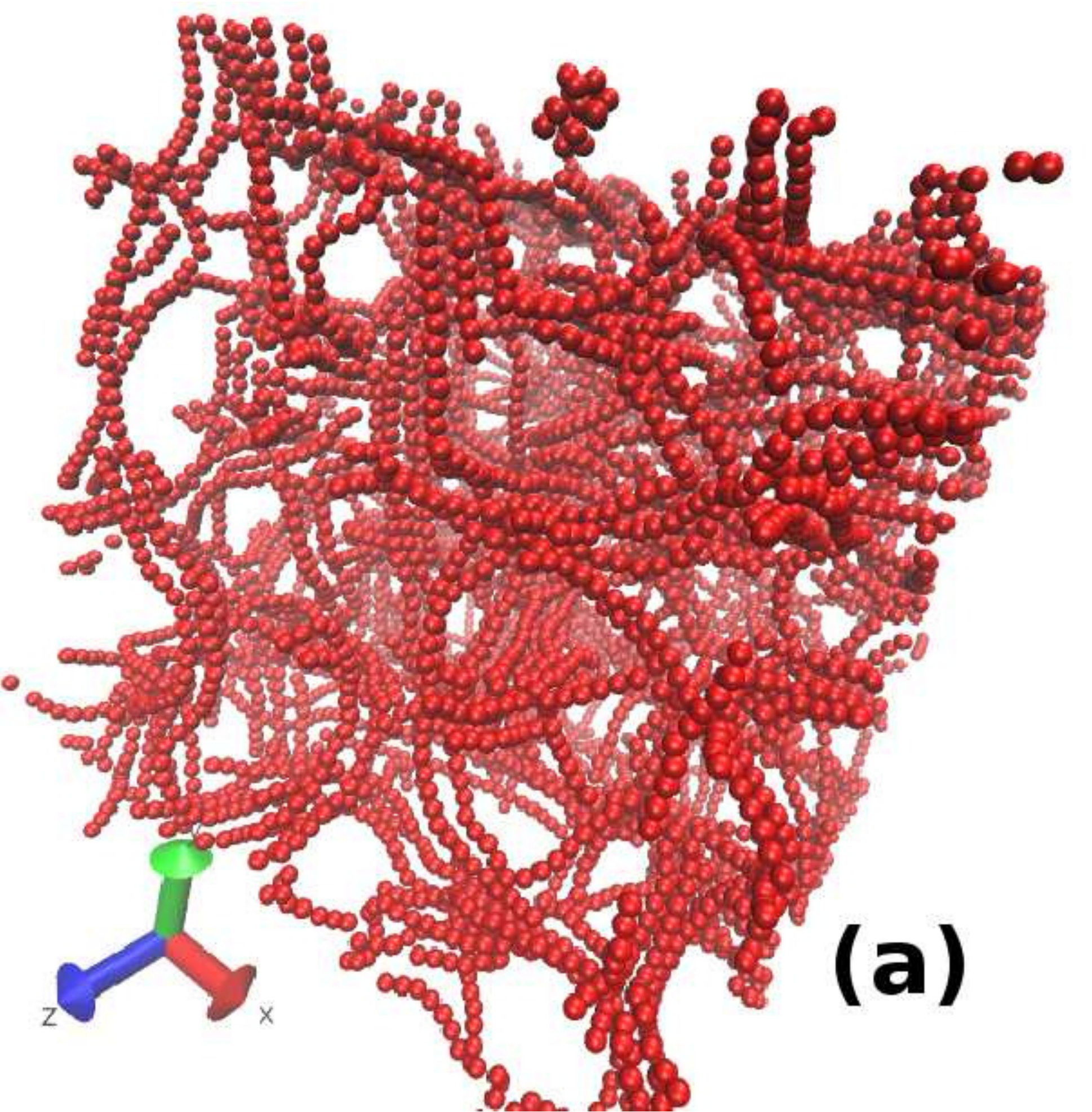}
\includegraphics[scale=0.2]{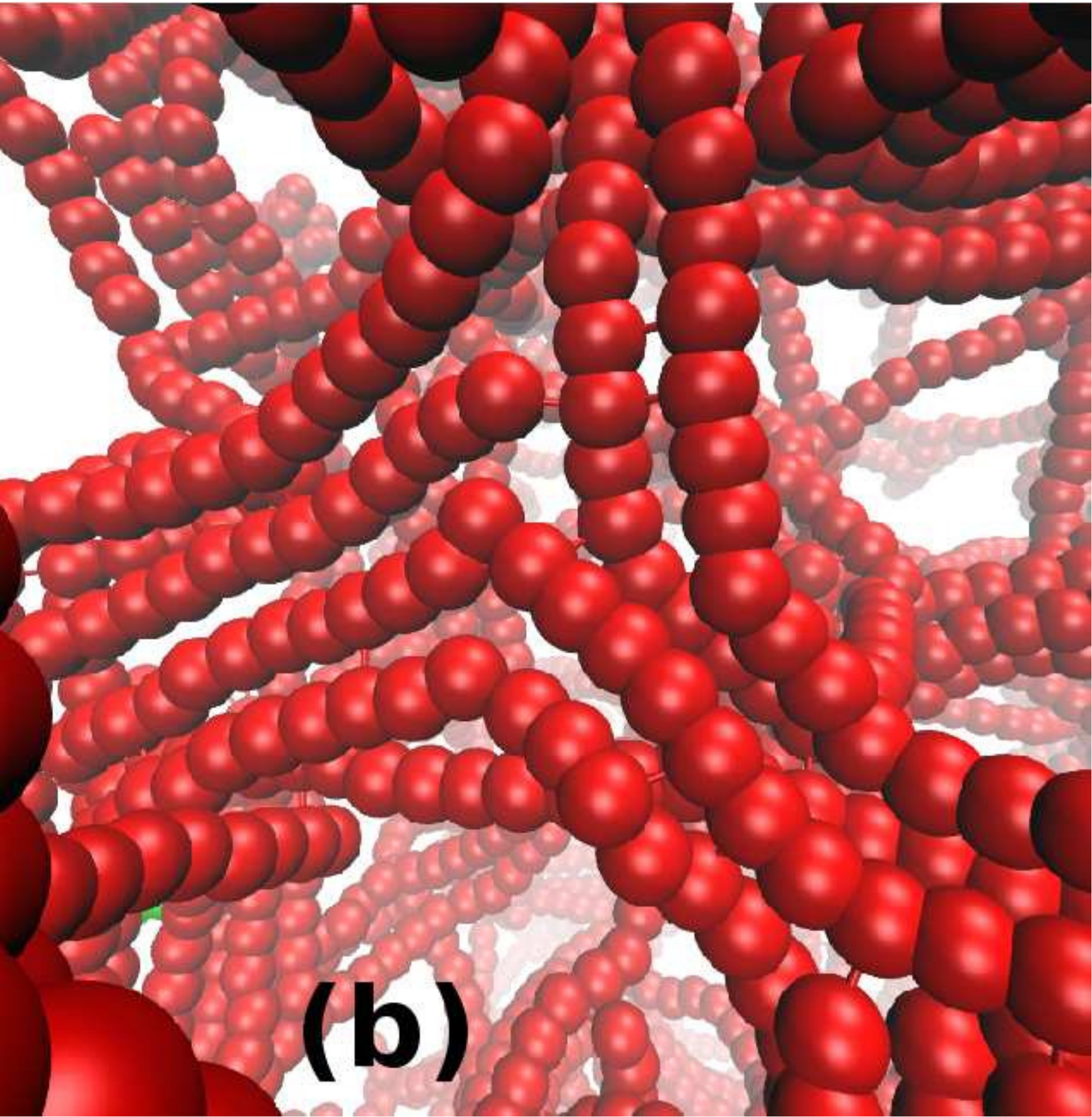}
\caption{The representative snapshot in (a) shows only the monomer configurations 
from a NP + micellar-polymer system  for  $\sigma_{4n}=2.5\sigma$ for the lowest 
value of monomer number density $\rho_m=0.037\sigma^{-3}$ for simulations performed in 
 a larger box size of $ 60\times 60\times 60\sigma^3$. 
This monomer-network is similar to the monomer-network structures obtained in a smaller box size 0f 
$30\times 30\times 30\sigma^3$ as shown in Fig.\ref{low_dens}. A magnified image of one of the 
network junction is shown in figure (b). Irrespective of the value of $\sigma_{4n}$, all the network 
structures obtained for $\rho_m=0.037\sigma^{-3}$ produce statistically similar monomeric network structures 
(with the same periodicity). }
\label{node}
\end{figure}

\begin{figure*}
\centering
\includegraphics[scale=0.21]{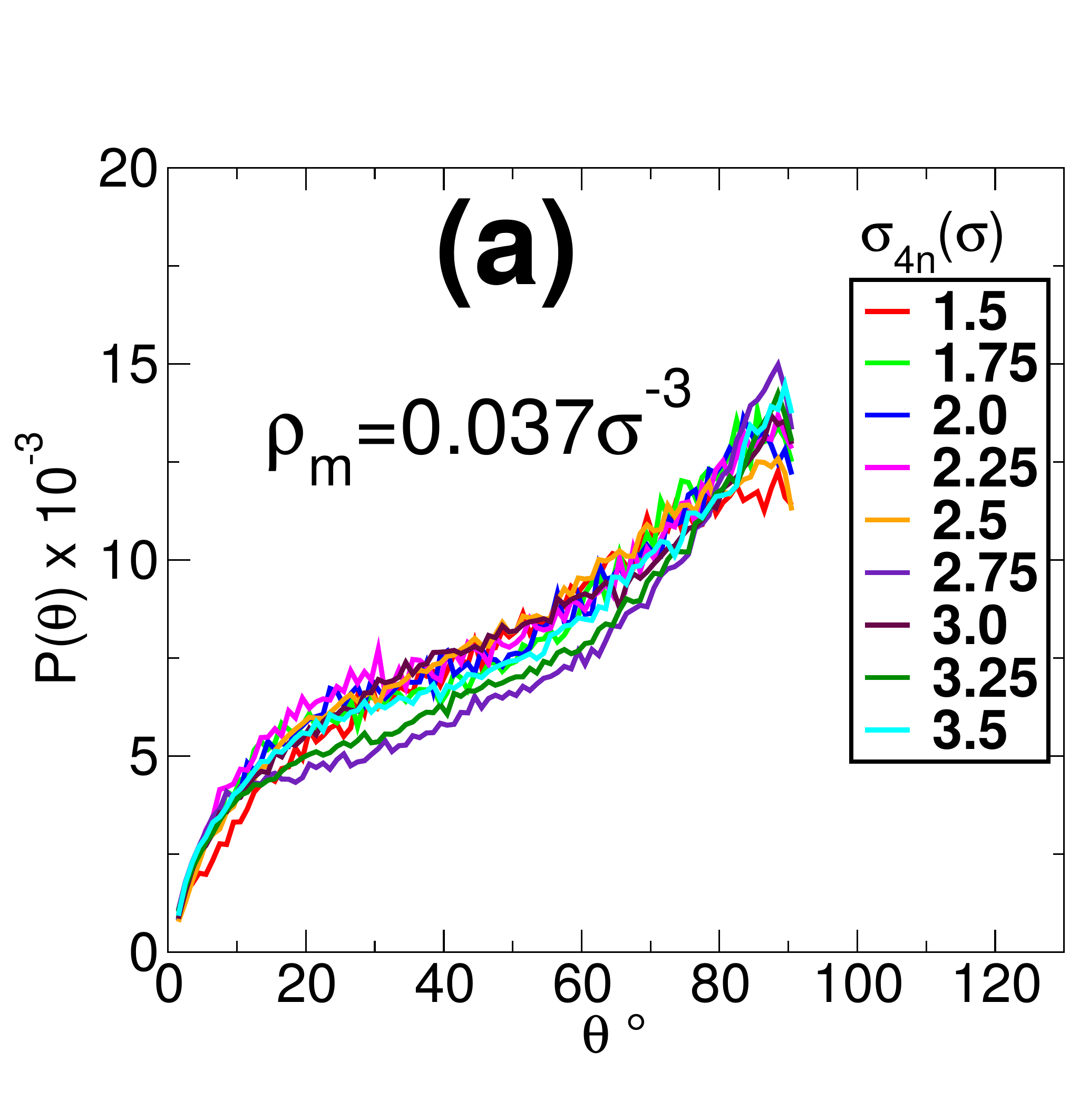}
\includegraphics[scale=0.21]{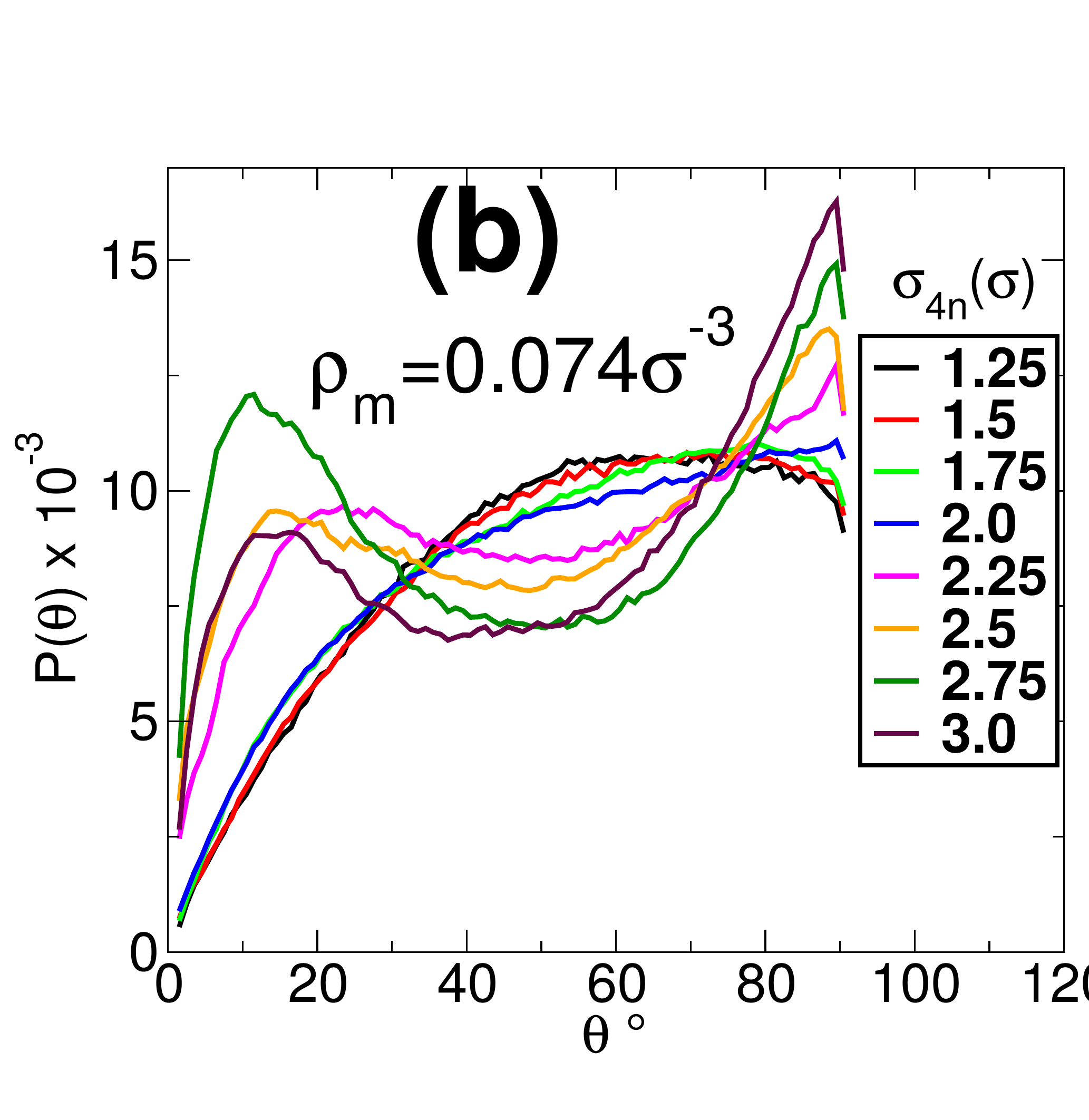}
\includegraphics[scale=0.21]{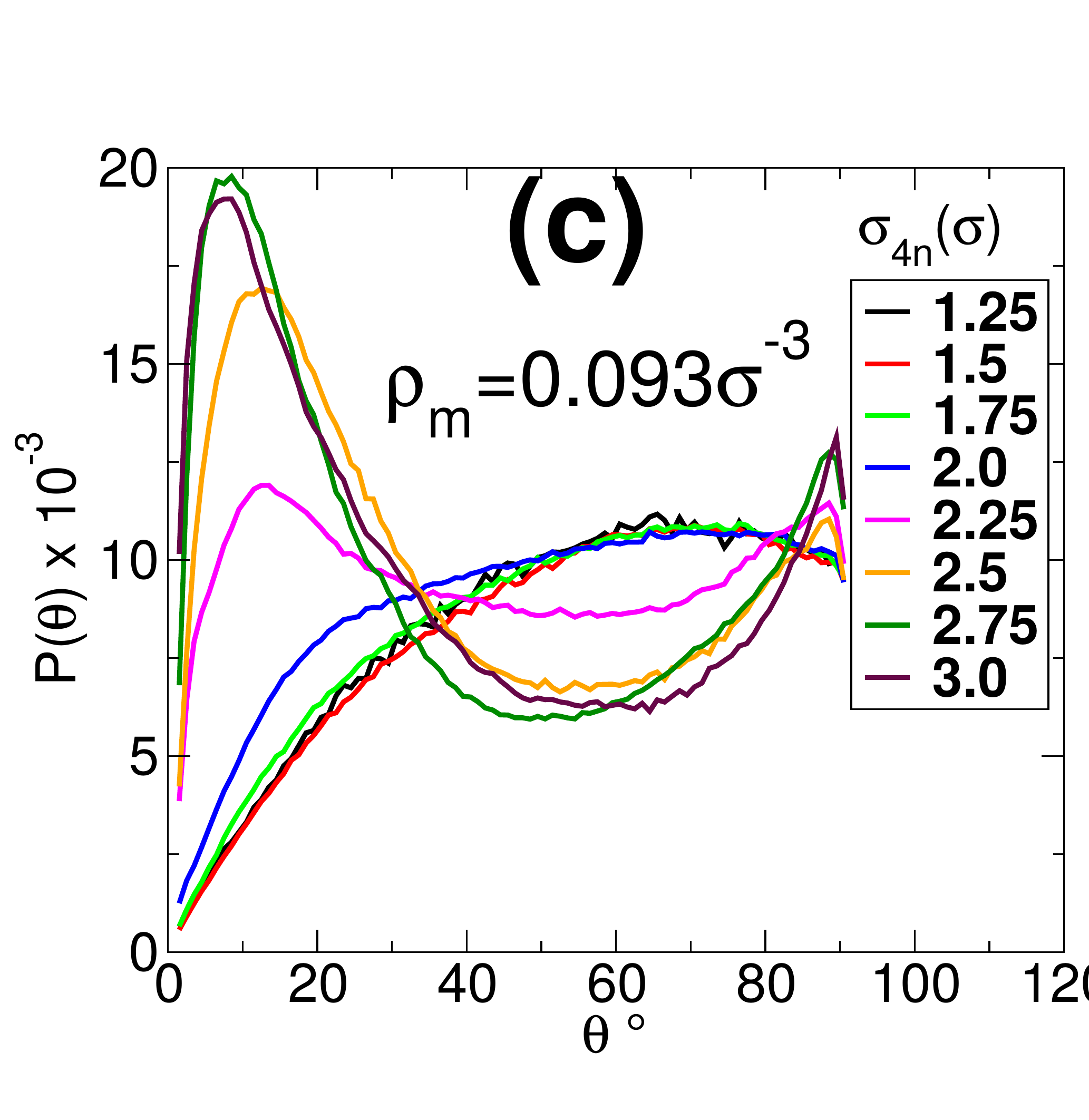}
\includegraphics[scale=0.21]{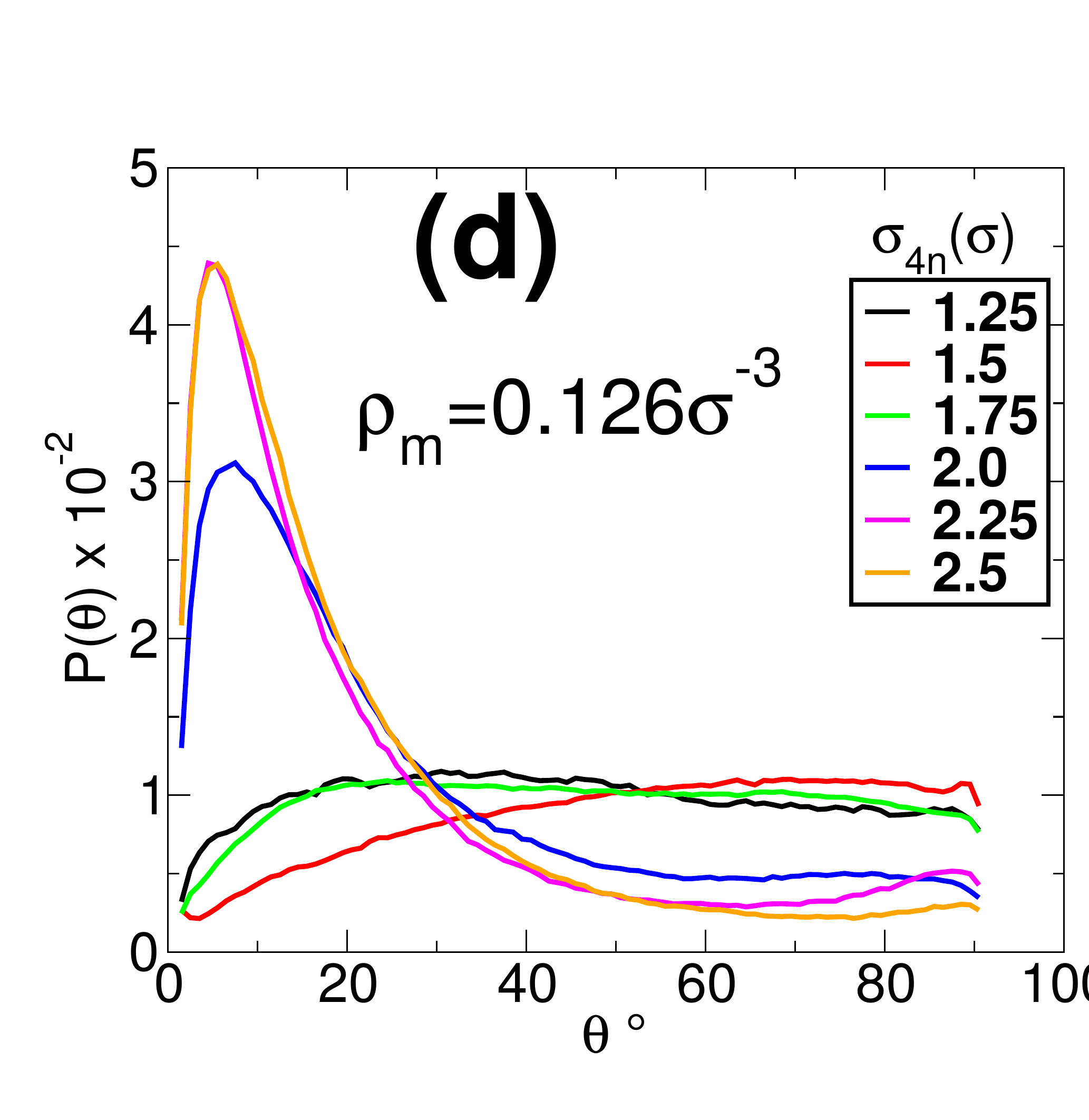}
\caption{The figure shows the distribution of angles subtended between pairs of micellar chains at
 different values of monomer densities, $ \rho_m$ = (a) $0.037\sigma^{-3}$, (b) $0.074\sigma^{-3}$, (c) $0.093\sigma{-3}$ and 
(d) $0.126\sigma^{-3}$, respectively. Each figure shows graphs for different values of $\sigma_{4n}$. 
For the lowest micellar density (a), it shows a peak at $90^{\circ}$ angle, hence confirming that 
many of the polymers lie to each other, as seen in snapshots of Fig.\ref{node}(b). Plots in (b) and (d) show 
two peaks one around $10^{\circ}$ and the other around $90^\circ$. The peak around $10^\circ$ corresponds 
to the formation of aligned domains of micellar chains which result in sheet-like structures, while the peak 
around $90^\circ$ shows the perpendicular arrangement of these sheet-like domains. The figure (d) shows 
a peak for a parallel arrangement of micellar chains which in turn results in the formation of rod-shaped
aggregates of nanoparticles.}
\label{ang_dist}
\end{figure*}

For the lowest micellar density considered $\rho_m=0.037\sigma^{-3}$, only the network-like structures 
are observed for the range of values of $\sigma_{4n}$ considered here (except for the case of 
$\sigma_{4}=1.25\sigma$). To ensure that the network architecture observed in the snapshots in 
Fig.\ref{low_dens} is not an artefact of the small box size, these structures were reproduced in a larger 
box size of $60 \times 60\times 60\sigma^3$ for all values of $\sigma_{4n}$ considered for 
$\rho_m=0.037\sigma^{-3}$. A representative snapshot for $\sigma_{4n}=2.5\sigma$ and $\rho_m=0.037\sigma^{-3}$ 
is shown in fig. \ref{node}(a). The figure shows that the snapshot have statistically similar network of 
micellar chains as in the smaller box size shown in Fig.\ref{low_dens}. The Fig.\ref{node}(b) shows a 
magnified image of one of the junctions of the network.

As seen in the snapshots of Figures.\ref{int_dens1}(h), \ref{int_dens2}(h) and \ref{high_dens}(g)\&(h), 
the anisotropy of the NP clusters depends on the micellar density. Moreover, the nanoparticle clusters that 
form a percolating network as in Figs.\ref{low_dens}(e), \ref{int_dens1}(e), \ref{int_dens2}(e) and 
\ref{high_dens}(e) also differ in their anisotropy thereby resulting in different pore shapes. Since, 
the structure of the nanoparticles also effects the arrangement of chains, or vice versa, the arrangement of 
micellar chains can be used to indirectly interpret the anisotropy of the NP structures. The distribution of 
angle between micellar chains $P(\theta)$ is analyzed and the plots are shown in Fig.\ref{ang_dist} for 
$\rho_m= 0.037\sigma^{-3}, 0.074\sigma^{-3}, 0.093\sigma^{-3}$, and $0.126\sigma^{-3}$ from (a) to (d), 
respectively. It is calculated by, 

\begin{equation}
P(\theta) = \frac{N(\theta)}{\sum_{\theta}N(\theta)}
\end{equation}

where, $N(\theta)$ is the total number of pair of chains at an angle $\theta$ and the angle $\theta$ between two chains is calculated by using the largest eigenvectors of the corresponding gyration tensors of the chains as follows,

\begin{equation}
\theta = \cos^{-1}\frac{\vec{e_1}.\vec{e_2}}{|\vec{e_1}||\vec{e_2}|}
\end{equation}

where, $\vec{e_1}$ and $\vec{e_2}$ are the largest eigenvectors of the gyration tensors of the two chains.

 The distribution $P(\theta)$ is normalized for different values of $\sigma_{4n}$ by dividing it by the 
number of all possible combinations of the pair of chains. Each plot in Fig.\ref{ang_dist} shows the 
distribution of angles between micellar chains for different values of $\sigma_{4n}$ as indicated in the 
figures. The lowest density of micellar system considered $\rho_m=0.037\sigma^{-3}$ shows a high value of 
the distribution around 90$^\circ$ for all values of $\sigma_{4n}$. This shows the perpendicular 
arrangement of micellar chains at the network junction as can be seen in the Fig.\ref{node}(b). 
The similar correlation plots in Fig.\ref{dens_corr} and the same distribution function for angles 
in Fig.\ref{ang_dist}(a) for $\rho_m=0.037\sigma^{-3}$ confirms the similarly periodic architecture of 
monomers networks for all $\sigma_{4n}$ as shown in Fig.\ref{low_dens}. The distribution plots for the 
monomer densities $\rho_m=0.074\sigma^{-3}$ and $0.0931\sigma^{-3}$ in figures \ref{ang_dist}(b) and (c) 
shows that after a particular value of $\sigma_{4n}$, there appear two peaks around $10^\circ$ and 
$90^\circ$ indicating a preference for a parallel and a perpendicular arrangement of micellar chains, 
respectively. These are the densities where NPs form sheet-like structures. The parallel arrangement 
indicates that the chains within an aligned micellar domain form sheet-like structures. The peak around 
$90^\circ$ in the distribution plot corresponds to the perpendicular arrangement of these sheet-like domains 
of aligned micellar chains. We find the formation of sheet-like domains to be unexpected and interesting 
and the reason for this will be explained later in the text. All the neighbouring NP sheets as shown in 
Figure. \ref{int_dens1} and \ref{int_dens2} are arranged such that their planes are perpendicular to each 
other. Finally for the highest micellar density considered $\rho_m=0.126\sigma^{-3}$, for 
$\sigma_{4n} \geq 2\sigma$, there appears a high peak around $10^\circ$. Here, the presence of peaks only 
around $10^\circ$ is indicative of the nematic ordering of micellar chains and correspondingly we observe 
rodlike structure of NPs. The height of the peaks at $90^\circ$ gradually reduces from (b) to (d) and 
resulting in only peaks for parallel arrangement in (d). This is indicative of the change in the anistropy 
of NP clusters. This can be confirmed by calculating the shape anisotropy $S_{AN}$ of NP clusters using the 
following formula,

\begin{equation}
S_{AN} = 1 - 3\frac{\lambda_1\lambda_2+\lambda_2\lambda_3+\lambda_1\lambda_3}{(\lambda_1+\lambda_2+\lambda_3)^2}
\label{eq_anis}
\end{equation}

where, $\lambda_1,\lambda_2,\lambda_3$ are the eigen values of the gyration tensor of the NP cluster.
Using the above formula, the average shape anisotropy of the nanoparticle clusters shown in snapshots of 
Figs.\ref{int_dens1}(h), \ref{int_dens2}(h) and \ref{high_dens}(h) is calculated to be $0.165,0.211$ and 
$0.414$, respectively.  This shows that the shape anisotropy of the NP clusters increases with increase 
in micellar density and the shape of the NP clusters is varying from sheet-like to rod-like structures.
 This shows that the anisotropy of the NP structures increases with increase in micellar density with 
nanostructures varying from sheets to rodlike structures.

\subsection{The morphological changes : our understanding}

     The purpose of the introduction of the parameter $\sigma_{4n}$ is to control the minimum approaching 
distance between micelles and NPs, but it also influences the arrangement and the number of NPs in the 
system and hence affects the effective volume of micelles. An increase in the value of $\sigma_{4n}$ 
indicates an increase in the effective volume of micelles. We remind the reader that, as discussed in the 
beginning of the section (refer Fig.\ref{eff_vol}), the effective volume of micelles also gets modified 
by the density of NPs present in the system. In the presence of a large number of NPs (lower value of 
$\sigma_{4n}$), there are relatively more contacts between monomers and NPs, which leads to a higher value 
of the effective volume of monomers. Moreover, the number of contacts also depend on the assembly 
architecture and the relative organization between NPs and micellar chains. Therefore, the effective 
volume of micelles not only depends on the value of $\sigma_{4n}$ but also on the NP density as well as 
the morphology and corresponding monomer-NP contacts. Hence, the effect of the transformation from 
network-like structure to non-percolating clusters can be understood in terms of change in the effective 
volume of micelles. 

From the experimental perspective, different values of $\sigma_{4}$ would correspond to micellar chains 
made up of different chemical compositions such that they have different diameters of micellar chains. 
Here, it is kept constant ($\sigma_4=1.75\sigma$). Now, with a given value of $\sigma_4$ and micellar density, 
it is the parameter $\sigma_{4n}$ that effectively determines the space available for the incoming NPs 
inside the simulation box. Experimentally, in the micelle-NP mixture, the effective repulsion between micellar 
chains of a particular composition (with a particular value of $\sigma_4$) and NPs can be tuned in to 
change the value of $\sigma_{4n}$ and thereby the effective volume available to NPs. That, in turn, will 
determine the morphology of NP mesostructures formed in the presence of a background micellar matrix. 

A characteristic quantity that changes with the density of monomers is the average length of monomer chains. 
The average chain length of WLMs increases with increase in the value of monomer density. In presence of a 
large number of NPs (in case of system spanning network of NPs), monomers are more packed.  This is also seen from 
the behaviour of $g(r)$ in Fig.\ref{corr_mono}) and can be compared to the systems with a lower number of NPs 
which result in  non-percolating clusters.   Therefore, the effect of the change of NP morphology from 
network to non-percolating clusters is expected to be seen in the value of average chain length of micelles. 
We first discuss the behaviour of the average length of micellar chains with a change in $\sigma_{4n}$, 
because, that will help us quantitatively better appreciate the behaviour of the effective volume 
fraction of micelles.

\begin{figure}
\centering
\includegraphics[scale=0.3]{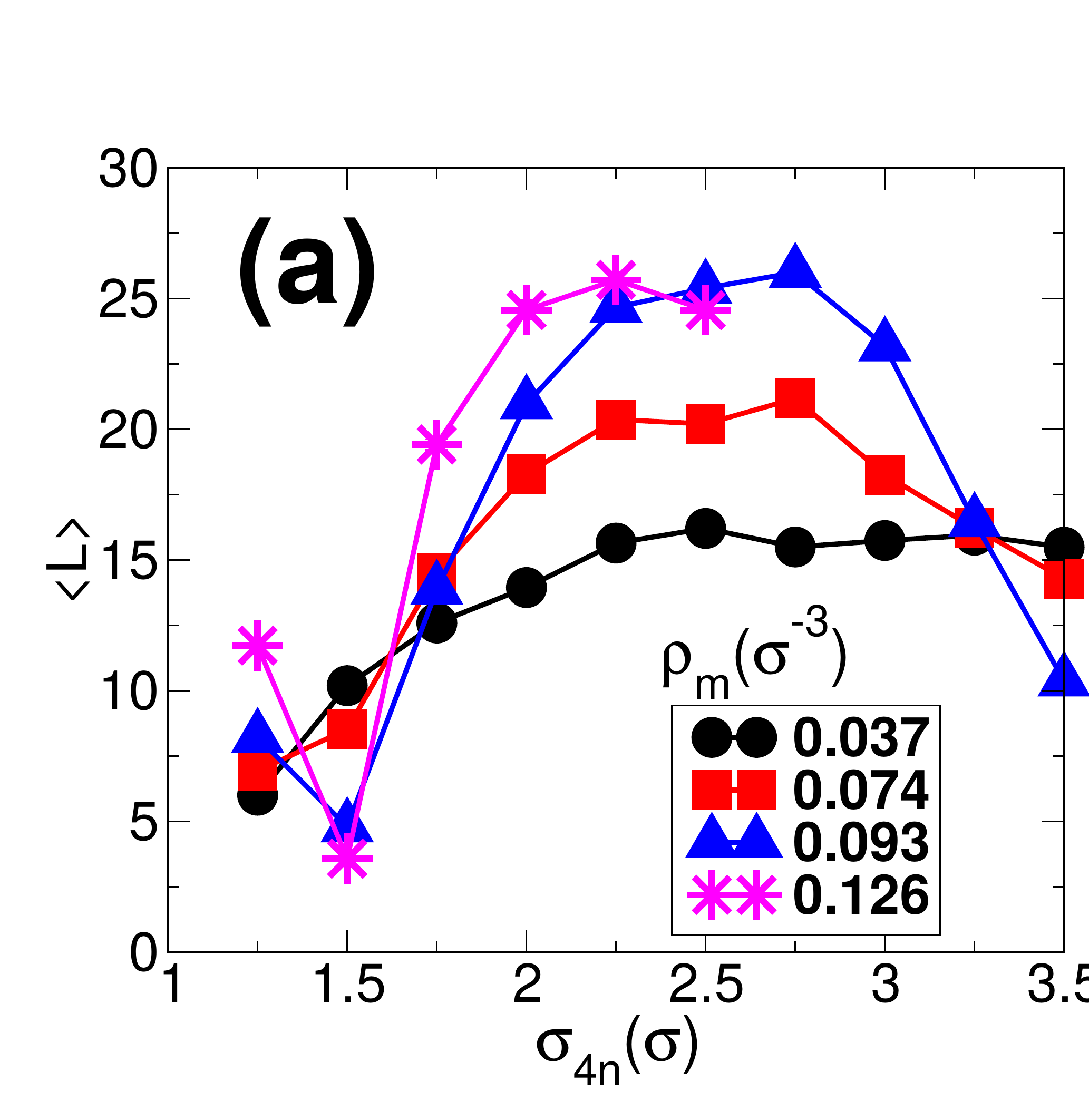}
\includegraphics[scale=0.3]{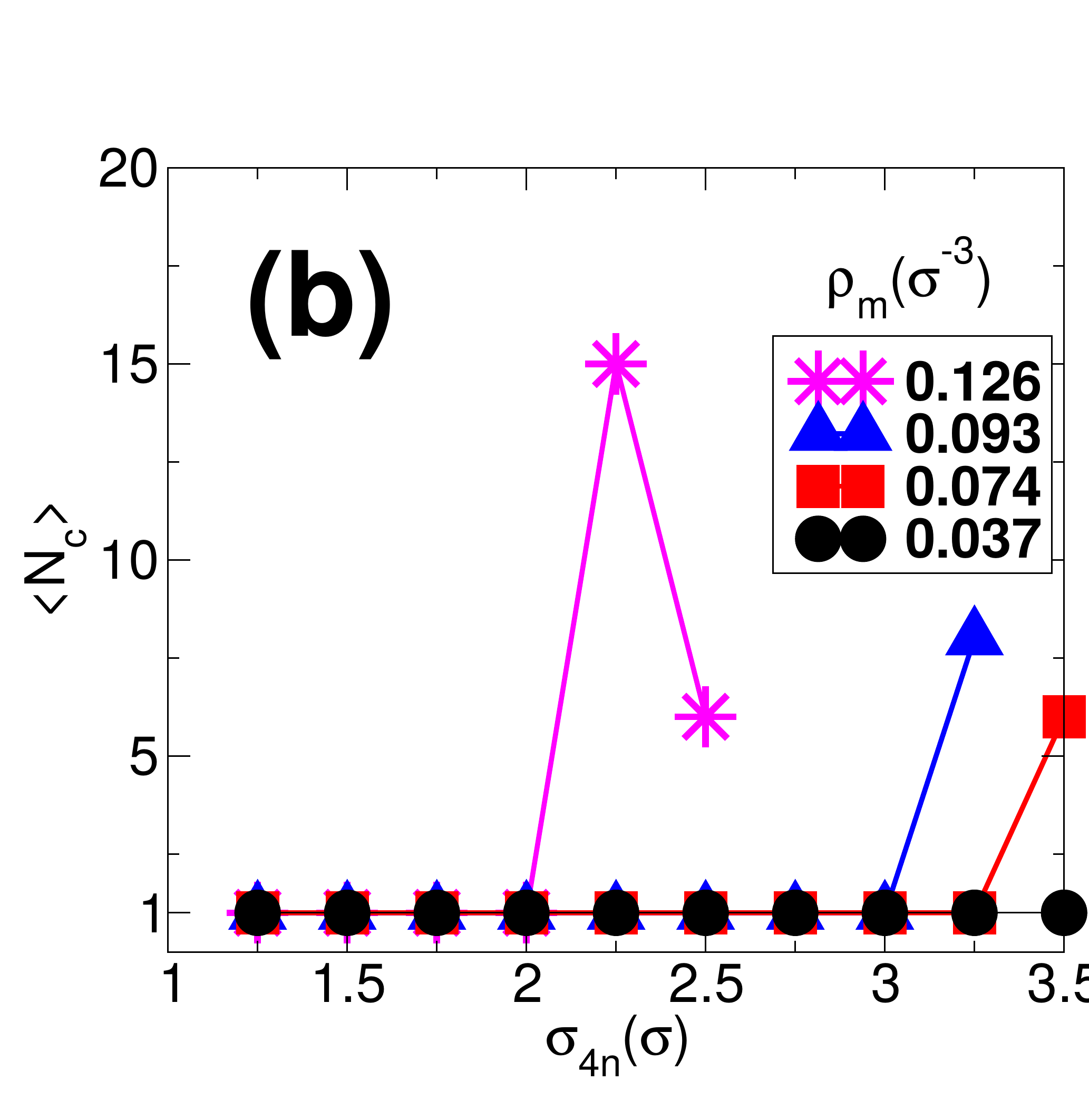}
\caption{(colour online) Figure (a) shows the variation in the average length $\langle L \rangle$ of micellar chains with 
$\sigma_{4n}$ where, L is the number of monomers in a particular chain. The figure (b) shows the plot for 
number of nanoclusters $\langle N_c \rangle$ in the system. The quantity $\langle L \rangle$ shows a decrease 
in its value at a higher 
value of $\sigma_{4n}$ where nanoparticle network breaks into non-percolating clusters. The quantity $<N_c>$ 
shows a jump  from its value of $1$ to a larger value when, the nanoparticle network breaks into 
non-percolating clusters.The nanoparticles aggregates with $<N_c>=1$ spans the length of the simulation box 
as seen in the snapshots and also confirmed by calculating the largest eigen value of the gyration tensor 
for these aggregates.}
\label{len}
\end{figure}

\begin{figure*}
\centering
\includegraphics[scale=0.28]{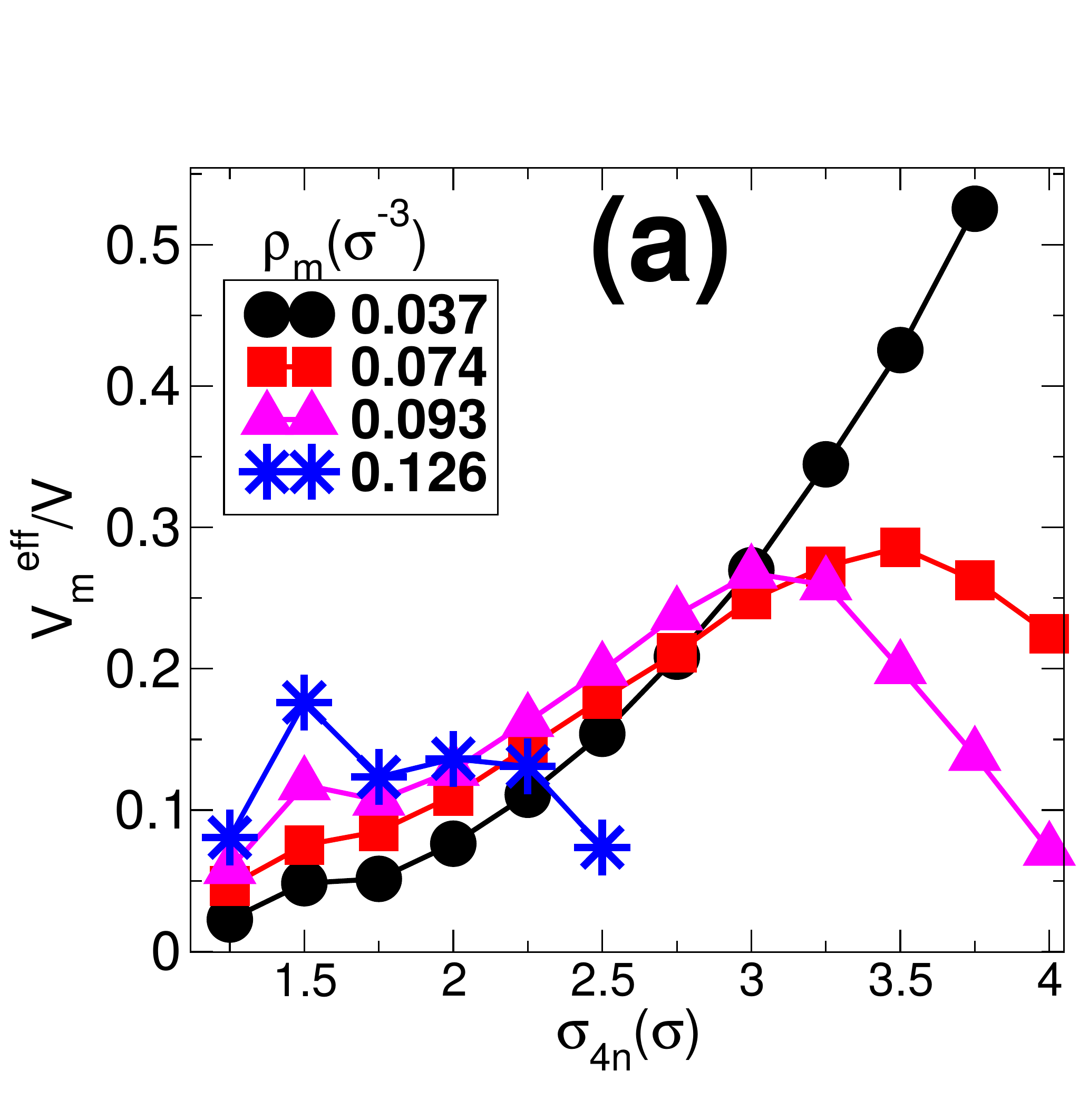}
\includegraphics[scale=0.28]{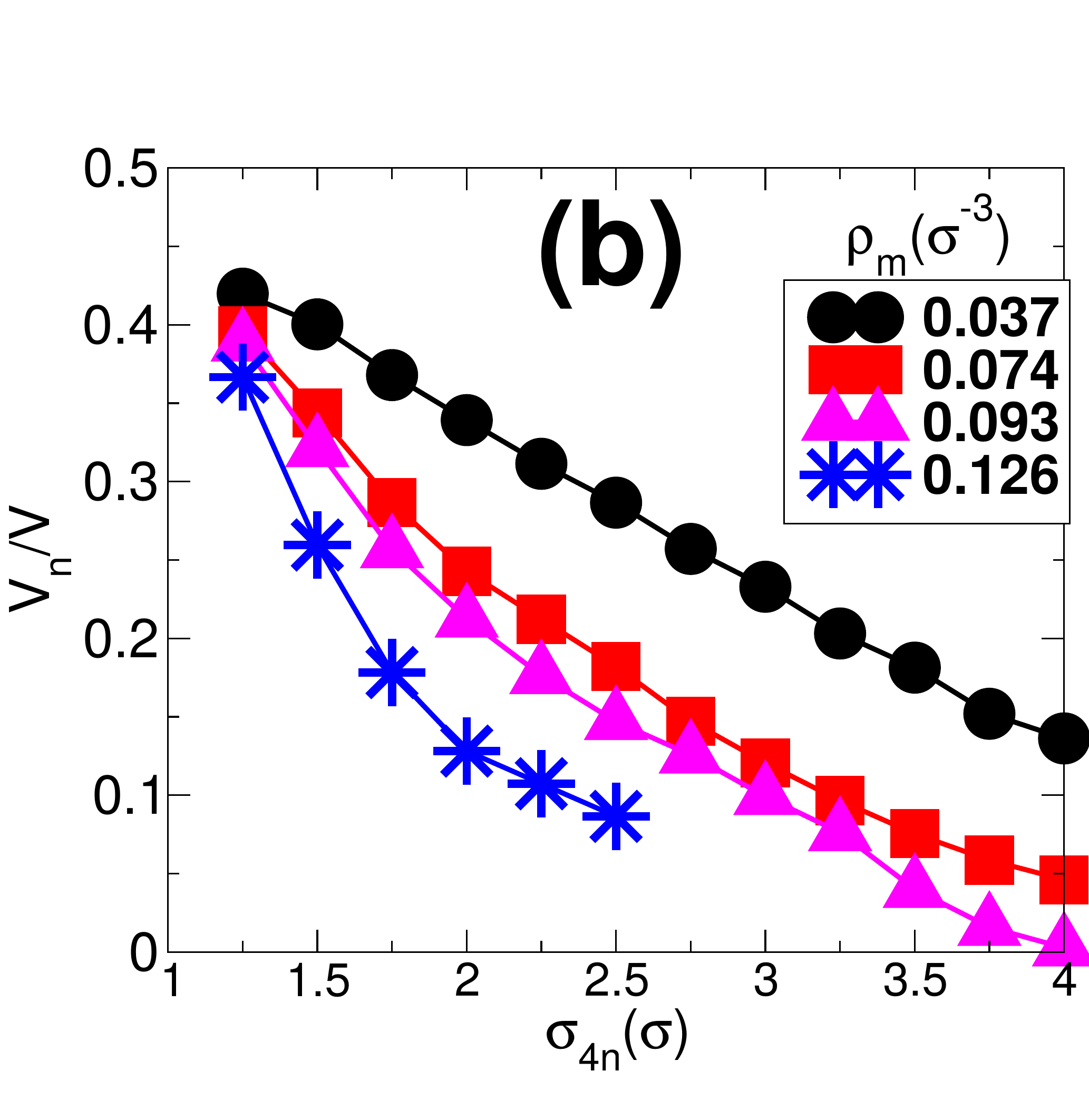}
\includegraphics[scale=0.28]{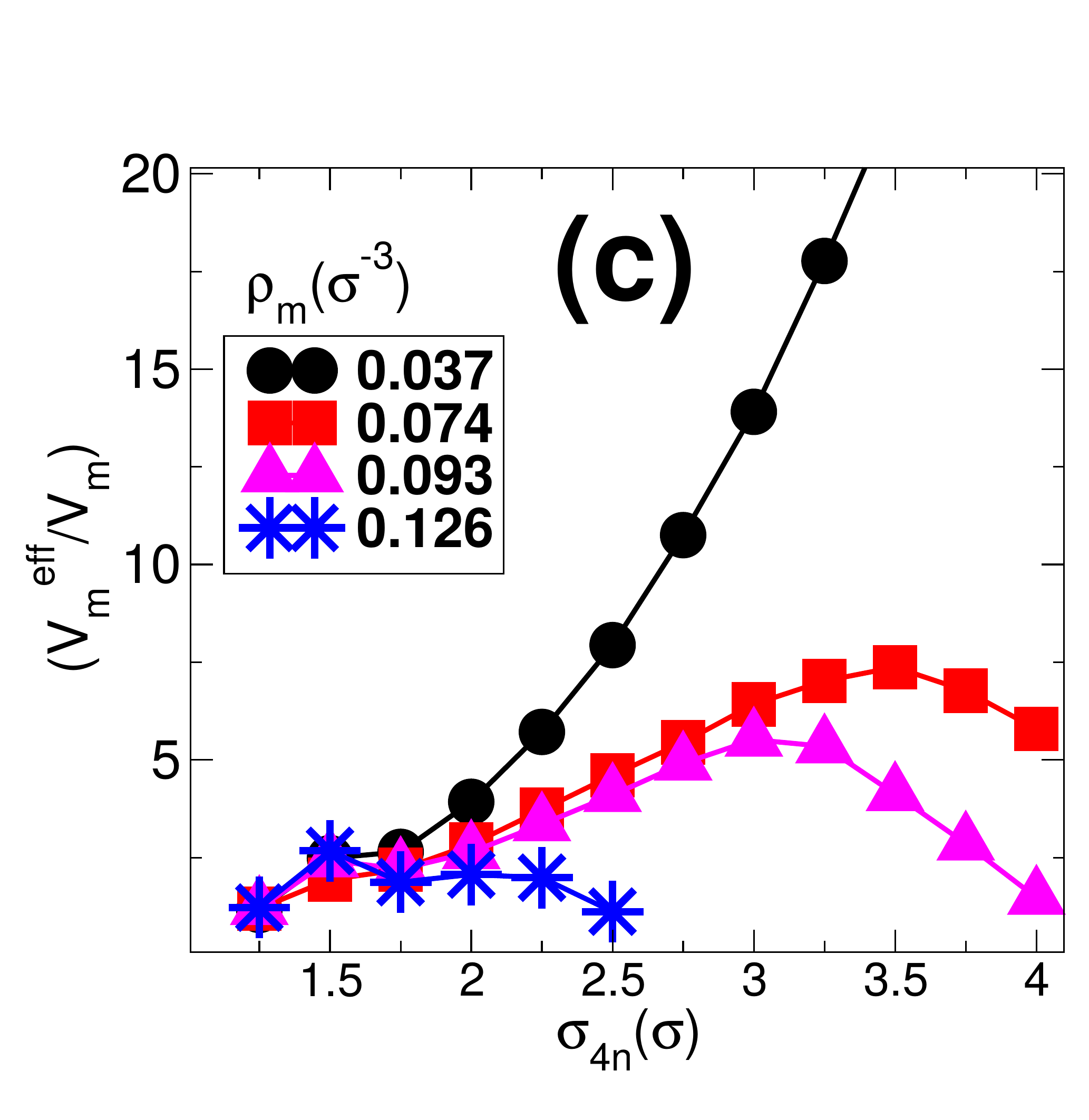}
\caption{(colour line) The figure shows the behaviour of (a) effective monomer volume fraction 
(${V_m}^{eff}/V$), (b) nanoparticle volume fraction ($V_n/V$) and  (c) the ratio of effective monomer volume 
to the  monomer volume (${V_m}^{eff}/{V_m}$), with change in the value of $\sigma_{4n}$, where 
$V_m=\rho_m 4/3\pi(\sigma/2)^3V$. Each figure shows graphs for four different values of number density of 
monomers i.e. $\rho_m=0.037\sigma^{-3}, 0.074\sigma^{-3},0.093\sigma^{-3}$ and $0.125\sigma^{-3}$ as indicated 
by the symbols. With an increase in the value of $\sigma_{4n}$, the nanoparticle volume fraction in (b) 
decreases while the effective monomeric volume fraction in (a) increases. At a higher value of 
$\sigma_{4n}$, the monomeric volume fraction shows a decrease in its value indicating a change of 
nanoparticle structure from network to non-percolating clusters. In figure (c), the graph represents the 
variation of the ratio of the effective monomer volume fraction to the monomer volume fraction. In other 
words, it represents the extent of the effect of the repulsive interactions on effective monomer volume 
fraction. A high value of ${V_m}^{eff}/{V_m}$ indicates that the presence of a large number of interacting 
pairs having repulsive interactions $V_4$ and $V_{4n}$, which have a cutoff distances $\sigma_4$ and 
$\sigma_{4n}$, respectively. The graph in (c) shows high values for intermediate values of $\sigma_{4n}$, 
where the system shows the formation of a network-like structure of nanoparticles. However, for very low 
values or high values of $\sigma_{4n}$, the ratio $V_m/{V_m}^{eff}$ shows low values ($ \sim 0.1$ for 
$\rho_m=0.126\sigma^{-4}$). This shows that the constituent particles in the system are away from each 
other, i.e., at a distance greater than the cutoff distances of the repulsive interactions $V_4$ and $V_{4n}$, 
for very low or higher values of $\sigma_{4n}$.}
\label{vol_frac}
\end{figure*}

    The behaviour of average micellar chain length is shown in Fig.\ref{len}(a). Different symbols indicate 
different micellar number densities. The average length at $\sigma_{4n}=1.25\sigma$ increases with increase in 
the number density of monomers. This is also consistent with the snapshots shown in Fig.\ref{crystal}. 
In that figure, we observe longer chains for higher values of monomer number densities. An increase in the 
value of $\sigma_{4n}$ from $1.25\sigma$ to $1.5\sigma$, reverses this behaviour. This is due to the fact 
that, at $\sigma_{4n}=1.5\sigma$, the micellar chains undergo a change in their arrangement from dispersed 
to network-like structure with clusters of micellar chains. This transition is such that the network-like 
structure of NPs has the same period for all micellar densities (refer Fig.\ref{dens_corr} and the 
corresponding snapshots in Fig.\ref{low_dens}, \ref{int_dens1}(a), \ref{int_dens2}(a) and \ref{high_dens}(a)). 
At $\sigma_{4n}$ $=1.5\sigma$, with the same periodicity of NP networks (refer Fig.\ref{dens_corr}), 
micellar chains with a high $\rho_m$ are more "crushed" (having smaller chains) compared to the systems with
lower density. Therefore, denser micellar systems have lower average length of micellar chains for this 
particular value of $\sigma_{4n}$. This can also be 
realised by comparing the snapshots shown in Fig.\ref{low_dens}(a), \ref{int_dens1}(a), \ref{int_dens2}(a) 
and \ref{high_dens}(a). With further increase in the value of $\sigma_{4n}$, the NP networks start breaking. 
Due to the presence of higher number of monomer-nanoparticle contacts in the systems with higher monomer 
densities, the effect of the increase in the value of $\sigma_{4n}$ is also higher. Therefore, for the 
same increase in the value of $\sigma_{4n}$ beyond $1.5\sigma$, the decrease in the number of nanoparticles 
is more for a higher value of $\rho_m$. Hence, the nanoparticle network is ``{\em more broken}'' for higher $\rho_m$.
Hence, the average length reverses its behaviour  at $1.75\sigma$, and the average length again increases 
with $\rho_m$ as was seen for the $\sigma_{4n}=1.25\sigma$ system.

As explained earlier, the effective polymeric chain volume depends on the combination of the value of $\sigma_{4n}$, 
the NP number density and more importantly, the number of contacts with NP and monomers. 
Therefore, at a higher value of $\sigma_{4n}$ ($> 2.5\sigma$  when the introduction of NPs 
inside the system becomes relatively difficult), the effective micellar volume decreases and hence a decrease 
in the average length of micellar chains can be seen. This decrease in the average length shows the extent of 
the breaking of NP network, and the breaking of network-like structures to non-percolating clusters of NPs 
is recorded once the average length of chains start to decrease. This occurs at $\sigma_{4n}=2.25\sigma, 
3.25\sigma$ and $3.5\sigma$ for the values of micellar densities $\rho_m=0.126\sigma^{-3}, 0.093\sigma^{-3}$ 
and $0.074\sigma^{-3}$, respectively. Thus, the behaviour of the average length of micelles clearly indicates 
the existence of two points of structural changes. The first one is for the change from dispersed state to 
clustering of micellar chains (from $\sigma_{4n}=1.25\sigma$ to $\sigma_{4n}=1.5\sigma$) and the other is for 
the breaking of NP network into non-percolating clusters at some higher value of $\sigma_{4n}$ depending 
on micellar density. 

\begin{figure}
\centering
\includegraphics[scale=0.2]{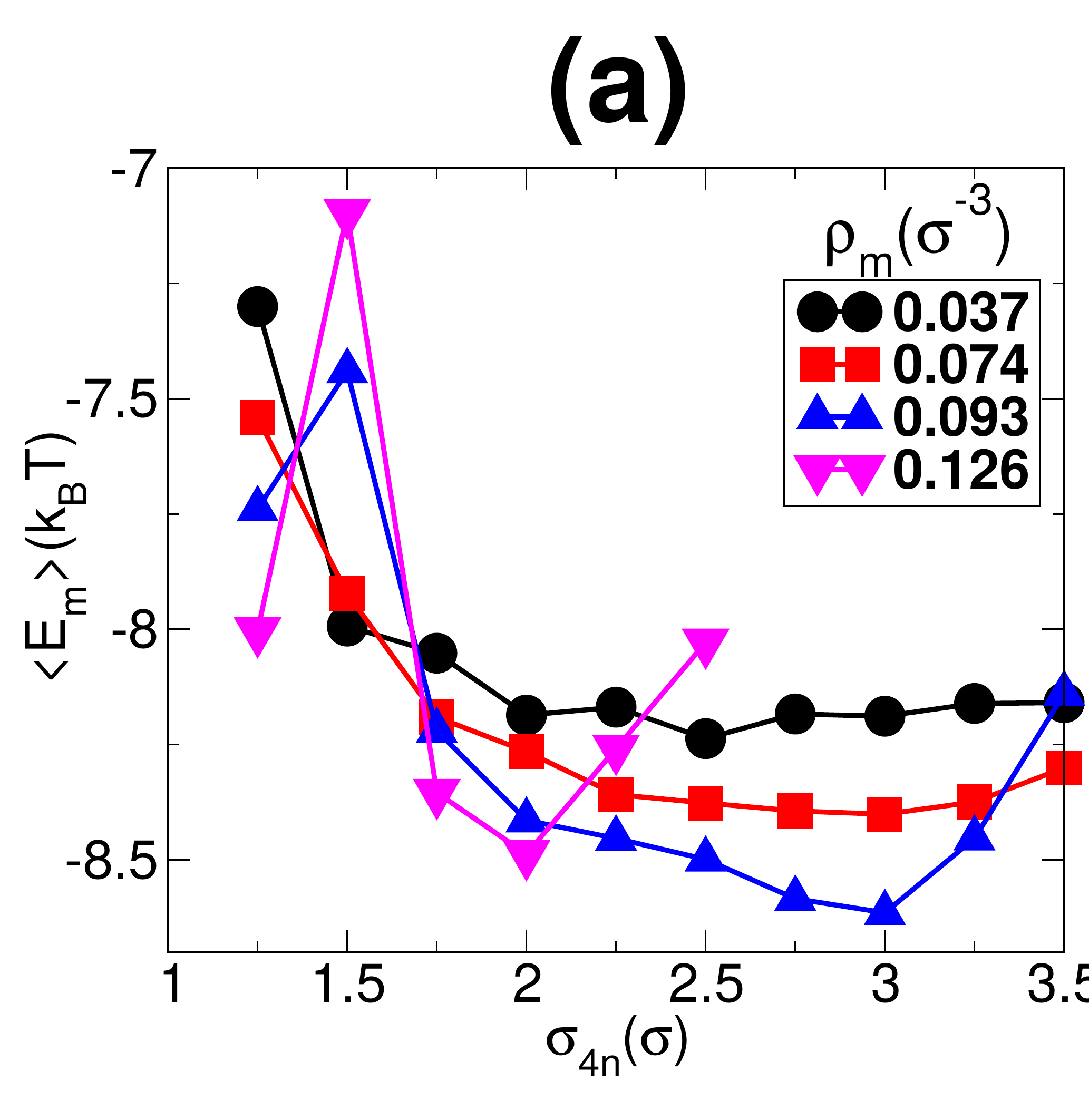}\\
\includegraphics[scale=0.2]{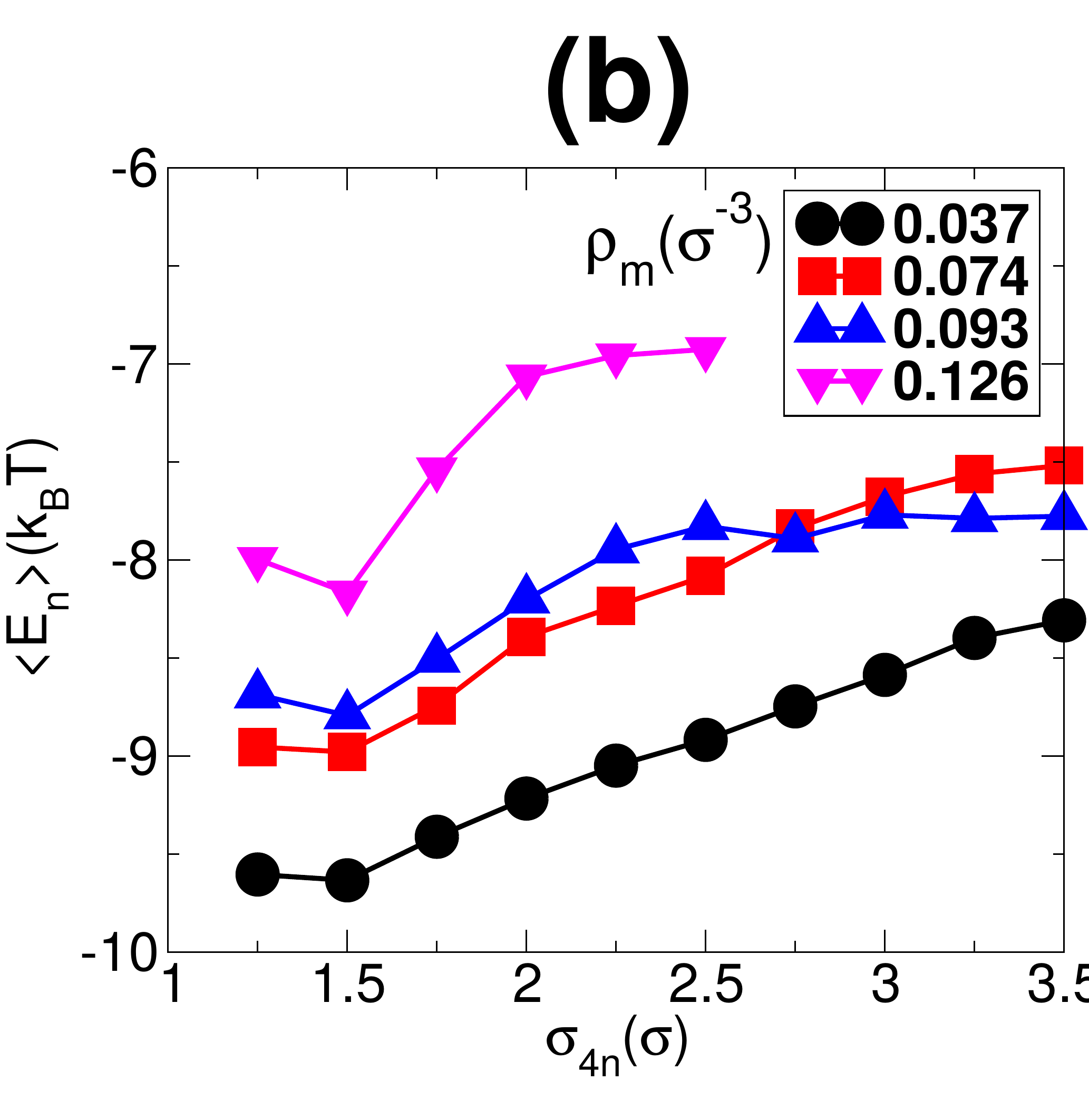}\\
\includegraphics[scale=0.2]{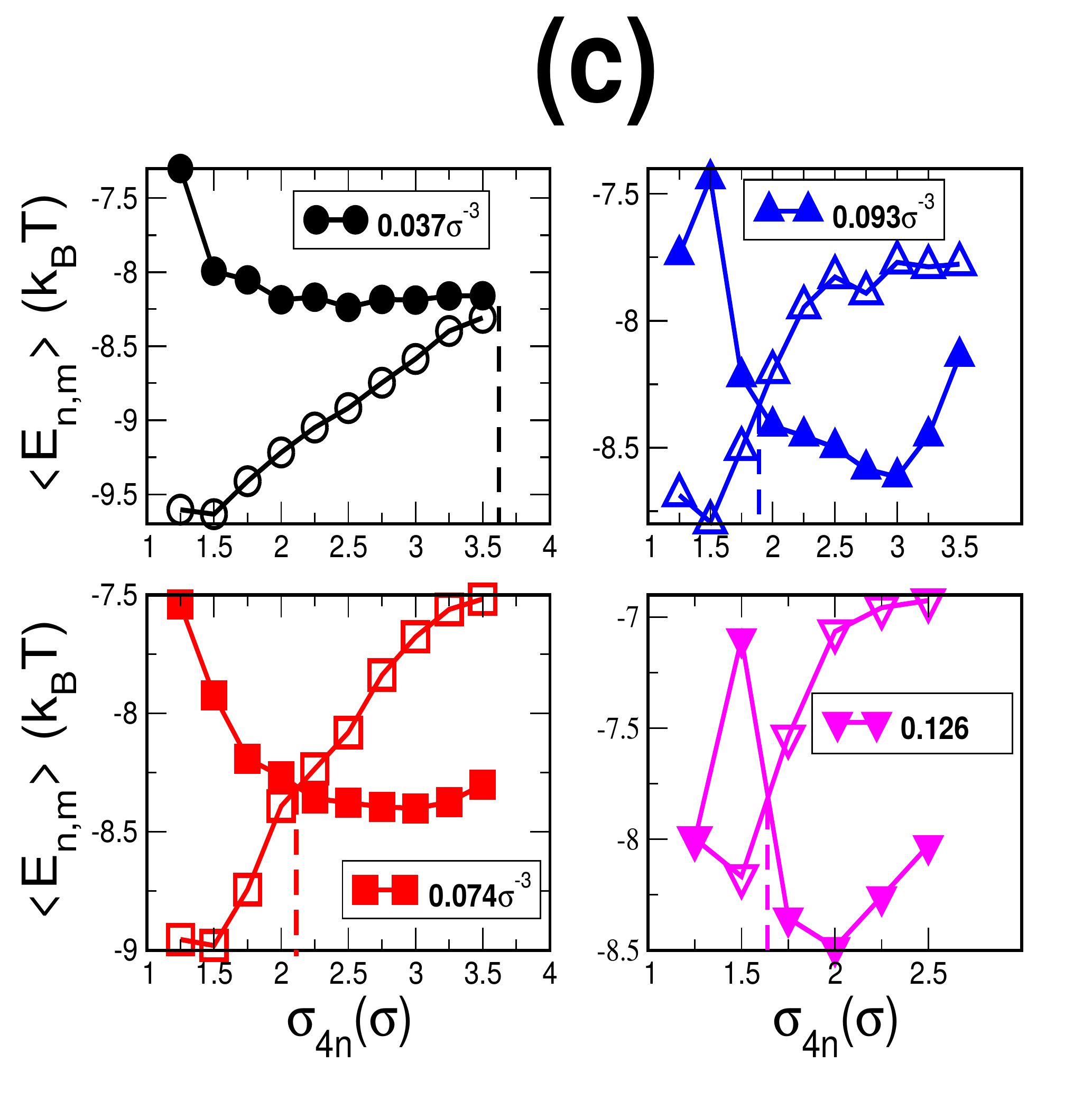}\\
\caption{The Figure shows graphs for variation in the average potential energy (PE) of (a) monomers, (b) nanoparticles and 
for different values of monomer densities. Plot (c) shows the monomer and NP PEs in the same plot, but each 
subfigure shows data for a particular value of $\rho_m$. With an increase in the value of $\sigma_{4n}$, the 
nanoparticle energy decreases while the energy of micelles increases. At a higher value of $\sigma_{4n}$, 
the monomer volume fraction shows a decrease in its value marking the change from network-like structure to 
non-percolating clusters of nanoparticles. In (c), the  lines showing variation in monomer PE (filled symbols)
and PE of nanoparticles (empty symbols) intersect at some intermediate value of $\sigma_{4n}$. 
The change from network to non-percolating clusters of the nanoparticle takes place at a value of 
$\sigma_{4n}$ higher than the $\sigma_{4n}$-value at which the intersection occurs. The points of intersections of 
the energies of monomers and nanoparticles for different values of monomer densities are indicated by the 
vertical lines which get shifted to lower values of $\sigma_{4n}$ with the increase in micellar densities. 
Therefore, with an increase in the micellar densities the point of change of nanoparticle morphology from 
network to non-percolating clusters also gets shifted to a lower value of $\sigma_{4n}$.}
\label{en_nano_mono}
\end{figure}

Since the change in average length indirectly indicates a change in effective volume of micelles, a decrease 
in the effective volume of micelles is expected around the point of transformation. To confirm these observations, 
the behaviour of the volume fractions of the constituents and the average energies of NPs and monomers are
examined, and the plots are shown in Figs.\ref{vol_frac} and \ref{en_nano_mono}, respectively.

The Fig.\ref{vol_frac} shows (a) the plot of the effective monomer volume fraction ${V_m}^{eff}/V$ (b) NP volume 
fraction $V_n/V$ and (c) the ratio of effective volume fraction of monomers to monomers volume fraction 
${V_m}^{eff}/V_m$. Change in the value of $\sigma_{4n}$ from $1.25\sigma$ to $1.5\sigma$ corresponds to the 
formation of clusters of micellar chains from a dispersed state as discussed before. These clusters of micellar 
chains join to form a system spanning network-like structure. With further increase in $\sigma_{4n}$, 
the effective volume fraction of monomers increases at first but then  starts to decrease 
after a certain value of $\sigma_{4n}$ ($> 2.5\sigma$) for the highest three number densities. 
The decrease in the value of the effective volume fraction of monomers corresponds to the 
transition from the network-like structure of nanoparticles to non-percolating clusters. This happens at 
$\sigma_{4n}=2.25\sigma$ for $\rho_m=0.126\sigma^{-3}$, $\sigma_{4n}= 3.25\sigma$ for $\rho_m= 0.093\sigma^{-3}$ 
and at $\sigma_{4n}=3.5\sigma$ for $\rho_m=0.074\sigma^{-3}$. This change in nanoparticle structure is not 
observed for the lowest density of micelles for the values of $\sigma_{4n}$ considered here. Hence, 
${V_m}^{eff}/V_m$ keeps on increasing for $\rho_m=0.037\sigma^{-3}$. In Fig.\ref{vol_frac}(b), the 
nanoparticle volume fraction decreases with increase in $\sigma_{4n}$ as introducing nanoparticles in 
the system becomes increasingly difficult with increasing $\sigma_{4n}$. 

The ratio of the effective volume fraction of monomers to their 
actual volume fraction [refer Fig.\ref{vol_frac}(c)] shows the extent of the effect of $\sigma_{4n}$ on the 
effective volume fraction of micelles. Its behaviour is exactly similar to the behaviour of the effective 
volume of monomers. The behaviour of the ratio ${V_m}^{eff}/V_m$  indicates the role of $\sigma_{4n}$ and 
micellar density in the arrangement of constituent particles and hence the morphology of the system. 
A high value of ${V_m}^{eff}/V_m$ is obtained when there exists a clustering of micellar chains as a 
consequence of an increase in $\sigma_{4n}$. When $(V_m^{eff}/V_m)\approx 1$, it shows a dispersed state of 
micellar chains.

\begin{figure*}
\centering
\includegraphics[scale=0.4]{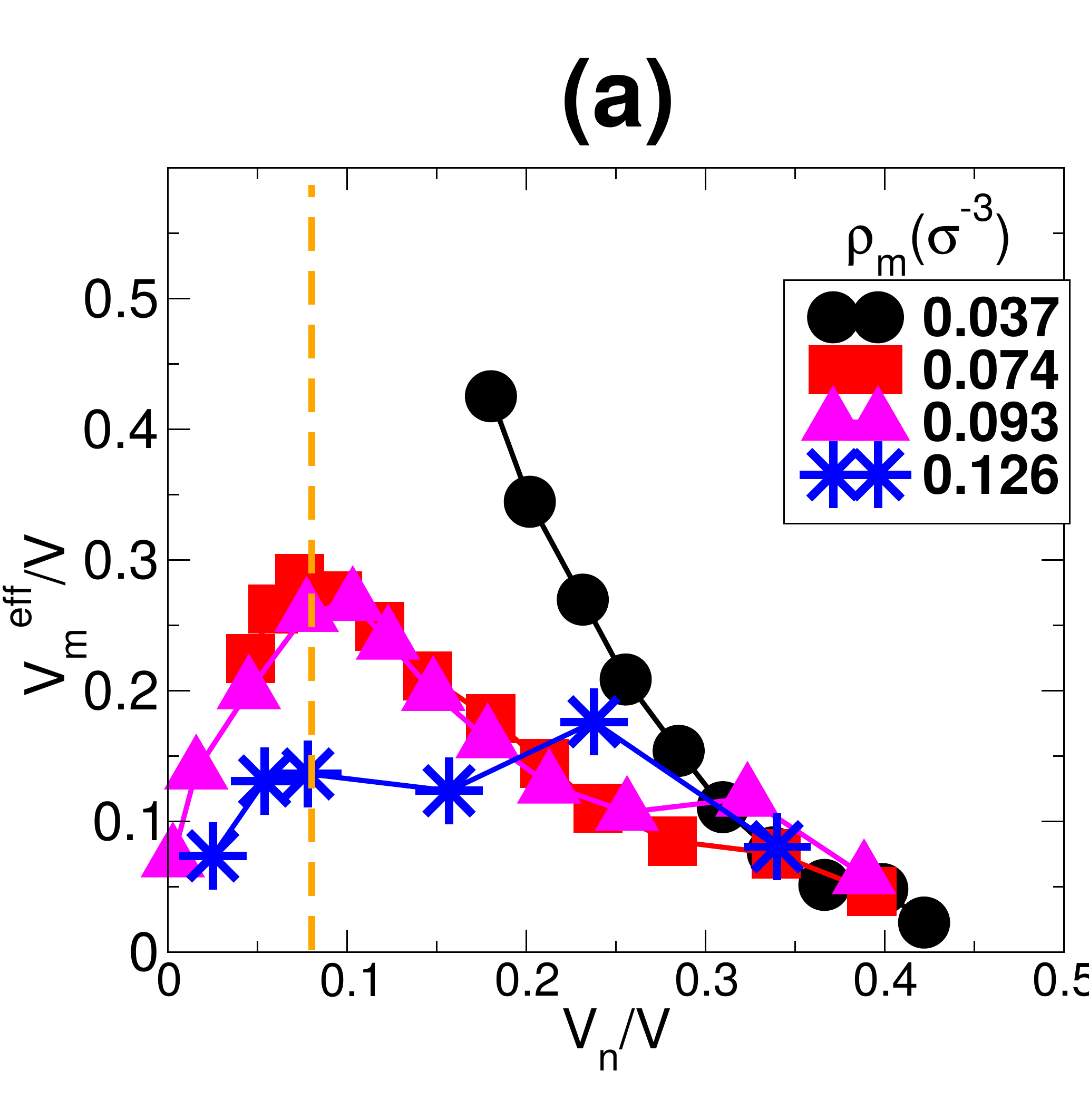}
\includegraphics[scale=0.4]{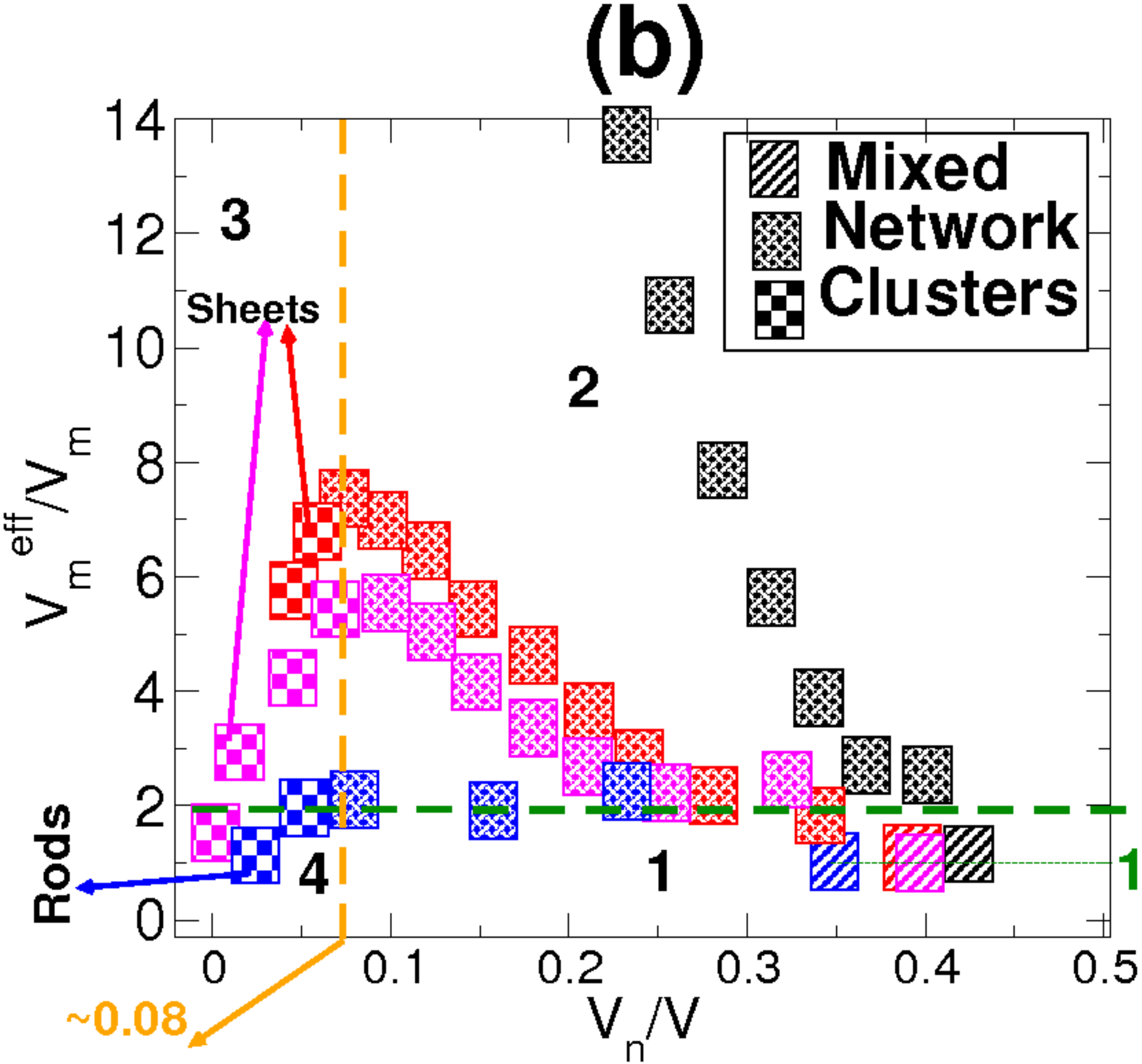}
\caption{(a) The figure shows the effective volume fraction ${V_m}^{eff}/V$ of equilibrium polymers (micelles)
versus the nanoparticle volume fraction $V_n/V$. The vertical broken line (orange) separates the values of $V_n/V$
which correspond to  formation of network morphology of NPs and micelles (right of the vertical line) from
the $V_n/V$ values which correspond to the non-percolating clusters of NPs (left of vertical line).
The exceptions to network morphology are the points with highest nanoparticle volume fraction for each $rho_m$. 
These correspond to the mixed phase of NPs and micellar chains.
(b) The figure shows the replot of the figure shown in (a) 
with the variable shown in Y-axis replaced by the ratio of ${V_m}^{V_m}$ and the symbols indicating 
the morphology of the system. 
These two lines divide the graph into four different regions marked as 1, 2, 3 and 4. The points in region-1 
are for $\sigma_{4n}=1.25\sigma$ which form a dispersed state (no clustering) of micellar chains (all joined 
by a (thin-green) horizontal line intersecting the Y-axis at $\approx 1$). Region-2 shows the systems with 
network-like structures of both nanoparticles and micellar chains. Regions-3 belongs to the systems 
forming non-percolating clusters of nanoparticles in between the clusters of micellar chains. Finally, the 
region-4 depicts the systems with non-percolating clusters of nanoparticles in the background matrix of 
dispersed micellar chains.}
\label{phase}
\end{figure*}

       The Fig.\ref{en_nano_mono} shows the graphs of average energy per particle of (a) monomers 
$\langle E_m\rangle$ and (b) NPs $\langle E_n\rangle$. The plot in (c) shows the average energy per particle for
 both monomers (filled symbols) and NPs (empty symbols), with data for different values of densities shown 
in separate graphs. For the change in $\sigma_{4n}$ from $1.25\sigma$ to $1.5\sigma$, the $\langle E_m\rangle$ 
shows an increase in its value for $\rho_m=0.126\sigma^{-3}$ and $\rho_m=0.093\sigma^{-3}$ while, it decreases 
for $\rho_m=0.074\sigma^{-3}$ and $\rho_m=0.037\sigma^{-3}$. This change corresponds to the transition of 
polymeric chains from dispersed state to the network of its clusters. We can consider that these 
networks are formed in between 
the network of nanoparticle aggregates. This is because  the volume fraction of nanoparticles is higher 
than the effective volume fraction of monomers. The network of monomer chains show same periodicity of the
 structure for all $\rho_m$ (refer Fig.\ref{dens_corr}) at $\sigma_{4n}=1.5\sigma$. Therefore, the monomers 
with number densities $\rho_m=0.126\sigma^{-3}$ and $\rho_m=0.093\sigma^{-3}$ are "crushed" in between 
nanoparticle clusters forming smaller chains compared to the systems with number densities 
$\rho_m=0.037\sigma^{-3}$ and $\rho_m=0.074\sigma^{-3}$ (refer Fig.\ref{len}). 

The existence of smaller chains in the box increases the values of both $V_2$ and $V_4$ for 
$\rho_m=0.126\sigma^{-3}$ and $\rho_m=0.093\sigma^{-3}$. 
Hence, the energy of monomers $\langle E_m\rangle$ shows an increase in its value for $\rho_m=0.126\sigma^{-3}$ 
and $\rho_m=0.093\sigma^{-3}$ but decreases for $\rho_m=0.037\sigma^{-3}$ and $\rho_m=0.074\sigma^{-3}$, 
for the same change in value of $\sigma_{4n}$ from $1.25\sigma$ to $1.5\sigma$. For the same change in 
$\sigma_{4n}$ (from $1.25\sigma$ to $1.5\sigma$), the  energy of NPs $\langle E_n\rangle$ show a decrease 
in their value. This decrease in nanoparticle energy is a result of an increase in the packing of 
nanoparticles due to increase in $\sigma_{4n}$. 

With further increase in $\sigma_{4n}$, $\langle E_m\rangle$ decreases while $\langle E_n\rangle$ 
shows an increase. This decrease in the $\langle E_m\rangle$ is due to the increase in the effective 
volume fraction of monomers whcih results  in an increased chain length.  However, $\langle E_m\rangle$ again 
shows an increase after $\sigma_{4n}=2\sigma$, $\sigma_{4n}=3\sigma$ and $\sigma_{4n}=3.25\sigma$ for 
$\rho_m=0.126\sigma^{-3}$, $\rho_m=0.093\sigma^{-3}$ and $\rho_m=0.074\sigma^{-3}$, respectively. 
This increase in the value of $\langle E_m\rangle$ corresponds to the breaking of nanoparticle network 
into non-percolating clusters that results in the decrease in micellar chain length (refer Figs.\ref{len} 
and \ref{vol_frac}(a)). Thus the dependence  of average monomer energy on $\rho_m$ changes  for the change in 
$\sigma_{4n}$ from $1.25\sigma$ to $1.5\sigma$, this is a consequence of  the change from dispersed state 
of chains of monomers to network-like structure. 

Moreover, the transition from nanoparticle network to non-percolating clusters is also indicated 
by an increase in $\langle E_m\rangle$ for higher values of $\sigma_{4n} > 1.5\sigma$. 
In Fig.\ref{en_nano_mono}(c) shows  the variation in $E_m$ and $E_n$ with $\sigma_{4n}$ on the same
plot for each value of $\rho_m$, the points of intersection of 
both the energies are shown by vertical lines (dashed). As the NP network gradually breaks with increase 
in $\sigma_{4n}$, the NP energy decreases and energy of monomers increases.  At a value of $\sigma_{4n}$,
just higher than the point of intersection of these two energy plots,  the nanoparticle network 
breaks into individual clusters. With the decrease in micellar density $\rho_m$, the point of 
intersection of the two energies gets shifted to higher values of $\sigma_{4n}$. Therefore, the value 
of $\sigma_{4n}$ at which the NP network breaks into non-percolating clusters also gets shifted to 
higher values with the decrease in micellar density.

All these observations can be well explained in a plot of effective micellar volume fraction 
versus NP volume fraction which is shown in Fig.\ref{phase}(a). Different monomer number densities 
are represented by different symbols. The orange coloured dashed vertical line  divides the parameter region 
where NP network structures are observed from the region of parameters where NPs form non-percolating 
clusters. All the points to the left of this line show the systems with non-percolating clusters of 
NPs, while all the points to the right show the percolating network-like structures of both NPs and 
polymeric chains. The rightmost point for each case of micellar density (the maximum value of NP volume 
fraction for each $\rho_m$) belongs to the system which has  polymeric chains dispersed in between NPs. 
In this plot, the regions having the points showing uniformly dispersed and network-like structures 
of micellar chains are not well separated and they overlap. If we replot this graph with the variable 
along the Y-axis replaced by ${V_m}^{eff}/V_m$, then the regions with the dispersed and the network 
of clusters of micellar chains can also be captured and both the morphological transformations can 
be clearly marked on the graph. This is shown in Fig.\ref{phase}(b). 

\begin{figure*}
\centering
\includegraphics[scale=0.16]{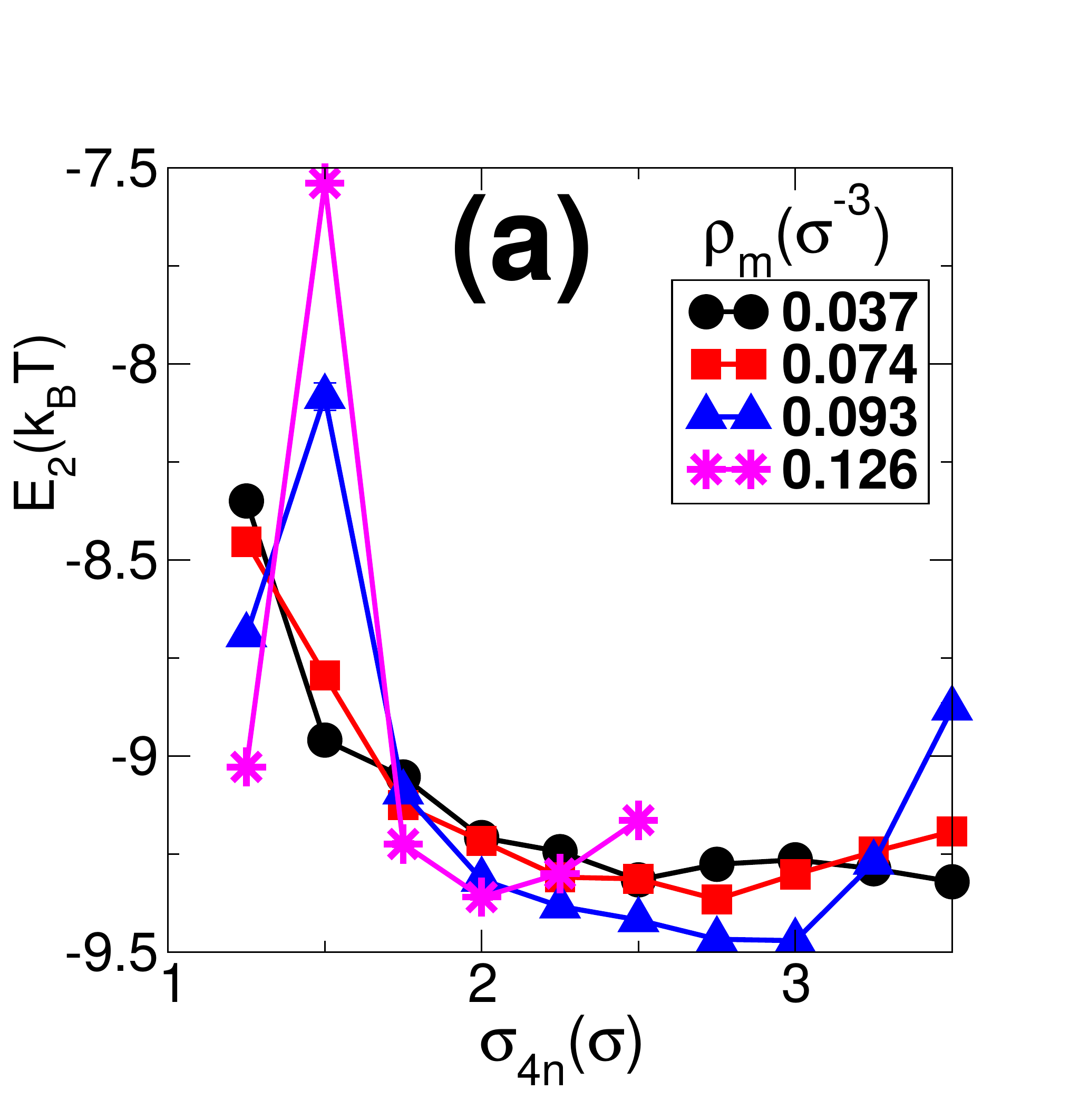}
\includegraphics[scale=0.16]{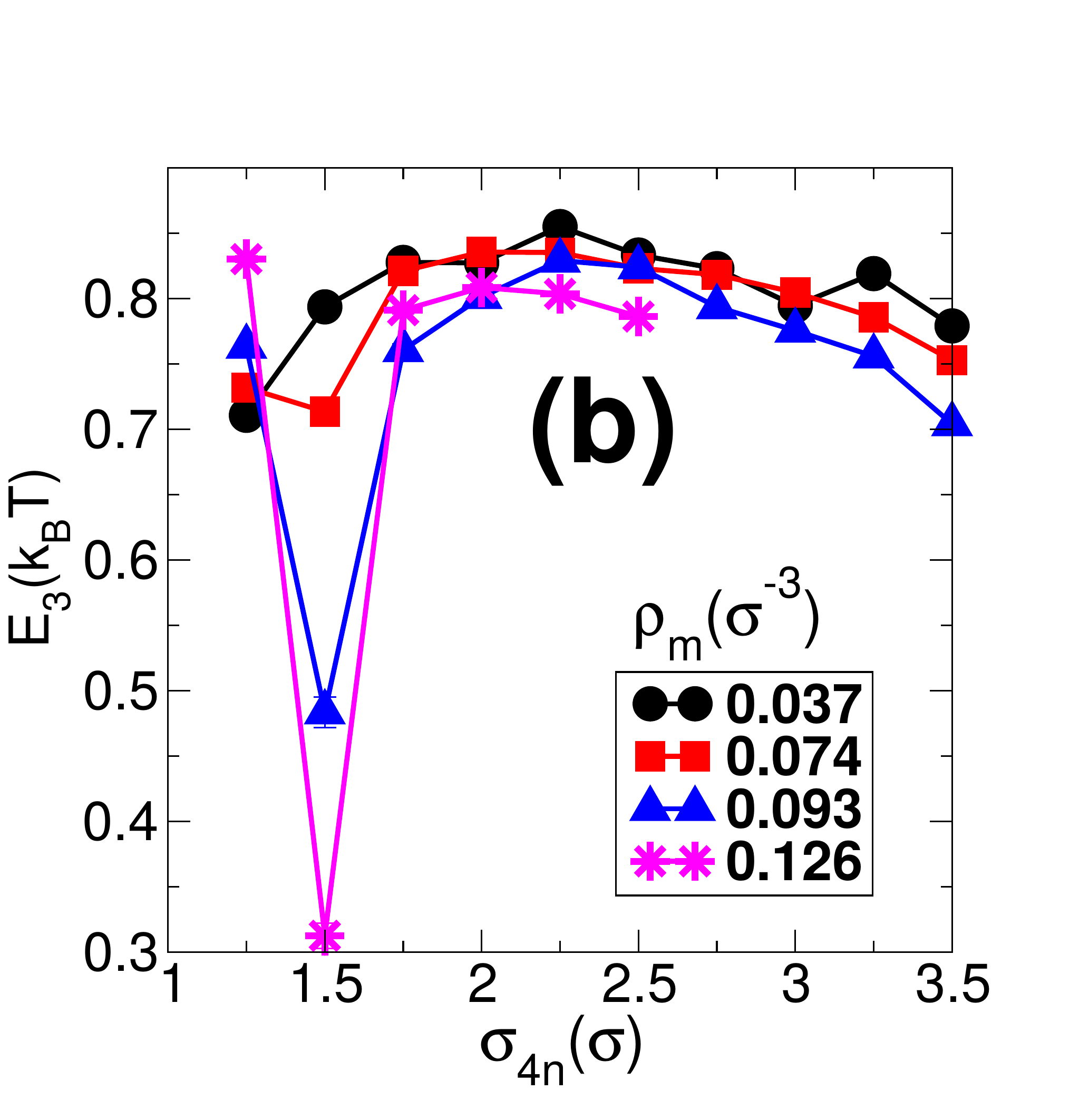}
\includegraphics[scale=0.16]{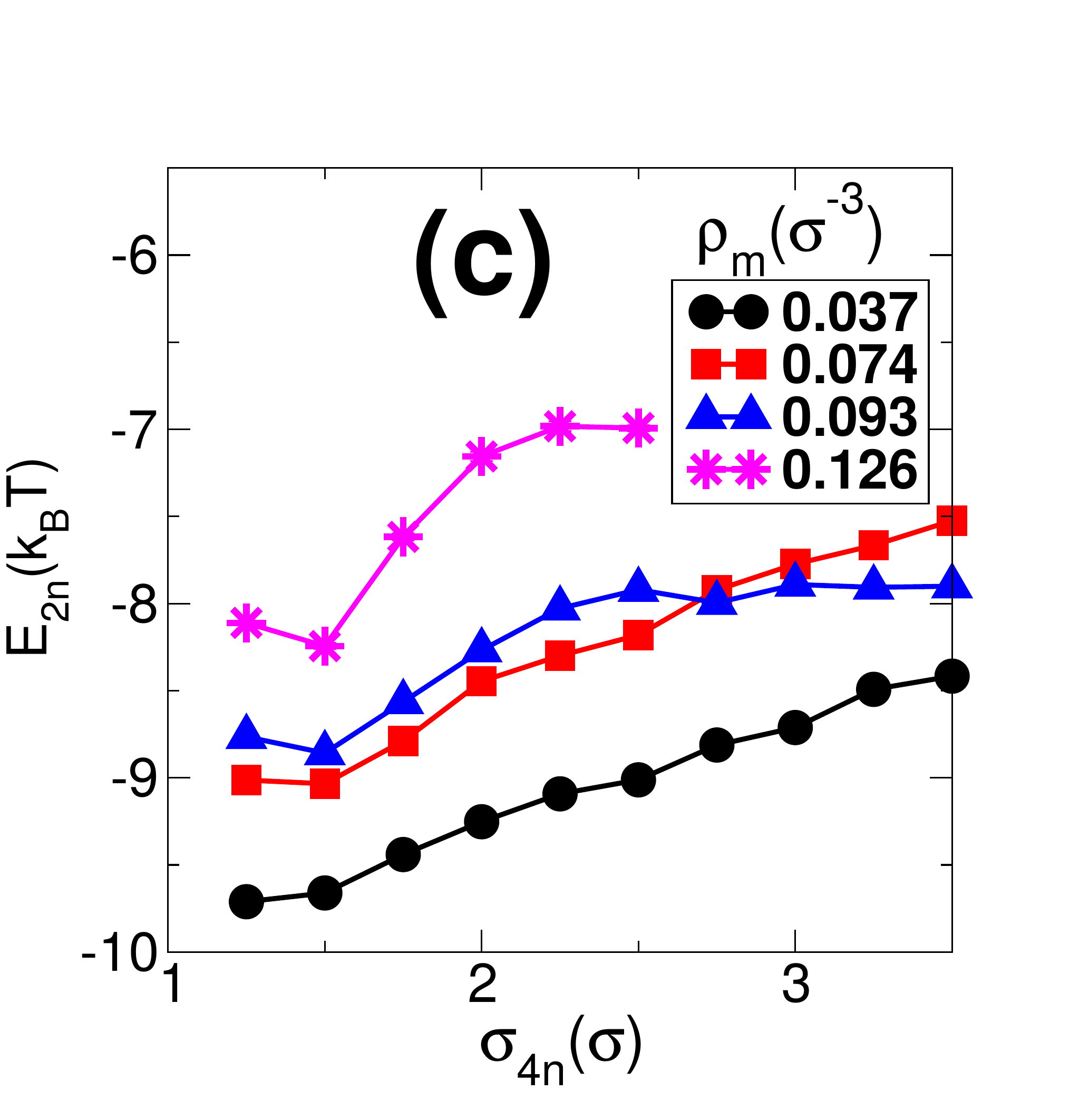}
\includegraphics[scale=0.16]{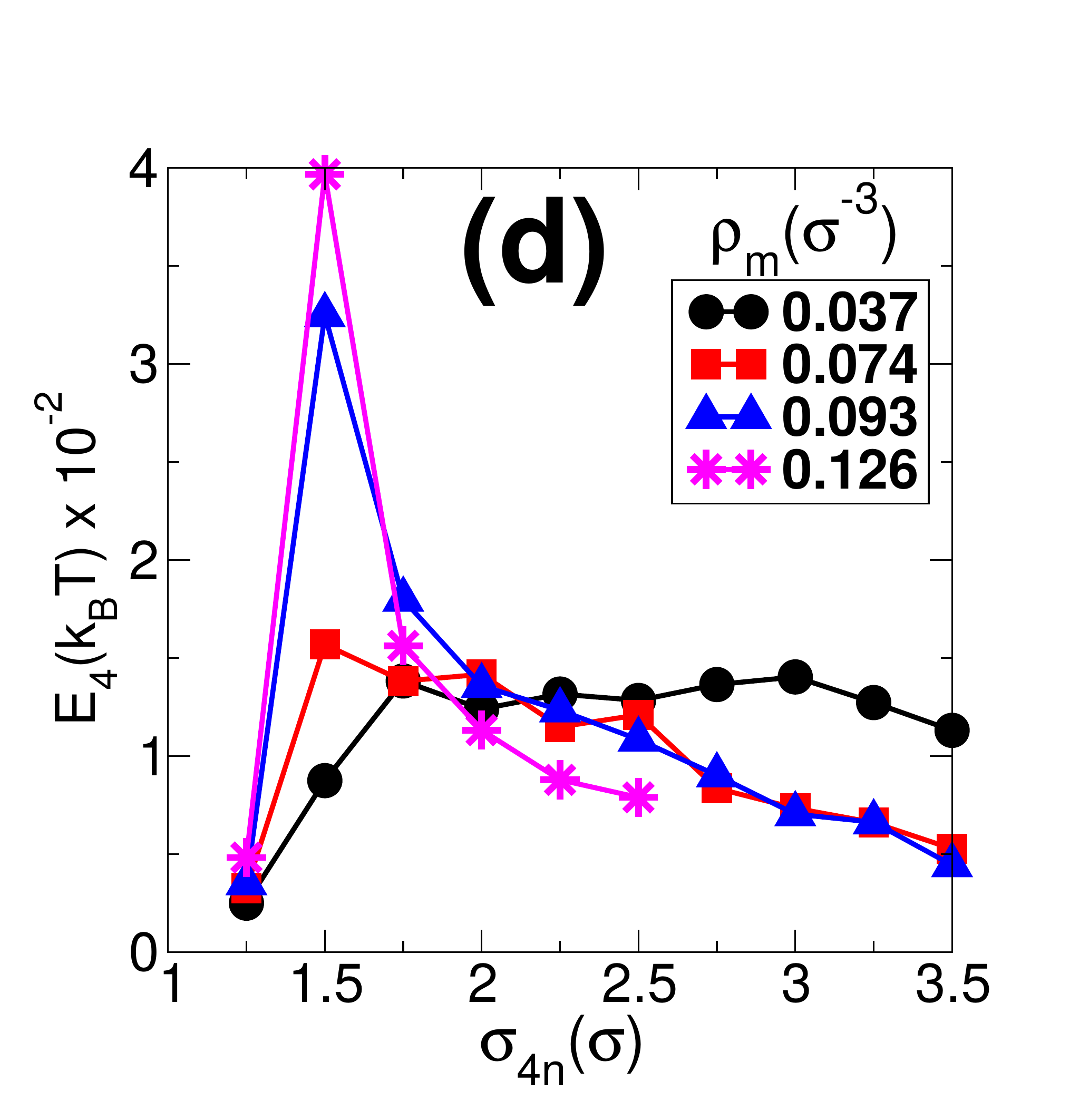}
\includegraphics[scale=0.16]{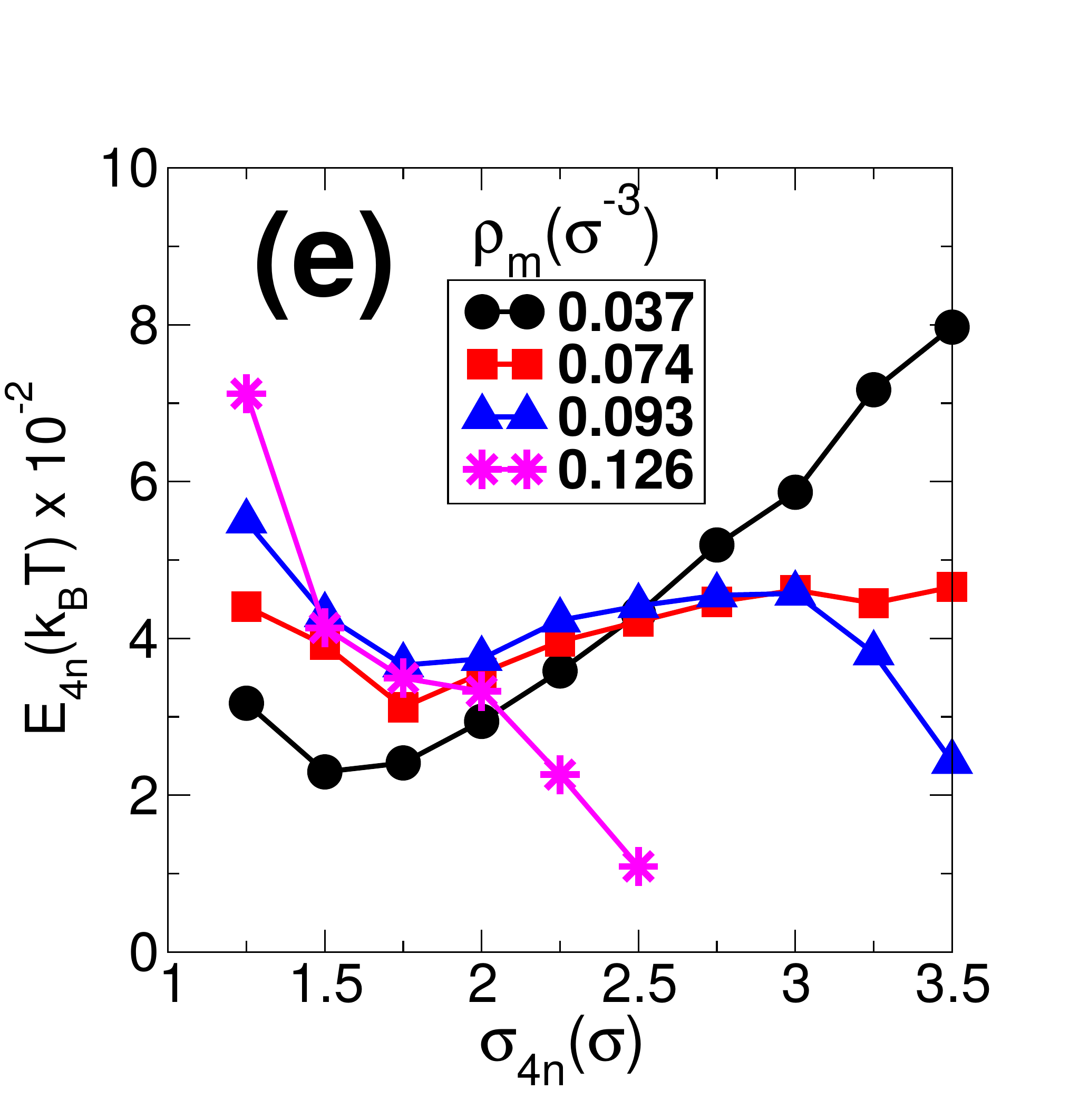}
%\decoRule
\caption{The figure shows the variations of the potential energies due to (a) $ V_2, (b) V_3, (c) V_{2n}, 
(d) V_{4}$ and (e) $V_{4n}$  with the excluded volume parameter $\sigma_{4n}$. The potentials 
are explained in the modeling section. All the points are represented along with the error bars for $10$ 
independent runs. The error bars are too low to be visible for all the plots.}
\label{pots}
\end{figure*}

In Fig.\ref{phase}(b), the ratio ${V_m}^{eff}/V_m$ shows the extent of the effect of change in 
$\sigma_{4n}$ on the effective volume of micelles (refer Fig.\ref{vol_frac}(c)). The different 
morphologies of the system are indicated by the different symbols. For any symbol, the different colours 
correspond to  the different micellar densities. At $\sigma_{4n}=1.25\sigma$ and for all the values 
of $\rho_m$, the value of the ratio ${V_m}^{eff}/V_m$ remains $\approx 1$. All such points 
are joined by a horizontal line (thin-broken-green) intersecting the Y-axis at $\approx 1$. An increase 
in the value of $\sigma_{4n}$ from $1.25\sigma$ to $1.5\sigma$ leads to the onset of formation of 
clusters of micellar chains. All such points with a network-like structure of polymeric chains are 
indicated by the corresponding symbol spanning over a large parameter space in the graph. All the points that 
show a network-like structure of clusters of micellar chains occurs above a value of 
${V_m}^{eff}/V_m\approx 2$ that is shown by a dashed horizontal line (green). Therefore, this line 
approximately marks the change from a uniformly dispersed state to a network-like structure of micellar 
chains.  With further increase in $\sigma_{4n}$, the NP volume fraction decreases and effective monomer 
volume fraction increases. At a even higher value of $\sigma_{4n}$ (depending on $\rho_m$), the NP 
networks break to form non-percolating clusters. This is marked by an decrease in the value of 
the effective volume of micelles and all such points are shown by the corresponding symbol for 
non-percolating clusters of NPs. Depending on the number density of monomers, the morphology of these 
non-percolating clusters changes from sheets to rodlike structures (with increasing $\rho_m$). 
For all the values of $\rho_m$, the change from network-like morphology to non-percolating clusters of 
NPs is observed to occur at a value of NP volume fraction $\approx 0.08$. This is marked by a dashed 
vertical line (orange).

The combination of the vertical line (orange-broken) and the horizontal line (green-broken) divides 
the space into four different regions which are marked as 1, 2, 3 and 4. Each region defines the 
structural morphology of the system as follows; \\
Region 1: Uniformly mixed state of micellar chains and nanoparticles with micellar chains dispersed 
in between nanoparticles. \\
Region 2: Both the polymeric chains and the nanoparticles form network-like structures which 
interpenetrate each other. \\
Region 3: Individual clusters of nanoparticles in between the clusters of micellar chains. \\
Region 4: In this region all the points show the value of $\frac{{V_m}^{eff}}{V_m} \approx 1$, 
therefore, the monomer chains are in a dispersed state. However, the nanoparticles form non-percolating 
clusters.  Therefore, in this region, non-percolating clusters of nanoparticles are present in the 
background of dispersed chains of micelles. \\

  All the observations discussed in this paper for the NP-micellar system can be verified by examining 
the behaviour of the potential energies involved in the system. The plots of these potential energies 
$E_2,E_3,E_{2n},E_4,E_{4n}$ are shown in Figs.\ref{pots}(a), (b), (c), (d) and (e) due to the potentials
 $ V_2, V_3, V_{2n}, V_{4}$ and $V_{4n}$, respectively. All the potential energy values are averaged over 10 
independent runs with $(15-20) \times 10^5$ iterations for each run and are plotted with an error bar. 
The error bars are too small to be visible. Moreover, these potential energies are normalized by dividing them 
with the number of all combinations of the particles that are interacting with the concerned potential. 
A higher value of $E_2$ indicates a higher average length of the chains and vice versa. The behaviour of 
$E_2$ in Fig.\ref{pots}(a) shows an increase in its value for a change of $\sigma_{4n}=1.25\sigma$ to 
$1.5\sigma$ for $\rho_m=0.126\sigma^{-3}$ and $\rho_m=0.093\sigma^{-3}$ indicating scission of chains.
 Apart from these two points $E_2$ decreases with increase in $\sigma_{4n}$ for all monomer densities 
indicating the increase in chain lengths. However, with further increase in $\sigma_{4n}$, it shows a 
decrease in its value (except $\rho_m=0.037\sigma^{-3}$). This happens at $\sigma_{4n}=2.25\sigma$ for 
$\rho_m=0.126\sigma^{-3}$ and at $\sigma_{4n}=3.25\sigma$ for $\rho_m=0.093\sigma^{-3}$ and 
$0.074\sigma^{-3}$. Thus, it confirms the behaviour of the average length of chains shown in Fig.\ref{len}.

Figure \ref{pots}(b) shows the behaviour of $E_3$ with $\sigma_{4n}$ for different values of 
$\rho_m$. Longer chains will give rise to a large number of 
bonded triplets along a chain resulting in a higher value of $E_3$ while, smaller chains will result 
in a lower value of $E_3$. Therefore, the decrease in the value of $E_3$ for change of 
$\sigma_{4n}=1.25\sigma$ to $1.5\sigma$ for $\rho_m=0.126\sigma^{-3}$ and $\rho_m=0.093\sigma^{-3}$ is 
indicative of the formation of smaller chains at $1.5\sigma$. Moreover, with further increase in the value 
of $\sigma_{4n}$, $E_3$ shows relatively higher values indicating the presence of longer chains. 
The interaction potential energies between NPs $E_{2n}$, shown in Fig.\ref{pots}(c), also shows a decrease at 
$1.5\sigma$ indicating an increase in the packing of nanoparticles in spite of the decrease in its 
number density. It then shows an increase in its value at a higher value of $\sigma_{4n}$ due to a 
decrease in its number density.

     The potential energies $E_4$ in Fig.\ref{pots}(d) which shows the repulsive interaction between micellar 
chains shows an increase in its value when the micellar chains transforms to a network-like structure 
at $\sigma_{4n}=1.5\sigma$ from a dispersed state at $\sigma_{4n}=1.25\sigma$. The quantity $E_4$ then 
decreases indicating that the average distance between micellar chains increases (chains get out of 
the range of the potential $E_4$) as $\sigma_{4n}$ increases. The repulsive interaction energy
between NPs and micellar chains $E_{4n}$ shown in Fig.\ref{pots}(e) also shows the two points of 
morphological transformations, first a decrease in $E_{4n}$ at $\sigma_{4n}=1.5\sigma$ indicating 
the change in the structure of micellar chains from dispersed to clusters that join to form network-like 
structure. The decrease in $E_{4n}$ is due to the re-organization of micellar chains to form clusters 
that reduces the number of monomer-NPs contacts. Then, it increases (or remains constant for 
$\rho_m=0.126\sigma^{-3}$), but again show a decrease in its value when the morphology of NPs changes 
from a network-like structure to non-percolating clusters. This decrease is due to the low density 
of NPs present in the system at high $\sigma_{4n}$. The values of the repulsive potential energies $E_4$ and 
$V_{4n}$ are of the order of $10^{-2}k_BT$, thereby, indicating that the rearrangement of the system 
architecture with the increase in $\sigma_{4n}$ is to avoid the repulsive interactions within the system.

     Thus, the plots of $E_4$ and $E_{4n}$ with very low values (of the order of $10^{-2}k_BT$) points to 
the emergence of different kind of self-assembled structures as a result of the system trying to decrease 
the interfacial or repulsive interactions. The repulsive interaction potential between NPs and micellar 
chains $V_{4n}$ will form a phase separated state at higher densities in order to reduce the repulsion 
between them. But the presence of the steep repulsive interaction potential within the micellar chains ($V_4$)
themselves evokes a competition between the two potentials $V_4$ and $V_{4n}$. Micellar chains 
aligned parallely with
 the average distance between chains $r<\sigma_4$, lead to a higher value of the potential energy due 
to $V_4$. Therefore, reducing the value of $V_4$ as well as $V_{4n}$  not only requires a reduction of 
the interface area between micelles and NPs but also encourages micellar chains to arrange such that they 
are out of the range of the repulsive interaction $V_4$ acting between themselves. 
For a system with low monomer density 
(Fig.\ref{low_dens}), the perpendicular arrangement between micellar chains which meet at the junctions of 
the network (refer Fig.\ref{node}(b) and fig.\ref{ang_dist}) is a consequence of the requirement to reduce 
the repulsive interaction between them. By remaining at an angle of $90^\circ$, the possibility of the chains
 of monomers coming in contact with each other is very low. With an increased density of micelles 
(Fig.\ref{int_dens1} and \ref{int_dens2}), the average length of the micellar chains also increases. 
Therefore, for higher monomer density if the same structures as shown in figure \ref{low_dens} is 
maintained, it will lead to an increased width of the clusters of micellar chains in the network 
resulting in a high repulsive energy due to $V_4$. Hence, the micellar chains arrange in an aligned fashion 
forming thin sheet-like domains and relatively away from each other in order to reduce values of $V_4$.
The steep increase of the potential $V_4$ with $r_4$ discourages those Monte Carlo steps which 
bring the monomers from adjacent chains just below $\sigma_4$, and thus the values of $E_4$ remains 
much lower than $k_BT$.  Micellar chains which form  sheet like domains have relatively lower number 
of neighbouring chains compared to chains in a hexagonally packed arrangement. Moreover, the 
arrangement of these sheetlike planes at an angle perpendicular 
to neighbouring domains (as noted in Figure. \ref{ang_dist}(b) and (c)) also seems to be the part of the 
strategy to lower the repulsive potential between micellar chains by reducing the possibility of contact 
between the domains. At even higher number densities of micelles, the chains become very long and formation 
of a sheet-like structure would lead to an increase in the interfacial area between micellar sheets and NPs.
 This in turn will result in an increase potential energy due to $V_{4n}$. Therefore, in this case, the
 NPs prefer to  nematically arrange themselves with the polymeric chains maintaining a distance 
$r>1.75\sigma$ from each other to be out of the range of the repulsive potential $V_4$.

\section{Discussion and Conclusion:}

In this paper, we examine in detail the behavior of mixtures attractive nanoparticle (NPs) and 
equilibrium polymers (EPs), examples 
of which are worm-like micellar systems, with the variation of number of monomer number density 
and the excluded volume parameter (EVP) between micellar chains and nanoparticles. The density of monomers which 
self-assemble to form equilibrium polymers affect the self organization of aggregates of NPs, which in 
turn affect the affect the organization and length of the self-assembled polymeric chains. One obtains 
a number of interesting equilibrium or kinetically arrested configurations of NP morphologies, starting
from a mixed phase of NPs and EPs, when the size of NPs are relatively small compared the width of micellar chains.
For higher values of the EVP, the morphology of the NP aggregates varies from porous percolating connected network of NPs,
where the thickness of the pores can be suitably controlled by the EVP and number density of monomers, 
to disconnected aggregates of NPs with varying shape anisotropy values. The NPs
aggregate between clusters of semi-flexible polymeric chains, hence instead of nucleating to spherically symmetric 
aggregates they develop into anisotropic rod like aggregates whih grow parallel to the allignment of backgroun 
semi-flexible micellar matrix. We establish that the anisotropy of the NP clusters is governed by the
density of the equilibrium polymeric matrix. Furthermore, the steep repulsive potential 
between neighboring micellar chains,  and its contribution relative to repulsive potential between chains and NPs,
can lead to the formation of planar sheet like arrangement of micellar chains
due to self-avoidance, which in turn lead to planar sheets of NP-aggregates. Apriori, 
this was was an unexpected result for us.  This and the other morphologies obtained are examples of the 
production of nanostructures via synergistic interactions of the equilibrium polymeric matrix and the NPs. 
The value of the volume fraction of NPs  for the transformation from network-like morphology to 
non-percolating clusters of NPs is approximately identified as $\approx 0.08$. We summarize the different 
kinetically arrested morphologies obtained in a (non-equilibrium) phase diagram with the effective volume fraction of EPs and 
the  volume fraction of NPs on the two axes. There have been experimental studies \cite{wagner} which explore 
the fundamental aspects of Wormlike micelle and NP interactions, where they see ``double network'' of micelles
and NPs. But there the underlying interactions are expected to be different, with chains getting attached to NPs to 
to form a network of micelles. We have purely repulsive interaction between chains and NPs.

Understanding the underlying physical mechanism and energetics of the  different morphologies obtained due to 
synergistic interactions between self-assembling polymers and attractive NPs is expected to help the experimental
scientist fine tune the relevant parameters to control  the yield of different porous nano-structures with
required surface area and porosity.  The micelles can  be dissovled away once the NPs aggregate and self-organize 
to the suitable morphology.  This might have significant relevance, for example, in the design of 
batteries materials and super-capacitors, where one needs to balance between large surface area (to increase charge 
storage) as well as large pores to enable quick dynamics of ions during the charging and discharging process,
which affects power density. On the other hand, the NPs could also be possibly dissolved away after the 
self assembled polymers are made into gels by addition of suitable additives to enable the experimentalist to 
obtain polymeric sponges as has been demonstrated using ice-templating techniques \cite{guru1,guru2}. 
We expect the EP-NP morphologies to be kinetically arrested phase separating states, except when the value 
of $\sigma_{4n} =1.25 \sigma$ when we get a thermodynamically stable mixed state. We use grand-canonical Monte
Carlo scheme to introduce (and remove) NPs at randomly chosen positions in the box, this might seem difficult 
to realize experimentally in a relatively dense polymeric matrix. However, one could possibly have background 
solution in which the monomers self assemble to form equilibrium polymers, and then appopriately chemical reagents 
could be added so initiate nucleation of nano-particles out of the solution which would then aggregate and
self-organize into the suitable morphologies depending on the synegistic interactions with the background 
polymeric matrix.

\section{Acknowledgement}

    We acknowledge the helpful discussion with Dr.Arijit Bhattacharyya, Dr.Guruswamy Kumaraswamy 
and Dr. Deepak Dhar. We thank the computational facility provided National Param Super Computing 
(NPSF) CDAC, India for use of Yuva cluster, the compute cluster in IISER-Pune, funded by DST, 
India by Project No. SR/NM/NS-42/2009.  We acknowledge funding and the use of a cluster bought 
using DST-SERB grant no. EMR/2015/000018 to A. Chatterji. AC acknowledges funding support by 
DST Nanomission, India under the Thematic Unit Program (grant no.: SR/NM/TP-13/2016).

\bibliographystyle{apsrev4-1}
\bibliography{paper_today1}
\end{document}